\documentclass[10pt,aps,a4paper,prd,floatfix,onecolumn,notitlepage,nofootinbib]{revtex4-2}

\usepackage[left=2cm,right=2cm,top=2cm,bottom=2cm]{geometry}

\usepackage[utf8]{inputenc}
\usepackage[british,UKenglish,USenglish,american]{babel}
\usepackage{amsmath}

\usepackage{graphicx}
\usepackage{subcaption}
\usepackage{booktabs}

\usepackage{setspace}

\usepackage{slashed}

\usepackage{mathrsfs}
\usepackage{amssymb}
\usepackage{bbm}

\usepackage{nicefrac}

\usepackage{microtype}

\usepackage{parskip}

\usepackage{physics}

\usepackage{xcolor}

\newcommand\numberthis{\addtocounter{equation}{1}\tag{\theequation}}

\DeclareMathOperator{\diag}{diag}
\DeclareMathOperator{\sgn}{sgn}
\DeclareMathOperator{\trf}{{\tr}_{\mathnormal{f}}}
\DeclareMathOperator{\trc}{{\tr}_{\mathnormal{c}}}
\DeclareMathOperator{\trD}{{\tr}_{\mathrm{D}}}
\DeclareMathOperator{\trx}{{\tr}_{\mathnormal{x}}}

\newcommand\operator{\hat{\mathcal{O}}}

\newcommand{\p}{\prime}

\renewcommand{\bar}{\overline}

\usepackage{bm}
\renewcommand{\vec}{\bm}

\newcommand{\e}{\mathrm{e}}

\newcommand{\nfermi}{f_{\mathrm{F}}}

\newcommand{\pair}{\mathrm{p}}
\newcommand{\scat}{\mathrm{s}}

\newcommand{\LHS}{\mathrm{LHS}}
\newcommand{\RHS}{\mathrm{RHS}}
\newcommand{\Linear}[1]{\mathrm{L}\left[#1\right]}
\newcommand{\Quadratic}[1]{\mathrm{Q}\left[#1\right]}

\newcommand{\CPV}{\mathscr{P}}

\newcommand{\sumeta}{\sum_{\eta=\pm 1}}

\newcommand{\TCoeff}[3]{\,T_{#1}^{#2#3}\,}

\newcommand{\depsilondEEMin}{E_\mathrm{min}^{(\dd{\varepsilon}\dd{E})}}

\newcommand{\depsilondEEMax}{E_\mathrm{max}^{(\dd{\varepsilon}\dd{E})}}

\newcommand{\depsilondEepsilonMin}{\varepsilon_\mathrm{min}^{(\dd{\varepsilon}\dd{E})}}

\newcommand{\depsilondEepsilonMax}{\varepsilon_\mathrm{max}^{(\dd{\varepsilon}\dd{E})}}

\newcommand{\dEdepsilonEMin}{E_\mathrm{min}^{(\dd{E}\dd{\varepsilon})}}

\newcommand{\dEdepsilonEMax}{E_\mathrm{max}^{(\dd{E}\dd{\varepsilon})}}

\newcommand{\dEdepsilonepsilonMin}{\varepsilon_\mathrm{min}^{(\dd{E}\dd{\varepsilon})}}

\newcommand{\dEdepsilonepsilonMax}{\varepsilon_\mathrm{max}^{(\dd{E}\dd{\varepsilon})}}

\begin{document}
\raggedbottom

\title{
A new approach to the 3-momentum regularization of the in-medium one and two fermion line integrals with applications to cross sections in the Nambu--Jona-Lasinio model
}

\author{Renan Câmara Pereira}
\email{renan.pereira@student.uc.pt}
\affiliation{CFisUC - Center for Physics of the University of Coimbra, Department of Physics, University of Coimbra, 3004-516 Coimbra, Portugal}

\author{João Moreira}
\email{jmoreira@uc.pt}
\affiliation{CFisUC - Center for Physics of the University of Coimbra, Department of Physics, University of Coimbra, 3004-516 Coimbra, Portugal}
\affiliation{Departamento de Física, Universidade da Beira Interior, 6201-001 Covilhã, Portugal}

\author{Pedro Costa}
\email{pcosta@uc.pt}
\affiliation{CFisUC - Center for Physics of the University of Coimbra, Department of Physics, University of Coimbra, 3004-516 Coimbra, Portugal}

\author{Constança Providência}
\email{cp@uc.pt}
\affiliation{CFisUC - Center for Physics of the University of Coimbra, Department of Physics, University of Coimbra, 3004-516 Coimbra, Portugal}

\begin{abstract}
We propose the 3-momentum sphere intersection regularization applied to the one and two fermion line integrals at finite temperature and chemical potential. The quark-antiquark polarization function in this new regularization approach is equivalent to the usual 3-momentum regularization, when the absolute value of the external 3-momentum of the polarization is zero. Additionally, it respects the particle-antiparticle symmetry of meson states in the Nambu$-$Jona-Lasinio (NJL) model for all values of temperature and chemical potential. Without this symmetry, in-medium cross sections calculated in the 3-momentum regularized NJL model are not consistent. In order to demonstrate the difference between the usual 3-momentum regularization with the one proposed in this work, we study the quark-quark and quark-antiquark cross sections in both regularization schemes. To this end we use the standard $SU(3)$ NJL model, with four and six quark interactions. We observe major quantitative and qualitative differences when comparing quark-quark cross sections in both schemes. The quark-antiquark cross sections, on the other hand, are very similar in both regularizations, owning to the equivalence between the regularizations when the absolute value of the external 3-momentum is zero. 
\end{abstract}

\maketitle

\section{Introduction}

The calculation of one loop integrals, at finite temperature and chemical potential, is an essential part of the study of in-medium quantum field theory.  These integrals arise naturally in calculations performed under the framework of effective theories of quantum chromodynamics (QCD), like Nambu$-$Jona-Lasinio (NJL) type models \cite{Rehberg:1995kh,Costa:2003uu,Costa:2005cz,Nagahiro:2006dr,Costa:2008dp,Contrera:2009hk,Inagaki:2010nb,Blanquier:2014kja,Blaschke:2016sqn,Costa:2019bua}. In the NJL model, the one and two fermion line integrals occupy a central role in the calculation of meson propagators, meson masses, cross sections, or thermodynamic quantities, like pressure, energy density and entropy density. Moreover, quark-quark and quark-antiquark cross sections, derived from the NJL model, can be used to estimate the quark relaxation times, a vital part of several theoretical predictions about the in-medium behavior of transport coefficients like the shear and bulk viscosities, as well as, the electrical and thermal conductivities of strongly interacting matter \cite{Rehberg:1995nr,Rehberg:1996vd,Zhuang:1995uf,Sasaki:2008um,Marty:2013ita,Soloveva:2020hpr}.

Due to the non-perturbative characteristics of QCD at low energies, phenomenological models, like the NJL model, become an essential tool to make qualitative predictions about QCD at finite temperature and density, specially when the goal is to study the behavior of the quark gluon plasma at temperatures and chemical potentials for which the \textit{ab initio} approaches of QCD, like lattice gauge theory or perturbative based calculations, are not applicable \cite{Vogl:1991qt,Klevansky:1992qe,Hatsuda:1994pi,Buballa:2003qv}. Hence, making sure that the calculation of one loop integrals, within the scope of NJL-type models, is made in a symmetry preserving manner, is extremely important in order to make better qualitative and quantitative predictions while also assuring that one is not introducing non-physical features in the phenomenological calculation. 

In this work we follow the terminology given in Ref. \cite{Rehberg:1995nr}, and denote the one fermion line integral by $A$ and the two fermion line integral by $B_0$. Here, we will not study the practical implementation of integrals with a fermion line counting higher then two. As it is well known, both of these integrals are divergent and some regularization technique must be employed, becoming an essential part of the calculations \cite{Rehberg:1995nr,Ripka:1997zb}. When studying finite temperature and chemical potential systems, the 3-momentum regularization is widely applied \cite{Rehberg:1995nr}. In this framework the integration over the internal 3-momentum of quarks, $\vec{p}$, is carried out up to a maximum value of its absolute value, $\abs{\vec{p}}<\Lambda$, with $\Lambda$ the 3-momentum cutoff of the model \cite{Ripka:1997zb}. The usual way in which the 3-momentum regularization is performed breaks a symmetry of the system: the quark-antiquark polarization loop, which mixes light and strange quarks, is not symmetric with respect to the interchange of quarks \cite{Rehberg:1996vd}. This originates from a shift of variables used in order to simplify the evaluation of the $B_0$ integral \cite{Rehberg:1995nr,Rehberg:1996vd}.

One of the main issues with this feature is that physical scattering processes which ought to be identical for all values of temperature and chemical potential, become different. A clear example is the cross sections of the quark-quark processes, $us \to us$ and $su \to su$ (the same occurs for $ds \to ds$ and $sd \to sd$). Considering the $us \to us$ scattering in the $u$-channel, the exchanged mesons are the pseudoscalar $K^+$ and the scalar $\kappa^+$ while, in the $t$-channel, the exchanged mesons are the flavorless, neutral mesons $\eta$, $\eta'$, $\pi_0$, $\sigma$, $f_0$ and $a_0^0$. For the $su \to su$ process the corresponding antimesons are involved in each channel. Using a theoretical setup in which the quark-quark processes, $us \to us$ and $su \to su$, yield different results is an indication that the set of meson propagators used in the calculations do not display the expected particle-antiparticle symmetry of meson states. One way to solve this problem is to follow the recipe introduced in Ref. \cite{Rehberg:1996vd} and, when appropriate, replace the two fermion line integral present in the quark-antiquark polarization loop function, $B_0=B_0 \qty[M_u,M_s,T,\mu_u,\mu_s, k_0, \abs{\vec{k}}] $, with an average, $B_0=\frac{1}{2}\qty( B_0 \qty[M_u,M_s,T,\mu_u,\mu_s, k_0, \abs{\vec{k}}] + B_0 \qty[M_s,M_u,T,\mu_s,\mu_u, k_0, \abs{\vec{k}}] )$. Here $M_f$ and $\mu_f$ are the effective mass and effective chemical potential of the quark of flavor $f$, $T$ is the temperature and  $k_0$, $\abs{ \vec{k} }$ refer to the temporal-like and spacial-like parts of the external momentum, respectively. In the physical scenario where isospin symmetry does not occur (up and down quarks have different bare quark masses) or if there is an asymmetry caused by non-degenerate up and down quark chemical potentials (for example, when considering matter in beta equilibrium), several other scattering processes are also affected, such as the $ud \to ud$ and $du \to du$ processes.

In this work we propose a novel approach to the 3-momentum regularization, namely the 3-momentum sphere intersection regularization scheme. In this new regularization scheme, we regularize the one loop integrals based on the number of quark propagators present in the integrand. Each quark propagator has a 3-momentum which must be smaller than the 3-momentum cutoff,  $\abs{\vec{p}}<\Lambda$. In the 3-momentum sphere intersection regularization, the individual momentum of each quark propagators must be smaller than the cutoff, even when external momenta are considered. As we will discuss in this paper, this can be achieved by considering that the integration over the momentum in the $N$ fermion line integrals must be evaluated inside the region defined by the intersection of $N$ spheres. Furthermore, in the case in which there are no external momenta applied to the $N$ fermion line integrals, the spheres completely intersect and the regularization is equivalent to the usual 3-momentum regularization. Interestingly, if the spheres do not intersect, the integral is automatically zero. As we will discuss, this happens for the meson propagators: if the momentum of the meson is larger than $2\Lambda$, the spheres corresponding to each quark propagator, inside the quark-antiquark polarization function, no longer intersect and the polarization function is automatically zero. With this approach to the  3-momentum regularization, all the symmetries of the system are maintained and the previously discussed problems with the scattering processes do not occur.

This paper is organized as follows. In Section \ref{theOneandTwoFermionLineIntegrals}, we derive the one and two fermion line integrals from the NJL model for any number of scalar and pseudoscalar quark-quark interactions. Such is accomplished in the mean field approximation by writing the linear and quadratic expansions of the NJL lagrangian. We also show how the expansion used to derive the meson propagators from the quadratic NJL model yields the same meson propagators as the ones obtained via the commonly employed Bethe-Salpeter equation in the random phase approximation. Additionally, the 3-momentum sphere intersection is introduced and discussed. The formalism developed here can be used to incorporate multi-quark interactions, such as the eight quark proposed in \cite{Osipov:2005tq,Osipov:2006ns}, or explicitly chiral symmetry breaking interactions, as those in \cite{Osipov:2012kk, Osipov:2013fka}. In Sections \ref{two_fermion_one_loop_integral} and \ref{one_fermion_one_loop_integral}, we evaluate the two fermion line integral, $B_0$, and the one fermion line integral, $A$, within the new regularization scheme. In Section \ref{crossSectionsNJLModel} the formalism to evaluate quark-quark and quark-antiquark cross sections is laid out and a specific version of the $SU(3)$ NJL model, with four and six quark interactions is introduced in order to study the effect of using both regularizations on the different scattering processes. The numerical results regarding quark-quark and quark-antiquark scattering processes for different scenarios of temperature and chemical potential are presented and discussed. Finally, in Section \ref{conclusions}, conclusions are formulated and additional research is suggested.

\section{The one and two fermion line integrals from NJL-type models}
\label{theOneandTwoFermionLineIntegrals}

The one fermion line integral, $A$, and the two fermion line integral, $B_0$, at one loop level, arise when studying NJL-type models. In this section we show one way to derive the definitions of these integrals from the NJL model and discuss regularization strategies, namely a new way to regularize the quark-antiquark polarization function.

Consider a general model describing  $N_f$ flavors of fermions including different types of scalar and pseudoscalar interactions and the standard free Dirac Lagrangian. This can be written as:
\begin{align}
\mathcal{L} [\bar{\psi}, \psi] = 
\bar{\psi} \qty( i\slashed{\partial} - m ) \psi + \mathcal{L}_\mathrm{int} \qty[ s_a, p_a ] .
\label{general_NJL_Lagrangian}
\end{align}
Here, $m=\diag  \qty{ m_0, m_1, \ldots , m_{N_f-1} } $ is the quark current mass matrix, the term $\mathcal{L}_\mathrm{int}$ contains several types of quark-quark interactions and is written as a function of quark bilinear operators, $s_a=\bar{\psi} \lambda_a \psi$ and $p_a=\bar{\psi} i \gamma^5 \lambda_a \psi$, where $\lambda_a$ are the $N_f^2$ matrices that span the U$(N_f)$ algebra. The usual 4-quark scalar-pseudoscalar interaction, for example, can be written as, $\mathcal{L}_{4q} \propto (\bar{\psi} \lambda_a \psi)^2 + (\bar{\psi} i \gamma_5 \lambda_a \psi)^2 = s_a^2 + p_a^2$ (Einstein summation convention is used).

Using this Lagrangian density, one can obtain the generating functional of the theory and derive all quantities of interest like the quark propagator, the thermodynamic potential and (anti)quark-(anti)quark cross sections (here and henceforth we are referring to quark-quark, quark-antiquark and antiquark-antiquark processes). However, the presence of more than two quark-quark interactions at the Lagrangian level renders impossible the exact integration over the quark fields \cite{Reinhardt:1988xu}. This is a natural difficulty when dealing with interacting field theories and some approximation must be employed in order to deal with the functional integrations and calculate the previously mentioned quantities.

In this work, we will use the linear and quadratic expansions of the Lagrangian, allied to the mean field approximation. The one fermion line integral, $A$, arises in the linear expansion of the NJL Lagrangian, namely in the calculation of the quark propagator and thermodynamic quantities. The two fermion line integral, $B_0$, emerges in the quadratic expansion and is essential in the calculation of the quark-antiquark polarization function, which is a fundamental step in the calculation of meson mass behavior and the cross sections. In the following we  show how these integrals arise from the linear and quadratic expansions of the Lagrangian defined in Eq. (\ref{general_NJL_Lagrangian}).

\subsection{The linear Lagrangian}

As already discussed, in order to obtain the quark propagator and thermodynamics of the model defined in Eq. (\ref{general_NJL_Lagrangian}), one can consider the linear expansion of the Lagrangian, allied to the mean field approximation. In this approximation one transforms all quark-quark interactions into quadratic quark interactions (linear in bilinear quark operators) by the introduction of auxiliary fields. Terms which have a larger than linear dependency in the quantum fluctuations are neglected. Using this technique, one can write a generating functional integrand that is quadratic in the quark fields.

Consider a quark bilinear operator, $\operator = \bar{\psi} \hat{\Gamma} \psi$, where $\hat{\Gamma}$ is an operator that acts on some internal quark index, like color or flavor. Without loss of generality, the bilinear can be written as a sum between its mean field value and a small fluctuation, $\operator = 
\expval*{ \operator } + \delta \operator$. In order to obtain the linear expansion of the product between $N-$bilinear operators, one only keeps terms that are linear in the fluctuations, $\delta \operator$, i.e., terms equal or superior to $(\delta \operator )^2$ are neglected. As a matter of fact, one can prove that the linear product between $N=n+1$ commuting operators can be written using the following formula (see Appendix \ref{linear_expansion_product_N_operators} for the proof) \cite{CamaraPereira:2020rtu,Pereira:2021xxv}:
\begin{align}
\Linear{ \prod_{i=1}^{n+1} \operator_i } 
& =  
\qty(
\sum_{i=1}^{n+1} \frac{ \operator_i }{ \expval*{ \operator_i } }
- n  
)
\prod_{j=1}^{n+1} \expval*{ \operator_j }  .
\label{linear_product_operators}
\end{align}
As an example, the linear product between two and three commuting operators, can directly be obtained from this equation:
\begin{align*}
\Linear{ \operator_1 \operator_2 }
& =
-\expval*{\operator_1} \expval*{\operator_2} + 
\expval*{\operator_2} \operator_1 + 
\expval*{\operator_1} \operator_2 ,
\numberthis
\\
\Linear{ \operator_1 \operator_2 \operator_3 }
& =
-2 
\expval*{\operator_1} 
\expval*{\operator_2} 
\expval*{\operator_3} + 
\operator_1 \expval*{\operator_2} \expval*{\operator_3} + \expval*{\operator_1} \operator_2 \expval*{\operator_3} + \expval*{\operator_1} \expval*{\operator_2} \operator_3 .
\numberthis
\end{align*}

Finally, using this approximation, one can obtain a Lagrangian density, $\mathcal{L}_{\mathrm{L}}$, that is linear in the quark bilinear operators,
\begin{align}
\mathcal{L}_{\mathrm{L}}
\qty[\bar{\psi},\psi, \expval{\vec{s}}, \expval{\vec{p}}]
=
\bar{\psi} 
\qty( i\slashed{\partial} - M \qty[m,\expval{\vec{s}}, \expval{\vec{p}}] ) \psi 
-  
\mathcal{U}\qty[ \expval{\vec{s}}, \expval{\vec{p}} ] .
\end{align}
Here, $\expval*{ \vec{s} }= \{\expval*{ s_0 }, \ldots, \expval*{ s_{N_f^2-1} }\}$ and $\expval*{ \vec{p} }= \{ \expval*{ p_0 }, \ldots, \expval*{ p_{N_f^2-1}}  \}$ are the sets of scalar and pseudoscalar mean fields obtained after performing the linearization procedure. These fields correspond to the expectation values of the $N_f^2$ scalar and pseudoscalar quark bilinear operators, $s_a=\bar{\psi} \lambda_a \psi$ and $p_a=\bar{\psi} i \gamma^5 \lambda_a \psi$, respectively ($a=0,\ldots,N_f^2-1$). The effective mass matrix, $M$ and the mean field potential $U$ are functions of these sets. Using this quantity, one can find the generating functional of the model, within this approximation, to be given by:
\begin{align*}
\mathcal{Z}_{\mathrm{L}} 
\qty[\expval{\vec{s}}, \expval{\vec{p}},\bar{\eta},\eta] 
& \propto
\int
\mathcal{D} \bar{\psi}
\mathcal{D} \psi
\exp
\qty[
i \int \dd[4]{x} 
\qty( 
\mathcal{L}_{\mathrm{L}}
\qty[\bar{\psi},\psi,\expval{\vec{s}}, \expval{\vec{p}}] 
+ \bar{\psi}\eta 
+ \bar{\eta}\psi  )
]
\\
& \propto
\exp \qty[ 
-i \mathcal{U} \, V 
]
\int
\mathcal{D} \bar{\psi}
\mathcal{D} \psi
\exp
\qty[
i \int \dd[4]{x}  
\qty( 
\int \dd[4]{y} 
\bar{\psi} [x]
\,
\mathcal{G}_q^{-1} \qty[x,y]
\,
\psi [y]
+ \bar{\psi} [x] \eta 
+ \bar{\eta}\psi [x] )
] ,
\numberthis
\end{align*}
where $\eta$ and $\bar{\eta}$ are the quark and antiquark sources, $V$ is the four-dimensional volume and the inverse quark propagator, $\mathcal{G}_q^{-1}$, was defined as:
\begin{align*}
\mathcal{G}_q^{-1} \qty[x,y]
& =
\qty( 
i\slashed{\partial} - M 
)
\delta^4 \qty(x-y)  .
\numberthis
\label{quarkPropagator}
\end{align*}
For simplicity, we omitted the mean field dependencies present on the quark effective mass, $M = M [m,\expval{\vec{s}}, \expval{\vec{p}}]$.

The quark fields can then be integrated out, and one can define the effective action of the model from the generating functional (with vanishing sources), $\Gamma\qty[ \expval{\vec{s}}, \expval{\vec{p}} ]  = - i \ln \qty[ \mathcal{Z}_{\mathrm{L}} 
\qty[ \expval{\vec{s}}, \expval{\vec{p}} ] ]$, which only depends on the sets of mean fields, $\expval{\vec{s}}$ and $\expval{\vec{p}}$. Such dependence must be removed by requiring the effective action to be an extreme with respect to these mean fields i.e., $\fdv*{ \Gamma\qty[\expval{\vec{s}}, \expval{\vec{p}}] }{\expval{s_a}}=\fdv*{ \Gamma\qty[\expval{\vec{s}}, \expval{\vec{p}}] }{\expval{p_a}}=0$. At this point one has access to the thermodynamic potential from which several thermodynamic variables of interest, such as pressure, energy density, entropy and speed of sound, can be evaluated.

Finally, the one fermion, one loop integral, $A$, can be defined by considering the quark condensate of flavor $i$ \cite{Klevansky:1992qe}:
\begin{align*}
\expval*{ \bar{\psi}_i \psi_i }
& =
-i \tr \mathcal{G}_q^i \qty[x,x]
\\
& =
-4 i N_c M_i \int\frac{ \dd[4]{p} }{ \qty( 2\pi )^4 }
\frac{1}{ p^2 - M_i^2 }
\\
& =
\frac{ N_c }{ 4 \pi^2 } M_i A \qty[M_i] .
\numberthis
\label{quarkCondensate}
\end{align*}
The one fermion, one loop integral in Minkowski spacetime, $A \qty[M_f]$, is given by:
\begin{align}
A \qty[M_i]
=
- 16 \pi^2 i
\int_{\mathrm{Reg}} \frac{ \dd[4]{p} }{ \qty( 2\pi )^4 }
\frac{1}{ p^2 - M_i^2 }.
\label{A_integral_definition}
\end{align}
Here, $M_i$ is the $i$-quark effective mass. The inclusion of temperature and chemical potential will be done in Section \ref{one_fermion_one_loop_integral}. Since this integral is divergent and some regularization scheme must be applied, we explicitly wrote the label $\mathrm{Reg}$.

\subsection{The quadratic Lagrangian}
\label{quadraticLagrangian}

The two fermion line integral, $B_0$, can be found when considering the quadratic expansion of the Lagrangian given in Eq. (\ref{general_NJL_Lagrangian}). Such expansion allows one to calculate several interesting properties, namely: meson propagators, meson masses, decay rates and (anti)quark-(anti)quark cross sections.

The approach to write a Lagrangian that is quadratic in the quark bilinear operators (fourth order in the quark fields) is similar to the one used in the previous section to obtain the linear Lagrangian. Firstly, as before, the operator is written as the sum,  $\operator = 
\expval*{ \operator } + \delta \operator$. However, to get the quadratic expansion of the product between $N$-bilinear operators, only up to quadratic terms in the fluctuations are kept and higher order terms are neglected. As before, instead of doing such expansion by hand, there is a formula from which one can readily obtain the quadratic product between $N=n+2$ commuting operators, with $n \geq 1$ (see Appendix \ref{quadratic_expansion_product_N_operators} for the proof). It is given by \cite{CamaraPereira:2020rtu,Pereira:2021xxv}:
\begin{align*}
\Quadratic{ \prod_{i=1}^{n+2} \operator_i }
& =  
\bigg(
\frac{1}{2}
\sum_{i=1}^{n+2}
\sum_{j=1}^{n+2}
\frac{ \operator_i  }{ \expval*{ \operator_i } }
\frac{ \operator_j }{ \expval*{ \operator_j } }
\qty( 1 - \delta_{ij} )
- n  
\sum_{i=1}^{n+2}
\frac{ \operator_i }{ \expval*{ \operator_i } }
+\frac{n}{2} \qty(n+1)
\bigg)
\prod_{k=1}^{n+2} \expval*{ \operator_k }  .
\numberthis
\end{align*}
As an example, the quadratic expansion of the product between three and four operators, are given by:
\begin{align*}
\Quadratic{ 
\operator_1
\operator_2
\operator_3
}
= &
-\expval*{\operator_2} \expval*{\operator_3} \operator_1
-\expval*{\operator_1} \expval*{\operator_3} \operator_2
-\expval*{\operator_1} \expval*{\operator_2} \operator_3
\\
& 
+\expval*{\operator_1} \operator_2 \operator_3
+\expval*{\operator_2} \operator_1 \operator_3
+\expval*{\operator_3} \operator_1 \operator_2
+\expval*{\operator_1} \expval*{\operator_2} \expval*{\operator_3} ,
\numberthis
\\
\Quadratic{ 
\operator_1
\operator_2
\operator_3
\operator_4
}
= &
- 2 \expval*{\operator_2} \expval*{\operator_3} \expval*{\operator_4} \operator_1 
- 2 \expval*{\operator_1} \expval*{\operator_3} \expval*{\operator_4} \operator_2 
\\
& 
- 2 \expval*{\operator_1} \expval*{\operator_2} \expval*{\operator_4} \operator_3 
- 2 \expval*{\operator_1} \expval*{\operator_2} \expval*{\operator_3} \operator_4 
\\
& 
+ \expval*{\operator_2} \expval*{\operator_4} \operator_1 \operator_3 
+ \expval*{\operator_1} \expval*{\operator_4} \operator_2 \operator_3
+ \expval*{\operator_2} \expval*{\operator_3} \operator_1 \operator_4 
\\
& 
+ \expval*{\operator_1} \expval*{\operator_3} \operator_2 \operator_4 
+ \expval*{\operator_1} \expval*{\operator_2} \operator_3 \operator_4
+ \expval*{\operator_3} \expval*{\operator_4} \operator_1 \operator_2 
\\
& 
+ 3 \expval*{\operator_1} \expval*{\operator_2} \expval*{\operator_3} \expval*{\operator_4} .
\numberthis
\end{align*}

Using the quadratic expansion above and neglecting second order terms which mix scalar and pseudoscalar quark bilinear operators ($s_a  p_b$ and $p_a s_b$)\footnote{This follows from the consideration that only the scalar condensates get a non-vanishing ground state and pseudoscalar condensates are zero.}, the quadratic Lagrangian density, $\mathcal{L}_{\mathrm{Q}}$, (fourth order in the quark fields) can be written as:
\begin{align}
\mathcal{L}_{\mathrm{Q}}
\qty[\bar{\psi},\psi,\expval{\vec{s}}, \expval{\vec{p}}]
=
\overline{\psi} \qty(i\slashed{\partial} - m ) \psi
+ s_a S_{ab} s_b
+ p_a P_{ab} p_b
+ s_a S_a
+ p_a P_a
+ R.
\label{quadraticLagrangianExpansion}
\end{align}
Here, $S_{ab}=S_{ab}[\expval{\vec{s}}, \expval{\vec{p}}]$ and $P_{ab}=P_{ab}[\expval{\vec{s}}, \expval{\vec{p}}]$ gather the coefficients that arise after the quadratic expansion of the operators, which are quadratic in the variables $s_a=\bar{\psi} \lambda_a \psi$ and $p_a=\bar{\psi} i \gamma^5 \lambda_a \psi$, respectively. They are the so-called meson projectors for the scalar modes $(S_{ab})$ and the pseudoscalar modes $(P_{ab})$. The first and zeroth order contributions are gathered in the coefficients $S_a=S_a[\expval{\vec{s}}, \expval{\vec{p}}]$, $P_a=P_a[\expval{\vec{s}}, \expval{\vec{p}}]$ and $R=R[\expval{\vec{s}}, \expval{\vec{p}}]$, respectively. 

The generating functional for the system, can be written as:
\begin{align}
\mathcal{Z}_\mathrm{Q} 
\qty[ \expval{\vec{s}}, \expval{\vec{p}} , \eta, \bar{\eta} ]  
\propto 
\int \mathcal{D}\overline{\psi} \mathcal{D} \psi \exp 
\qty{  
i \int \dd[4]{x} 
\qty( 
\mathcal{L}_{\mathrm{Q}}
\qty[\bar{\psi},\psi,\expval{\vec{s}}, \expval{\vec{p}}]
+ \overline{\psi} \eta  
+ \bar{\eta} \psi 
)   
}. 
\end{align}

In order to integrate out the quark fields, one can bosonize the model by introducing auxiliary meson fields,  $\sigma_a$ and $\pi_a$, through Hubbard–Stratonovich transformations \cite{PhysRevLett.3.77,Stratonovich,Ebert:1997fc}. For the $s_a$ contributions, consider the following Gaussian integral:
\begin{align}
\exp
\qty{ 
i \int \dd[4]{x} \qty( s_a S_{ab} s_b )  
}
& \propto
\int \prod_c \mathcal{D}\sigma_c
\exp\qty{ i \int \dd[4]{x} 
\qty(
- \frac{1}{4} \sigma_a \qty( S ^{-1})_{ab} \sigma_b 
-s_a \sigma_a
) 
} .
\end{align}
Considering a change of variables on the functional integral as: $\sigma_a \to \sigma_a + \qty(S_a + \Delta_a)$  where $\Delta_a = M_a - m_a$ i.e., $\lambda_a \Delta_a = M - m$, and multiplying both sides by $\exp\qty{
i \int \dd[4]{x} \qty( s_a S_a ) 
}$, we can write\footnote{Here we used the fact that the matrix $S$ is symmetric.}:
\begin{align*}
\exp
\bigg\{
i \int \dd[4]{x} 
\qty( s_a S_{ab} s_b + s_a S_a ) 
\bigg\}
&
\\
 \propto
\int \prod_c \mathcal{D}\sigma_c 
\exp\bigg\{ &
i \int \dd[4]{x} 
\bigg(
- \frac{1}{4} \sigma_a \qty( S ^{-1})_{ab} \sigma_b 
- \frac{1}{2} \sigma_a \qty( S^{-1})_{ab} \qty(S_b + \Delta_b)
- s_a  \qty( \sigma_a + \Delta_a )
\bigg) 
\bigg\}  .
\numberthis
\end{align*}
With this particular change of variables we are considering the possibility of scalar condensation in the vacuum. In this work we do not consider the possibility of pseudoscalar condensation. In this case, for the $p_a$ contributions, the approach is similar except that the change of variables is simply $\pi_a \to \pi_a + P_a $.  Using these identities, the action becomes quadratic in the fermion fields and one can integrate out the quark fields to yield:
\begin{align*}
\mathcal{Z}_\mathrm{Q} 
\qty[ \expval{\vec{s}}, \expval{\vec{p}} ]
& \propto
\int 
\prod_a 
\mathcal{D}\sigma_a 
\mathcal{D} \pi_a 
\exp 
\qty{
i \mathcal{S}_\mathrm{Q}
\qty[ \vec{\sigma}, \vec{\pi} , \expval{\vec{s}}, \expval{\vec{p}} ]
} .
\numberthis
\label{Zeff_after_quark_integration}
\end{align*}
Here, $\mathcal{S}_\mathrm{Q}
\qty[ \vec{\sigma}, \vec{\pi} , \expval{\vec{s}}, \expval{\vec{p}} ]$ is the action of the system and $\vec{ \sigma }= \{\sigma_0, \ldots, \sigma_{N_f^2-1}\}$ and $\vec{ \pi }= \{ \pi_0 , \ldots,  \pi_{N_f^2-1}  \}$ are the sets of scalar and pseudoscalar auxiliary fields (respectively) introduced with the Hubbard–Stratonovich transformation. The action of the system is explicitly given by:
\begin{align*}
\mathcal{S}_\mathrm{Q}
\qty[ \vec{\sigma}, \vec{\pi}, \expval{\vec{s}}, \expval{\vec{p}} ]
 =
-i 
\Tr \ln 
\big[
i\slashed{\partial} - M
& - \lambda_a\sigma_a - i\gamma_5 \lambda_a \pi_a
\big]
\\ 
+ \int \dd[4]{x}
\bigg(
&- \frac{1}{4} \sigma_a \qty( S^{-1})_{ab} \sigma_b 
- \frac{1}{4} \pi_a \qty( P^{-1} )_{ab} \pi_b 
\\
&- \frac{1}{2} \sigma_a \qty( S^{-1})_{ab} \qty(S_b + \Delta_b)
- \frac{1}{2} \pi_a \qty( P^{-1} )_{ab} P_b 
+ R 
\bigg) .
\numberthis
\label{quadratic_action}
\end{align*}

In the above expressions, $\Tr$ is the usual notation for the total trace over continuous spacetime indexes and discrete internal indexes such as color, flavor and spinor, $\Tr \equiv \trx  \trc \trf \trD $  and the functional trace of an operator, $\hat{O}$, is defined by: $\trx \hat{O} 
=
\int \dd[d]{x}
\bra{ x }
\hat{O}
\ket{ x }$ \cite{BALL19891,Klevansky:1992qe}. Here, we will use the functional formalism introduced in Refs. \cite{Novikov:1983gd,Fraser:1984zb,Aitchison:1985pp}, where $\hat{O}$ is considered an operator acting directly on an fictitious Hilbert space, $\qty{ \ket{x} }$, with the position ($\hat{x}_\mu$) and momentum operators ($\hat{p}_\mu$) obeying the usual commutation relation, $\qty[\hat{x}_\mu,\hat{p}_\nu] = i\delta_{\mu \nu}$ \cite{Novikov:1983gd,Fraser:1984zb,Aitchison:1985pp,BALL19891}.

The gap equations of the	 model can be obtained by evaluating the path integral over the auxiliary meson fields, $\sigma_a$ and $\pi_a$ in the stationary phase (or saddle point) approximation \cite{Vogl:1991qt,Hatsuda:1994pi,Ebert:1997fc,Ripka:1997zb,Oertel:2000jp}, which is equivalent to the mean field or Hartree approximation. It amounts to requiring the action of the model to be stationary with respect to the auxiliary meson fields,
$
\left.\fdv*{ \mathcal{S}_\mathrm{Q} }{ \sigma_a}\right|_{\qty{\sigma^{\mathrm{st}}_a,\pi^{\mathrm{st}}_a}}=
\left.\fdv*{ \mathcal{S}_\mathrm{Q} }{ \pi_a }\right|_{\qty{\sigma^{\mathrm{st}}_a,\pi^{\mathrm{st}}_a}}=0$. Here, we note that we already allowed for a non-vanishing expectation value of the field $\sigma_a$ in the change of variables, $\sigma_a \to \sigma_a + \qty(S_a + \Delta_a)$, and thus are requiring that the stationary field configurations must vanish: $\sigma^{\mathrm{st}}_a=\pi^{\mathrm{st}}_a=0$. Using these requirements, one can derive the self-consistent gap equations of the model which relates the quarks effective mass, $M$, and the expectation values of the quark bilinear operators, $\expval{\vec{s}}$ and $\expval{\vec{p}}$.  This approach leads exactly to the same gap equations as the ones obtained using the linear Lagrangian in the mean field approximation presented in the previous section. Indeed, for a specific NJL Lagrangian, one can check this equivalence explicitly. This correspondence might be related the exact Hubbard–Stratonovich transformation, used to transform the quadratic action (in terms of quark bilinear operators) into a linear action. A proper demonstration, between the equivalence of the gap equations derived in the linear and quadratic Lagrangians (alongside the saddle point approximation for the auxiliary meson fields), for any interactions terms, is left as future work.

The quark-antiquark polarization function, at finite chemical potential, can be derived from this formalism by considering the expansion of the fermionic determinant to second order in the mesonic auxiliary fields, $\sigma_a$ and $\pi_a$. Finite chemical potential can be included in the calculation by making the following substitution, $\partial_\alpha \to D_\alpha = \partial_\alpha - i \mu \delta_{\alpha0}$, with $\mu$ the quark chemical potential matrix in flavor space \cite{Klevansky:1992qe}. This matrix can be considered to be diagonal with components, $\mu=\diag\qty( \mu_0, \mu_1, \ldots , \mu_{N_f-1})$. Next, we consider the Taylor series expansion of the logarithm of a matrix in Eq. (\ref{quadratic_action}). We write \cite{Klevansky:1992qe,Nikolov:1996jj,Costa:2002gk,Costa:2003uu,Blaschke:2013zaa,Blaschke:2017boi,HaberNotesMatrixExponentialLogarithm}:
\begin{align*}
\Tr \ln 
\big[
i\slashed{D} - M
- \lambda_a\sigma_a - i\gamma_5 \lambda_a \pi_a
\big]
& =
\Tr \ln 
\big[
i\slashed{D} - M
\big]
+
\Tr \ln 
\qty[
1
+
\qty[i\slashed{D} - M]^{-1}
\bigg(- \lambda_a\sigma_a - i\gamma_5 \lambda_a \pi_a \bigg)
]
\\
& = 
\Tr \ln 
\big[
i\slashed{D} - M
\big]
-
\sum_{n=1}^{\infty} \Tr \qty[ T_n ] .
\numberthis
\label{seriesExpansionFermionicDeterminant}
\end{align*}
In the above, the logarithm of a matrix was written using its defining Taylor series expansion (assuming that the series converges) \cite{HaberNotesMatrixExponentialLogarithm}:
\begin{align}
\ln \qty[X] & =
\sum_{n=1}^{\infty}
\frac{ \qty(-1)^{n+1} }{ n }
\qty( X - I )^n .
\end{align}
A convergence criterion can be written as: $\abs{ X - I }<1$, with $\abs{Y}$ a proper way to evaluate the norm of the matrix $Y$ \cite{HaberNotesMatrixExponentialLogarithm}. Additionally, the operator $T_n$ is defined as:
\begin{align}
T_n 
= 
\frac{ 1 }{ n }
\qty( 
\qty[i\slashed{D} - M]^{-1}
\qty(\lambda_a\sigma_a + i\gamma_5 \lambda_a \pi_a )
)^n .
\end{align}

We are interested in evaluating the second order term in the expansion, $T_2$, which is an essential part in the definition of the inverse meson propagators. To this end, we define the operators, $\mathcal{G}=\qty[i\slashed{D} - M]^{-1}$, $\sigma = \lambda_a \sigma_a$ and $\pi = \lambda_a \pi_a$. We highlight that $\mathcal{G}$ is a matrix in flavor space due to the structure of the effective quark mass and effective chemical potential matrices, $M=\diag\qty( M_0, M_1, \dots, M_{N_f-1} )$, and $\mu=\diag\qty( \mu_0, \mu_1, \ldots , \mu_{N_f-1})$. With this notation, we can write:
\begin{align*}
T_2 
& = 
\frac{1}{2}
\mathcal{G} \qty(\sigma + i\gamma_5 \pi )
\mathcal{G} \qty(\sigma + i\gamma_5 \pi )
\\
& = 
T_2^{ \sigma \sigma } +
T_2^{ \sigma \pi    } + 
T_2^{ \pi    \sigma } +
T_2^{ \pi    \pi    } .
\numberthis
\end{align*}
Only the traces of the first and fourth operators are nonzero, $T_2^{ \sigma \sigma }$ and $T_2^{ \pi \pi }$. The trace over spinor indexes of the ``mixed'' contributions involving the identity matrix $I_4$ and the $\gamma^5$ matrix vanish. Without loss of generality, we can deal with the quantity, $T_2^{ \phi \phi } \qty[ \Gamma_1, \Gamma_2 ]
=
\frac{1}{2}
\mathcal{G} \Gamma_1 \phi 
\mathcal{G} \Gamma_2 \phi  
$, with $\phi$ standing in for $\sigma$ or $\pi$. It yields $T_2^{ \sigma \sigma }$ when $\Gamma_1 = \Gamma_2 = I_4$ ($I_4$ is the four-dimensional unity matrix) and $T_2^{ \pi \pi }$ in the case in which $\Gamma_1 = \Gamma_2 = i\gamma_5$. The operator $\Gamma$ only has discrete Dirac indexes. The fields $\phi$ and the quantity $\mathcal{G}$ both have discrete flavor indexes as well as continuous spacetime indexes but are structureless in color space. Additionally, the operator $\mathcal{G}$ also carries Dirac indexes. As a result of this internal structure, the evaluation of the traces over these different vector spaces has to be done taking into account the noncommutative nature of these operators, i.e., one has to take into account that, $ \qty[\mathcal{G},\Gamma] \neq 0$ in spinor space, $ \qty[\mathcal{G},\phi] \neq 0$ both for flavor and continuous spacetime indexes, while the remaining commutation relations are zero (for instance, $\qty[\Gamma,\phi]=0$ in flavor and spinor spaces). The way in which one evaluates the traces can lead to different mathematical expansions. As we will explain more thoroughly later, we are interested in obtaining an expansion of the fermionic determinant which yields the same result for the meson propagators as the ones obtained via Bethe-Salpeter equation in the random phase approximation (or, equivalently, the ring approximation \cite{Hatsuda:1994pi}). We will make further comments about deriving the meson propagators from the fermionic determinant in Section \ref{comments_expansion_fermionic_determinant}.

Lets start by evaluating the trace over flavor degrees of freedom, $\trf \qty[ \mathcal{G} \Gamma_1 \phi \mathcal{G} \Gamma_2 \phi   ]$. To this end we introduce the matrices $\hat{e}_{AB}$, with indexes in the range $A,B=\qty{0,1,\ldots,N_f-1}$ and matrix elements given by: $\qty(\hat{e}_{AB})_{ij}=\delta_{Ai}\delta_{Bj}$. Explicitly, they are given by:
\begin{align*}
\hat{e}_{00} & =
\mqty(
1 & 0  & \ldots & 0 \\ 
0 & 0  & \ldots & 0 \\ 
\vdots & \vdots  & \ddots & \vdots    \\
0 & 0 & \ldots & 0
) 
, \;
\hat{e}_{01}  =
\mqty(
0 & 1  & \ldots & 0 \\ 
0 & 0  & \ldots & 0 \\ 
\vdots & \vdots  & \ddots & \vdots    \\
0 & 0 & \ldots & 0
) 
, \;
\ldots
, \;
\hat{e}_{N_f-1 N_f-1} =
\mqty(
0 & 0  & \ldots & 0 \\ 
0 & 0  & \ldots & 0 \\ 
\vdots & \vdots  & \ddots & \vdots    \\
0 & 0 & \ldots & 1
) .
\numberthis
\end{align*}
Any $N_f \times N_f$ complex matrix, $C$, can be written as a linear combination of the $\hat{e}_{AB}$ matrices alongside $N_f^2$ complex coefficients, $c_{AB}$, as: $C = c_{AB} \hat{e}_{AB}$. These matrices are linearly independent since the zero matrix can only be obtained if all coefficients vanish. Additionally, one can check that the trace between two $\hat{e}_{AB}$ matrices is, $ \tr \qty[ \hat{e}_{AB} \hat{e}_{CD} ] = \delta_{AD}\delta_{BC} $. The $N_f^2$ generators of the $U(N_f)$ group, $\lambda_a$, can be written in terms of this new basis. Using the completeness relation we can write: $\lambda_a = R_{AB}^{a} \hat{e}_{AB}$. Multiplying both sides by $\hat{e}_{CD}$ and taking the trace, one can get the expression for the $R_{AB}^a$ coefficients:
\begin{align}
R_{AB}^a = \tr\qty[ \lambda_a \hat{e}_{BA}  ].
\label{R_coefficients}
\end{align}
Using this new set of matrices, the trace over flavor space is:
\begin{align}
\trf
\qty[
\mathcal{G} \Gamma_1 \phi \mathcal{G} \Gamma_2 \phi  
]
& =
R_{ij}^{a}
R_{ji}^{b}
\qty( \mathcal{G} )_{ii} \Gamma_1 \phi_a 
\qty( \mathcal{G} )_{jj} \Gamma_2 \phi_b .
\end{align}
Here, $a,b,i,j=\{0,1,\ldots,N_f-1\}$ and the coefficients $R^a_{AB}$ are given by Eq. (\ref{R_coefficients}). Also, we used the fact that the operator $\mathcal{G}$ is diagonal in flavor space, $\mathcal{G}=\diag\qty( \mathcal{G}_0, \mathcal{G}_1, \ldots, \mathcal{G}_{N_f-1} )$. For the $N_f=3$ case, which will be studied later in this work, the nine generators of the $U(3)$ group are constituted by the scaled unit matrix plus the eight Gell-Mann matrices and one can identify the $\qty(\mathcal{G})_{ii}$ elements with the up, down and strange quark propagators, i.e., $i=0=u$ (up), $i=1=d$ (down) and $i=2=s$ (strange) and $\mathcal{G}=\diag\qty( \mathcal{G}_u, \mathcal{G}_d, \mathcal{G}_s )$.

Due to the lack of structure in color space, the trace over such degrees of freedom yields a simple $N_c$ factor. The full trace over spacetime and discrete indexes is given by:
\begin{align*}
\Tr \qty[ T_2^{ \phi \phi } \qty[ \Gamma_1, \Gamma_2 ] ]
& =
R_{ij}^{a}
R_{ji}^{b} 
\frac{N_c}{2}
\trD
\int \dd[4]{x}
\bra{ x }
\qty( \mathcal{G} )_{ii} \Gamma_1 \phi_a 
\qty( \mathcal{G} )_{jj} \Gamma_2 \phi_b 
\ket{ x } .
\numberthis
\label{expression_for_trx}
\end{align*}
Evaluating it, yields the following result:
\begin{align*}
\Tr \qty[ T_2^{ \phi \phi } \qty[ \Gamma_1, \Gamma_2 ] ]
& =
R_{ij}^{a}
R_{ji}^{b} 
\frac{N_c}{2}
\int_{\mathrm{Reg}} \frac{ \dd[4]{k} }{ \qty(2\pi)^4 }
\phi_a \qty[ k ] \phi_b \qty[ - k ] 
\int_{\mathrm{Reg}[ k ]}  \frac{ \dd[4]{p} }{ \qty(2\pi)^4 }
\trD
\qty[
\frac{1}{ \slashed{p}_i - M_i } 
\Gamma_1 
\frac{1}{ (\slashed{p}_j - \slashed{k}) - M_j } 
\Gamma_2
]
\\
& =
-
\frac{ i }{2}
R_{ij}^{a}
R_{ji}^{b} 
\int_{\mathrm{Reg}} \frac{ \dd[4]{k} }{ \qty(2\pi)^4 }
\phi_a \qty[ k ] \phi_b \qty[ - k ] 
\Pi_{ij} \qty[\Gamma_1, \Gamma_2, k]
\\
& =
-
\frac{ i }{2} 
\int_{\mathrm{Reg}} \frac{ \dd[4]{k} }{ \qty(2\pi)^4 }
\phi_a \qty[ k ] 
\, 
\Pi_{ab} \qty[\Gamma_1, \Gamma_2, k] 
\, 
\phi_b \qty[ - k ] . 
\label{traceT2PhiPhi}
\numberthis
\end{align*}
Here, we defined $\slashed{p}_f = \gamma^0( p_0 + \mu_f ) - \vec{\gamma} \cdot \vec{p}$, with $\mu_f$ the chemical potential of the quark with flavor $f$ \cite{Klevansky:1992qe}. Furthermore, we applied regularization to both momentum integrations, with a small caveat: we allowed the regularization of the innermost momentum integration ($p$) to be a function of the outermost momentum ($k$). The reason for this will be discussed later. The quark-antiquark polarization function with flavor indexes, $\Pi_{ij}\qty[\Gamma_1, \Gamma_2, k]$ and its analogue, $\Pi_{ab}\qty[\Gamma_1, \Gamma_2, k]$, with indexes coming from the generators of the $U(N_f)$  group, were defined in this expression as:
\begin{align}
\Pi_{ij} \qty[\Gamma_1, \Gamma_2, k]
& =
i N_c
\int_{\mathrm{Reg}[ k ]}  \frac{ \dd[4]{p} }{ \qty(2\pi)^4 }
\frac
{ \trD \qty[ ( \slashed{p}_i + M_i  ) \Gamma_1 
((\slashed{p}_j - \slashed{k}) + M_j) \Gamma_2  ] }
{ (p_i^2 - M_i^2)((p_j - k)^2 - M_j^2) } ,
\label{quarkAntiquarkPolarization_flavorIndexes}
\\
\Pi_{ab} \qty[\Gamma_1, \Gamma_2, k]
& =
R_{ij}^{a} R_{ji}^{b} \Pi_{ij} \qty[k] .
\label{quarkAntiquarkPolarization_gellmannIndexes}
\end{align}
The only trace left to calculate is over Dirac indexes. This can be achieved by choosing $\Gamma_1$ and $\Gamma_2$. As previously stated, only the cases $\Gamma_1=\Gamma_2=I_4$ and $\Gamma_1=\Gamma_2=i \gamma^5$ are nonzero. Thus, we define the scalar quark-antiquark polarization function as $\Pi_{ij}^S [M_i, M_j, \mu_i, \mu_j, k]  = \Pi_{ij} \qty[I_4, I_4, k]$, and the pseudoscalar quark-antiquark polarization function by $\Pi_{ij}^P [M_i, M_j, \mu_i, \mu_j, k]  = \Pi_{ij} \qty[i\gamma^5, i\gamma^5, k]$. Evaluating the trace over spinor space yields:
\begin{align*}
\Pi_{ij}^{S|P} [M_i, M_j, \mu_i, \mu_j, k] 
=
-\frac{ N_c }{ 8 \pi^2 }
\bigg\{
\mathcal{A} & [M_i, \mu_i, k] + \mathcal{A} [M_j, \mu_j, k]  
\\
& +
\qty( \qty( M_i \pm M_j )^2 - \qty( \mu_i - \mu_j + k_0 )^2 + \vec{k}^2 )
B_0 [M_i, M_j, \mu_i, \mu_j, k] 
\bigg\}.
\label{scalarPseudoscalar_polarization_def}
\numberthis
\end{align*}
In the right-hand side of this equation, the positive sign $(+)$ is used for the scalar function, $\Pi_{ij}^S$, while, the negative sign $(-)$, is used for the pseudoscalar quark-antiquark polarization function, $\Pi_{ij}^P$. Here, we also defined the two fermion line integral, $B_0$, in Minkowski spacetime:
\begin{align}
B_0 [M_i, M_j, \mu_i, \mu_j, k] 
& =
- 16 \pi^2 i
\int_{\mathrm{Reg}[ k ]} \frac{ \dd[4]{p} }{ \qty(2\pi)^4 }
\frac{ 1 }
{ 
( \qty(p_0 + \mu_i)^2 - \vec{p}^2 - M_i^2)
((p_0 + \mu_j - k_0)^2 - (\vec{p} - \vec{k})^2 - M_j^2) 
}  .
\label{B0_definition}
\end{align}
The function $\mathcal{A}$ is defined by:
\begin{align}
\mathcal{A} [M_i, \mu_i, k] 
& =
- 16 \pi^2 i
\int_{\mathrm{Reg}[ k ]} \frac{ \dd[4]{p} }{ \qty(2\pi)^4 }
\frac{ 1 }{ \qty(p_0 + \mu_i)^2 - \vec{p}^2 - M_i^2 }  .
\label{Ak_integral_definition}
\end{align}
The integrand is exactly the same as the one present in the definition of the one fermion line integral $A$ (see Eq. (\ref{A_integral_definition})). However, the function $\mathcal{A}$ carries the same momentum dependent regularization as the integral $B_0$. In the special case where the same regularization is applied to both $A$ and $\mathcal{A}$, then $A=\mathcal{A}$. Hence, the integral $A$, defined in Eq. (\ref{A_integral_definition}) and the integral $\mathcal{A}$, defined above, are exactly the same quantity but, within different regularizations.

Finally, we can define the meson propagators of the model. Using the series expansion of Eq. (\ref{seriesExpansionFermionicDeterminant}) alongside the trace calculation of Eq. (\ref{traceT2PhiPhi}). The second order term, proportional to $\sigma_a\sigma_b$ and $\pi_a\pi_b$, of the quadratic action, $\mathcal{S}_{ \mathrm{Q} }$, is given by:
\begin{align*}
\mathcal{S}_{ \mathrm{Q} }^{ (2) }
\qty[ \vec{\sigma}, \vec{\pi}, \expval{\vec{s}}, \expval{\vec{p}} ]
= &  
-
\frac{ 1 }{2} 
\int_{\mathrm{Reg}} \frac{ \dd[4]{k} }{ \qty(2\pi)^4 }
\,
\sigma_a \qty[ k ] 
\qty(
\frac{1}{2}
\qty( S^{-1})_{ab}
-
\Pi_{ab}^S \qty[k] 
)
\sigma_b \qty[ - k ]
\\
& 
-
\frac{ 1 }{2} 
\int_{\mathrm{Reg}} \frac{ \dd[4]{k} }{ \qty(2\pi)^4 }
\, 
\pi_a \qty[ k ] 
\qty(
\frac{1}{2}
\qty( P^{-1})_{ab}
-
\Pi_{ab}^P \qty[k] 
)
\pi_b \qty[ - k ] .
\numberthis
\label{actionSecondOrderTerm}
\end{align*}
Using this result we can define the inverse scalar and pseudoscalar meson propagators, $D_S^{-1}$ and $D_P^{-1}$, respectively. They are defined by:
\begin{align}
\qty( D_S^{-1} )_{ab} \qty[k] 
& =
\frac{1}{2} \qty( S^{-1})_{ab} - \Pi_{ab}^S \qty[k]  ,
\label{scalarInverseMesonPropagator}
\\
\qty( D_P^{-1} )_{ab} \qty[k] 
& =
\frac{1}{2} \qty( P^{-1} )_{ab} - \Pi_{ab}^P \qty[k]  .
\label{pseudoscalarInverseMesonPropagator}
\end{align}
As we will see later, the meson propagators are essential in the calculation of quark-quark and quark-antiquark cross sections.

\subsubsection{Equivalence with the Bethe-Salpeter equation in the random phase approximation}
\label{equivalenceBetheSalpeterRPA}

\begin{figure}[ht!]
\centering
\includegraphics[scale=1.06]
{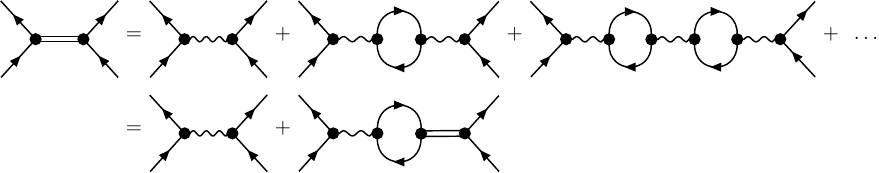}
\caption{Bethe-Salpeter equation in the random phase approximation (RPA) \cite{Vogl:1991qt,Klevansky:1992qe,Oertel:2000jp,Buballa:2003qv}. In the first line the transfer matrix, $\mathcal{T}$, is written as an infinite sum over quark-antiquark polarization loops, $\Pi$. In the second line, the equation is expressed in a self-consistent way. The double lines represents the meson propagator and the solid lines the quark propagators. The wiggly line between two black dots is used to signal the tensor nature of the NJL scattering kernel, $\mathcal{K}$, due to its structure in flavor, color and Dirac spaces.}
\label{bethe_salpeter_diagram}
\end{figure}

\begin{figure}[ht!]
\centering
\includegraphics[scale=1.3]
{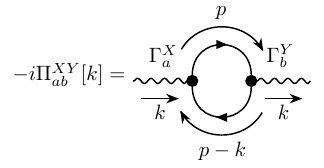}
\caption{Diagram representing the scalar/pseudoscalar quark-antiquark polarization function \cite{Klevansky:1992qe}. The solid lines represent quark propagators and the wiggly line with a black dot represents $\Gamma_a^X$, with $\Gamma_a^S=\lambda_a$, for scalar modes and $\Gamma_a^P=i\gamma^5 \lambda_a$, for pseudoscalar modes.}
\label{polarization_function_diagram}
\end{figure}

Another commonly employed method to obtain the meson propagators involves the Bethe-Salpeter equation, which describes the bound states of two body quantum systems. In the so-called random phase approximation, this equation is given by (see Fig. \ref{bethe_salpeter_diagram} for its diagrammatic representation) \cite{Vogl:1991qt,Klevansky:1992qe}:
\begin{align*}
i \mathcal{T} \qty[k]
& =
i \mathcal{K} 
+
i \mathcal{K} 
\Big( -i \Pi \qty[k] \Big)
i \mathcal{K} 
+
i \mathcal{K} 
\Big( -i \Pi \qty[k] \Big)
i \mathcal{K} 
\Big( -i \Pi \qty[k] \Big)
i \mathcal{K} 
+ \ldots
\\
& =
i \mathcal{K} 
+
i \mathcal{K} 
\Big( -i \Pi \qty[k] \Big)
i \mathcal{T} \qty[k] .
\numberthis
\label{betheSalpeterEquation}
\end{align*}
Here, $\mathcal{T}$ is the so-called quark-antiquark scattering matrix, $\mathcal{K}$ is the scattering kernel and $\Pi \qty[k]$ is the quark-antiquark polarization loop (see Fig. \ref{polarization_function_diagram}). These quantities contain several indexes including color, flavor and spinor.

The Bethe-Salpeter equation in the random-phase approximation, applied to NJL-type models in the mean field approximation (or Hartree approximation \cite{Vogl:1991qt,Klevansky:1992qe,Hatsuda:1994pi,Oertel:2000jp,Buballa:2003qv}), has been extensively applied in the literature to study in-medium meson behavior, and to evaluate quark-quark and quark-antiquark cross sections \cite{Bernard:1987sg,Vogl:1991qt,Klevansky:1992qe,Hatsuda:1994pi,Rehberg:1995kh,Zhuang:1995uf,Rehberg:1995nr,Rehberg:1996vd,Oertel:2000jp,Buballa:2003qv,Saionji:2022sxn}. Such standard approach \cite{Oertel:2000jp}, is known to be a symmetry preserving one, being compatible with chiral symmetry and respecting the Goldstone theorem and the Gell-Mann$-$Oakes$-$Renner relation \cite{Vogl:1991qt,Klevansky:1992qe,Hatsuda:1994pi,Nikolov:1996jj,Oertel:2000jp,Buballa:2003qv}. In this section we highlight the equivalence between this method and the functional expansion derived previously. To this end, one can use Eq. (\ref{betheSalpeterEquation}), to derive a self-consistent expression for the scalar and pseudoscalar meson propagator, $D_{ab}^{S|P}$. Consider the transfer matrix written in terms of the meson propagator: $\mathcal{T} = D^{XY}_{ab} \qty( \Gamma_a^X \otimes \Gamma_b^Y )$, with $\Gamma_a^X$ carrying the tensor structure of the vertex, $\Gamma_a^S = \lambda_a$ and $\Gamma_a^P = i \gamma^5 \lambda_a$ \cite{Vogl:1991qt,Klevansky:1992qe,Oertel:2000jp}. In a theory with only four-Fermi interactions (like the one considered in this work, see the quadratic Lagrangian in Eq. (\ref{quadraticLagrangianExpansion})), the scattering kernel is given by: $\mathcal{K} = 2 g_{ab}^{XY} \qty( \Gamma_a^X \otimes \Gamma_b^Y )$, with $g_{ab}^{XY} = g_{ab}^{X} \, \delta_{XY}$ and $g_{ab}^{S} = S_{ab}$ and $g_{ab}^{P} = P_{ab}$ for scalar and pseudoscalar degrees of freedom, respectively \cite{Vogl:1991qt,Klevansky:1992qe,Oertel:2000jp}. The factor of $2$ comes from the Feynman rules and is due to the application of Wick’s theorem \cite{Vogl:1991qt,Klevansky:1992qe}. The polarization loop, $\Pi_{ab}^{XY}$ is the quark-antiquark polarization loop defined by \cite{Vogl:1991qt,Klevansky:1992qe,Oertel:2000jp,Buballa:2003qv}:
\begin{align}
\Pi_{ab}^{XY} \qty[k]
& =
i
\int  \frac{ \dd[4]{p} }{ \qty(2\pi)^4 }
\tr
\qty[
\frac{ \slashed{p} + M }{ p^2 - M^2 } 
\Gamma_a^X 
\frac{ (\slashed{p} - \slashed{k}) + M  }{ ({p} - {k})^2 - M^2 } 
\Gamma_b^Y
] .
\end{align}
Here, the trace is made over flavor, color and Dirac discrete indexes, i.e., $\tr \equiv \trD \trc \trf$. After evaluating the trace, this expression agrees with the polarization loop obtained via functional methods, defined in Eqs. (\ref{quarkAntiquarkPolarization_flavorIndexes}) and (\ref{quarkAntiquarkPolarization_gellmannIndexes}). A diagrammatic representation of the quark-antiquark polarization function is shown in Fig. \ref{polarization_function_diagram}. Using these definitions, one can write an equation for the meson propagator as:
\begin{align*}
D_{ab}^{XY} \qty( \Gamma_a^X \otimes \Gamma_b^Y )
=
2 g_{ab}^{XY} \qty( \Gamma_a^X \otimes \Gamma_b^Y )
+
2 g_{ac}^{XK} 
\Pi_{cd}^{KL}
D_{db}^{LY}
\qty( \Gamma_a^X \otimes \Gamma_b^Y ) .
\numberthis
\end{align*}
Since the matrix structure is completely contained in the $\Gamma_a^X$ quantities, we can write an equation for the scalar function, $D_{ab}^{XY}$:
\begin{align*}
D_{ab}^{XY}
=
2 g_{ab}^{XY} 
+
2 g_{ac}^{XK} 
\Pi_{cd}^{KL}
D_{db}^{LY} .
\numberthis
\end{align*}
Considering $D_{ab}^{XY}$ as matrix elements, we can write the above equation in matrix form \cite{Vogl:1991qt,Klevansky:1992qe,Saionji:2022sxn}, $D = 2 g + 2 g \Pi D $, which can be solved for the meson propagator, $D$, to yield \cite{Vogl:1991qt,Klevansky:1992qe,Buballa:2003qv,Saionji:2022sxn}:
\begin{align*}
D & = \qty[ 1 - 2 g \Pi ]^{-1} 2 g .
\numberthis
\end{align*}
Writing the flavor indexes explicitly, we can write the scalar and pseudoscalar meson propagators, $D^{S|P}_{ab}$ as:
\begin{align}
D^{S|P}_{ab}
=
\qty[ \frac{1}{2} \qty(S|P^{-1})_{ab} -  \Pi^{S|P}_{ab} ]^{-1} .
\numberthis
\end{align}
Hence, the meson propagators obtained using the Bethe-Salpeter equation, in the random phase approximation, applied to the four-fermion Lagrangian density of Eq. (\ref{quadraticLagrangianExpansion}), are identical to the ones obtained from the functional expansion presented earlier in this work, see Eqs. (\ref{scalarInverseMesonPropagator}) and (\ref{pseudoscalarInverseMesonPropagator}).

\subsubsection{Comments on the expansion of the fermionic determinant}
\label{comments_expansion_fermionic_determinant}

In the aforementioned calculations, the meson propagators were acquired through a specific expansion method applied to the fermionic determinant. Our approach involved utilizing the series representation of the logarithm of a matrix to derive the expansion. However, the expansion obtained is not unique and our particular choice is rooted on the fact that we wanted to derive the same expression for the meson propagators as the one obtained using the Bethe-Salpeter equation in the random phase approximation \cite{Klevansky:1992qe,Rehberg:1996vd,Rehberg:1995nr,Hatsuda:1994pi,Vogl:1991qt,Lutz:1992dv,Oertel:2000jp}.

In a different expansion, one could have introduced in the calculation kinetic terms for the meson fields. Such step could have been made when evaluating the functional trace over the continuous spacetime indexes. Since the operator $\qty( \mathcal{G} )_{jj}$ and the fields $\phi_a$ do not commute, instead of directly evaluating the trace in Eq. (\ref{expression_for_trx}), one could evaluate the trace over these degrees of freedom independently, by moving all the functions of momentum operators to the left and all functions of the position operator to the right, as introduced in Refs. \cite{Novikov:1983gd,Fraser:1984zb,Aitchison:1985pp,BALL19891}. This can be accomplished by using the following identity \cite{Fraser:1984zb,Aitchison:1985pp}:
\begin{align}
\phi \mathcal{G} = 
\mathcal{G} \phi
+ 
\mathcal{G}^2 \qty[ \mathcal{G}^{-1} , \phi ]
+ 
\mathcal{G}^3 \qty[ \mathcal{G}^{-1} , \qty[ \mathcal{G}^{-1} , \phi ] ]
+
\ldots
\label{operator_identity_commutators}
\end{align}
As a side note, the first term in this identity diverges, requiring regularization, while the following terms are not divergent \cite{Ripka:1997zb}. By substituting $\mathcal{G}=\qty( \mathcal{G} )_{jj}$ and $\phi=\phi_a$ one can understand that the tower of iterated commutators are proportional to field derivatives like $\partial_\mu \phi_a$, $p \cdot \partial \phi_a$, $\Box \phi_a$ , etc. Following the use of this identity, one can write Eq. (\ref{expression_for_trx}) as:
\begin{align*}
\Tr \qty[ T_2^{ \phi \phi } \qty[ \Gamma_1, \Gamma_2 ] ]
& =
R_{ij}^{a}
R_{ji}^{b} 
\frac{N_c}{2}
\trD
\int \dd[4]{x}
\bra{ x }
\qty( \mathcal{G} )_{ii} \Gamma_1 \qty( \mathcal{G} )_{jj} \Gamma_2 
\ket{ x }
\phi_a (x) \phi_b (x) 
+ \text{$\phi$ derivative terms} .
\numberthis
\end{align*}
Hence, in such expansion, the action of the model would contain kinetic terms for the meson fields, $\sigma_a$ and $\pi_a$, not present in the our previous calculation. Additionally, only one fermion line quark loops would contribute to the calculation without any contribution coming from the two fermion line integral, $B_0$. 

We would also like to highlight the fact that such result is only obtained if one starts the evaluation of the total trace with the trace over flavor space. Starting the trace evaluation by applying directly the identity of Eq. (\ref{operator_identity_commutators}), as a first step in the trace calculation leads to an infinite tower of terms proportional to iterated commutators between the mass matrix $M$ and the fields, $\qty[M,\phi]$, $\qty[M, \qty[M,\phi]]$, \ldots Of course, in the case in which the mass matrix is diagonal and degenerate in flavor space, $M= M_q \diag\qty( 1, 1, \dots, 1 )$, all these contribution vanish (with $M_q$ the effective quark mass). However in the case in which this matrix is not diagonal and degenerate, this cancellation does not occur. This is the case when considering the physical scenario of $N_f=3$ in which the strange quark mass is considerably higher than the up and down quark masses. Furthermore, when considering finite chemical potential, there is an extra matrix contribution to the $\mathcal{G}^{-1}$ operator, in the form of an effective quark chemical potential matrix, $\mu=\diag\qty( \mu_0, \mu_1, \ldots , \mu_{N_f-1} )$. Just like the mass case, in physical scenarios in which this quantity is not diagonal and degenerate an infinite sequence of iterated commutators between $\mu$ and the fields appears, $\qty[\mu,\phi]$, $\qty[\mu, \qty[\mu,\phi]]$, \ldots Finally, if one wishes to allow for the possibility of pseudoscalar condensation in the medium, the operator $\mathcal{G}^{-1}$ will certainly contain contributions which are not diagonal and degenerate.

To end this section we would like to point out another technique that can be used to extract the meson propagators from the fermionic determinant, the so-called heat kernel expansion \cite{BALL19891,EBERT1986453,alkofer1995chiral,PhysRevD.33.3645,Osipov:2000rg,Osipov:2001bj,Salcedo:2001qp,OSIPOV201750,OSIPOV200481,Vassilevich:2003xt,Dunne:2007rt}. Such method has been widely studied and developed in the context of two and three flavor NJL-type models, including different types of quark interactions at the Lagrangian level \cite{Osipov:2001bj,Osipov:2001nx,Osipov:2000rg,OSIPOV200481,Moreira:2014qna}. When applying this expansion to such type of models, one considers a Wick rotation of the operator inside the fermionic determinant in order to make it possible to obtain its Schwinger proper-time representation. In general, one can write \cite{OSIPOV200481}:
\begin{align}
\ln \abs{ \det \qty[D] }
=
- \frac{1}{2} \int_0^\infty \frac{ \dd{t} }{ t } \rho \qty[t, \Lambda^2] \Tr \qty[ \e^{- t D_\mathrm{E}^\dagger D_\mathrm{E} } ]  .
\end{align}
Here, $\rho \qty[t, \Lambda^2]$ is a function, often called regulator or kernel, which incorporates the necessary regularization \cite{Osipov:2000rg}, $\Tr\qty[\ldots]$ stands for the total trace and the operator $D_\mathrm{E}$, is the Wick rotated Dirac operator in the presence of background fields $\sigma=\sigma_a \lambda_a$ and $\pi = \pi_a \lambda_a$. Additionally, $D_\mathrm{E}^\dagger D_\mathrm{E} = M^2 - \partial^2 + Y$, with $M=M_a \lambda_a$ the effective quark mass matrix and $Y$ contains only contributions coming from the background fields \cite{OSIPOV200481,Osipov:2001bj}. In the case in which the quark mass matrix is not diagonal and degenerate, it does not commute with the rest of the operator and a resummation technique was developed in order to solve this difficulty \cite{Osipov:2001bj,Osipov:2001nx,Osipov:2000rg}.

Applying the heat kernel expansion to Eq. (\ref{Zeff_after_quark_integration}), would lead to a different model, with different meson propagators when compared to our approach, which are equivalent to the ones obtained via the Bethe-Salpeter equation in the random phase approximation. Thus, it could be insightful to evaluate the NJL quark-quark and quark-antiquark cross sections from meson propagators derived from a heat kernel expansion. The results of such study could then be directly compared to the ones obtained in this work. This could be used to assess the impact of considering different approximations to extract meson propagators from a purely fermionic theory as is the case of the NJL model.

\subsection{The 3-momentum sphere intersection regularization}
\label{regularization}

As previously established, in a 4-dimensional spacetime, the one and the two fermion lines integrals ($A$ and $B_0$) are divergent quantities. Other integrals with higher fermion line count, like the three fermion line integral ($C_0$) are, on the other hand, convergent. When faced with divergent integrals, some regularization procedure must be applied. Then, the regularization procedure becomes part of the model in question \cite{Ripka:1997zb} and should be carried out in a symmetry preserving manner. The primary objective is to maintain the expected physical properties of the model unchanged after applying the regularization.

To attain such goal, a plethora of techniques was developed over the years: the 4-momentum regularization, the 3-momentum regularization, dimensional regularization, Pauli-Villars regularization, implicit regularizations and the proper time regularization \cite{1990ZPhyA,Kahana:1992jm,Nikolov:1996jj,Ripka:1997zb,Inagaki:2011uj,Inagaki:2013hya,Kohyama:2015hix,TorresBobadilla:2020ekr}. The choice of regularization is of extreme importance since it introduces a scale, which informs about the range of applicability of a particular model \cite{Ripka:1997zb}. Such choice must also take into account the physical scenario under study. For instance, in quantum field theory at finite temperature and density, a compatible regularization scheme must be chosen, e.g., if a full sum over Matsubara frequencies is performed, one is implicitly considering the full range of the integration over the temporal component of the momentum.

The fact that the $A$ and $B_0$ are divergent, poses the following question: in a model calculation, does one simply regularize the diverging integrals, or is the regularization scheme applied to all integrals? For instance, in the NJL model, thermodynamic quantities like pressure and energy density contain both the well known divergent zero energy mode term (which corresponds to the vacuum contribution), as well as, convergent contributions (which are temperature and chemical potential dependent and can be regarded as in-medium contributions). The entropy density, in the same model, is defined by convergent contributions. In cases like this, one must choose where to apply the regularization (see for instance \cite{Xue_2022}). Indeed, in some works which only deal with the one fermion line integral, only the vacuum contribution of the $A$ integral is regularized \cite{Costa:2009ae}. One can also find studies where the convergent contribution of the $A$ integral is not regularized, but the convergent contribution of the $B_0$ integral is, effectively mixing regularization strategies \cite{Hansen:2019lnf,Costa:2020dgc,Soloveva:2020hpr,Xue_2022}. In this work, as it was performed in Ref. \cite{Rehberg:1995nr}, we will use the Matsubara frequency summation to introduce temperature and apply the 3-momentum regularization to the remaining integrals. Furthermore, we choose to consider the regularization as part of the model, affecting all contributions, divergent and convergent alike.

\begin{figure*}[ht!]
\begin{subfigure}[b]{0.4\textwidth}
\includegraphics[width=\textwidth]
{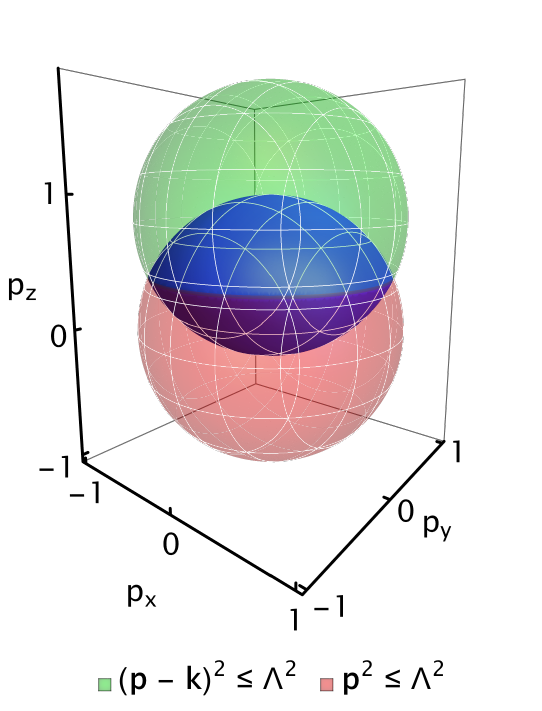}
\label{sphereSphereIntersection3D}
\end{subfigure}
\begin{subfigure}[b]{0.46\textwidth}
\includegraphics[width=\textwidth]
{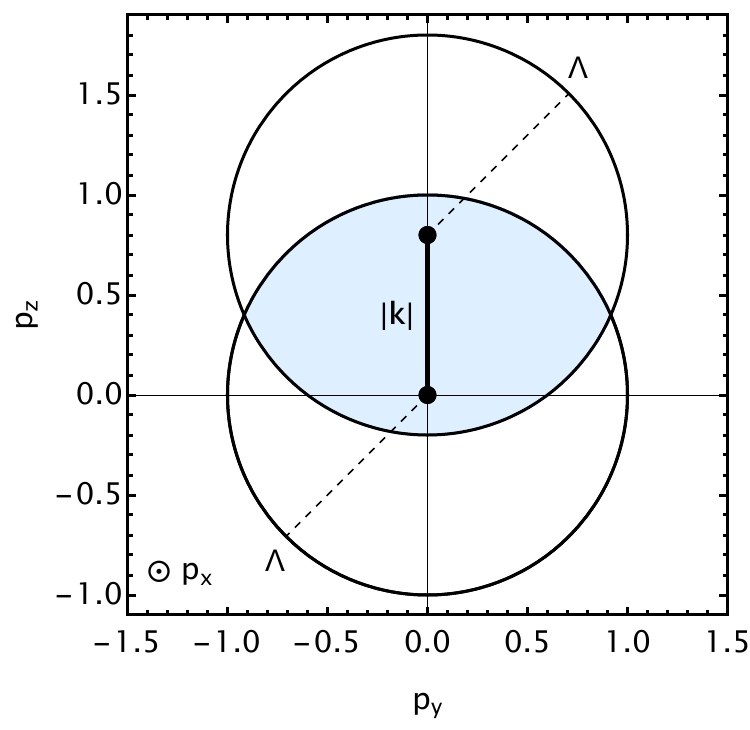}
\label{integration_region_B0}
\end{subfigure}
\caption{Region of integration in the 3-momentum sphere intersection regularization scheme, used to regularize the quark-antiquark polarization function, $\Pi_{ij}^{S|P}$. In the left panel: a particular 3-dimensional view of the intersection of two spheres. The red sphere is centered at the origin (defined by the equation $\vec{p}^2 \leq \Lambda^2$) and the green sphere is distanced $\abs{ \vec{k} }$ from it, in the $p_z$ direction (defined by the equation $\qty( \vec{p} - \vec{k} )^2 \leq \Lambda^2$). The blue region corresponds to the intersection between these two spheres and is designated by $\mathbb{I}_{\abs{\vec{k}}}^\Lambda$. In the right panel, we show the $p_y-p_z$ cross section of the same integration region.}
\label{integration_region}
\end{figure*}

At this point we introduce and discuss the 3-momentum sphere intersection regularization which will be applied to the quark-antiquark polarization function $\Pi_{ij}^{S|P}$. In the 3-momentum regularization, the quark propagator  is defined up to the maximum allowed momentum, the cutoff, $\Lambda$. In such scheme, a quark can have any 3-momentum inside a sphere of radius $\Lambda$. In the definition of the scalar and pseudoscalar polarization functions written in Eqs. (\ref{quarkAntiquarkPolarization_flavorIndexes}) and (\ref{quarkAntiquarkPolarization_gellmannIndexes}), there are two quark propagators which must be integrated over a loop, one with momenta $p$ and another with $p-k$, which carries an insertion from an external momentum $k$. In order to truly restrict the 3-momentum of the quarks, both momenta must be smaller than the cutoff, $\Lambda$. This means that the absolute value of the quark loop 3-momentum, $\abs{\vec{p}}$, present in the integral of the polarization loop functions, must obey: $ 0 \leq \abs{ \vec{p} } \leq \Lambda \wedge 0 \leq \abs{ \vec{p} - \vec{k} } \leq \Lambda $. Such restriction can be visualized as the intersection between two spheres. Let $\mathbb{S}^\Lambda_0$ and $\mathbb{S}_{\abs{\vec{k}}}^\Lambda$ be spheres of radius $\Lambda$, where the first is centered at the origin ($0$) and the second at a point distanced $\abs{\vec{k}}$ from the origin. We denote the intersection of these spheres\footnote{The volume of the intersection between two spheres of radius $\Lambda$, distanced k between each other, can be calculated to yield, $V_{ \mathbb{I}_{\abs{\vec{k}}}^\Lambda } = \frac{1}{12} \pi \qty(2\Lambda-k)^2 \qty(4\Lambda+k) $. In the $k \to 0$ limit one trivially recovers the volume of a sphere of radius $\Lambda$, $V_{ \mathbb{I}_0^\Lambda } = \frac{4 }{3} \pi \Lambda^3 $. } by $\mathbb{I}_{\abs{\vec{k}}}^\Lambda =\mathbb{S}_0^\Lambda \cap \mathbb{S}_{\abs{\vec{k}}}^\Lambda$. Hence, the integrals present in the quark-antiquark polarization functions will be regularized by the intersection of these two spheres of radii $\Lambda$, distanced by $\abs{\vec{k}}$. An interesting property of the 3-momentum sphere intersection regularization is that it becomes the usual 3-momentum regularization for zero external momentum: in such case both spheres completely intersect each other. This means that the one fermion line integral present in the quark-antiquark polarization function $\mathcal{A}[ M_i,\abs{\vec{k}} ]$, regularized with the sphere intersection scheme, degenerates to the usual one fermion line integral $A[M_i]$ in the $\abs{\vec{k}} \to 0$ limit, i.e., $A[M_i]=\mathcal{A}[ M_i,0 ]$. In Fig. \ref{integration_region}, we show the region of integration which regularizes the quark-antiquark polarization function, $\Pi_{ij}^{S|P}$. In the left panel of this figure, we show a particular 3-dimensional view of this integration region which corresponds to a ``convex lens shaped'' volume, defined as $\mathbb{I}_{\abs{\vec{k}}}^\Lambda$. In the right panel of the same figure, we show the two-dimensional cross section of the region in the $p_y-p_z$ plane.

Another interesting property of this regularization scheme is its behavior for $\abs{\vec{k}} \to \infty$. As a matter of fact, for $\abs{\vec{k}} \geq 2 \Lambda$, the spheres no longer intersect, the integration region disappears, the integrals $\mathcal{A}$ and $B_0$ are automatically zero and the quark-antiquark polarization functions vanish. So, the asymptotic behavior of the meson propagators are given by:
\begin{align}
D^{S|P}
& =
\qty[ \frac{1}{2} \qty(S|P)^{-1} - \Pi^{S|P} ]^{-1}
\underset{ \abs{\vec{k}} \to \infty }{ \Rightarrow }
D_{ab}
=
2 \qty(S|P)_{ab} .
\end{align}
Thus, for large momentum, the meson propagators converge to twice the meson projectors ($2S_{ab}$ and $2P_{ab}$), which are temperature and chemical potential dependent quantities.

\begin{figure*}[ht!]
\includegraphics[width=0.4\textwidth]
{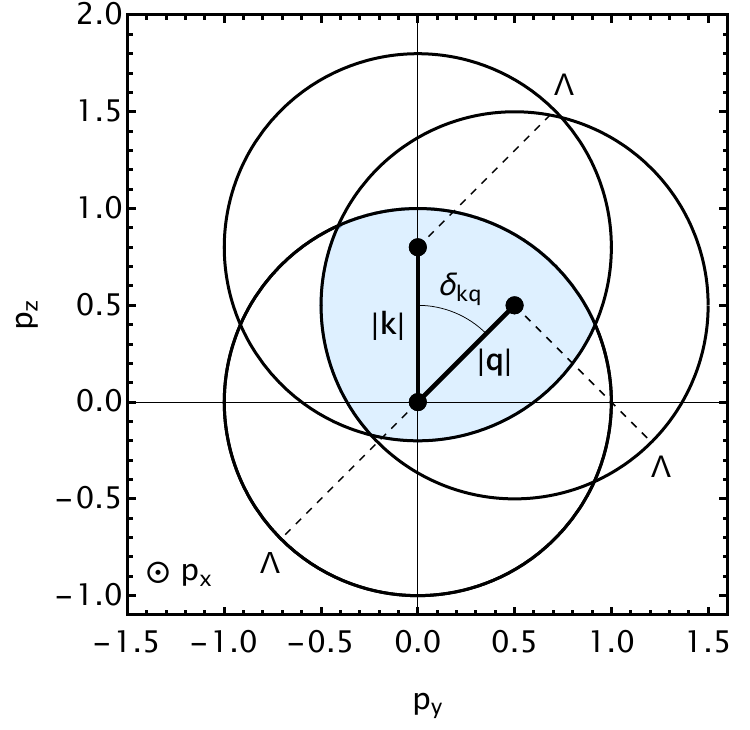}
\caption{
Cross section in the $p_y-p_z$ plane of the region of integration for the three fermion line integral, $C_0$, in the 3-momentum sphere intersection regularization scheme. }
\label{integration_region_C0}
\end{figure*}

This novel approach to the 3-momentum regularization, can also be extended to integrals with three or more quark propagators present in the integrand: in such cases the region of integration is the intersection between the appropriate number of spheres. For example, for the three fermion line integral, $C_0$, three quark propagators would be present, meaning that the integral would be calculated over the intersection between three spheres of radii $\Lambda$. One sphere located at the origin, other distanced $\abs{\vec{k}}$ from the origin and another at distance  $\abs{\vec{q}}$ from the origin, with an angle $\delta_{\vec{k} \vec{q}}$ between the 3-vectors $\vec{k}$ and $\vec{q}$. In Fig. \ref{integration_region_C0}, we show the  $p_y-p_z$ cross section of the integration region for the three fermion line integral, within the sphere intersection regularization scheme. We leave the details of the explicit calculation of the $C_0$ integral, within this new regularization, to future work.

\section{The in-medium two fermion, one loop integral}
\label{two_fermion_one_loop_integral}

In order to introduce finite temperature, one first considers a Wick rotation to Euclidean spacetime, and then applies the Matsubara formalism (see for instance Ref. \cite{Kapusta:Book}). Through this process Eq. (\ref{B0_definition}) yields:
\begin{align*}
B_0 \qty[M_i,M_j,T,\mu_i,\mu_j, i\nu_m, \abs{\vec{k}}] 
=
16\pi^2 T \sum^{+\infty}_{n=-\infty}
\int_{\mathrm{Reg}[ \abs{\vec{k}} ]}   
\frac{ \dd[3]{\vec{p}} }{ \qty( 2\pi )^3 }
&
\frac{1}{ \qty( i\omega_n + \mu_i)^2 - E\qty[M_i,\vec{p}]^2 }
\times
\\
& \qquad 
\frac{1}{ \qty( i\omega_n + \mu_j - i\nu_m)^2 - E[M_j,\vec{p}-\vec{k}]^2 } ,
\numberthis
\label{B0def}
\end{align*}
with $\omega_n=(2n +1)\pi T$ denoting the Matsubara frequency for fermions and $\nu_m$ the Matsubara frequency for the external particle. Here, there are two distinct chemical potentials, $\mu_i$ and $\mu_j$, each corresponding to different fermion masses, $M_i$ and $M_j$ and $E \qty[ M, \vec{p}]$ is the usual dispersion relation:
\begin{align}
E \qty[ M, \vec{p}] = \sqrt{ \vec{p}^2 + M^2 }.
\label{energyEquation}
\end{align}
Applying a Wick rotation to the external momentum, one can perform an analytical continuation as, $i\nu_m \to k_0$.

One very important symmetry of this function arises when considering a shift in the Matsubara frequency as: $\omega_n \to  -\omega_n + \nu_m$, which is allowed in the case of this infinite sum. Applying this substitution and performing the aforementioned Wick rotation, one can write:
\begin{align*}
B_0 \qty[M_i,M_j,T,\mu_i,\mu_j, k_0, \abs{\vec{k}}] 
= &
B_0 \qty[M_j,M_i,T,-\mu_j,-\mu_i, k_0, \abs{\vec{k}}] .
\numberthis
\label{symmetryB0}
\end{align*}
Hence, the two fermion, one loop integral $B_0$, is invariant with respect to the interchange of the masses, $M_i \leftrightarrow M_j$, and chemical potentials, albeit with an extra minus sign, $\mu_i \rightarrow -\mu_j$ and $\mu_j \rightarrow -\mu_i$.  Such symmetry can also be understood as exchanging the fermion in the upper line with the antifermion in the lower line in the loop diagram shown in Fig. \ref{polarization_function_diagram}, where time flows from left to right.

When dealing with the integration over the 3-momentum, $\vec{p}$, it will become necessary to introduce an infinitesimal shift in the complex plane, $\pm i\epsilon$ with $\epsilon>0$, in order to deform the contour around singularities. As discussed in Ref. \cite{Rehberg:1995nr}, one can regard the masses as having an infinitesimal complex contribution, $M^2 \to M^2-i\epsilon$, or consider other recipes, such as a shift in the zero component of the external momentum, $k_0 \to k_0 \pm i\epsilon$ \cite{Fujii:2004jt}. The way in which the $\pm i\epsilon$ term is introduced in the calculation has an impact in a very important symmetry property of the $B_0$ function. In practice, choosing different recipes amounts to different time orderings of the single particle Green function. However, for the purposes of calculating meson masses in the NJL model and (differential) cross sections, the chosen $\pm i\epsilon$ recipe will not change the results. The reason for this ``recipe invariance'' will be discussed later.

Considering the usual shift in the masses, $M^2 \to M^2-i\epsilon$ (see Ref. \cite{Rehberg:1995nr}), the $B_0$ function takes the form:
\begin{align*}
B_0 \qty[M_i,M_j,T,\mu_i,\mu_j, k_0, \abs{\vec{k}}] 
=
16\pi^2 T \sum^{+\infty}_{n=-\infty}
\int_{\mathrm{Reg}[ \abs{\vec{k}} ]}   
\frac{ \dd[3]{\vec{p}} }{ \qty( 2\pi )^3 }
&
\frac{1}{ \qty( i\omega_n + \mu_i)^2 - E\qty[M_i,\vec{p}]^2 + i\epsilon }
\times
\\
& \qquad 
\frac{1}{ \qty( i\omega_n + \mu_j - k_0)^2 - E[M_j,\vec{p}-\vec{k}]^2 + i\epsilon  }.
\numberthis
\end{align*}
Applying the simple change of variable, $\omega_n \rightarrow -\omega_n$, in the Matsubara sum, one can establish the following connection:
\begin{align}
B_0 \qty[M_i,M_j,T,\mu_i,\mu_j, k_0, \abs{\vec{k}}] 
& =
B_0 \qty[M_i,M_j,T,-\mu_i,-\mu_j, -k_0, \abs{\vec{k}}] .
\label{symm_symmetric}
\end{align}
Hence, for this choice of shift, the $B_0$ function is symmetric with respect to the zero component of the external momentum, $k_0$. As discussed above, another way to introduce the shift is to consider it in the external momentum, $k_0 \to k_0 \pm i\epsilon$. In such case one can write:
\begin{align*}
B_0 \qty[M_i,M_j,T,\mu_i,\mu_j, k_0, \abs{\vec{k}}] 
=
16\pi^2 T \sum^{+\infty}_{n=-\infty}
\int_{\mathrm{Reg}[ \abs{\vec{k}} ]}   
\frac{ \dd[3]{\vec{p}} }{ \qty( 2\pi )^3 }
&
\frac{1}{ \qty( i\omega_n + \mu_i)^2 - E\qty[M_i,\vec{p}]^2 }
\times
\\
& \qquad 
\frac{1}{ \qty( i\omega_n + \mu_j - k_0 - i\epsilon)^2 - E[M_j,\vec{p}-\vec{k}]^2  }.
\numberthis
\end{align*}
In this case, a different symmetry relation is obtained, involving the complex conjugate of the $B_0$ function:
\begin{align}
B_0 \qty[M_i,M_j,T,\mu_i,\mu_j, k_0, \abs{\vec{k}}] 
& =
B_0 \qty[M_i,M_j,T,-\mu_i,-\mu_j, -k_0, \abs{\vec{k}}]^\ast .
\label{symm_antisymmetric}
\end{align}
We highlight that, differently from the previous symmetry, the masses and chemical potentials do not switch positions, one simply changes, $\mu_i \to -\mu_i$, $\mu_j \to -\mu_j$ and $k_0 \to -k_0$. In this work we will use the previously introduced mass shift however, to simplify the notation, we will only write the contribution $-i\epsilon$ explicitly, when it is necessary. In later sections we will discuss why this symmetry is essential in order to maintain the symmetry between meson and anti-meson pairs and, consequently, provide a proper framework to evaluate the (differential) cross sections of different quark-quark and quark-antiquark processes.

The infinite sum over the Matsubara frequencies can be performed analytically, for example, by converting it into a contour integration on the complex plane. In such approach, the Matsubara sum is written as a sum of residues which, in turn, can be written as a sum of contour integrations where each contour surrounds a simple pole located at a particular Matsubara frequency. The contours can be deformed and the Cauchy residue theorem applied in order to obtain a result for the original Matsubara sum \cite{Nieto:1993pr}. One gets:
\begin{align*}
B_0 \qty[M_i,M_j,T,\mu_i,\mu_j, k_0, \abs{\vec{k}}] 
 =
16\pi^2 
\int_{\mathrm{Reg}[ \abs{\vec{k}} ]}
\frac{ \dd[3]{\vec{p}} }{ \qty( 2\pi )^3 }
\Bigg\{
& -
\frac{1}{ 4 E_1 }
\frac
{ 1 - 2 \nfermi\qty[ E_1-\mu_i,T] }
{ \qty( E_1 - \lambda )^2 - E_2^2  }
\\
& \qquad -
\frac{1}{ 4 E_1 }
\frac
{ 1 - 2 \nfermi\qty[E_1+\mu_i,T] }
{ \qty( E_1 + \lambda )^2 - E_2^2  }
\\
& \qquad \qquad -
\frac{1}{ 4 E_2 }
\frac
{ 1 - 2 \nfermi\qty[E_2-\mu_j,T] }
{ \qty( E_2 + \lambda )^2 - E_1^2  }
\\
& \qquad \qquad \qquad -
\frac{1}{ 4 E_2 }
\frac
{ 1 - 2 \nfermi\qty[E_2+\mu_j,T] }
{ \qty( E_2 - \lambda )^2 - E_1^2  }
\Bigg\},
\numberthis
\end{align*}
where, $\nfermi$ is the Fermi-Dirac distribution function, given by:
\begin{align}
\nfermi \qty[ E , T ]
& =
\frac{1}{ \e^{E/T} + 1 },
\label{fermiDiracDistribution}
\end{align}
and, for simplicity, we have defined:
\begin{align}
E_1 & = E\qty[M_i,\vec{p}],
\label{E1_def}
\\
E_2 & = E[M_j,\vec{p}-\vec{k}],
\label{E2_def}
\\
\lambda & = k_0 + \mu_i - \mu_j .
\label{lambda_def}
\end{align}

\begin{figure*}[ht!]
\includegraphics[width=0.5\textwidth]
{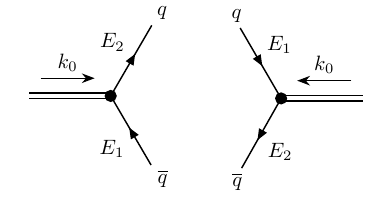}
\caption{Pair creation/annihilation processes \cite{Fujii:2004jt}.}
\label{pair_visual_representation}
\end{figure*}

\begin{figure*}[ht!]
\includegraphics[width=0.5\textwidth]
{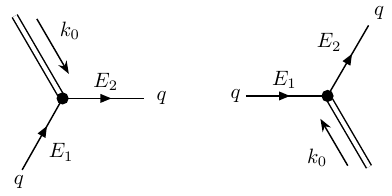}
\caption{Scattering processes: absorption and emission \cite{Fujii:2004jt,Mustafa:2022got}.}
\label{scat_visual_representation}
\end{figure*}

As introduced in \cite{Fujii:2004jt,Yamazaki:2012ux} it is quite useful to write this function using the so-called pair creation/annihilation, $B_{0,\pair}$, and scattering contributions, $B_{0,\scat}$. In Figs. \ref{pair_visual_representation} and \ref{scat_visual_representation}, we show schematic diagrams representing the pair creation/annihilation and scattering contributions, respectively. For a deeper explanation about all these different processes see Ref. \cite{Mustafa:2022got}. Using this separation, we can write:
\begin{align}
B_0 \qty[M_i,M_j,T,\mu_i,\mu_j, k_0, \abs{\vec{k}}]  =
B_{0,\pair} \qty[M_i,M_j,T,\mu_i,\mu_j,k_0,\abs{\vec{k}}] +
B_{0,\scat} \qty[M_i,M_j,T,\mu_i,\mu_j,k_0,\abs{\vec{k}}]  ,
\label{B0_components_def}
\end{align}
where, each component is given explicitly by:
\begin{align}
B_{0,\pair} \qty[M_i,M_j,T,\mu_i,\mu_j,k_0,\abs{\vec{k}}] 
& =
16\pi^2 
\sumeta
\int_{\mathrm{Reg}[ \abs{\vec{k}} ]}   
\frac{ \dd[3]{\vec{p}} }{ \qty( 2\pi )^3 }
\frac{ 1 }{ 4 E_1 E_2 }
\frac
{ 1 - \nfermi\qty[ E_1 + \eta \mu_i, T ] - \nfermi\qty[ E_2 - \eta \mu_j, T ] }
{ E_2 + E_1 + \eta \lambda  - i \epsilon } ,
\label{B0_pair_def}
\\
B_{0,\scat} \qty[M_i,M_j,T,\mu_i,\mu_j,k_0,\abs{\vec{k}}] 
& =
-
16\pi^2 
\sumeta
\int_{\mathrm{Reg}[ \abs{\vec{k}} ]}   
\frac{ \dd[3]{\vec{p}} }{ \qty( 2\pi )^3 }
\frac{ 1 }{ 4 E_1 E_2 }
\frac
{ \nfermi\qty[ E_1 - \eta \mu_i, T ] - \nfermi\qty[ E_2 - \eta \mu_j, T ] }
{ E_2 - E_1 + \eta \lambda  -  \sgn\qty[E_2 - E_1] i \epsilon  } .
\label{B0_scat_def}
\end{align}
Here, we introduced the mass shift, $M^2 \to M^2-i\epsilon$ \cite{Rehberg:1995nr}, which will be necessary in order to use the Sokhotski–Plemelj theorem\footnote{Introducing the mass shift and expanding the difference $E_2[\epsilon]-E_1[\epsilon]$ around zero yields: 
$
E_2[\epsilon]-E_1[\epsilon]
=
\qty(E_2-E_1)
- i \epsilon
\frac{ \qty(E_2-E_1) }{ 2 E_1 E_2 }
+
O\qty[\epsilon^2].
$ Since $E_1,E_2>0$, the sign of the first order term will depend on the difference of these energies and the denominator is absorbed by the infinitesimal $\epsilon$.
}. The function $\sgn[x]$ returns the sign of the quantity $x$, i.e., $\sgn[x]=+1$ if $x>0$ and $\sgn[x]=-1$ if $x<0$.

In the vacuum, the scattering contribution, $B_{0,\scat}$, vanishes and only $B_{0,\pair}$ will contribute: in the vacuum there is no medium for scattering to occur, only the creation and annihilation processes contribute. This vacuum contribution is the only one which diverges. For cases with non-zero temperature and/or density the scattering term will contribute. In the later case, for $T=0$ and finite chemical potential, one can introduce a Fermi momentum and easily perform two of the three integrations.

\subsection{Practical implementation}
\label{practical_implementation_B0}

\subsubsection{ The $\abs{\vec{k}}>0$ case}
\label{practical_implementation_knot0}

Lets discuss the implementation of the $B_0$ integral for the case where $\abs{\vec{k}}>0$. Since the integrand does not depend on the azimuthal angle, the integration over such variables is trivial, yielding a factor of $2\pi$. The remaining 2-dimensional integration is more complicated. One way to simplify the calculations and perform such integration, is to consider a change of variables. Here, we extend the approach made for the degenerate mass case ($M_i=M_j$) that was introduced in \cite{Yamazaki:2012ux} and also applied in \cite{CamaraPereira:2020ipu}, by taking into account different masses. Consider the following change of variables:
\begin{align}
E & = \frac{ 1 }{2} \qty(  E_2 + E_1 ) ,
\label{E_B0_COV}
\\
\varepsilon & = E_2 - E_1 .
\label{epsilon_B0_COV}
\end{align}
From these definitions one has the following relations: $E_1 = E - \nicefrac{\varepsilon}{2}$ and $E_2 = E + \nicefrac{\varepsilon}{2}$. Looking at Eq. (\ref{epsilon_B0_COV}), the quantity $\varepsilon$ can be both positive or negative while, checking Eq. (\ref{E_B0_COV}), $E$ is always a positive quantity. We also point out that the quantities $E_1$ and $E_2$ are always positive since they are energies.

The integration measure will be changed as $\dd{ E }
\dd{ \varepsilon }
 =
\abs{ \det[ J ] } 
\dd{ \abs{\vec{p}} }
\dd{ \theta } $, where $J= J \qty[ \abs{\vec{p}}, \theta ]  =\frac{\partial(E,\varepsilon)}{\partial(\abs{\vec{p}}, \theta )}$, is the Jacobian matrix. Performing the derivatives, one can write:
\begin{align}
\det[ J ]
=
\abs{ \vec{k} }
\frac{ \vec{p}^2 \sin[\theta] }{ E_1 E_2 } .
\label{jacobianCOV_knot0}
\end{align}
From the above one can realize that this change of variables is only valid for the $\abs{ \vec{k} }>0$. We treat the case $\abs{ \vec{k} }=0$ as a special case since it is simpler and a different change of variables must be used. 

Considering the 3-momentum sphere intersection regularization, $\mathrm{Reg}[ \abs{\vec{k}} ] \to \mathbb{I}_{\abs{\vec{k}}}^\Lambda$ and performing the proposed change of variables, one can write the integrals as:
\begin{align*}
B_{0,\pair} \qty[M_i,M_j,T,\mu_i,\mu_j,k_0,\abs{\vec{k}}] 
& = 
\frac{ 1 }{ 2\abs{ \vec{k} } }
\sumeta
\int_{ \mathbb{I}_{\abs{\vec{k}}}^\Lambda }
\frac{\dd{\phi}}{2\pi}
\dd{ E }
\dd{ \varepsilon }
\frac
{ 
1 - 
\nfermi\qty[ E - \nicefrac{\varepsilon}{2} + \eta \mu_i, T ] - 
\nfermi\qty[ E + \nicefrac{\varepsilon}{2} - \eta \mu_j, T ] 
}
{ E + \nicefrac{ \eta \lambda }{ 2 } - i \epsilon } ,
\numberthis
\label{B0pairGeneralDef}
\\
B_{0,\scat} \qty[M_i,M_j,T,\mu_i,\mu_j,k_0,\abs{\vec{k}}] 
& =
-
\frac{ 1 }{ \abs{ \vec{k} } }
\sumeta
\int_{ \mathbb{I}_{\abs{\vec{k}}}^\Lambda }
\frac{\dd{\phi}}{2\pi}
\dd{ E }
\dd{ \varepsilon }
\frac
{ 
\nfermi\qty[ E - \nicefrac{\varepsilon}{2} - \eta \mu_i, T ] - 
\nfermi\qty[ E + \nicefrac{\varepsilon}{2} - \eta \mu_j, T ] }
{ \varepsilon + \eta \lambda  -  \sgn\qty[\varepsilon] i \epsilon  } .
\numberthis
\label{B0scattGeneralDef}
\end{align*}
The integration over the azimuthal angle yields $2\pi$ and the integration over the remaining variables corresponds to evaluating the integral over the 2-dimensional cross section of the intersection between two spheres (see Fig. \ref{integration_region}). We have written the 2-dimensional integration without referring the specific order of integration. In fact it is extremely useful to consider both order of integrations. For the pair creation/annihilation contribution, $B_{0,\pair}$, we will use the order, $\int_{ \dEdepsilonEMin }^{ \dEdepsilonEMax } 
\dd{E} 
\int_{ \dEdepsilonepsilonMin }^{ \dEdepsilonepsilonMin } 
\dd{\varepsilon}$, while for the scattering contribution, $B_{0,\scat}$, the order $\int_{ \depsilondEepsilonMin }^{ \depsilondEepsilonMax }
\dd{\varepsilon}
\int_{ \depsilondEEMin }^{ \depsilondEEMax } 
\dd{E}$ will be used. In principle, if the integrand is a well behaved function, switching the order of integration yields the same results \cite{nahin2014inside}. The reason for using the two possible integration orders is twofold. First, it allows to perform one of the integrations analytically, resulting in only one numerical integration in order to evaluate the $B_0$ function. Second, it allows for a straightforward isolation of the simple poles present at $E=\nicefrac{-\eta \lambda}{2}$ and $\varepsilon=-\eta \lambda$, allowing the use of the appropriate numerical methods to evaluate such type of integrals. For convenience, we define the functions $g_\pair$ and $g_\scat$ as:
\begin{align}
g_\pair \qty[T,\mu_i,\mu_j,\eta,E,\varepsilon] 
& =
1 - 
\nfermi\qty[ E - \nicefrac{\varepsilon}{2} + \eta \mu_i, T ] - 
\nfermi\qty[ E + \nicefrac{\varepsilon}{2} - \eta \mu_j, T ] , 
\\
g_\scat \qty[T,\mu_i,\mu_j,\eta,E,\varepsilon] 
& =
\nfermi\qty[ E - \nicefrac{\varepsilon}{2} - \eta \mu_i, T ] - 
\nfermi\qty[ E + \nicefrac{\varepsilon}{2} - \eta \mu_j, T ] .
\end{align}

For the pair creation/annihilation contribution, we write:
\begin{align*}
B_{0,\pair} \qty[M_i,M_j,T,\mu_i,\mu_j,k_0,\abs{\vec{k}}] 
& = 
\frac{ 1 }{ 2\abs{ \vec{k} } }
\sumeta
\int_{ \dEdepsilonEMin }^{ \dEdepsilonEMax } 
\dd{E} 
\frac
{ 1 }{ E + \nicefrac{ \eta \lambda }{ 2 } -  i \epsilon } 
\int_{ \dEdepsilonepsilonMin }^{ \dEdepsilonepsilonMax }
\dd{\varepsilon}
g_\pair \qty[T,\mu_i,\mu_j,\eta,E,\varepsilon]
\\
& = 
\frac{ 1 }{ 2\abs{ \vec{k} } }
\sumeta
\int_{ \dEdepsilonEMin }^{ \dEdepsilonEMax } 
\dd{E} 
\frac
{ G_\pair \qty[M_i,M_j,T,\mu_i,\mu_j,\eta,\abs{\vec{k}},E] }
{ E + \nicefrac{ \eta \lambda }{ 2 } - i \epsilon }  .
\numberthis
\end{align*}
While, for the scattering contribution, one gets:
\begin{align*}
B_{0,\scat} \qty[M_i,M_j,T,\mu_i,\mu_j,k_0,\abs{\vec{k}}] 
& =
-
\frac{ 1 }{ \abs{ \vec{k} } }
\sumeta
\int_{ \depsilondEepsilonMin }^{ \depsilondEepsilonMax }
\dd{\varepsilon}
\frac
{ 1 }{ \varepsilon + \eta \lambda -  \sgn\qty[\varepsilon] i \epsilon }
\int_{ \depsilondEEMin }^{ \depsilondEEMax } 
\dd{E} 
g_\scat \qty[T,\mu_i,\mu_j,\eta,E,\varepsilon] 
\\
& =
-
\frac{ 1 }{ \abs{ \vec{k} } }
\sumeta
\int_{ \depsilondEepsilonMin }^{ \depsilondEepsilonMax }
\dd{\varepsilon}
\frac
{ G_\scat \qty[M_i,M_j,T,\mu_i,\mu_j,\eta,\abs{\vec{k}},\varepsilon]  }
{ \varepsilon + \eta \lambda -  \sgn\qty[\varepsilon] i \epsilon }
.
\numberthis
\end{align*}
Both integration regions and all the functions that define them, $\dEdepsilonEMin$, $\dEdepsilonEMax$, $\dEdepsilonepsilonMin$, $\dEdepsilonepsilonMax$, $\depsilondEEMin$, $\depsilondEEMax$, $\depsilondEepsilonMin$, $\depsilondEepsilonMax$, are explicitly derived in Appendix \ref{integration_region_B0COV}. We point out that the integration endpoints will be functions of several variables, such as $M_i$, $M_j$, $\abs{\vec{k}}$ and $\Lambda$. However, here, such dependencies are omitted for simplicity. The functions $G_\pair$ and $G_\scat$ have been defined as the innermost integration for each contribution, separately, i.e.:
\begin{align}
G_\pair \qty[M_i,M_j,T,\mu_i,\mu_j,\eta,\abs{\vec{k}},E] 
& = 
\int_{ \dEdepsilonepsilonMin }^{ \dEdepsilonepsilonMax }
\dd{\varepsilon}
g_\pair \qty[T,\mu_i,\mu_j,\eta,E,\varepsilon] ,
\\
G_\scat \qty[M_i,M_j,T,\mu_i,\mu_j,\eta,\abs{\vec{k}},\varepsilon]
& =
\int_{ \depsilondEEMin }^{ \depsilondEEMax } 
\dd{E} 
g_\scat \qty[T,\mu_i,\mu_j,\eta,E ,\varepsilon] .
\end{align}
The reason for this separations is that these integrals can be computed analytically, simplifying the calculation of the $B_0$ function. For the pair creation/annihilation contribution, $G_\pair$, one gets\footnote{Numerically, for low temperatures, it can be useful to make use of the Puiseux series expansion of $\ln[1+x]$ at $x \to \infty$: $\ln[1+x] = \ln[x] + 1/x - 1/(2x^2) + 1/(3x^3) - 1/(4x^4) + O[1/x^5]$.}:
\begin{align*}
G_\pair \qty[M_i,M_j,T,\mu_i,\mu_j,\eta,\abs{\vec{k}},E] 
= 
\dEdepsilonepsilonMax - \dEdepsilonepsilonMin
& + 
2 T\ln \qty[ 1 + \e^{ \qty( -E + \frac{1}{2}\dEdepsilonepsilonMin -\eta \mu_i )/T } ] 
\\ & -
2 T\ln \qty[ 1 + \e^{ \qty( -E - \frac{1}{2}\dEdepsilonepsilonMin +\eta \mu_j )/T } ] 
\\ & -
2 T\ln \qty[ 1 + \e^{ \qty( -E + \frac{1}{2}\dEdepsilonepsilonMax -\eta \mu_i )/T } ] 
\\ & +
2 T\ln \qty[ 1 + \e^{ \qty( -E - \frac{1}{2}\dEdepsilonepsilonMax +\eta \mu_j )/T } ] .
\numberthis
\label{GPair}
\end{align*}
For the scattering contribution, one gets:
\begin{align*}
G_\scat \qty[M_i,M_j,T,\mu_i,\mu_j,\eta,\abs{\vec{k}},\varepsilon]
= 
& -
T \ln \qty[ 1 + \e^{ \qty( -\depsilondEEMin - \frac{1}{2}\varepsilon + \eta \mu_j )/T } ] 
\\
& +
T \ln \qty[ 1 + \e^{ \qty( -\depsilondEEMin + \frac{1}{2}\varepsilon + \eta \mu_i )/T } ] 
\\
& -
T \ln \qty[ 1 + \e^{ \qty( -\depsilondEEMax + \frac{1}{2}\varepsilon + \eta \mu_i )/T } ] 
\\
& +
T \ln \qty[ 1 + \e^{ \qty( -\depsilondEEMax - \frac{1}{2}\varepsilon + \eta \mu_j )/T } ]  .
\numberthis
\label{GScat}
\end{align*}

In the case where a singularity resides within the integration bounds, the integration is made using the Sokhotski–Plemelj theorem and the $B_0$ function presents both real and imaginary components. The real part of the pair creation/annihilation contribution, $B_{0,\pair}$, and the real part of the scattering contribution, $B_{0,\scat}$, are given by:
\begin{align*}
\Re 
\Big[
B_{0,\pair} \qty[M_i,M_j,T,\mu_i,\mu_j,k_0,\abs{\vec{k}}] 
\Big]
& = 
\frac{ 1 }{ 2\abs{ \vec{k} } }
\sumeta 
\CPV
\int_{ \dEdepsilonEMin }^{ \dEdepsilonEMax } 
\dd{E} 
\frac
{ G_\pair \qty[M_i,M_j,T,\mu_i,\mu_j,\eta,\abs{\vec{k}},E] }
{ E + \nicefrac{ \eta \lambda }{ 2 } } ,
\numberthis
\\
\Re 
\Big[
B_{0,\scat} \qty[M_i,M_j,T,\mu_i,\mu_j,k_0,\abs{\vec{k}}] 
\Big]
& = 
-
\frac{ 1 }{ \abs{ \vec{k} } }
\sumeta
\CPV
\int_{ \depsilondEepsilonMin }^{ \depsilondEepsilonMax }
\dd{\varepsilon}
\frac
{ G_\scat \qty[M_i,M_j,T,\mu_i,\mu_j,\eta,\abs{\vec{k}},\varepsilon]  }
{ \varepsilon + \eta \lambda } .
\numberthis
\end{align*}
The symbol $\CPV$ stands for Cauchy principal value. The respective imaginary contributions are:
\begin{align*}
\Im 
\Big[
B_{0,\pair} &\qty[M_i,M_j,T,\mu_i,\mu_j,k_0,\abs{\vec{k}}] 
\Big]
 = 
\\
&
\frac{ \pi }{ 2\abs{ \vec{k} } }
\sumeta 
\left\{
\begin{array}{lr}
G_\pair \qty[M_i,M_j,T,\mu_i,\mu_j,\eta,\abs{\vec{k}}, \nicefrac{ -\eta \lambda }{ 2 }], & \text{if } \nicefrac{ -\eta \lambda }{ 2 } \in \qty[\dEdepsilonEMin, \dEdepsilonEMax] ,
\\
0, & \text{otherwise } , 
\end{array}
\right.
\numberthis
\label{imagPartPair}
\\
\Im 
\Big[
B_{0,\scat} &\qty[M_i,M_j,T,\mu_i,\mu_j,k_0,\abs{\vec{k}}] 
\Big]
 = 
\\
-
&\frac{ \pi  }{ \abs{ \vec{k} } }
\sumeta 
\left\{
\begin{array}{lr}
\sgn\qty[ -\eta \lambda  ] 
G_\scat \qty[M_i,M_j,T,\mu_i,\mu_j,\eta,\abs{\vec{k}},-\eta \lambda ] , & \text{if } -\eta \lambda \in \qty[\depsilondEepsilonMin, \depsilondEepsilonMax] ,
\\
0, & \text{otherwise } .
\end{array}
\right.
\numberthis
\label{imagPartScat}
\end{align*}

If one considers the external momentum shift, $k_0 \to k_0 \pm i\epsilon$, the real part of both contributions, $\Re 
\qty[B_{0,\pair}]$ and $\Re 
\qty[B_{0,\scat}]$, remains unchanged, however, the overall sign in front of the imaginary contributions is altered. The imaginary contributions are changed as:
\begin{align}
G_\pair \qty[M_i,M_j,T,\mu_i,\mu_j,\eta,\abs{\vec{k}}, \nicefrac{ -\eta \lambda }{ 2 }] 
& \to 
\mp \eta G_\pair \qty[M_i,M_j,T,\mu_i,\mu_j,\eta,\abs{\vec{k}}, \nicefrac{ -\eta \lambda }{ 2 }] ,
\label{momentumShiftChangeIma1}
\\
\sgn\qty[ -\eta \lambda  ] 
G_\scat \qty[M_i,M_j,T,\mu_i,\mu_j,\eta,\abs{\vec{k}},-\eta \lambda ] 
& \to 
\mp \eta G_\scat \qty[M_i,M_j,T,\mu_i,\mu_j,\eta,\abs{\vec{k}}, -\eta \lambda] .
\label{momentumShiftChangeImag2}
\end{align}

Finally, the $B_0$ function can be written using the real and imaginary parts of the pair creation/annihilation and scattering contributions as (see Eq. (\ref{B0_components_def})):
\begin{align*}
B_{0} \qty[M_i,M_j,T,\mu_i,\mu_j,k_0,\abs{\vec{k}}] 
=&
\Re 
\bigg[
B_{0,\pair} \qty[M_i,M_j,T,\mu_i,\mu_j,k_0,\abs{\vec{k}}] 
+
B_{0,\scat} \qty[M_i,M_j,T,\mu_i,\mu_j,k_0,\abs{\vec{k}}] 
\bigg]
\\
&
+i
\Im
\bigg[
B_{0,\pair} \qty[M_i,M_j,T,\mu_i,\mu_j,k_0,\abs{\vec{k}}] 
+
B_{0,\scat} \qty[M_i,M_j,T,\mu_i,\mu_j,k_0,\abs{\vec{k}}] 
\bigg] .
\numberthis
\label{B0_defined_terms_Re_and_IM}
\end{align*}

\subsubsection{ The $\abs{\vec{k}}=0$ case}

The $\abs{\vec{k}}=0$ limiting case cannot be calculated from the case $\abs{\vec{k}}>0$. As it was explained earlier (see Section \ref{practical_implementation_knot0}), the Jacobian determinant used when making the change of variables is zero for $\abs{\vec{k}}=0$ (see Eq. (\ref{jacobianCOV_knot0})). Hence, the case $\abs{\vec{k}}=0$ must be treated separately. Differently from the previous case, our calculation for the case $\abs{\vec{k}}=0$ completely agrees with the one found in \cite{Rehberg:1995nr}. In this case there is no separation between the spheres and the region of integration is simply the one inside a sphere of radius $\abs{\vec{p}} \leq \Lambda$, the usual 3-momentum regularization. Nonetheless, for completeness, we will briefly show the explicit calculation.

It is easier to consider a particular change of variables for the integral $B_{0,\pair}$ and a different one for the integral $B_{0,\scat}$. Furthermore, in this special case the integrand does not have angular dependencies, meaning that the integral over the azimuthal angle, $\varphi$, and polar angle, $\theta$, yield an overall factor of $4\pi$ and an integration over the norm of the 3-momentum, $\abs{\vec{p}}$.

Lets start by discussing the $\abs{\vec{k}}=0$ case of the pair creation/annihilation contribution, $B_{0,\pair}$ by considering the $\abs{\vec{k}}=0$ limit of Eq. (\ref{B0_pair_def}). Solving the equation, $E=\qty(E_2 + E_1)/2=\qty(E\qty[M_i,{\vec{p}}]+E\qty[M_j,{\vec{p}}])/2$, for the variable $\abs{\vec{p}}$, one can consider a change of variables from $\abs{\vec{p}} \to E$. Using the positive solution, one gets:
\begin{align}
\abs{\vec{p}} 
=
\frac{1}{4} 
\sqrt{ 
\frac{
\qty( 2 E - M_i + M_j ) 
\qty( 2 E - M_i - M_j ) 
\qty( 2 E + M_i - M_j ) 
\qty( 2 E + M_i + M_j ) 
}{ E^2 }
} 
=
P \qty[M_i,M_j,E]
.
\end{align}
The integration measure, $\dd{ \abs{\vec{p}} }$, changes as, $\dd{ \abs{\vec{p}} }
\frac{ \vec{p}^2 }{ E_1 E_2  }
=
\dd{ E }
\frac{ P \qty[M_i,M_j,E] }{ E }$ and the integral is written as:
\begin{align}
B_{0,\pair} \qty[M_i,M_j,T,\mu_i,\mu_j,k_0, 0] 
& =
\sumeta
\int_{E_0}^{E_\Lambda}
\dd{ E }
\frac{ P \qty[M_i,M_j,E] }{ E }
\frac
{ 1 - \nfermi[ \bar{E_1} + \eta \mu_i, T ] - \nfermi[ \bar{E_2}  - \eta \mu_j, T ] }
{ E + \nicefrac{ \eta \lambda }{ 2 }  - i \epsilon } .
\end{align}
Here, $E_0$ and $E_\Lambda$ are the endpoints obtained after the change of variables and are given by:
\begin{align}
E_0 & = 
\frac{1}{2}\qty(E\qty[M_i,0]+E\qty[M_j,0])
=
\frac{1}{2}\qty( M_i + M_j ) , 
\\
E_\Lambda & = 
\frac{1}{2}\qty(E\qty[M_i,\Lambda]+E\qty[M_j,\Lambda])
=
\frac{1}{2}\qty( 
\sqrt{ \Lambda^2 + M_i^2 } + \sqrt{ \Lambda^2 + M_j^2 } 
).
\end{align}
For simplicity, we introduced the functions $\bar{E_1}$ and $\bar{E_2}$, which are the functions defined in Eqs. (\ref{E1_def}) and (\ref{E2_def}), written in terms of $E$ instead of $\vec{p}$. They are explicitly given by:
\begin{align}
\bar{E_1} & = \bar{E_1}\qty[M_i,M_j,E] = 
2 E - 
\frac{1}{4} 
\sqrt{\frac{\qty(4 E^2 - M_i^2 + M_j^2)^2}{E^2}} ,
\\
\bar{E_2} & = \bar{E_2}\qty[M_i,M_j,E] = 
2 E - 
\frac{1}{4} 
\sqrt{\frac{\qty( 4 E^2 + M_i^2 - M_j^2)^2}{E^2}} .
\end{align}

As for the case $\abs{\vec{k}}>0$, one can use the Sokhotski–Plemelj theorem to write the real and imaginary parts, of the $\abs{\vec{k}}=0$ limit of the $B_0$ function. We write:
\begin{align*}
\Re 
\Big[
& B_{0,\pair} \qty[M_i,M_j,T,\mu_i,\mu_j,k_0,0] 
\Big]
 =
\sumeta 
\CPV 
\int_{E_0}^{E_\Lambda}
\dd{ E }
\frac{ P \qty[M_i,M_j,E] }{ E }
\frac
{ 1 - \nfermi[ \bar{E_1} + \eta \mu_i, T ] - \nfermi[ \bar{E_2}  - \eta \mu_j, T ] }
{ E + \nicefrac{ \eta \lambda }{ 2 } } ,
\numberthis
\\
\Im 
\Big[
& B_{0,\pair} \qty[M_i,M_j,T,\mu_i,\mu_j,k_0,0] 
\Big]
 = 
\\
& \qquad \qquad
\pi
\sumeta 
\left\{
\begin{array}{lr}
\frac{ P \qty[M_i,M_j, \nicefrac{ -\eta \lambda }{ 2 }] }{ \nicefrac{ -\eta \lambda }{ 2 } }
\qty( 1 - \nfermi[ \bar{E_1} + \eta \mu_i, T ] - \nfermi[ \bar{E_2}  - \eta \mu_j, T ] ) , & \text{if } \nicefrac{ -\eta \lambda }{ 2 } \in \qty[E_0,E_\Lambda] ,
\numberthis
\\
0, & \text{otherwise } .
\end{array}
\right.
\end{align*}
Again, by considering the external momentum shift, $k_0 \to k_0 \pm i\epsilon$, instead of the mass shift, only the imaginary part is changed with the following substitution: $P \qty[M_i,M_j, \nicefrac{ -\eta \lambda }{ 2 }]\to - \eta P \qty[M_i,M_j, \nicefrac{ -\eta \lambda }{ 2 }]$.

Finally, we deal with the scattering contribution, $B_{0,\scat}$ applying the $\abs{\vec{k}}=0$ limit to Eq. (\ref{B0_scat_def}). For this case we consider a change of variables based on solving the equation, $\varepsilon=E_2 - E_1=E\qty[M_j,{\vec{p}}]-E\qty[M_i,{\vec{p}}]$ for the absolute value of the momentum, $\abs{\vec{p}}$ and get:
\begin{align}
\abs{\vec{p}} 
=
\frac{1}{2} 
\sqrt{ 
\frac{
\qty( \varepsilon - M_i + M_j ) 
\qty( \varepsilon - M_i - M_j ) 
\qty( \varepsilon + M_i - M_j ) 
\qty( \varepsilon + M_i + M_j ) 
}{ \varepsilon^2 }
} 
=
P \qty[M_i,M_j,\varepsilon] .
\end{align}
In this case, the integration measure after the change of variables is, 
$\frac{ \dd{ \abs{\vec{p}} }\vec{p}^2 }{ E_1 E_2  }
=
-
\frac{ \dd{ \varepsilon } P \qty[M_i,M_j,\varepsilon] }{ \varepsilon }$ and the integral can then be written as:
\begin{align}
B_{0,\scat} \qty[M_i,M_j,T,\mu_i,\mu_j,k_0, 0] 
& =
2
\sumeta
\int_{\varepsilon_0}^{\varepsilon_\Lambda}
\dd{ \varepsilon }
\frac{ P \qty[M_i,M_j,\varepsilon] }{ \varepsilon } 
\frac
{ \nfermi[ \hat{E_1} - \eta \mu_i, T ] - \nfermi[ \hat{E_2} - \eta \mu_j, T ] }
{ \varepsilon + \eta \lambda  -  \sgn\qty[\varepsilon] i \epsilon  }  .
\end{align}
As before, the functions, $\hat{E_1}$ and $\hat{E_2}$ are the functions defined in Eqs. (\ref{E1_def}) and (\ref{E2_def}), written in terms of $\varepsilon$ instead of $\vec{p}$ and are given by:
\begin{align}
\hat{E_1} & = \hat{E_1}\qty[M_i,M_j,\varepsilon] =  
\frac{1}{2} 
\sqrt{\frac{\qty( \varepsilon^2 - M_i^2 + M_j^2)^2}{\varepsilon^2}}
- \varepsilon ,
\\
\hat{E_2} & = \hat{E_2}\qty[M_i,M_j,\varepsilon] = 
\frac{1}{2} 
\sqrt{\frac{\qty( \varepsilon^2 + M_i^2 - M_j^2)^2}{\varepsilon^2}}
+ \varepsilon  .
\end{align}
The integration endpoints are defined as:
\begin{align}
\varepsilon_0 & = 
E\qty[M_j,0]-E\qty[M_i,0]
=
M_j - M_i , 
\\
\varepsilon_\Lambda & = 
E\qty[M_j,\Lambda]-E\qty[M_i,\Lambda]
=
\sqrt{ \Lambda^2 + M_j^2 } - \sqrt{ \Lambda^2 + M_i^2 } .
\end{align}
Following the previous steps, the real and imaginary parts can be written as:
\begin{align*}
\Re 
\Big[
& B_{0,\scat} \qty[M_i,M_j,T,\mu_i,\mu_j,k_0,0] 
\Big]
 = 
2
\sumeta 
\CPV 
\int_{\varepsilon_0}^{\varepsilon_\Lambda}
\dd{ \varepsilon }
\frac{ P \qty[M_i,M_j,\varepsilon] }{ \varepsilon } 
\frac
{ \nfermi[ \hat{E_1} - \eta \mu_i, T ] - \nfermi[ \hat{E_2} - \eta \mu_j, T ] }
{ \varepsilon + \eta \lambda  } , 
\numberthis
\\
\Im 
\Big[
& B_{0,\scat} \qty[M_i,M_j,T,\mu_i,\mu_j,k_0,0] 
\Big]
 = 
\\
& \qquad \qquad
2\pi
\sumeta 
\left\{
\begin{array}{lr}
\sgn\qty[ -\eta \lambda ]
\frac{ P \qty[M_i,M_j,-\eta \lambda] }{ -\eta \lambda } 
( \nfermi[ \hat{E_1} - \eta \mu_i, T ] - \nfermi[ \hat{E_2} - \eta \mu_j, T ] ) , & \text{if }  -\eta \lambda  \in \qty[\varepsilon_0,\varepsilon_\Lambda] ,
\numberthis
\\
0, & \text{otherwise } .
\end{array}
\right.
\end{align*}
In the momentum shift case, the imaginary part is modified as: $\sgn\qty[ -\eta \lambda ] P \qty[M_i,M_j,-\eta \lambda] \to -\eta P \qty[M_i,M_j,-\eta \lambda]$. The complete function, $B_0$, for the case with $\abs{\vec{k}}=0$, can then be obtained as before, using Eq. (\ref{B0_defined_terms_Re_and_IM}) by using the real and imaginary parts calculated above.

In the Appendix \ref{numericalResultsB0}, we show and discuss some numerical results from evaluating the two fermion line integral, $B_0$ for different values of temperature, chemical potential and external momentum by fixing a cutoff $\Lambda$ and the effective masses, $M_i$ and $M_j$.

\section{The in-medium one fermion, one loop integral}
\label{one_fermion_one_loop_integral}

As discussed in earlier sections, different versions of the one fermion line integral arise when making calculations in the context of the NJL model. On the one hand, it appears in the calculation of the quark condensate (integral $A$ defined in Eq. (\ref{A_integral_definition})) while, on the other hand, it contributes to the quark-antiquark polarization function (integral $\mathcal{A}$ defined in Eq. (\ref{Ak_integral_definition})). In this work, we use the 3-momentum sphere intersection regularization, implying that the integrals $A$ and $\mathcal{A}$ are different. The first, $A$, within this regularization, is integrated over the region of the intersection of one sphere centered in the origin and radius $\Lambda$, with itself, making it identical to the usual 3-momentum regularization.  The second, $\mathcal{A}$, is a contribution to the quark-antiquark polarization function, $\Pi_{ij}^{S|P}$, and its integral is calculated inside the region of intersection between two spheres, distanced $\abs{\vec{k}}$ from each other, just like the $B_0$ integral from the previous section.

Finite temperature can be introduced in the one fermion line integral, $A$ (and $\mathcal{A}$), in the same manner as for the two fermion, one loop integral, $B_0$, see Section \ref{two_fermion_one_loop_integral}. Applying these steps to Eq. (\ref{A_Klev_3mometum_integral}) yields:
\begin{align}
A[M_i,T,\mu_i] 
& = 
16\pi^2 T \sum^{+\infty}_{n=-\infty}
\int_{\mathrm{Reg}[ \abs{\vec{k}} ]}
\frac{ \dd[3]{\vec{p}} }{ \qty(2 \pi)^3 } 
\frac{1}{ \qty( i \omega_n + \mu_i )^2 - E\qty[M_i,\vec{p}]^2} . 
\end{align}
Here, $\mu_i$ is the chemical potential associated with the fermion of mass $M_i$. Performing the infinite sum over the allowed Matsubara frequencies yields the well known result:
\begin{align*}
A[M_i,T,\mu_i] 
& =
-16 \pi^2 
\int_{\mathrm{Reg}[ \abs{\vec{k}} ]}
\frac{ \dd[3]{\vec{p}} }{ \qty(2 \pi)^3 } 
\frac{1}{2 E\qty[M_i,\vec{p}] }
\qty(
1 - 
\nfermi \qty[ E\qty[M_i,\vec{p}] - \mu_i, T] - 
\nfermi \qty[ E\qty[M_i,\vec{p}] + \mu_i, T] 
) 
\\
& =
-4 	
\int_{0}^\Lambda 
\frac{ \dd{ \abs{\vec{p}} } }{ \sqrt{ \vec{p}^2 + M_i^2 } }
\qty(
1 - 
\nfermi \qty[ \sqrt{ \vec{p}^2 + M_i^2 } - \mu_i, T] - 
\nfermi \qty[ \sqrt{ \vec{p}^2 + M_i^2 } + \mu_i, T] 
) .
\numberthis
\label{A_Klev_3mometum_integral}
\end{align*}
Since the integrand only depends on $\abs{\vec{p}}$, the angular integrations are trivial, leaving only one integration to be performed numerically. From the above results, the divergence of the vacuum contribution to the integral is evident and the necessity to regularize the integral is explicit. In the vacuum, one gets the well known result:
\begin{align}
A[M_i,0,0] 
& =
- 2 \Lambda \sqrt{ \Lambda^2 + M_i^2 }
- M_i^2  
\ln 
\qty[
\frac
{ \qty( \Lambda - \sqrt{ \Lambda^2 + M_i^2 } )^2}
{ M_i^2 }
] .
\label{AKlev_vacuum_limit}
\end{align}

Now, we consider the one fermion line loop integral which arises in the second order contribution, $\mathcal{A}$ (see Eq. (\ref{Ak_integral_definition})). In the 3-momentum sphere intersection regularization scheme, we can write it as:
\begin{align*}
\mathcal{A}[M_i,T,\mu_i,\abs{\vec{k}}] 
& =
-16 \pi^2 
\int_{ \mathbb{I}_{\abs{\vec{k}}}^\Lambda } \frac{ \dd[3]{\vec{p}} }{ \qty(2 \pi)^3 } 
\frac{1}{2 E\qty[M_i,\vec{p}] }
\qty(
1 - 
\nfermi \qty[ E\qty[M_i,\vec{p}] - \mu_i, T] - 
\nfermi \qty[ E\qty[M_i,\vec{p}] +\mu_i, T] 
) .
\numberthis
\label{Ak_integral_3MSIR}
\end{align*}
As already mentioned, the region of integration ${ \mathbb{I}_{\abs{\vec{k}}}^\Lambda }$ is formed by the intersection of two spheres, dislocated on the $p_z$ axis (see Fig. \ref{integration_region}). In order to simplify the calculation, one can shift the region of integration in order to have the center of the intersection coinciding with the origin. This can be done by performing the following change of variables: $\vec{p} \to \vec{p} + \nicefrac{\vec{k}}{2}$. Using this transformation, the integral becomes:
\begin{align*}
\mathcal{A}[M_i,T,\mu_i,\abs{\vec{k}}] 
& =
-16 \pi^2 
\int_{ {\mathbb{I}^\p}_{\abs{\vec{k}}}^\Lambda } \frac{ \dd[3]{\vec{p}} }{ \qty(2 \pi)^3 } 
\frac{1}{2 E\qty[M_i, \vec{p} + \frac{\vec{k}}{2}]  }
\\
& \qquad  \qquad
\times
\qty(
1 - 
\nfermi \qty[ E\qty[M_i, \vec{p} + \frac{\vec{k}}{2}]  - \mu_i, T] - 
\nfermi \qty[ E\qty[M_i, \vec{p} + \frac{\vec{k}}{2}]  +\mu_i, T] 
) .
\numberthis
\end{align*}
Here, ${ {\mathbb{I}^\p}_{\abs{\vec{k}}}^\Lambda }$ is the new integration region constituted by the intersection of two spheres, one centered at $\qty(p_x,p_y,p_z)=\qty(0,0,\nicefrac{\abs{\vec{k}}}{2})$ and other centered at $\qty(p_x,p_y,p_z)=\qty(0,0,\nicefrac{-\abs{\vec{k}}}{2})$, with the respective centers distanced by $\abs{\vec{k}}$. Interestingly, after performing the change of variables to center the region of integration and integrating over the azimutal angle, we are left with an extra dependence on the polar angle, $\theta$, making this integral slightly more complicated to calculate when compared to its analogue with only one sphere, $A$, given in Eq. (\ref{A_Klev_3mometum_integral}). Using the steps shown in Appendix \ref{integration_region_Rij} (see Eqs. (\ref{general_integrand_p_theta_3MSIR})-(\ref{pMax_general_integrand_p_theta_3MSIR})), we can write this integral, in the vacuum limit, as:
\begin{align*}
\mathcal{A}[M_i,T,\mu_i,\abs{\vec{k}}] 
& =
2
\int_0^{ \Lambda -\frac{ \abs{\vec{k}} }{2} } \dd{\abs{\vec{p}}}
\vec{p}^2
\int_{ 1 }^{ -1 } 
\frac{ 
\dd{\alpha}
}
{ 
\sqrt{ \vec{p}^2 + \frac{\vec{k}^2}{4} + \vec{p} \abs{\vec{k}} \alpha + M_i^2 }
}
\\
& \qquad \qquad  +
2
\int_{ \Lambda -\frac{ \abs{\vec{k}} }{2} }^{ \sqrt{ \Lambda ^2 - \frac{ \vec{k}^2 }{4} } } \dd{\abs{\vec{p}}}
\vec{p}^2
\int_{ \frac
{\Lambda ^2 - \frac{\vec{k}^2}{4} - \vec{p}^2}
{\abs{\vec{k}} \abs{\vec{p}}} }^{ -\frac
{\Lambda ^2 - \frac{\vec{k}^2}{4} - \vec{p}^2}
{\abs{\vec{k}} \abs{\vec{p}}} } 
\frac{ 
\dd{\alpha}
}
{ 
\sqrt{ \vec{p}^2 + \frac{\vec{k}^2}{4} + \vec{p} \abs{\vec{k}} \alpha + M_i^2 }
}
\\
& =
-\frac{ 1 }{ 3 \abs{\vec{k}} }
\Bigg\{
\vec{k}^2 
\qty(
\sqrt{ ( \Lambda - \abs{\vec{k}} )^2 + M_i^2 }
- 3 \sqrt{ \Lambda^2 + M_i^2 }
)
\\
& \qquad \qquad 
+ \abs{\vec{k}} \Lambda  
\qty( 
\sqrt{ ( \Lambda - \abs{\vec{k}} )^2 + M_i^2 } + 
3 \sqrt{ \Lambda^2 + M_i^2 } )
\\
& \qquad \qquad 
+ 2 
\qty( \Lambda^2 + M_i^2 ) 
\qty( 
\sqrt{ \Lambda^2 + M_i^2 } - 
\sqrt{ ( \Lambda - \abs{\vec{k}} )^2 + M_i^2 } 
)
\\
& \qquad \qquad 
+ 3 \abs{\vec{k}} M_i^2 
\ln 
\qty[
\frac{1}{M_i^2} 
\qty( \Lambda - \sqrt{ \Lambda^2 + M_i^2 } ) 
\qty( \Lambda - \abs{\vec{k}} - \sqrt{ ( \Lambda - \abs{\vec{k}} )^2 + M_i^2 }  
)
]
\Bigg\} .
\numberthis
\label{AKlev_vacuum_result}
\end{align*}
In the $\abs{\vec{k}} \to 0$ case, the two spheres in the regularization completely intersect and one gets the expected result given in Eq. (\ref{AKlev_vacuum_limit}), i.e., $\mathcal{A}[M_i,0,0,0] = A[M_i,0,0] $. Of course, this equivalence also holds at finite temperature and chemical potential as long as $\abs{\vec{k}} = 0$.

At finite temperature and chemical potential it is not possible to write this quantity as an 1-dimensional integral using this approach. However, from the two fermion line integral $B_0$, we know that it is possible to further simplify this quantity, making an appropriate change of variables. For simplicity, we will use exactly the same change of variables performed in the case of the $B_0$ integral, with $\abs{\vec{k}}>0$. Thus, considering the original form of the $\mathcal{A}$ integral (see Eq. (\ref{Ak_integral_3MSIR})), we multiply the integrand by one written as $1=E\qty[M_j, \vec{p} - \vec{k}]/E\qty[M_j, \vec{p} - \vec{k}]=E_2/E_2$ (see Eq. (\ref{E2_def})) and employ the change of variables used in Section \ref{practical_implementation_knot0} (see Eqs. (\ref{E_B0_COV})$-$(\ref{jacobianCOV_knot0})) in order to write:
\begin{align}
\mathcal{A}[M_i,T,\mu_i,\abs{\vec{k}}] 
& = 
-
\frac{2}{ \abs{\vec{k}} }
\sumeta
\int_{ \mathbb{I}_{\abs{\vec{k}}}^\Lambda }
\frac{\dd{\phi}}{2\pi}
\dd{ E }
\dd{ \varepsilon }
\qty( E + \frac{\varepsilon}{2} )
\qty(
\frac{1}{2}
+
\nfermi \qty[ E - \frac{\varepsilon}{2} + \eta \mu_i, T]
).
\end{align}
We highlight that this integral does not depend on the value of $M_j$ introduced when one multiplies the integrand by $E_2/E_2$. Indeed, any value for $M_j$ can be used in the calculation and this apparent extra parameter only arises in the calculation to allow us to use the formalism developed earlier in order to simplify the calculations. By analogy with the $B_0$ integral, we separate this integral into two contributions:
\begin{align}
\mathcal{A}[M_i,T,\mu_i,\abs{\vec{k}}] 
& = 
\mathcal{A}_\pair [M_i,T,\mu_i,\abs{\vec{k}}] 
+
\mathcal{A}_\scat [M_i,T,\mu_i,\abs{\vec{k}}] 
\end{align} 
where each contribution is given by:
\begin{align}
\mathcal{A}_\pair [M_i,T,\mu_i,\abs{\vec{k}}] 
& = 
-
\frac{2}{ \abs{\vec{k}} }
\sumeta
\int_{ \dEdepsilonEMin }^{ \dEdepsilonEMax }
\dd{ E }
E
\,
G_{E} \qty[M_i,T,\mu_i,\eta,\abs{\vec{k}},E] , 
\\
\mathcal{A}_\scat [M_i,T,\mu_i,\abs{\vec{k}}] 
& =
-
\frac{2}{ \abs{\vec{k}} }
\sumeta
\int_{ \depsilondEepsilonMin }^{ \depsilondEepsilonMax } 
\dd{ \varepsilon }
\varepsilon
\,
G_{\varepsilon} \qty[M_i,T,\mu_i,\eta,\abs{\vec{k}},\varepsilon]  . 
\end{align}
Here, we defined:
\begin{align*}
G_{E} \qty[M_i,T,\mu_i,\eta,\abs{\vec{k}},E] 
& = 
\int_{ \dEdepsilonepsilonMin }^{ \dEdepsilonepsilonMax }
\dd{\varepsilon}
\qty(
\frac{1}{2}
+
\nfermi \qty[ E - \frac{\varepsilon}{2} + \eta \mu_i, T]
) ,
\numberthis
\\
G_{\varepsilon} \qty[M_i,T,\mu_i,\eta,\abs{\vec{k}},\varepsilon] 
& =
\int_{ \depsilondEEMin }^{ \depsilondEEMax } 
\dd{E} 
\qty(
\frac{1}{2}
+
\nfermi \qty[ E - \frac{\varepsilon}{2} + \eta \mu_i, T]
) .
\numberthis
\end{align*}
As a result of making the appropriate change of variables, these integrals can be evaluated analytically to yield:
\begin{align*}
G_{E} \qty[M_i,T,\mu_i,\eta,\abs{\vec{k}},E] 
= 
\frac{1}{2}\qty(\dEdepsilonepsilonMax-\dEdepsilonepsilonMin) 
& -
2 T \ln \qty[ \e^{ \qty(\frac{1}{2}\dEdepsilonepsilonMax - E - \eta \mu_i)/T } + 1] 
\\
& +
2 T \ln \qty[ \e^{ \qty(\frac{1}{2}\dEdepsilonepsilonMin - E - \eta \mu_i)/T } + 1] ,
\numberthis
\\
G_{\varepsilon} \qty[M_i,T,\mu_i,\eta,\abs{\vec{k}},\varepsilon] 
 =
\frac{1}{2}\qty(\depsilondEEMax-\depsilondEEMin) 
& +
T \ln \qty[ \e^{ \qty( \frac{1}{2}\varepsilon -\depsilondEEMax - \eta \mu_i)/T } + 1]
\\
& -
T \ln \qty[ \e^{ \qty( \frac{1}{2}\varepsilon -\depsilondEEMin - \eta \mu_i)/T } + 1] .
\numberthis
\end{align*}
In this way, the one fermion line integral, in the 3-momentum sphere intersection regularization, can be written as a sum of one dimensional integrals. Taking the vacuum limit, one can also perform the remaining integrations in order to obtain Eq. (\ref{AKlev_vacuum_result}).

\section{Cross sections for quark-quark and quark-antiquark processes in the NJL model}
\label{crossSectionsNJLModel}

The goal of this section is to evaluate the cross sections of the NJL model and show the influence of the regularization scheme proposed in this work versus the 3-momentum regularization used in other works. We chose the cross sections to show the differences between regularization schemes because they constitute a physical quantity which is dependent on the absolute value of external momentum, $\abs{\vec{k}}$, present in the quark-antiquark polarization function $\Pi_{ij}^{S|P}$. As already discussed, the 3-momentum sphere intersection regularization scheme is dependent on the value of   $\abs{\vec{k}}$ and, for  $\abs{\vec{k}}=0$, one gets the usual 3-momentum regularization. Evaluating thermodynamic quantities (at mean field level, there are no external momenta), or meson masses (usually calculated at rest with  $\abs{\vec{k}}=0$) would not yield any differences when compared to the usual 3-momentum regularization with a cutoff in all integrals.

\subsection{The model}

In order to calculate the cross sections for quark-quark and quark-antiquark scatterings, we will use the following $SU(3)$ NJL Lagrangian:
\begin{align*}
\mathcal{L} = 
\bar{\psi} \qty(i\slashed{\partial}-m) \psi
+ \frac{G}{2}
\Big(
\qty(\bar{\psi} \lambda_a \psi)^2 + 
(\bar{\psi} i \gamma^5 \lambda_a \psi)^2 
\Big)
+ 8 \kappa 
\Big(
\det\qty[ \bar{\psi} P_R \psi ] 
+ 
\det\qty[ \bar{\psi} P_L \psi ]  
\Big) .
\numberthis
\label{NJL_Lagrangian}
\end{align*}
Here, $\psi^T= \qty(\psi_u,\psi_d,\psi_s)$ and $m=\diag  \qty{ m_u, m_d, m_s } $ is the quark current mass matrix. This popular model includes, besides the free Dirac Lagrangian, a chirally symmetric set of interactions with a spontaneous chiral symmetry broken ground state in the vacuum. The choice of model is rooted in the goal to compare our results with other results found in the literature, where the regularization of the one fermion and and two fermion line integrals are performed in a different way than the one proposed here \cite{Rehberg:1995nr}. Hence, we only include in the Lagrangian the necessary four quark-quark scalar-pseudoscalar interaction and the 't Hooft interaction \cite{Rehberg:1995nr,Rehberg:1995kh}. However, the formalism developed earlier allows for a straightforward extension to include, at the Lagrangian level, multi-scalar and multi-pseudoscalar quark interactions, such as eight quark interactions or explicit chiral symmetry breaking interactions \cite{Osipov:2013fka,Moreira:2014qna,Morais:2017bmt,CamaraPereira:2020rtu}. Indeed, in the future, we will consider an extended version of the NJL model which includes eight-quark interactions in order to study the influence of considering these higher quark interactions in the transport properties of the NJL model.

The central motivation behind the NJL model is the incorporation of a mechanism which induces spontaneous chiral symmetry breaking. In its original form this is achieved through the inclusion of a local four-fermion contact interaction term. One of the drawbacks of the model in this formulation is the fact that it exhibits a axial $U_A(1)$ symmetry which is not observed in Nature (there are no degenerate mirror hadrons with opposite parity). The fact that no corresponding Nambu$-$Goldstone boson is observed led to the so called ``$U_A(1)$ problem'' which can be solved by the anomalous breaking of this symmetry through instanton effects \cite{'tHooft:1976up,tHooft:1976snw}. This mechanism can be incorporated in the model with the inclusion of the 't Hooft determinant $2 N_f-$fermion interaction term \cite{Kunihiro:1987bb,Bernard:1987sg, Reinhardt:1988xu}. It should be noted that in the three flavor version of the model this has been shown to lead to vacuum stability problems which, in turn, can be solved with the inclusion of eight quark interactions \cite{Osipov:2005tq, Osipov:2006ns}. In the present work, however, we will not consider these in order to enable a straightforward comparison of our results with the ones found in the literature.

Using the linear expansion of the Lagrangian, written in Eq. (\ref{NJL_Lagrangian}), as explained in previous sections, one can readily obtain the gap equations for this model:
\begin{align}
M_i & = m_i 
- 2 G \expval*{ \bar{\psi}_i \psi_i }
- 2 \kappa \expval*{ \bar{\psi}_j \psi_j } \expval*{ \bar{\psi}_k \psi_k } .
\end{align}
Here, $i\ne j\ne k\in \{u,d,s\}$. The value of the quark effective masses, at finite temperature and chemical potential, can be obtained by solving these equations self-consistently and the quark condensates, $\expval*{ \bar{\psi}_i \psi_i }$, can be calculated from Eq. (\ref{quarkCondensate}). These gap equations can be obtained by calculating the thermodynamic potential of the model, $\Omega$, and requiring it to be thermodynamically consistent. This amounts to requiring the potential to be stationary with respect to the mean fields, $\expval*{s_a}$, i.e., $\fdv*{\Omega}{ \expval*{s_a} }=0$.

As pointed out earlier, and has it will become clear in following sections, to calculate the cross sections for different processes, one will also need to evaluate different meson propagators at finite temperature and density. The scalar and pseudoscalar meson propagators, $D_S$ and $D_P$, were defined in Eqs. (\ref{scalarInverseMesonPropagator}) and (\ref{pseudoscalarInverseMesonPropagator}), respectively. The meson propagators are written in terms of the interaction-independent quark-antiquark polarization function, $\Pi_{ij}^{S|P}$, and of the interaction-dependent meson projectors $S_{ab}$ and $P_{ab}$. As discussed in previous sections, the polarization function is independent of the type of interactions included at the Lagrangian level. As a matter of fact we derived it in Section \ref{quadraticLagrangian}, without making any choice regarding the quark interaction Lagrangian. The meson projectors, on the other hand, are completely dependent on the type of interactions considered at the Lagrangian (as explained in Section \ref{quadraticLagrangian}) and are defined as the coefficients in front of the quadratic products, $s_a s_b=(\bar{\psi} \lambda_a \psi)(\bar{\psi} \lambda_b \psi)$ and $p_a p_b=(\bar{\psi} i \gamma^5 \lambda_a \psi)(\bar{\psi} i \gamma^5 \lambda_b \psi)$. As discussed, these coefficients arise when one makes the quadratic expansion of the product between quark bilinear operators (see Eq. (\ref{quadraticLagrangianExpansion})). The interaction-dependent meson projectors, $S_{ab}$ and $P_{ab}$, for the Lagrangian defined in Eq. (\ref{NJL_Lagrangian}), are given by:
\begin{align*}
S_{ab}
& = 
\frac{1}{2}
\qty(
G \delta_{ab} 
+
\frac{3\kappa}{2} A_{abc} \expval{s_c} 
),
\numberthis
\\
P_{ab}
& = 
\frac{1}{2}
\qty(
G \delta_{ab} -
\frac{ 3\kappa }{2} A_{abc} \expval{s_c}
) .
\numberthis
\end{align*}
Here, $A_{abc}$ is given by:
\begin{align*}
A_{abc} = 
\frac{2}{3} d_{abc} +
\sqrt{ \frac{2}{3} } 	
\qty(
\delta_{a0} \delta_{b0} \delta_{c0} -
\delta_{a0} \delta_{bc}
-
\delta_{b0} \delta_{ca} -
\delta_{c0} \delta_{ab}
) .
\numberthis
\label{eq:Aabc}
\end{align*} 
The quantities $d_{abc}$ are the symmetric structure constants of the unitary group $U(3)$. As previously introduced, the expectation value $\expval*{s_a}=\expval*{\bar{\psi} \lambda_a \psi}$ and we consider that there is no pseudoscalar condensation, i.e., $\expval*{p_a}=\expval*{\bar{\psi} i \gamma^5 \lambda_a \psi}=0$. The non-zero scalar projectors, $S_{ab}$, are explicitly given by:
\begin{align*}
{S}_{00}
& = 
\frac{1}{2}
\qty( 
G
+
\frac{2}{3} \kappa \qty( \expval*{ \bar{\psi}_u \psi_u } + \expval*{ \bar{\psi}_d \psi_d } + \expval*{ \bar{\psi}_s \psi_s } )
) ,
\\
S_{11} & = S_{22} = S_{33} =
\frac{1}{2}
\qty( G - \kappa \expval*{ \bar{\psi}_s \psi_s } ) ,
\\
S_{44} & = S_{55} = S_{66} = S_{77} =
\frac{1}{2}
\qty( G - \kappa \expval*{ \bar{\psi}_d \psi_d } ),
\\
S_{88} & =
\frac{1}{2}
\qty( G  - \frac{\kappa}{3} \qty( 2 \expval*{ \bar{\psi}_u \psi_u } + 2 \expval*{ \bar{\psi}_d \psi_d } - \expval*{ \bar{\psi}_s \psi_s } )
) ,
\\
S_{03} & = S_{30} =  
-\frac{\kappa}{2\sqrt{6}} \qty( \expval*{ \bar{\psi}_u \psi_u } - \expval*{ \bar{\psi}_d \psi_d } ) ,
\\
S_{08} & = S_{80} =  
- \frac{ \sqrt{2} \kappa }{12} \qty( \expval*{ \bar{\psi}_u \psi_u } + \expval*{ \bar{\psi}_d \psi_d } - 2\expval*{ \bar{\psi}_s \psi_s } ) .
\numberthis
\end{align*}
The non-zero pseudoscalar coefficients, $P_{ab}$, are:
\begin{align*}
P_{00} 
& = 
\frac{1}{2}
\qty( 
G
- \frac{2}{3} \kappa \qty( \expval*{ \bar{\psi}_u \psi_u } + \expval*{ \bar{\psi}_d \psi_d } + \expval*{ \bar{\psi}_s \psi_s } )
) ,
\\
P_{11} & = P_{22} = P_{33} =
\frac{1}{2}
\qty( G + \kappa \expval*{ \bar{\psi}_s \psi_s } ) ,
\\
P_{44} & = P_{55} = P_{66} = P_{77} =
\frac{1}{2}
\qty( G + \kappa \expval*{ \bar{\psi}_d \psi_d } ),
\\
P_{88} & =
\frac{1}{2}
\qty( G + \frac{\kappa}{3} \qty( 2 \expval*{ \bar{\psi}_u \psi_u } + 2 \expval*{ \bar{\psi}_d \psi_d } - \expval*{ \bar{\psi}_s \psi_s } )
) ,
\\
P_{03} & = P_{30} =  
\frac{\kappa}{2\sqrt{6}} ( \expval*{ \bar{\psi}_u \psi_u } - \expval*{ \bar{\psi}_d \psi_d } ) ,
\\
P_{08} & = P_{80} =  
\frac{ \sqrt{2} \kappa }{12} \qty( \expval*{ \bar{\psi}_u \psi_u } + \expval*{ \bar{\psi}_d \psi_d } - 2\expval*{ \bar{\psi}_s \psi_s } ) .
\numberthis
\end{align*}
These are exactly the same as the one obtained in Ref. \cite{Rehberg:1996vd} with a small caveat: in the NJL Lagrangian density defined in Eq. (\ref{NJL_Lagrangian}), the four quark scalar pseudoscalar interaction is proportional to $G/2$ while, in Ref. \cite{Rehberg:1996vd}, it is proportional to $G$.

At this point, one can explicitly write down the expressions for the meson propagators, defined in Eqs. (\ref{scalarInverseMesonPropagator}) and (\ref{pseudoscalarInverseMesonPropagator}). The indexes of the inverse meson propagators, $\qty( D_S^{-1} )_{ab}$ and $\qty( D_P^{-1} )_{ab}$, run from 0 to 8, the number of generators of the $U(3)$ group. Thus, these objects can be considered as two matrices of dimension $9 \times 9$. It is quite useful to write the scalar, $\Pi_{ab}^{S}$ and pseudoscalar, $\Pi_{ab}^{P}$, polarization functions in the quark flavor basis, i.e., switch from $a,b=\qty{0,\ldots,8}$ to $\Pi_{ij}^{S|P}$, with $i,j=\qty{u,d,s}$. Using Eq. (\ref{quarkAntiquarkPolarization_gellmannIndexes}), we can write the scalar and pseudoscalar polarization functions in the quark flavor basis. Explicitly, all non-zero elements are given by (for the sake of simplicity, we omit the scalar and pseudoscalar labels, ${S|P}$ ):
\begin{align*}
\Pi_{00} & = \nicefrac{2}{3} \qty( \Pi_{uu} + \Pi_{dd} + \Pi_{ss} ) ,
\\
\Pi_{03} & = \Pi_{30} = \sqrt{\nicefrac{2}{3}} \qty( \Pi_{uu} - \Pi_{dd} ) ,
\\
\Pi_{08} & = \Pi_{80} = \nicefrac{\sqrt{2}}{3} \qty( \Pi_{uu} + \Pi_{dd} - 2 \Pi_{ss} ) ,
\\
\Pi_{11} & = \Pi_{22} = \Pi_{ud} + \Pi_{du},
\\
\Pi_{12} & = -\Pi_{21} = i \qty( \Pi_{ud} - \Pi_{du} ),
\\
\Pi_{33} & = \Pi_{uu} + \Pi_{dd},
\\
\Pi_{38} & = \Pi_{83} = \nicefrac{1}{\sqrt{3}} \qty( \Pi_{uu} - \Pi_{dd} ) ,
\\
\Pi_{44} & = \Pi_{55} = \Pi_{us} + \Pi_{su},
\\
\Pi_{45} & = -\Pi_{54} = i \qty( \Pi_{us} - \Pi_{su} ),
\\
\Pi_{66} & = \Pi_{77} = \Pi_{ds} + \Pi_{sd},
\\
\Pi_{67} & = -\Pi_{76} = i \qty( \Pi_{ds} - \Pi_{sd} ),
\\
\Pi_{88} & = \nicefrac{1}{3} \qty( \Pi_{uu} + \Pi_{dd} + 4 \Pi_{ss} ) .
\numberthis
\end{align*}
All the other elements are zero. The polarization functions written in terms of the flavor indexes are not invariant under index exchange, i.e., $\Pi_{ud}$ and $\Pi_{du}$, $\Pi_{us}$ and $\Pi_{su}$ or $\Pi_{sd}$ and $\Pi_{ds}$ are not necessarily equal at finite temperature and chemical potential. Indeed, it is this difference which breaks the particle/antiparticle symmetry in the in-medium meson behavior \cite{Costa:2019bua}. For instance the charged kaons behave differently at finite chemical potential due to the mass difference between the quark and anti-quark particles which constitute the kaons: at finite chemical potential, the bath is more populated by particles than anti-particles.

Additionally, in this model, the inverse meson propagators are block matrices. Thus, we can write the second order term of the quadratic expansion of the effective action in Eq. (\ref{actionSecondOrderTerm})  in different contributions. There will be a sector related to the indexes $a,b=1,2$, $a,b=4,5$, $a,b=6,7$ and $a,b=0,3,8$. In other models, with different physical considerations (like considering pion/kaon condensation) or interaction terms at the Lagrangian level, this may not be possible. The sector with indexes $a,b=1,2$ are related to the charged pion propagators, $D_{ \pi^+ | \pi^- }$. Consider the matrices $\lambda_\pm = \nicefrac{1}{\sqrt{2}} \qty( \lambda_1 \pm i \lambda_2 ) $ with $\lambda_\pm$ creating a quark bilinear with the quantum numbers of the $\pi^\mp$, i.e., $\pi^\pm = \bar{\psi} \lambda_\mp \psi $. Using these identities one can write the $\pi_{1|2}$ fields in terms of $\pi_\pm$. Thus, one can write the $a,b=1,2$ sector of the quadratic effective action of the model (defined in Eq. (\ref{actionSecondOrderTerm})) as:
\begin{align*}
\mathcal{S}_{ \mathrm{Q} }^{ (2) }
& = 
-
\frac{ 1 }{2} 
\int_{\mathrm{Reg}} \frac{ \dd[4]{k} }{ \qty(2\pi)^4 }
\bigg\{
\pi^+ [k]
\qty(
\frac{   \qty( D_P^{-1} )_{11} }{2} + 
\frac{ i \qty( D_P^{-1} )_{12} }{2} - 
\frac{ i \qty( D_P^{-1} )_{21} }{2} + 
\frac{   \qty( D_P^{-1} )_{22} }{2}
)
\pi^- [-k]
\\
& \qquad \qquad  \qquad \quad \;
+
\pi^- [k]
\qty(
\frac{   \qty( D_P^{-1} )_{11} }{2} -
\frac{ i \qty( D_P^{-1} )_{12} }{2} + 
\frac{ i \qty( D_P^{-1} )_{21} }{2} + 
\frac{   \qty( D_P^{-1} )_{22} }{2}
)
\pi^+ [-k]
+ \ldots
\bigg\}
\\
& = 
-
\frac{ 1 }{2} 
\int_{\mathrm{Reg}} \frac{ \dd[4]{k} }{ \qty(2\pi)^4 }
\bigg\{
\pi^+ [k]
\qty(
D_{ \pi^- }^{-1}
)
\pi^- [-k]
+
\pi^- [k]
\qty(
D_{ \pi^+ }^{-1}
)
\pi^+ [-k]
+ \ldots
\bigg\} .
\numberthis
\end{align*}
Here, we have defined the inverse propagators of the charged pions as:
\begin{align}
\qty( D_{ \pi^+ | \pi^- } )^{-1}
& =
\frac{   \qty( D_P^{-1} )_{11} }{2} \mp
\frac{ i \qty( D_P^{-1} )_{12} }{2} \pm
\frac{ i \qty( D_P^{-1} )_{21} }{2} + 
\frac{   \qty( D_P^{-1} )_{22} }{2}  .
\end{align}
In this model, the inverse model propagator with indexes $a,b=1,2$ is given, in matrix form, by:
\begin{align}
\qty( D_P^{-1} )_{ab} 
=
\frac{1}{2}
\mqty(
P_{11} & 0  
\\ 
0 & P_{11}
)^{-1} 
-
\mqty(
\Pi_{ud}^P + \Pi_{du}^P & i \Pi_{ud}^P - i \Pi_{du}^P 
\\ 
i \Pi_{ud}^P - i \Pi_{du}^P  & \Pi_{ud}^P + \Pi_{du}^P
) .
\end{align}
We can then find that the charged pion propagators, $D_{ \pi^+ | \pi^- }$, are given by;
\begin{align}
D_{ \pi^+ | \pi^- }
&=
\frac{ 2 P_{11} }{ 1 - 4 P_{11} \Pi_{ ud|du }^P } .
\numberthis
\label{chargedPionPropagator}
\end{align}
Following these steps, one can find the remaining charged pseudoscalar meson propagators including, $K^\pm$, $K^0$ and $\bar{K^0}$. They correspond to the $a,b=4,5$ and $a,b=6,7$ sectors of the pseudoscalar meson propagators matrix. They are given by:
\begin{align}
D_{ K^+ | K^- }
& =
\frac{ 2 P_{44} }{ 1 - 4 P_{44} \Pi_{ us|su }^P } ,
\numberthis
\\
D_{ K^0 | \bar{K^0} }
& =
\frac{ 2 P_{66} }{ 1 - 4 P_{66} \Pi_{ ds|sd }^P } .
\numberthis
\end{align}
Likewise, the charged scalar particles can be obtained in the same manner. Their expressions are akin to the ones written above with the substitution of pseudoscalar quantities to scalar ones. The sector $a,b=0,3,8$, corresponds to the diagonal Gell-Mann matrices plus the scaled identity matrix, meaning they correspond to mesons constituted by linear combinations of $\bar{u} u$, $\bar{d}d$ and $\bar{s} s$, i.e.,  the pseudoscalars $\pi^0$, $\eta$ and $\eta'$ and the scalars $a^0_0$, $f_0$ and $\sigma$ \cite{Klevansky:1992qe}. Due to this mixture, one must use a diagonalization procedure in order to obtain the expressions for the propagators of these particles \cite{Klevansky:1992qe}. For the purposes of calculating the cross sections of the NJL model, one does not have to calculate the propagators of these states. Indeed, as we will see, in the calculation of the differential cross sections, different mixtures of the elements of $D_{ab}$ (with $a,b=0,3,8$) will be used, depending on the particular quark-quark or quark-antiquark scattering process considered \cite{Rehberg:1996vd}.

\subsection{Cross sections}
\label{cross_sections}

In this section we lay out the analytical tools to calculate the differential cross section for several quark-quark and quark-antiquark processes within the context of the NJL model. Since these quantities depend on the meson propagator and furthermore, are functions of the external momentum of the propagator, they can be used to gauge the difference between using the regularization proposed in this work versus the usual regularization used in the literature. Hence, we will compare the cross section obtained in this work with the one provided in \cite{Rehberg:1996vd}. The different cross sections can be obtained by integrating over the differential cross sections \cite{Zhuang:1995uf,Rehberg:1996vd,Soloveva:2020hpr}.

\begin{figure}[ht!]
\begin{subfigure}[b]{0.32\textwidth}
\includegraphics[scale=1]
{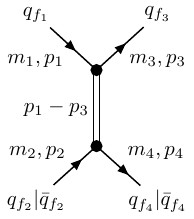}
\caption{$t$-channel}
\label{tChannel}
\end{subfigure}
\begin{subfigure}[b]{0.32\textwidth}
\includegraphics[scale=1]
{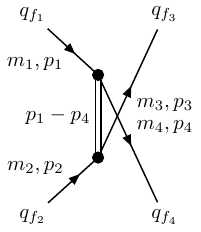}
\caption{$u$-channel}
\label{uChannel}
\end{subfigure}
\begin{subfigure}[b]{0.32\textwidth}
\includegraphics[scale=1]
{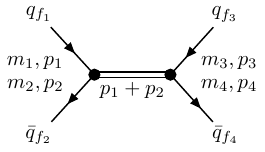}
\caption{$s$-channel}
\label{sChannel}
\end{subfigure}
\caption{The three possible Feynman diagrams involved in quark-quark and quark-antiquark scattering processes at leading order in an $1/N_c$ expansion of the transition amplitude \cite{Zhuang:1995uf} (time is considered in the horizontal axis). The $t$- and $u$-channels arise in the scattering between quarks and the $t$- and $s$-channels arise in the quark-antiquark scattering processes. 
}
\label{feynmanDiagramsForQuarkQuarkAndQuarkAntiquarkScatterings}
\end{figure}

As usual, in order to represent the different scattering processes, we will use the Mandelstam variables, $s$, $t$ and $u$. These are defined by:
\begin{align}
s & = \qty( p_1 + p_2 )^2,
\\
t & = \qty( p_1 - p_3 )^2,
\\
u & = \qty( p_1 - p_4 )^2,
\label{Mandelstam_variables}
\end{align}
with the four-momenta $p_i$ $i\in\left\{1,2,3,4\right\}$ ($i\in\left\{1,2\right\}$ incoming particles and $i\in\left\{3,4\right\}$ outgoing particles, see Fig. \ref{feynmanDiagramsForQuarkQuarkAndQuarkAntiquarkScatterings}). Momentum conservation implies the identity $s+t+u=\sum_{i=1}^4 m_i^2$, allowing one to use the relativistic kinematics of the two-particle scattering to write all quantities of interest in terms of the $s$- and $t$-channels, eliminating the need to a direct reference to the $u$-channel quantities. By definition, the differential cross section is $\dv*{\sigma_{ 12 \to 34} }{t}$. Thus integrating over the $t$-channel will yield the cross section of a given process as a function of the center of mass energy, $s$. In the center of mass frame it can be checked that the momentum of the exchanged particle, in the $t$-channel, is given by:
\begin{align}
k_0^{(t)}
& =
\frac{ \qty( 
m_1^2 - 
m_2^2 - 
m_3^2 + 
m_4^2 
) }{ 2\sqrt{s} } ,
\label{k0_tChannel}
\\
\abs{ \vec{k} }^{(t)}
& =
\sqrt{
\frac{ \qty( m_1^2 - m_2^2 - m_3^2 + m_4^2)^2 }{ 4s }
- t
} .
\label{absk_tChannel}
\end{align}
In the $u$-channel, the momentum components are:
\begin{align}
k_0^{(u)}
& =
\frac{ \qty( 
m_1^2 - 
m_2^2 +
m_3^2 -
m_4^2 
) }{ 2\sqrt{s} } ,
\label{k0_uChannel}
\\
\abs{ \vec{k} }^{(u)}
& =
\sqrt{
\frac{ \qty( m_1^2 - m_2^2 + m_3^2 - m_4^2)^2 }{ 4s }
- m_1^2 - m_2^2 - m_3^2 - m_4^2 + s + t 
} .
\label{absk_uChannel}
\end{align}
Naturally, in the center of  mass reference frame, $k_0^{(s)}=\sqrt{s}$ and $\abs{ \vec{k} }^{(s)}=0$. The scattering angle, $\Theta$, in the center of mass reference frame, can be written in terms of the incoming/outgoing masses and the $s$- and $t$-channels. It can be found to be given by:
\begin{align}
\Theta
=
\arccos
\qty[
\frac{
\qty( m_1^2 - m_2^2 + s ) 
\qty( m_3^2 - m_4^2 + s )
- 2 s 
\qty( m_1^2 + m_3^2 - t )
}
{ 
\sqrt{ \lambda(s,m_1^2,m_2^2) } 
\sqrt{ \lambda(s,m_3^2,m_4^2) }
}
] ,
\label{scatteringAngle}
\end{align}
with $\lambda(a,b,c)=a^2+b^2+c^2-2ab-2ac-2bc$, the so-called triangle function. Thus, we can parametrize the differential cross sections in terms of the Mandelstam variables $s$ and $t$.





At leading order in an $1/N_c$ expansion of the transition amplitude matrix, $\mathcal{M}$, for the quark-quark scattering, only the $t$- and $u$-channels contribute while, for the quark-antiquark scattering, only the $t$- and $s$-channels are involved \cite{Zhuang:1995uf,Rehberg:1996vd}. The Feynman diagrams for the quark-quark and quark-antiquark scatterings can be observed in Fig. \ref{feynmanDiagramsForQuarkQuarkAndQuarkAntiquarkScatterings}. Taking into account all symmetries, at finite temperature and chemical potential there are a total of 6 different quark-quark processes, the corresponding 6 antiquark-antiquark processes and 15 different quark-antiquark processes. When dealing with exact isospin symmetry, at zero chemical potential, there would exist only 4 different quark-quark processes ($uu \to uu$, $ud \to ud$, $us \to us$, $ss \to ss$) and 7 different quark-antiquark processes ($u\bar{u} \to u\bar{u}$, $u\bar{u} \to d\bar{d}$, $u\bar{u} \to s\bar{s}$, $u\bar{d} \to u\bar{d}$, $u\bar{s} \to u\bar{s}$, $s\bar{s} \to u\bar{u}$, $s\bar{s} \to s\bar{s}$).

\begin{table*}[t!]
\begin{center}
\begin{tabular}{@{}lc@{}} 
\toprule
$12 \to 34$ process \hspace{2cm} & 
$ 
\qty(
\TCoeff{a}{f_3}{f_1}
D_{ab}^{P|S} [t]
\TCoeff{b}{f_4}{f_2}
)
$
\\
\midrule
$ud \to ud$ , $du \to du$   
&
$\frac{2}{3} D_{00}^{P|S} + \frac{2\sqrt{2}}{3} D_{08}^{P|S} - D_{33}^{P|S} + \frac{1}{3} D_{88}^{P|S}$  
\\[0.3cm]
$us \to us$ , $su \to su$  
& 
$
\frac{2}{3} D_{00}^{P|S} + \sqrt{\frac{2}{3}} D_{03}^{P|S} -
\frac{\sqrt{2}}{3} D_{08}^{P|S} - \frac{2}{\sqrt{3}} D_{38}^{P|S} - \frac{2}{3} D_{88}^{P|S}
$    
\\[0.3cm]
$ds \to ds$ , $sd \to sd$  
& 
$
\frac{2}{3} D_{00}^{P|S} - \sqrt{\frac{2}{3}} D_{03}^{P|S} -
\frac{\sqrt{2}}{3} D_{08}^{P|S} + \frac{2}{\sqrt{3}} D_{38}^{P|S} - \frac{2}{3} D_{88}^{P|S}
$    
\\[0.3cm]
$uu \to uu$  
& 
$
\frac{2}{3} D_{00}^{P|S} + 2\sqrt{\frac{2}{3}} D_{03}^{P|S} +
\frac{2\sqrt{2}}{3} D_{08}^{P|S} + D_{33}^{P|S} + \frac{2}{\sqrt{3}} D_{38}^{P|S} + \frac{1}{3} D_{88}^{P|S}
$ 
\\[0.3cm]
$dd \to dd$  
& 
$
\frac{2}{3} D_{00}^{P|S} - 2\sqrt{\frac{2}{3}} D_{03}^{P|S} +
\frac{2\sqrt{2}}{3} D_{08}^{P|S} + D_{33}^{P|S} - \frac{2}{\sqrt{3}} D_{38}^{P|S} + \frac{1}{3} D_{88}^{P|S}
$   
\\[0.3cm]
$ss \to ss$  
& 
$
\frac{2}{3} D_{00}^{P|S} - \frac{4\sqrt{2}}{3} D_{08}^{P|S} + \frac{4}{3} D_{88}^{P|S}
$  
\\
\bottomrule
\end{tabular}
\end{center}
\caption{$t-$channel propagator for each particular quark-quark process.}
\label{t_channel_prop_qq}
\end{table*}

\begin{table*}[t!]
\begin{center}
\begin{tabular}{@{}lc@{}} 
\toprule
$12 \to 34$ process \hspace{1cm} & 
$ 
\qty(
\TCoeff{a}{f_4}{f_1}
D_{ab}^{P|S} [ t ] 
\TCoeff{b}{f_3}{f_2}
)
$
\\
\midrule
$u \bar{d} \to u \bar{d}$ , $d \bar{u} \to d \bar{u}$  &
$\frac{2}{3} D_{00}^{P|S} + \frac{2\sqrt{2}}{3} D_{08}^{P|S} - D_{33}^{P|S} + \frac{1}{3} D_{88}^{P|S}$  
\\[0.3cm]
$u \bar{s} \to u \bar{s}$ , $s \bar{u} \to s \bar{u}$  & 
$
\frac{2}{3} D_{00}^{P|S} + \sqrt{\frac{2}{3}} D_{03}^{P|S} -
\frac{\sqrt{2}}{3} D_{08}^{P|S} - \frac{2}{\sqrt{3}} D_{38}^{P|S} - \frac{2}{3} D_{88}^{P|S}
$  
\\[0.3cm]
$d \bar{s} \to d \bar{s}$ , $s \bar{d} \to s \bar{d}$  & 
$
\frac{2}{3} D_{00}^{P|S} - \sqrt{\frac{2}{3}} D_{03}^{P|S} -
\frac{\sqrt{2}}{3} D_{08}^{P|S} + \frac{2}{\sqrt{3}} D_{38}^{P|S} - \frac{2}{3} D_{88}^{P|S}
$   
\\[0.3cm]
$u \bar{u} \to u \bar{u}$  & 
$
\frac{2}{3} D_{00}^{P|S} + 2\sqrt{\frac{2}{3}} D_{03}^{P|S} +
\frac{2\sqrt{2}}{3} D_{08}^{P|S} + D_{33}^{P|S} + \frac{2}{\sqrt{3}} D_{38}^{P|S} + \frac{1}{3} D_{88}^{P|S}
$   
\\[0.3cm]
$u \bar{u} \to d \bar{d}$  & 
$ D_{11}^{P|S} - i D_{12}^{P|S} + i D_{21}^{P|S} + D_{22}^{P|S} = 2 D_{\pi^+ | a_0^+}   $
\\[0.3cm]
$u \bar{u} \to s \bar{s}$  & 
$ D_{44}^{P|S} - i D_{45}^{P|S} + i D_{54}^{P|S} + D_{55}^{P|S} = 2 D_{K^+ | \kappa^+}   $
\\[0.3cm]
$d \bar{d} \to u \bar{u}$  & 
$ D_{11}^{P|S} + i D_{12}^{P|S} - i D_{21}^{P|S} + D_{22}^{P|S} = 2 D_{\pi^- | a_0^-}  $
\\[0.3cm]
$d \bar{d} \to d \bar{d}$  & 
$
\frac{2}{3} D_{00}^{P|S} - 2\sqrt{\frac{2}{3}} D_{03}^{P|S} +
\frac{2\sqrt{2}}{3} D_{08}^{P|S} + D_{33}^{P|S} - \frac{2}{\sqrt{3}} D_{38}^{P|S} + \frac{1}{3} D_{88}^{P|S}
$   
\\[0.3cm]
$d \bar{d} \to s \bar{s}$  & 
$ D_{66}^{P|S} - i D_{67}^{P|S} + i D_{76}^{P|S} + D_{77}^{P|S} = 2 D_{K^0 | \kappa^0}   $
\\[0.3cm]
$s \bar{s} \to u \bar{u}$  & 
$  D_{44}^{P|S} + i D_{45}^{P|S} - i D_{54}^{P|S} + D_{55}^{P|S} = 2 D_{K^- | \kappa^-} 	  $
\\[0.3cm]
$s \bar{s} \to d \bar{d}$  & 
$ D_{66}^{P|S} + i D_{67}^{P|S} - i D_{76}^{P|S} + D_{77}^{P|S} = 2 D_{\bar{K^0} | \bar{\kappa^0} }   $
\\[0.3cm]
$s \bar{s} \to s \bar{s}$  & 
$
\frac{2}{3} D_{00}^{P|S} - \frac{4\sqrt{2}}{3} D_{08}^{P|S} + \frac{4}{3} D_{88}^{P|S}
$  
\\
\bottomrule
\end{tabular}
\end{center}
\caption{$t-$channel propagator for each particular quark-antiquark process.}
\label{t_channel_prop_qqbar}
\end{table*}

\begin{table*}[t!]
\begin{center}
\begin{tabular}{@{}lc@{}} 
\toprule
$12 \to 34$ process \hspace{1cm} 
& 
$ 
\qty(
\TCoeff{a}{f_4}{f_1}
D_{ab}^{P|S} [u]
\TCoeff{b}{f_3}{f_2}
)
$
\\
\midrule
$ud \to ud$
& 
$ D_{11}^{P|S} - i D_{12}^{P|S} + i D_{21}^{P|S} + D_{22}^{P|S} = 2 D_{\pi^+ | a_0^+}   $
\\[0.3cm]
$du \to du$   
& 
$ D_{11}^{P|S} + i D_{12}^{P|S} - i D_{21}^{P|S} + D_{22}^{P|S} = 2 D_{\pi^- | a_0^-}   $
\\[0.3cm]
$us \to us$  
& 
$ D_{44}^{P|S} - i D_{45}^{P|S} + i D_{54}^{P|S} + D_{55}^{P|S} = 2 D_{K^+ | \kappa^+}   $
\\[0.3cm]
$su \to su$  
& 
$ D_{44}^{P|S} + i D_{45}^{P|S} - i D_{54}^{P|S} + D_{55}^{P|S} = 2 D_{K^- | \kappa^-}   $
\\[0.3cm]
$ds \to ds$  
& 
$ D_{66}^{P|S} - i D_{67}^{P|S} + i D_{76}^{P|S} + D_{77}^{P|S} = 2 D_{K^0 | \kappa^0}   $
\\[0.3cm]
$sd \to sd$  
& 
$ D_{66}^{P|S} + i D_{67}^{P|S} - i D_{76}^{P|S} + D_{77}^{P|S} = 2 D_{\bar{K^0} | \bar{\kappa^0} }   $
\\[0.3cm]
$uu \to uu$  
& 
$
\frac{2}{3} D_{00}^{P|S} + 2\sqrt{\frac{2}{3}} D_{03}^{P|S} +
\frac{2\sqrt{2}}{3} D_{08}^{P|S} + D_{33}^{P|S} + \frac{2}{\sqrt{3}} D_{38}^{P|S} + \frac{1}{3} D_{88}^{P|S}
$ 
\\[0.3cm]
$dd \to dd$  
& 
$
\frac{2}{3} D_{00}^{P|S} - 2\sqrt{\frac{2}{3}} D_{03}^{P|S} +
\frac{2\sqrt{2}}{3} D_{08}^{P|S} + D_{33}^{P|S} - \frac{2}{\sqrt{3}} D_{38}^{P|S} + \frac{1}{3} D_{88}^{P|S}
$   
\\[0.3cm]
$ss \to ss$  
& 
$
\frac{2}{3} D_{00}^{P|S} - \frac{4\sqrt{2}}{3} D_{08}^{P|S} + \frac{4}{3} D_{88}^{P|S}
$  
\\
\bottomrule
\end{tabular}
\end{center}
\caption{$u-$channel propagator for each particular particular quark-quark process.}
\label{u_channel_prop_qq}
\end{table*}

\begin{table*}[t!]
\begin{center}
\begin{tabular}{@{}lc@{}} 
\toprule
$12 \to 34$ process \hspace{1cm} & 
$ 
\qty(
\TCoeff{a}{f_2}{f_1}
D_{ab}^{P|S} [ s ] 
\TCoeff{b}{f_3}{f_4}
)
$
\\
\midrule
$u \bar{d} \to u \bar{d}$  &
$ D_{11}^{P|S} - i D_{12}^{P|S} + i D_{21}^{P|S} + D_{22}^{P|S} = 2 D_{\pi^+ | a_0^+}   $
\\[0.3cm]
$u \bar{s} \to u \bar{s}$  & 
$ D_{44}^{P|S} - i D_{45}^{P|S} + i D_{54}^{P|S} + D_{55}^{P|S} = 2 D_{K^+ | \kappa^+}   $
\\[0.3cm]
$d \bar{s} \to d \bar{s}$  & 
$ D_{66}^{P|S} - i D_{67}^{P|S} + i D_{76}^{P|S} + D_{77}^{P|S} = 2 D_{K^0 | \kappa^0}   $
\\[0.3cm]
$d \bar{u} \to d \bar{u}$  & 
$ D_{11}^{P|S} + i D_{12}^{P|S} - i D_{21}^{P|S} + D_{22}^{P|S} = 2 D_{\pi^- | a_0^-}  $
\\[0.3cm]
$s \bar{u} \to s \bar{u}$  & 
$  D_{44}^{P|S} + i D_{45}^{P|S} - i D_{54}^{P|S} + D_{55}^{P|S} = 2 D_{K^- | \kappa^-} 	  $
\\[0.3cm]
$s \bar{d} \to s \bar{d}$  & 
$ D_{66}^{P|S} + i D_{67}^{P|S} - i D_{76}^{P|S} + D_{77}^{P|S} = 2 D_{\bar{K^0} | \bar{\kappa^0} }   $
\\[0.3cm]
$u \bar{u} \to u \bar{u}$  & 
$
\frac{2}{3} D_{00,t}^{P|S} + 2\sqrt{\frac{2}{3}} D_{03,t}^{P|S} +
\frac{2\sqrt{2}}{3} D_{08,t}^{P|S} + D_{33,t}^{P|S} + \frac{2}{\sqrt{3}} D_{38,t}^{P|S} + \frac{1}{3} D_{88,t}^{P|S}
$   
\\[0.3cm]
$u \bar{u} \to d \bar{d}$ , $d \bar{d} \to u \bar{u}$  & 
$
\frac{2}{3} D_{00}^{P|S} + \frac{2\sqrt{2}}{3} D_{08}^{P|S} - D_{33}^{P|S} + \frac{1}{3} D_{88}^{P|S}
$  
\\[0.3cm]
$u \bar{u} \to s \bar{s}$ , $s \bar{s} \to u \bar{u}$  & 
$
\frac{2}{3} D_{00}^{P|S} + \sqrt{\frac{2}{3}} D_{03}^{P|S} - \frac{\sqrt{2}}{3} D_{08}^{P|S} - \frac{2}{\sqrt{3}} D_{38}^{P|S} - \frac{2}{3} D_{88}^{P|S}
$ 
\\[0.3cm]
$d \bar{d} \to d \bar{d}$  & 
$
\frac{2}{3} D_{00}^{P|S} - 2\sqrt{\frac{2}{3}} D_{03}^{P|S} +
\frac{2\sqrt{2}}{3} D_{08}^{P|S} + D_{33}^{P|S} - \frac{2}{\sqrt{3}} D_{38}^{P|S} + \frac{1}{3} D_{88}^{P|S}
$   
\\[0.3cm]
$d \bar{d} \to s \bar{s}$ , $s \bar{s} \to d \bar{d}$  & 
$
\frac{2}{3} D_{00}^{P|S} - \sqrt{\frac{2}{3}} D_{03}^{P|S} -
\frac{\sqrt{2}}{3} D_{08}^{P|S} + \frac{2}{\sqrt{3}} D_{38}^{P|S} - \frac{2}{3} D_{88}^{P|S}
$   
\\[0.3cm]
$s \bar{s} \to s \bar{s}$  & 
$
\frac{2}{3} D_{00}^{P|S} - \frac{4\sqrt{2}}{3} D_{08}^{P|S} + \frac{4}{3} D_{88}^{P|S}
$   
\\
\bottomrule
\end{tabular}
\end{center}
\caption{$s-$channel propagator for each particular quark-antiquark process.}
\label{s_channel_prop_qqbar}
\end{table*}

Starting with the quark-quark scatterings processes, the matrix elements in the $t$- and $u$-channels are formally given by  \cite{Zhuang:1995uf,Rehberg:1996vd}:
\begin{align*}
-i\mathcal{M}_t^{12 \to 34}
& = 
\delta_{c_1 c_3}
\delta_{c_2 c_4}
\bigg\{
\Big( \bar{u} [p_3] \TCoeff{a}{f_3}{f_1} u [p_1] \Big) 
\,
i D_{ab}^S [ t;  p_1-p_3] 
\,
\Big( \bar{u} [p_4]  \TCoeff{b}{f_4}{f_2} u [p_2] \Big)
\\
&
\qquad \qquad \qquad \qquad 
+
\Big( \bar{u} [p_3] i \gamma_5 \TCoeff{a}{f_3}{f_1} u [p_1] \Big)
\,
i D_{ab}^P [t; p_1-p_3] 
\,
\Big( \bar{u} [p_4] i \gamma_5 \TCoeff{b}{f_4}{f_2} u [p_2] \Big) 
\bigg\} ,
\numberthis
\\
-i\mathcal{M}_u^{12 \to 34}
& = 
\delta_{c_1 c_4}
\delta_{c_2 c_3}
\bigg\{
\Big( \bar{u} [p_4] \TCoeff{a}{f_4}{f_1} u [p_1] \Big)
\,
i D_{ab}^S [u; p_1-p_4] 
\,
\Big( \bar{u} [p_3] \TCoeff{b}{f_3}{f_2} u [p_2] \Big)
\\
&
\qquad \qquad \qquad \qquad 
+
\Big( \bar{u} [p_4] i \gamma_5 \TCoeff{a}{f_4}{f_1} u [p_1] \Big)
\,
i D_{ab}^P [u; p_1-p_4] 
\,
\Big( \bar{u} [p_3] i \gamma_5 \TCoeff{b}{f_3}{f_2} u [p_2] \Big) 
\bigg\} .
\numberthis
\end{align*}
Here, $u[p_i]$, $c_i$ and $f_i$ refer to the spinor, color and flavor of the respective $i-$th particle, with momenta $p_i$. The quantity $\TCoeff{a}{f_i}{f_j} = q_{f_i}^T \lambda_a q_{f_j} $, with $q_{f_i}^T = \qty{ \delta_{u f_i}, \delta_{d f_i}, \delta_{s f_i} }$, a vector that represents the flavor component of the Dirac spinor of the particle with momentum $p_i$. Following the calculation performed in Ref. \cite{Rehberg:1996vd}, one can arrive at the following matrix elements squared:
\begin{align*}
\frac{1}{4 N_c^2} \sum_{s,c} \abs{\mathcal{M}_t^{12 \to 34}}^2
& =
\abs{
\TCoeff{a}{f_3}{f_1}
D_{ab}^S [ t ] 
\TCoeff{b}{f_4}{f_2}
}^2
t^+_{13} t^+_{24}
+
\abs{
\TCoeff{a}{f_3}{f_1}
D_{ab}^P [ t ] 
\TCoeff{b}{f_4}{f_2}
}^2
t^-_{13} t^-_{24} ,
\numberthis
\\
\frac{1}{4 N_c^2} \sum_{s,c} \abs{\mathcal{M}_u^{12 \to 34}}^2
& =
\abs{
\TCoeff{a}{f_4}{f_1}
D_{ab}^S [ u ] 
\TCoeff{b}{f_3}{f_2}
}^2
u^+_{14} u^+_{23}
+
\abs{
\TCoeff{a}{f_4}{f_1}
D_{ab}^P [ u ] 
\TCoeff{b}{f_3}{f_2}
}^2
u^-_{14} u^-_{23} ,
\numberthis
\\
\frac{1}{4 N_c^2} \sum_{s,c} \mathcal{M}_t^{12 \to 34} {\mathcal{M}_u^{12 \to 34}}^*
& =\frac{1}{ 4 N_c }
\qty(
\TCoeff{a}{f_3}{f_1}
D_{ab}^S [ t ] 
\TCoeff{b}{f_4}{f_2}
)
\qty(
\TCoeff{a}{f_4}{f_1}
D_{ab}^S [ u ] 
\TCoeff{b}{f_3}{f_2}
)^*
\qty(
t^+_{13} t^+_{24}
- s^+_{12} s^+_{34}
+ u^+_{14} u^+_{23}
)
\\
& 
-
\frac{1}{ 4 N_c }
\qty(
\TCoeff{a}{f_3}{f_1}
D_{ab}^S [ t ] 
\TCoeff{b}{f_4}{f_2}
)
\qty(
\TCoeff{a}{f_4}{f_1}
D_{ab}^P [ u ] 
\TCoeff{b}{f_3}{f_2}
)^* 
\qty(
t^+_{13} t^+_{24}
- s^-_{12} s^-_{34}
+ u^-_{14} u^-_{23}
)
\\
& 
-
\frac{1}{ 4 N_c }
\qty(
\TCoeff{a}{f_3}{f_1}
D_{ab}^P [ t ] 
\TCoeff{b}{f_4}{f_2}
)
\qty(
\TCoeff{a}{f_4}{f_1}
D_{ab}^S [ u ] 
\TCoeff{b}{f_3}{f_2}
)^* 
\qty(
t^-_{13} t^-_{24}
- s^-_{12} s^-_{34}
+ u^+_{14} u^+_{23}
)
\\
& 
+
\frac{1}{ 4 N_c }
\qty(
\TCoeff{a}{f_3}{f_1}
D_{ab}^P [ t ] 
\TCoeff{b}{f_4}{f_2}
)
\qty(
\TCoeff{a}{f_4}{f_1}
D_{ab}^P [ u ] 
\TCoeff{b}{f_3}{f_2}
)^* 
\qty(
t^-_{13} t^-_{24}
- s^+_{12} s^+_{34}
+ u^-_{14} u^-_{23}
).
\numberthis
\end{align*}
Here, the sum is to be made over spin and color degrees of freedom. The quantities $s_{ij}^\pm$, $t_{ij}^\pm$ and $u_{ij}^\pm$ are given by:
\begin{align}
x^\pm_{ij}
& =
x - \qty( m_i \pm m_j )^2 ,
\label{xijPMdefinition}
\end{align}
with $x$ the appropriate Mandelstam variable, i.e., $x=s,t,u$. The propagators, in a given channel, are evaluated with the appropriate momenta using Eqs. (\ref{k0_uChannel})-(\ref{absk_uChannel}): $D_{ab}^{P|S} \qty[ t ] = D_{ab}^{P|S} \qty[ k_0^{(t)} , \abs{ \vec{k} }^{(t)} ]$ and $D_{ab}^{P|S} \qty[ u ]=D_{ab}^{P|S} \qty[ k_0^{(u)} , \abs{ \vec{k} }^{(u)} ]$.

For a given quark-quark scattering process, one can use the above equations in order to calculate the squared matrix elements $\abs{\mathcal{M}_t}^2$, $\abs{\mathcal{M}_u}^2$ and the interference term $\mathcal{M}_t \mathcal{M}_u^*$. To this end, for a particular quark-quark process, one must evaluate the different scalar and pseudoscalar propagators which are involved in such scattering. For the $t$-channel this means calculating $\TCoeff{a}{f_3}{f_1}
D_{ab}^{P|S} [t]
\TCoeff{b}{f_4}{f_2}$ while, for the $u$-channel, one has to calculate $\TCoeff{a}{f_4}{f_1}
D_{ab}^{P|S} [u]
\TCoeff{b}{f_3}{f_2}$. As it is evident, these quantities are process dependent, since one must specify the flavors of the incoming and outgoing particles. In Tables \ref{t_channel_prop_qq} and \ref{u_channel_prop_qq} we show the explicit form of these propagators for all quark-quark processes in the $t$- and $u$-channels, respectively.

Moving to the quark-antiquark scatterings processes, the matrix elements in the $s$- and $t$-channels are formally given by  \cite{Zhuang:1995uf,Rehberg:1996vd}:
\begin{align*}
-i\mathcal{M}_s^{12 \to 34} 
& = 
\delta_{c_1 c_2}
\delta_{c_3 c_4}
\sum_{a,b}
\bigg\{
\Big( \bar{v} [p_2] \TCoeff{a}{f_2}{f_1} u [p_1] \Big) 
\,
i D_{ab}^S [ s;  p_1 + p_2 ] 
\,
\Big( \bar{u} [p_3] \TCoeff{b}{f_3}{f_4} v [p_4] \Big)
\\
&
\qquad \qquad \qquad \qquad 
+
\Big( \bar{v} [p_2] i \gamma_5 \TCoeff{a}{f_2}{f_1} u [p_1] \Big)
\,
i D_{ab}^P [ s; p_1 + p_2 ]  
\,
\Big( \bar{u} [p_3] i \gamma_5 \TCoeff{b}{f_3}{f_4} v [p_4] \Big) 
\bigg\} ,
\numberthis
\\
-i\mathcal{M}_t^{12 \to 34}
& = 
\delta_{c_1 c_3}
\delta_{c_2 c_4}
\sum_{a,b}
\bigg\{
\Big( \bar{u} [p_4] \TCoeff{a}{f_4}{f_1} u [p_1] \Big) 
\,
i D_{ab}^S [ t;  p_1-p_3] 
\,
\Big( \bar{v} [p_3] \TCoeff{b}{f_3}{f_2} v [p_2] \Big)
\\
&
\qquad \qquad \qquad \qquad 
+
\Big( \bar{u} [p_4] i \gamma_5 \TCoeff{a}{f_4}{f_1} u [p_1] \Big)
\,
i D_{ab}^P [t; p_1-p_3] 
\,
\Big( \bar{v} [p_3] i \gamma_5 \TCoeff{b}{f_3}{f_2} v [p_2] \Big) 
\bigg\}.
\numberthis
\end{align*}
Here, the notations are the same as for the quark-quark scatterings with the addition that $v$ is the antiquark spinor. Again, following Ref. \cite{Rehberg:1996vd}, one can arrive at:
\begin{align*}
\frac{1}{4 N_c^2} \sum_{s,c} \abs{\mathcal{M}_s^{12 \to 34}}^2
& =
\abs{
\TCoeff{a}{f_2}{f_1}
D_{ab}^S [ s ] 
\TCoeff{b}{f_3}{f_4}
}^2
s^+_{12} s^+_{34}
+
\abs{
\TCoeff{a}{f_2}{f_1}
D_{ab}^P [ s ] 
\TCoeff{b}{f_3}{f_4}
}^2
s^-_{12} s^-_{34} ,
\numberthis
\\
\frac{1}{4 N_c^2} \sum_{s,c} \abs{\mathcal{M}_t^{12 \to 34}}^2
& =
\abs{
\TCoeff{a}{f_4}{f_1}
D_{ab}^S [ t ] 
\TCoeff{b}{f_3}{f_2}
}^2
t^+_{13} t^+_{24}
+
\abs{
\TCoeff{a}{f_4}{f_1}
D_{ab}^P [ t ] 
\TCoeff{b}{f_3}{f_2}
}^2
t^-_{13} t^-_{24} ,
\numberthis
\\
\frac{1}{4 N_c^2} \sum_{s,c} \mathcal{M}_s^{12 \to 34} {\mathcal{M}_t^{12 \to 34}}^*
& =\frac{1}{ 4 N_c }
\qty(
\TCoeff{a}{f_2}{f_1}
D_{ab}^S [ s ] 
\TCoeff{b}{f_3}{f_4}
)
\qty(
\TCoeff{a}{f_4}{f_1}
D_{ab}^S [ t ] 
\TCoeff{b}{f_3}{f_2}
)^*
\qty(
s^+_{12} s^+_{34}
- u^+_{14} u^+_{23}
+ t^+_{13} t^+_{24}
)
\\
& 
-
\frac{1}{ 4 N_c }
\qty(
\TCoeff{a}{f_2}{f_1}
D_{ab}^S [ s ] 
\TCoeff{b}{f_3}{f_4}
)
\qty(
\TCoeff{a}{f_4}{f_1}
D_{ab}^P [ t ] 
\TCoeff{b}{f_3}{f_2}
)^* 
\qty(
s^+_{12} s^+_{34}
- u^-_{14} u^-_{23}
+ t^-_{13} t^-_{24}
)
\\
& 
-
\frac{1}{ 4 N_c }
\qty(
\TCoeff{a}{f_2}{f_1}
D_{ab}^P [ s ] 
\TCoeff{b}{f_3}{f_4}
)
\qty(
\TCoeff{a}{f_4}{f_1}
D_{ab}^S [ t ] 
\TCoeff{b}{f_3}{f_2}
)^* 
\qty(
s^-_{12} s^-_{34}
- u^-_{14} u^-_{23}
+ t^+_{13} t^+_{24}
)
\\
& 
+
\frac{1}{ 4 N_c }
\qty(
\TCoeff{a}{f_2}{f_1}
D_{ab}^P [ s ] 
\TCoeff{b}{f_3}{f_4}
)
\qty(
\TCoeff{a}{f_4}{f_1}
D_{ab}^P [ t ] 
\TCoeff{b}{f_3}{f_2}
)^* 
\qty(
s^-_{12} s^-_{34}
- u^+_{14} u^+_{23}
+ t^-_{13} t^-_{24}
).
\numberthis
\end{align*}
As before, the quantities $\TCoeff{a}{f_2}{f_1}
D_{ab}^{P|S} [ s ] 
\TCoeff{b}{f_3}{f_4}$ and $\TCoeff{a}{f_4}{f_1}
D_{ab}^{P|S} [ t ] 
\TCoeff{b}{f_3}{f_2}$ have to be calculated for each quark-antiquark process and $D_{ab}^{P|S} \qty[ s ]=D_{ab}^{P|S} \qty[ \sqrt{s} , 0 ]$, $D_{ab}^{P|S} \qty[ t ]=D_{ab}^{P|S} \qty[ k_0^{(t)} , \abs{ \vec{k} }^{(t)} ]$. The results of such calculations are shown in Tables \ref{t_channel_prop_qqbar} and \ref{s_channel_prop_qqbar} for all processes in the $t$- and $s$-channels, respectively.

In the case of the quark-quark, (or antiquark-antiquark) scatterings in the $t$-channel (see Table \ref{t_channel_prop_qq} for quark-quark processes), since the incoming and outgoing particles are the same ($1 \leftrightarrow 3$ and $2 \leftrightarrow 4$), from conservation of electric and flavor charges at the vertexes we can deduce that a flavorless quark-antiquark combination of the same flavor is being exchanged. This means that  only mesons without electric charge and strangeness can be involved. Hence, one expects that only some combinations of the flavorless, neutral, pseudoscalar ($\eta$, $\eta'$ and $\pi_0$) and scalar ($\sigma$, $f_0$ and $a_0^0$) mesons will contribute to the quark-quark (antiquark-antiquark) scattering in the t-channel.

The same reasoning holds for the $t-$channel in quark-antiquark processes (see Table \ref{t_channel_prop_qqbar}) when the incoming and outgoing flavors remain unchanged. As there is direct correspondence ($1 \leftrightarrow 3$ and $2 \leftrightarrow 4$), with conservation of electric and flavor charges between incoming and outgoing particles,  the exchanged combination corresponds to a quark-antiquark pair of the same flavor and the exchanged meson must be flavorless. When the flavor of incoming and outgoing particles changes, which can only happen if the flavor of the two incoming is the same and different from that of the outgoing particles $f \bar{f} \to f\bar{f}'$, then the exchanged meson is given by $f\bar{f}'$ or $\bar{f}f'$, depending on the ordering.

In quark-quark (or antiquark-antiquark) processes in the $u-$channel (see Table \ref{u_channel_prop_qq} quark-quark processes) the exchanged meson, in the case of different flavors ($f f' \to f f'$), has a determined composition ($f\bar{f}'$ or $f'\bar{f}$ depending on the time ordering). In the case of equal flavors ($f f \to f f$) the same reasoning leads to a flavorless combination which corresponds to a mixture of the flavorless mesons.

For quark-antiquark processes in the $s-$channel (Table \ref{s_channel_prop_qqbar}) we have the case of equal flavors and an intermediate undetermined flavorless state (which implies a contribution coming from the exchange of all flavorless mesons). The outgoing particles can also be any flavorless combination (i.e. $f\bar{f} \to f'\bar{f'}$). In the case of different flavors the intermediate state is determined.

Finally, after evaluating the different matrix elements, one is able to calculate the differential cross sections for quark-quark and quark-antiquark processes. It is given by \cite{Zhuang:1995uf,Rehberg:1996vd,Soloveva:2020hpr}:
\begin{align*}
\dv{\sigma_{ 12 \to 34} }{t}
& =
\frac{1}{16\pi s^+_{12}s^+_{12}}
\frac{1}{ 4 N_c^2 }
\sum_{s,c}
\left|
\mathcal{M}_{s|u}^{12 \to 34} - \mathcal{M}_t^{12 \to 34}
\right|^2 .
\numberthis
\label{differentialCrossSectionDefinition}
\end{align*}
Here, $s^+_{12}$ can be calculated from Eq. (\ref{xijPMdefinition}). The total cross section in the medium is then given by \cite{Zhuang:1995uf,Rehberg:1996vd,Soloveva:2020hpr}:
\begin{align}
\sigma_{12 \to 34} 
\qty[T,\mu_u,\mu_d,\mu_s,s]
& =
\int^{t^+}_{t^-}\dd{t}
\dv{\sigma_{ 12 \to 34} }{t}
\qty(
1 - \nfermi\qty[ E_3\mp \mu_3, T]
)
\qty( 
1 - \nfermi\qty[ E_4\mp \mu_4, T]
).
\label{cross_section_definition}
\end{align}
Here, the quantities inside curved brackets represent the so-called Pauli blocking factor. This quantity is essential since it accounts for the occupation of particles in the outgoing state \cite{Zhuang:1995uf}. In this quantity, $\nfermi\qty[ E, T]$, is the Fermi-Dirac distribution, defined in Eq. (\ref{fermiDiracDistribution}). The $\mp$ signs in the Fermi distributions of the integrand refer to the particle and antiparticle nature of each outgoing particle (respectively). The integration limits can be calculated by finding the $t$ values corresponding to the maximum and minimum scattering angles. Hence, by considering $\cos[\Theta]=\pm1$ in Eq. (\ref{scatteringAngle}) and considering the physical cases in which $s \geq (m_1 + m_2)^2$ and $s \geq (m_3 + m_4)^2$, one can solve for $t^\pm$, to give:
\begin{align*}
t^\pm
& =
m_1^2 + m_3^2 - 
\frac{\qty( s + m_1^2 - m_2^2 )
\qty( s + m_3^2 - m_4^2 )}{2s} 
\pm 2\sqrt{
\frac{\qty(s + m_1^2 - m_2^2 )^2}{4s} - m_1^2
}
\sqrt{
\frac{\qty(s + m_3^2 - m_4^2 )^2}{4s} - m_3^2
}.
\numberthis
\label{crossSection_integration_endpoints}
\end{align*}
For cross sections corresponding to processes with identical incoming and outgoing particles ($uu \to uu$, $dd \to dd$ $ss \to ss$, $\bar{u}\,\bar{u} \to \bar{u}\,\bar{u}$, $\bar{d}\,\bar{d} \to \bar{d}\,\bar{d}$ and $\bar{s}\,\bar{s} \to \bar{s}\,\bar{s}$), this formula is not valid and one should use instead, $t^- = -\frac{1}{2}\qty( s - 4 m^2 )$ and $t^+=0$ \cite{Zhuang:1995uf}.

At this point we comment on the previously mentioned ``recipe invariance'' with respect to evaluating differential cross sections in the NJL model. In order to evaluate the $B_0$ function, necessary to calculate the meson propagators, we had to perform an infinitesimal shift in the complex plane, $\pm i\epsilon$ with $\epsilon>0$, deforming the contour of integration around singularities. As demonstrated in Section \ref{practical_implementation_B0}, taking the mass shift, $M^2 \to M^2-i\epsilon$, or the external momentum temporal component shift, $k_0 \to k_0 \pm i\epsilon$, amounts to different signs of the imaginary part of the $B_0$ function, see Eqs. (\ref{imagPartPair})$-$(\ref{momentumShiftChangeImag2}). Furthermore, it amounts to implying different symmetry properties, see Eqs (\ref{symm_symmetric}) and (\ref{symm_antisymmetric}). One can understand the ``recipe invariance'' of the differential cross sections by observing the expressions to calculate the different matrix elements squared, $\abs{\mathcal{M}_{s|u}}^2$, $\abs{\mathcal{M}_t}^2$, $\mathcal{M}_t \mathcal{M}_u^*$ and $\mathcal{M}_s \mathcal{M}_u^*$. In these equations, only the absolute value of meson propagators contribute. Even when dealing with the product between two different meson propagators, only the absolute value of the resulting complex number matters to the calculation. Since the meson propagators only acquire an imaginary part due to the imaginary part of the $B_0$ function, the sign or symmetry properties of the imaginary part of this function does not change the values of the matrix elements squared. In order to illustrate this point, for simplicity, lets consider the propagator of the positively charged pion, $D_{\pi^+}$. Using Eq. (\ref{chargedPionPropagator}), it can be written as:
\begin{align*}
D_{ \pi^+ }
& =
\frac
{ 
2 P_{11} - 8 P_{11}^2 
\qty( 
\Re[\Pi_{ud}^P] 
+ 
i \frac{N_c}{8\pi^2} \Delta_{ud} \Im[B_{0,ud}]
) 
}
{ 
\qty(1 - 4 P_{11} \Pi_{ud}^P)\qty(1 - 4 P_{11} \Pi_{ud}^P)^* 
} .
\numberthis
\end{align*}
Here, we have written the polarization function $\Pi_{ud}^P$ in terms of real and imaginary parts, $\Pi_{ud}^P=\Re[\Pi_{ud}^P] + i \Im[\Pi_{ud}^P]=\Re[\Pi_{ud}^P] - i \frac{N_c}{8\pi^2} \Delta_{ud} \Im[B_{0,ud}]$, with $\Delta_{ud}=( \qty( M_u - M_d )^2 - \qty( k_0 + \mu_u - \mu_d )^2 + \vec{k}^2 )
$ and $B_{0,ud}=B_0 \qty[M_u,M_d,T,\mu_u,\mu_d, k_0,\abs{\vec{k}}]$, see Eq. (\ref{scalarPseudoscalar_polarization_def}). The imaginary part of the propagator is:
\begin{align}
\Im[D_{\pi^+}]
& =
-
\qty(
\frac{N_c P_{11}^2 \Delta_{ud} }{\pi^2}
)
\frac
{ 
\Im[B_{0,ud}]
}
{ 
\qty(1 - 4 P_{11} \Pi_{ud}^P)\qty(1 - 4 P_{11} \Pi_{ud}^P)^* 
} .
\end{align}
Thus, the propagator only acquires a finite imaginary part due to the imaginary part of the $B_0$ function, as argued above. In the squared matrix elements, terms as $\qty|D_{\pi^+}|^2$ arise. This is a real quantity which can be written as:
\begin{align*}
\qty|D_{\pi^+}|^2
& = 
\qty(\Re \qty[D_{\pi^+}])^2
+
\qty(
\frac{N_c P_{11}^2 \Delta_{ud}  }{\pi^2}
)^2
\frac
{ 
\qty( \Im[B_{0,ud}] )^2
}
{ 
\qty(
\qty(1 - 4 P_{11} \Pi_{ud}^P)\qty(1 - 4 P_{11} \Pi_{ud}^P)^* 
)^2
} .
\numberthis
\end{align*}
The effect of choosing different shifting pole recipes amounts to a different overall sign of $\Im[B_{0,ud}]$. Since $\qty|D_{\pi^+}|^2$ only depends on the square of $\Im[B_{0,ud}]$, the overall sign of $\Im[B_{0,ud}]$ does not change the value of $\qty|D_{\pi^+}|^2$. Hence, as argued before, taking different recipes to shift the poles, does not change the values of the squared matrix elements and consequently of the (differential) cross sections. For other cases, not involving only charged meson propagators, the same reasoning and conclusions can be obtained, although leading to more extensive calculations.

In order to finish this section we would like to highlight on the importance of considering the regularization proposed in this work, in order to maintain symmetries of the system, specially the one related with changing the sign of the momentum's zero component, $k_0$, and of the chemical potentials of the two fermion line integral, $B_0$ (see Eqs. (\ref{symm_symmetric}) and (\ref{symm_antisymmetric})). 

Consider the quark-quark processes, $us \to us$ and $su \to su$. According to Table \ref{u_channel_prop_qq}, in the $u$-channel of the $us \to us$ scattering, the exchanged mesons are the pseudoscalar $K^+$ and the scalar $\kappa^+$ while, for $su \to su$ the exchanged mesons are their electrical negative versions, $K^-$ and $\kappa^-$. Of course, these particles constitute particle-antiparticle pairs, $K^+ / K^-$ and $\kappa^+ / \kappa^-$. For the $t$-channel, the exchanged mesons, in both processes, is some neutrally charged combination of the components $D_{ab}^{P|S}$, with $a,b={0,3,8}$. In this case, this chargeless meson is its own particle and antiparticle. Following the connection between these mesons, at finite temperature and zero chemical potential, one should be able to transform the propagators of each meson-antimeson pairs into each other by changing, in the meson propagators, the sign of the zero component of the momentum. At finite chemical potential this change must be accompanied by a change of the sign of the quarks chemical potential. Naturally, the two process mentioned above, $us \to us$ and $su \to su$, should have exactly the same differential cross sections, since the incoming and outgoing pairs of particles are equal. This is only possible if the ability to transform a meson in its corresponding antimeson, and vice-versa, is valid in the model calculation. This can be understood by explicitly writing the transition matrix elements squared for these two processes. For the $us \to us$ scattering, we write:
\begin{align*}
\frac{1}{4 N_c^2} \sum_{s,c} \abs{\mathcal{M}_t^{us \to us}}^2
& =
\abs{
D_{us}^{S}
}^2
t^+_{13} t^+_{24}
+
\abs{
D_{us}^{P}
}^2
t^-_{13} t^-_{24} ,
\numberthis
\\
\frac{1}{4 N_c^2} \sum_{s,c} \abs{\mathcal{M}_u^{us \to us}}^2
& =
\abs{
2 D_{\kappa^+}
}^2
u^+_{14} u^+_{23}
+
\abs{
2 D_{K^+}
}^2
u^-_{14} u^-_{23} ,
\numberthis
\\
\frac{1}{4 N_c^2} \sum_{s,c} \mathcal{M}_t^{us \to us} {\mathcal{M}_u^{us \to us}}^*
& =\frac{1}{ 4 N_c }
\qty(
D_{us}^{S}
)
\qty(
2 D_{\kappa^+}
)^*
\qty(
t^+_{13} t^+_{24}
- s^+_{12} s^+_{34}
+ u^+_{14} u^+_{23}
)
\\
& 
-
\frac{1}{ 4 N_c }
\qty(
D_{us}^{S}
)
\qty(
2 D_{K^+}
)^* 
\qty(
t^+_{13} t^+_{24}
- s^-_{12} s^-_{34}
+ u^-_{14} u^-_{23}
)
\\
& 
-
\frac{1}{ 4 N_c }
\qty(
D_{us}^{P}
)
\qty(
2 D_{\kappa^+}
)^* 
\qty(
t^-_{13} t^-_{24}
- s^-_{12} s^-_{34}
+ u^+_{14} u^+_{23}
)
\\
& 
+
\frac{1}{ 4 N_c }
\qty(
D_{us}^{P}
)
\qty(
2 D_{K^+}
)^* 
\qty(
t^-_{13} t^-_{24}
- s^+_{12} s^+_{34}
+ u^-_{14} u^-_{23}
).
\numberthis
\end{align*}
Here,
$D_{us}^{P|S}=\frac{2}{3} D_{00}^{P|S} + \sqrt{\frac{2}{3}} D_{03}^{P|S} -
\frac{\sqrt{2}}{3} D_{08}^{P|S} - \frac{2}{\sqrt{3}} D_{38}^{P|S} - \frac{2}{3} D_{88}^{P|S}$ and the propagators $D_{K^+}$ and $D_{\kappa^+}$ and the propagators for the propagators for $D_{K^+}$ and $D_{\kappa^+}$ are evaluated with the momenta, $\left.k_0^{(u)}\right|_{us \to us}
=
\qty( 
M_u^2 - 
M_s^2
)/ \sqrt{s} $ and $\left.\abs{ \vec{k} }^{(u)}\right|_{us \to us}$ (see Eqs. (\ref{k0_uChannel}) and (\ref{absk_uChannel})). Likewise, one can write the same quantities for the $su \to su$ scattering process:
\begin{align*}
\frac{1}{4 N_c^2} \sum_{s,c} \abs{\mathcal{M}_t^{su \to su}}^2
& =
\abs{
D_{us}^{S}
}^2
t^+_{13} t^+_{24}
+
\abs{
D_{us}^{P}
}^2
t^-_{13} t^-_{24} ,
\numberthis
\\
\frac{1}{4 N_c^2} \sum_{s,c} \abs{\mathcal{M}_u^{su \to su}}^2
& =
\abs{
2 D_{\kappa^-}
}^2
u^+_{14} u^+_{23}
+
\abs{
2 D_{K^-}
}^2
u^-_{14} u^-_{23} ,
\numberthis
\\
\frac{1}{4 N_c^2} \sum_{s,c} \mathcal{M}_t^{su \to su} {\mathcal{M}_u^{su \to su}}^*
& =\frac{1}{ 4 N_c }
\qty(
D_{us}^{S}
)
\qty(
2 D_{\kappa^-}
)^*
\qty(
t^+_{13} t^+_{24}
- s^+_{12} s^+_{34}
+ u^+_{14} u^+_{23}
)
\\
& 
-
\frac{1}{ 4 N_c }
\qty(
D_{us}^{S}
)
\qty(
2 D_{K^-}
)^* 
\qty(
t^+_{13} t^+_{24}
- s^-_{12} s^-_{34}
+ u^-_{14} u^-_{23}
)
\\
& 
-
\frac{1}{ 4 N_c }
\qty(
D_{us}^{P}
)
\qty(
2 D_{\kappa^-}
)^* 
\qty(
t^-_{13} t^-_{24}
- s^-_{12} s^-_{34}
+ u^+_{14} u^+_{23}
)
\\
& 
+
\frac{1}{ 4 N_c }
\qty(
D_{us}^{P}
)
\qty(
2 D_{K^-}
)^* 
\qty(
t^-_{13} t^-_{24}
- s^+_{12} s^+_{34}
+ u^-_{14} u^-_{23}
).
\numberthis
\end{align*}
In this case, the propagators $D_{K^-}$ and $D_{\kappa^-}$ are evaluated with the momenta, $\left.k_0^{(u)}\right|_{su \to su}
=
\qty( 
M_s^2 - 
M_u^2
)/\sqrt{s} $ and $\left.\abs{ \vec{k} }^{(u)}\right|_{su \to su}$. Since $\left.k_0^{(u)}\right|_{su \to su}=-\left.k_0^{(u)}\right|_{us \to us}$ (and the three momenta are identical, $\left.\abs{ \vec{k} }^{(u)}\right|_{us \to us}=\left.\abs{ \vec{k} }^{(u)}\right|_{su \to su}$) we can write, $D_{K^-} \qty[\left.\abs{ \vec{k} }^{(u)}\right|_{su \to su}]=D_{K^-} \qty[-\left.\abs{ \vec{k} }^{(u)}\right|_{us \to us}]$. From the Feynman diagrammatic point of view, a propagating particle, with a negative zero momentum component, is indistinguishable from its antiparticle, propagating with the same positive zero momentum component. Thus, we can substitute, in the above expression, $D_{K^-} \qty[\left.\abs{ \vec{k} }^{(u)}\right|_{su \to su}]=D_{K^+} \qty[\left.\abs{ \vec{k} }^{(u)}\right|_{us \to us}]$. The same can be performed for the $D_{\kappa^-}$ propagator. Hence, the set of matrix elements squared for the $us \to us$ and $su \to su$ processes are identical, as they should be. However, such is true, if and only if the particle/antiparticle symmetry is respected in the model calculation. In the model in question, this can only happen if the $B_0$ function is symmetric with respect to changing the sign of the zero component of the momentum alongside switching the masses and chemical potentials of the particles. As we discussed in Section \ref{two_fermion_one_loop_integral}, these are symmetries of the two fermion, one loop integral $B_0$ (see Eqs. (\ref{symmetryB0})$-$\ref{symm_antisymmetric}) however, they may be broken in the practical implementation of this function. In the usual 3-momentum regularization approach (see Ref. \cite{Rehberg:1995nr}), this symmetry is broken due to a shift in a variable applied to the integral which introduces a symmetry breaking term which is proportional to $1/\Lambda$ \cite{Rehberg:1996vd}. As discussed in Ref. \cite{Rehberg:1995nr}, this is a well known shortcoming of using the usual 3-momentum cutoff regularization in order to evaluate the two fermion, one loop integral $B_0$ with different quark masses, specially affecting the meson propagators in the kaonic sector. At zero chemical potential, this unwanted feature does not affect the pionic sector or the isosinglet mesons $\eta$ and $\eta'$ since their meson propagators are calculated in the isotopic limit with quark-antiquark polarizations with the degenerate masses, $\Pi_{uu}^{P|S}$, $\Pi_{dd}^{P|S}$, $\Pi_{ss}^{P|S}$, $\Pi_{ud}^{P|S}$ and $\Pi_{du}^{P|S}$. It also does not affect the evaluation of quantities at zero external 3-momentum, $\abs{\vec{k}}=0$, such as meson masses. In the evaluation of cross sections, however, this is not the case as the dependency at finite values of $\abs{\vec{k}}$ is extremely relevant.

At finite chemical potential, the pionic sector can also be affected by the breaking of this symmetry. This is the case if one considers systems with isospin asymmetry, such as neutron star matter, or systems with charge fractions $Y_Q=\rho_Q/\rho_B$ different than 0.5 (here, $\rho_B$ and $\rho_Q$ are the baryon and electrical charge densities, respectively). For instance, the case of $Y_Q=0.4$ is extremely important to study heavy-ion collisions and core-collapse supernova matter \cite{Costa:2020dgc} or the case of $Y_Q=0.0$, used to study neutron star matter \cite{glendenning2012compact,CamaraPereira:2016chj}.  Using the regularization proposed in this work however, fixes the problem and the particle/antiparticle symmetry is respected.

A different approach to evaluate the polarization function, without using the 3-momentum sphere intersection regularization introduced in this work, is to use the Pauli-Villars regularization. We leave the implementation of such regularization scheme, its implications on the calculation of cross sections in the NJL model and its comparison to the 3-momentum sphere intersection regularization as future work.

\subsection{Numerical results}

In this section we show the results of calculating the cross sections of some quark-quark and quark-antiquark scattering processes, considering the 3-momentum sphere intersection regularization procedure introduced in this work versus the usual 3-momentum regularization. The parameters set for the Lagrangian density defined in Eq. (\ref{NJL_Lagrangian}) are the same as in  Ref. \cite{Rehberg:1995kh} so as to reproduce the results from Ref. \cite{Rehberg:1996vd} when using the usual 3-momentum regularization.

The set of parameters used in this work can be found in Table \ref{ParameterSets}, including the light and strange quark current masses, $m_l=m_u=m_d$ and $m_s$, the 3-momentum cutoff, $\Lambda$ and the interaction couplings $G$ and $\kappa$. Solving the gap equations of the NJL model in the vacuum, for this parameter set, yields a light and strange effective masses, $M_l=M_u=M_d$, $M_s$, and light and strange quark condensates, $\expval*{\bar{\psi}_l \psi_l}^{1/3}$ and $\expval*{\bar{\psi}_s \psi_s}^{1/3}$. These results can be found in Table \ref{quarkQuantities}.


\begin{table*}[ht!]
\begin{center}
\begin{tabular*}{0.25\textwidth}{@{\extracolsep{\fill}}lr@{}}
\toprule
\multicolumn{1}{l}{Parameter} & 
\multicolumn{1}{r}{ Value } 
\\
\midrule
$m_l$ [MeV]  & $5.500$ 
\\
$m_s$ [MeV]  & $140.700$
\\
$\Lambda~\qty[\text{MeV}]$   & $602.300$ 
\\
$G \Lambda^2$   & $3.670$
\\
$\kappa \Lambda^5$   & $-12.360$
\\
$G~\left[\text{GeV}^{-2}\right]$   & $\phantom{+}10.117$
\\
$\kappa~\qty[\text{GeV}^{-5}]$   & $-155.939$ 
\\
\bottomrule
\end{tabular*}
\end{center}
\caption{
Parameter set: current masses of the light ($m_l=m_u=m_d$), and strange quarks ($m_s$), 3-momentum cutoff ($\Lambda$) and dimensionless couplings for the NJL model including the four quark interactions, ($G\Lambda^2$) and the 't Hooft determinant interaction ($\kappa\Lambda^5$).
} 
\label{ParameterSets}
\end{table*}

\begin{table*}[ht!]
\begin{center}
\begin{tabular*}{0.25\textwidth}{@{\extracolsep{\fill}}lr@{}}
\toprule
\multicolumn{1}{l}{Quantity} & 
\multicolumn{1}{r}{ Value [MeV]} 
\\
\midrule
$M_l$  & $\phantom{+}367.648$ 
\\
$M_s$  & $\phantom{+}549.479$
\\
$ \expval*{\bar{\psi}_l \psi_l}^{1/3}$   & $-241.946$ 
\\
$ \expval*{\bar{\psi}_s \psi_s}^{1/3}$   & $-257.688$
\\
\bottomrule
\end{tabular*}
\end{center}
\caption{
Vacuum observables for the parameter set defined in Table \ref{ParameterSets}: effective quark masses, $M_l$, $M_s$ and quark condensates, $\expval*{\bar{\psi}_l \psi_l}^{1/3}$ (light) and $\expval*{\bar{\psi}_s \psi_s}^{1/3}$ (strange), in the vacuum.
} 
\label{quarkQuantities}
\end{table*}

The pseudoscalar and scalar meson masses, in the vacuum, can also be calculated for the chosen parameter set. To evaluate the mass for a given meson, we search for the $k_0$ value for which the respective inverse meson propagator is zero, at the center of mass of the meson, i.e., $D_{M}^{-1}\qty[k_0, \abs{\vec{k}}=0]=0$. In order to allow for unbound meson states with a finite width, $\Gamma$, we search for solutions of complex $k_0$. As usual this complex pole solution is decomposed as $k_0=M_M - i \frac{\Gamma}{2}$, with real values of $M_M$ and $\Gamma$. Thus, for the charged pions, for example, one has to find the pair of values $(M_M,\Gamma_M)$ for which the real and imaginary parts of the denominator of Eq. (\ref{chargedPionPropagator}) are zero. The pseudoscalar and scalar meson masses, and respective widths, can be found in Table \ref{MesonQuantities}. The leptonic decays of the pion and kaon, $f_{\pi^\pm}$ and $f_{K^\pm}$, can be estimated in the model following the procedure outlined, for instance, in Ref. \cite{Klevansky:1992qe}, yielding $ f_{\pi^\pm} = 92.39~\text{MeV}$ and $ f_{K^\pm} = 95.23 ~\text{MeV}$.

The masses of the pion and kaon mesons are well reproduced in this parametrization of the model (see Table \ref{MesonQuantities}). This should come as unsurprising since the parameter choice is determined by the imposition of the values for the masses of the pion and kaon along with the mass of the $\eta'$ and the pion leptonic decay, while keeping the light quarks current mass fixed at  5.5 MeV \cite{Rehberg:1995kh}. The inability of the model to provide an estimation closer to the empirical value of the ratio between the two leptonic decays (which is found to be $f_{\pi^\pm}/f_{K^\pm} = 1.1928(26)$ \cite{Tanabashi:2018oca})  may be improved upon by introducing explicit chiral symmetry breaking interactions \cite{Moreira:2014qna}.

From these results we can also observe that the pions, kaons, $\eta$ and $\sigma$ mesons are bound states in the vacuum ($\Gamma$ is numerically compatible with zero, $\Gamma \sim 0.00$) while, the remaining ones are unbounded, i.e., they display a finite width, $\Gamma_M$. Of course the in-medium behavior of the meson states at finite temperature and chemical potential can be very intricate and bound states can become unbounded and vice-versa. For phenomenological studies of the in-medium meson behavior, see \cite{Rehberg:1995kh,Costa:2003uu,Costa:2005cz,Nagahiro:2006dr,Costa:2008dp,Contrera:2009hk,Inagaki:2010nb,Blanquier:2014kja,Blaschke:2016sqn,Costa:2019bua}.

\begin{table*}[ht!]
\begin{center}
\begin{tabular*}{0.35\textwidth}{@{\extracolsep{\fill}}lcr@{}}
\toprule
\multicolumn{1}{l}{Meson, $M$} & 
\multicolumn{1}{c}{ $M_M$ [MeV]} &
\multicolumn{1}{r}{ $\Gamma_M$ [MeV]} 
\\
\midrule
$\pi^\pm,\pi^0$ & $\phantom{0}134.97$ & $\sim 0.00$ 
\\
$K^\pm,K^0,\bar{K^0}$  & $\phantom{0}497.67$ & $\sim 0.00$
\\
$ \eta $   & $\phantom{0}514.76$ & $\sim 0.00$ 
\\
$ \eta'$   & $\phantom{0}957.74$ & $268.52$
\\
$a_0^\pm, a_0^0$  & $\phantom{0}880.08$ & $251.89$
\\
$\kappa^\pm,\kappa^0,\bar{\kappa^0}$  & $1050.37$ & $240.57$
\\
$ \sigma $   & $\phantom{0}728.77$ & $\sim 0.00$ 
\\
$ f_0$   & $1198.25$ & $192.08$
\\
\bottomrule
\end{tabular*}
\end{center}
\caption{
Meson masses, $M_M$, and meson widths, $\Gamma_M$, in the vacuum, for the parameter set defined in Table \ref{ParameterSets}. For the mesons which display a width numerically compatible with $0$, we write $\sim 0.00$. 
} 
\label{MesonQuantities}
\end{table*}

As a matter of fact, a bound state can effectively melt at finite temperature. In the literature, the temperature at which this occurs, for a given meson, is termed the meson Mott temperature, $T_M^{\mathrm{Mott}}$ \cite{Costa:2019bua,Blaschke:2016sqn}. In the NJL model this temperature can be used as an effective description of a deconfined state \cite{Rehberg:1996vd}: if all meson states are unbounded the system should be in a regime where quark-quark and quark-antiquark scatterings can occur \cite{Rehberg:1996vd,Rehberg:1995kh}. In this work, following other works \cite{Rehberg:1996vd}, we will take the Mott temperature of the charged pions, $T_{\pi^\pm}^{\mathrm{Mott}}$, as an indications that it is safe to use the NJL model to evaluate the cross sections. We use use the Mott temperature of the pion because it corresponds to the lightest particle in the spectrum, representing the most stable state. In an improved model, one could use the Polyakov loop as an order parameter for the deconfining phase transition instead of using the Mott temperatures. The Polyakov loop can be incorporated in the NJL model by coupling the quark fields to a gluonic background in the temporal direction. Such extended model is called the Polyakov$-$Nambu$-$Jona-Lasinio model and it has been extensively used in the literature to explore the QCD phase diagram. Indeed, in Ref. \cite{Soloveva:2020hpr}, the Polyakov loop was partially used to study the quark-quark and quark-antiquark scattering at finite temperature and chemical potential. In such work, the Polyakov loop was only considered in the evaluation of the quark effective masses but its contribution to the quark-antiquark polarization functions was not considered. For instance in Ref. \cite{Costa:2019bua}, the meson behavior considering the effect of the Polyakov loop in the quark-antiquark polarization function was studied.

Since the quark-quark and quark-antiquark cross sections are highly dependent on the meson propagators, the incorporation of the Polyakov loop dynamics, at this level, can be very important to get better qualitative predictions from the calculation. The formalism developed here to evaluate the two fermion line integral ($B_0$), including the 3-momentum sphere intersection regularization, can readily be used to include the effect of a static background of gluons at the Lagrangian level. This will be done in future works.

\begin{table*}[ht!]
\begin{center}
\begin{tabular*}{0.25\textwidth}{@{\extracolsep{\fill}}lr@{}}
\toprule
\multicolumn{1}{l}{Meson, $M$} & 
\multicolumn{1}{r}{ $T^{\mathrm{Mott}}_M$ [MeV]} 
\\
\midrule
$\pi^\pm,\pi^0$ & $211.57$ 
\\
$K^\pm,K^0,\bar{K^0}$  & $210.02$ 
\\
$ \eta $   & $179.19$ 
\\
$ \eta'$   & $-$
\\
$a_0^\pm, a_0^0$  & $-$ 
\\
$\kappa^\pm,\kappa^0,\bar{\kappa^0}$  & $-$ 
\\
$ \sigma $   & $184.72$ 
\\
$ f_0$   & $-$ 
\\
\bottomrule
\end{tabular*}
\end{center}
\caption{
Mott temperatures for the meson states with zero width at zero chemical potential. Mesons that are resonant states at zero chemical potential are marked with $-$.
} 
\label{MottTemperatures}
\end{table*}

The reason behind the choice of those two values of temperature ($T=215$ MeV and $T=250$ MeV) is, first and foremost, for comparison purposes: this exact calculation was performed in Ref. \cite{Rehberg:1995nr}, allowing us to directly compare our calculation with results found in the literature. These values of temperature are slightly above the Mott temperature of the pions and kaons, the lightest meson states in this model. This ensures the validity of the calculation in a region expected to correspond to a deconfined state of matter. The Mott temperature, at zero chemical potential, of all meson states can be found in Table. \ref{MottTemperatures}.

In all figures concerning cross section for different processes, the minimum center of mass energy will be signaled by colored dots. Naturally, for a given process $12 \to 34$, such energy is given by the smallest of the two quantities, $s_{\mathrm{min}}= \mathrm{min} \qty[ (m_1+m_2)^2,(m_3+m_4)^2]$. The maximum center of mass energy, on the other hand, is imposed by considering the maximum possible energy for the process involving the lightest quarks in the system, with the highest possible 3-momentum, $s_\mathrm{max} = (\sqrt{\Lambda^2 + M_l^2}+\sqrt{\Lambda^2 + M_l^2})^2= 4\qty(\Lambda^2 + M_l^2)$ \cite{Rehberg:1996vd}. In our model calculations, with isospin symmetry and equal chemical potentials, this mass corresponds to the effective up and down quark masses, $M_l=M_u=M_d$.

\subsubsection{Comparing regularizations procedures at finite temperature and zero chemical potential}

\begin{figure*}[t!]
\begin{subfigure}[b]{0.4\textwidth}
\includegraphics[width=\textwidth]
{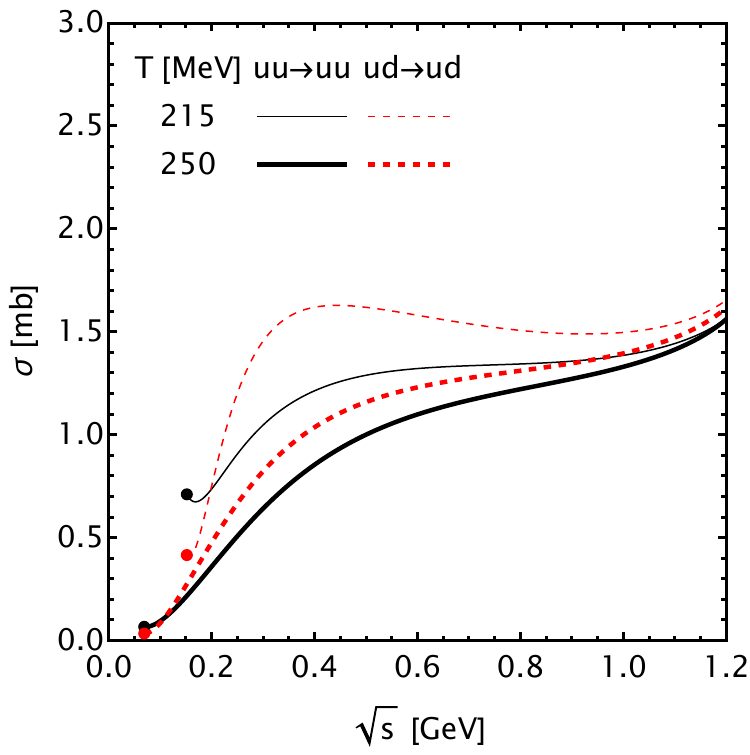}
\caption{}
\label{sigma_uuuu_udud_Klev}
\end{subfigure}
\begin{subfigure}[b]{0.4\textwidth}
\includegraphics[width=\textwidth]
{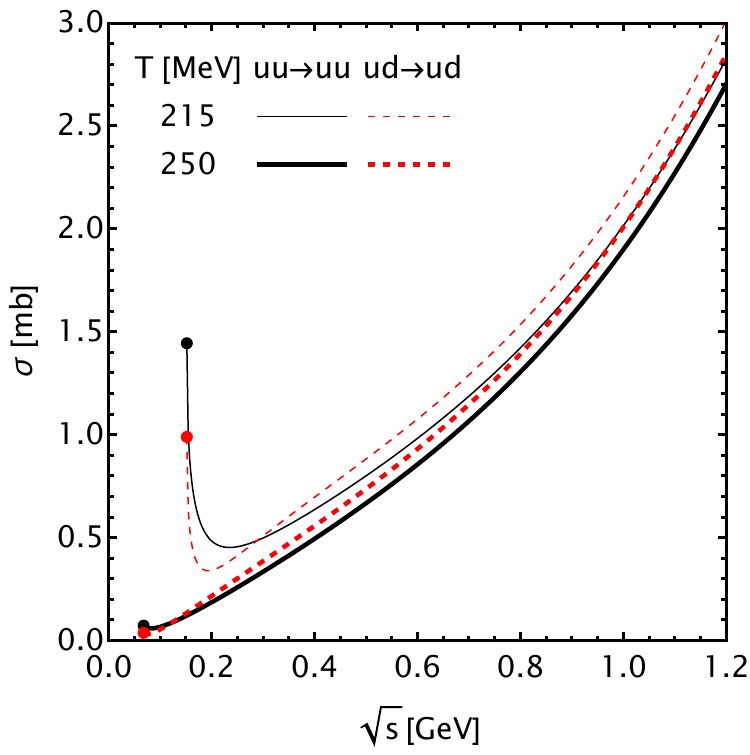}
\caption{}
\label{sigma_uuuu_udud_Yama}
\end{subfigure}
\\
\begin{subfigure}[b]{0.4\textwidth}
\includegraphics[width=\textwidth]
{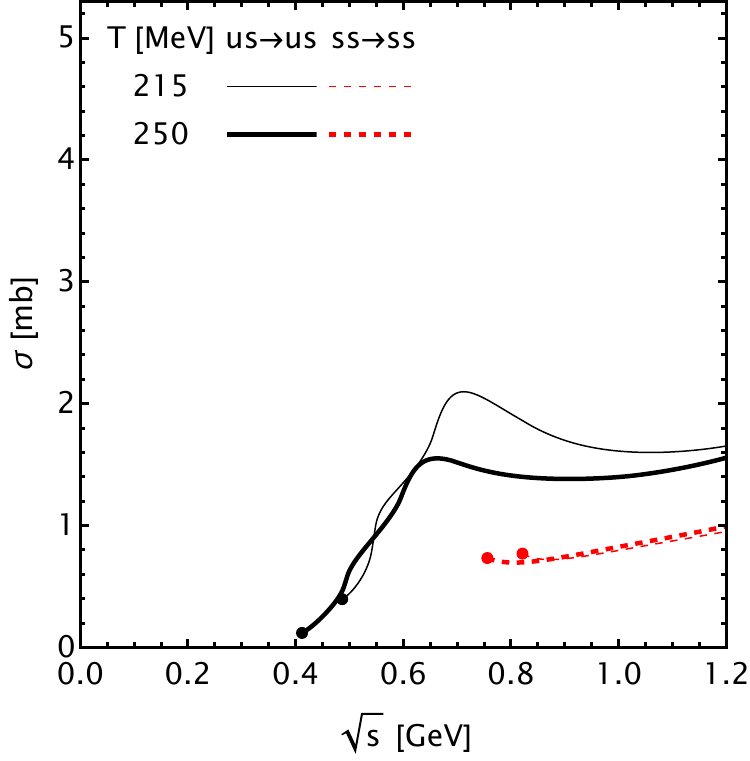}
\caption{}
\label{sigma_usus_ssssKlev}
\end{subfigure}
\begin{subfigure}[b]{0.4\textwidth}
\includegraphics[width=\textwidth]
{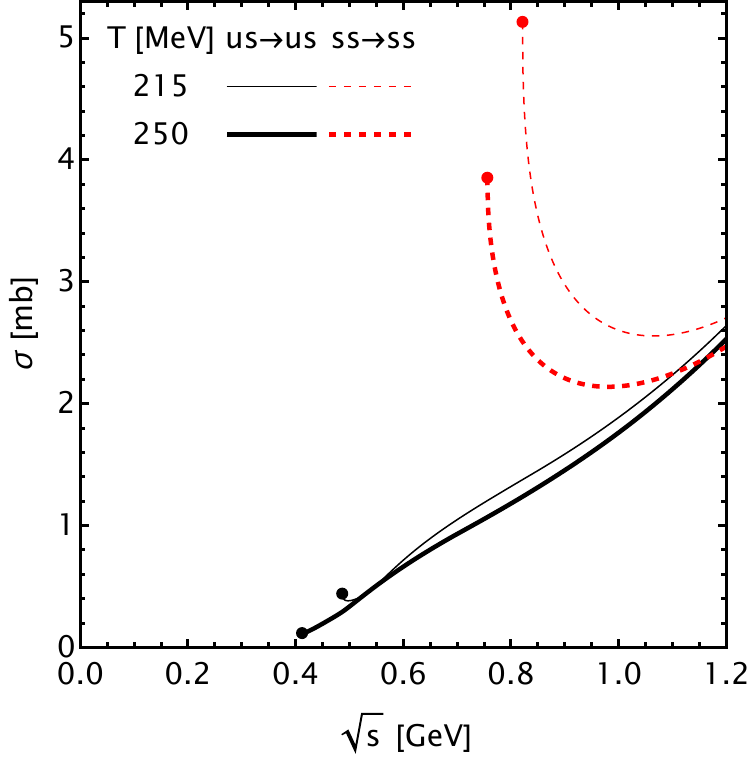}
\caption{}
\label{sigma_usus_ssssYama}
\end{subfigure}
\caption{Cross sections ($\qty[\sigma]=\mathrm{mb}$), as a function of the center of mass energy ($\qty[\sqrt{s}]=\mathrm{GeV}$), for the $u u\to u u$, $u d \to u d$, $us \to us$ and $ss \to ss$ processes for $T=215~\mathrm{MeV}$ and $T=250~\mathrm{MeV}$. On the left-hand side panel, \ref{sigma_uuuu_udud_Klev} and \ref{sigma_usus_ssssKlev}, we use the usual 3-momentum regularization proposed in Ref. \cite{Rehberg:1995nr} whereas, in the right-hand side panel, \ref{sigma_uuuu_udud_Yama} and \ref{sigma_usus_ssssYama}, we use the 3-momentum sphere intersection regularization, proposed in this work. The large dots mark the minimum center of mass energy requirements.
}
\label{sigma_uuuu_udud_usus_ssss}
\end{figure*}

\begin{figure*}[t!]
\begin{subfigure}[b]{0.4\textwidth}
\includegraphics[width=\textwidth]
{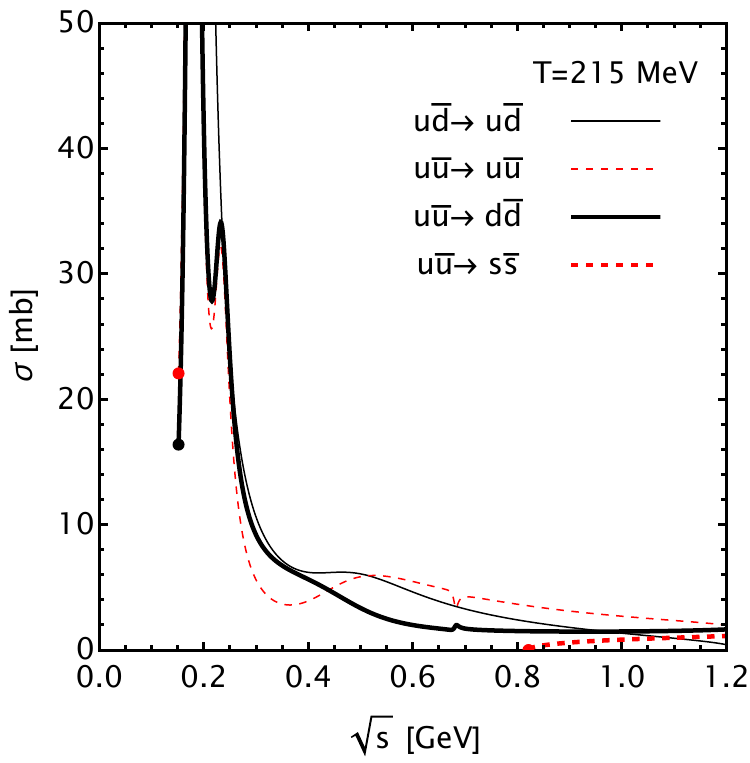}
\caption{}
\label{sigma_udbarudbar_uubaruubar_uubarddbar_uubarssbarT0215Klev}
\end{subfigure}
\begin{subfigure}[b]{0.4\textwidth}
\includegraphics[width=\textwidth]
{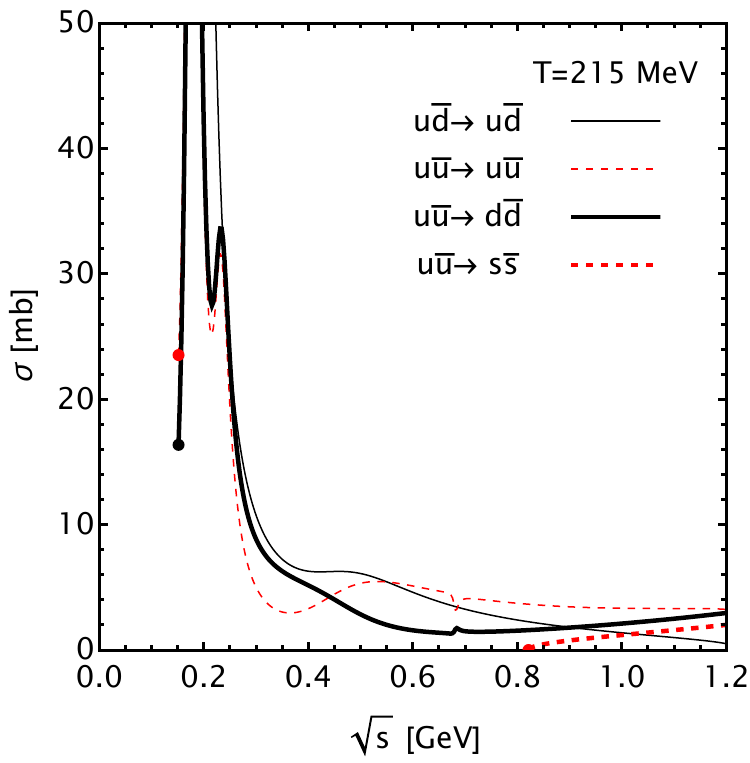}
\caption{}
\label{sigma_udbarudbar_uubaruubar_uubarddbar_uubarssbarT0215Yama}
\end{subfigure}
\\
\begin{subfigure}[b]{0.4\textwidth}
\includegraphics[width=\textwidth]
{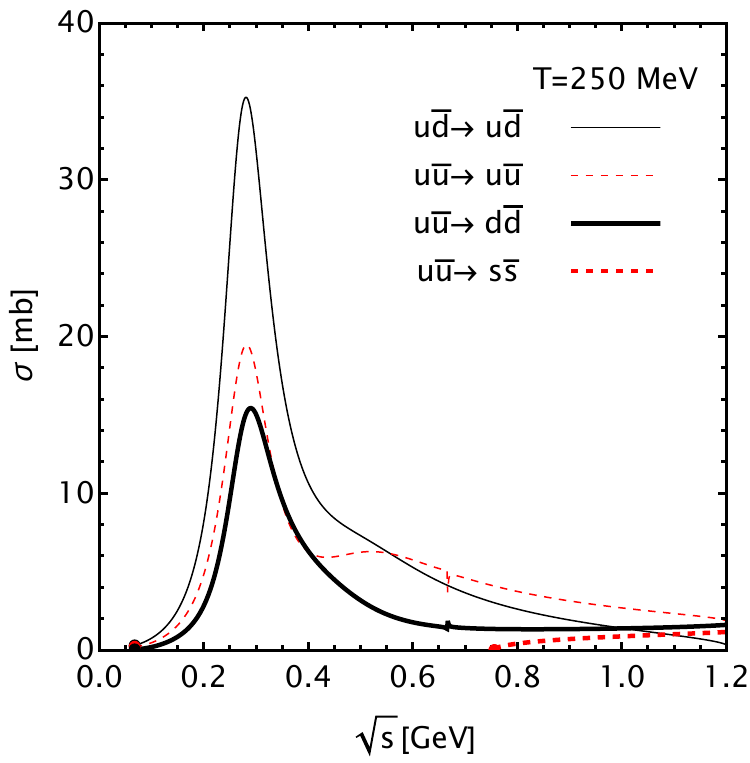}
\caption{}
\label{sigma_udbarudbar_uubaruubar_uubarddbar_uubarssbarT0250Klev}
\end{subfigure}
\begin{subfigure}[b]{0.4\textwidth}
\includegraphics[width=\textwidth]
{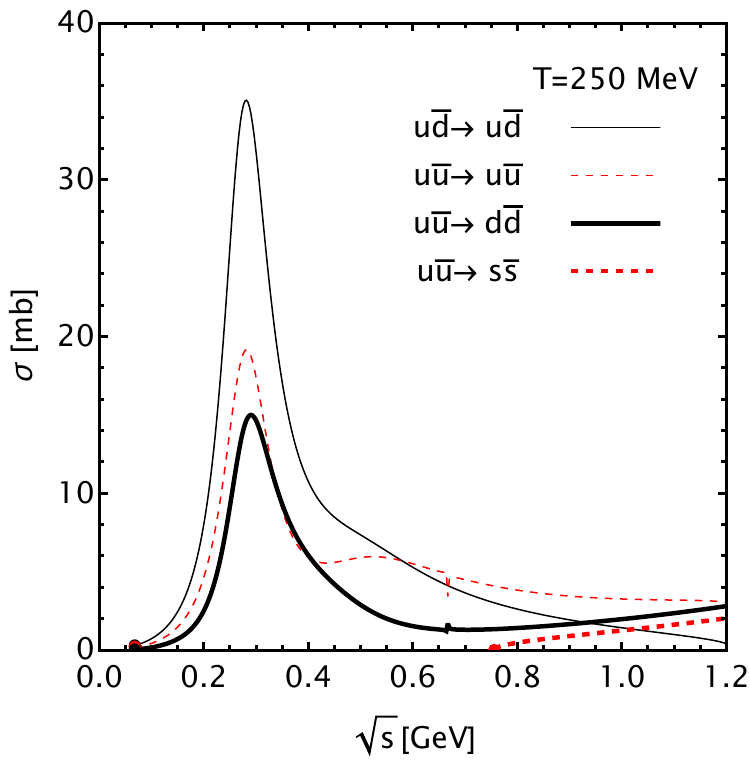}
\caption{}
\label{sigma_udbarudbar_uubaruubar_uubarddbar_uubarssbarT0250Yama}
\end{subfigure}
\caption{Cross sections ($\left[\sigma\right]=\mathrm{mb}$), as a function of the center of mass energy ($\left[\sqrt{s}\right]=\mathrm{GeV}$), for the $u \bar{d}\to u \bar{d}$, $u \bar{u} \to u \bar{u}$, $u \bar{u} \to d \bar{d}$ and $u \bar{u} \to s \bar{s}$ processes, for $T=215~\mathrm{MeV}$ in the top and $T=250~\mathrm{MeV}$ in the bottom panels. On the left-hand side panels, \ref{sigma_udbarudbar_uubaruubar_uubarddbar_uubarssbarT0215Klev} and \ref{sigma_udbarudbar_uubaruubar_uubarddbar_uubarssbarT0250Klev}, the usual 3-momentum regularization proposed in Ref. \cite{Rehberg:1995nr} is used whereas, in the right-hand side panel, \ref{sigma_udbarudbar_uubaruubar_uubarddbar_uubarssbarT0215Yama} and \ref{sigma_udbarudbar_uubaruubar_uubarddbar_uubarssbarT0250Yama}, the 3-momentum sphere intersection regularization, proposed in this work, is used. The large dots mark the minimum center of mass energy requirements.
}
\label{sigma_udbarudbar_uubaruubar_uubarddbar_uubarssbar}
\end{figure*}

In Fig. \ref{sigma_uuuu_udud_usus_ssss}, we show the cross section for the $u u\to u u$, $u d \to u d$, $us \to us$ and $ss \to ss$ quark-quark scattering processes, as a function of the center of mass energy ($\sqrt{s}$), for the two temperatures mentioned before, $T=215$ MeV and $T=250$ MeV. On the left-hand side panels,  \ref{sigma_uuuu_udud_Klev} and \ref{sigma_usus_ssssKlev}, we calculated the cross sections using the usual 3-momentum regularization proposed in Ref. \cite{Rehberg:1995nr} while, on the right-hand side panels, \ref{sigma_uuuu_udud_Yama} and \ref{sigma_usus_ssssYama}, we used the approach proposed in this work with the 3-momentum sphere intersection regularization. From this result we can clearly observe that using the two different regularization procedures result in very different cross sections for both $u u \to u u$ and $u d \to u d$ processes. The reason behind this huge difference can be traced back to the values of the mesons 3-momenta that are exchanged in each channel, for each scattering process. Since we are considering the limit of equal up and down quark masses, $M_u=M_d=M_l$, (and also equal chemical potentials, $\mu_u=\mu_d=\mu_s=0$), for both scatterings, the zero component of the mesons momentum in the $t$- and $u$-channels is $k_0=0$ while, the absolute value of the exchanged 3-momentum is $\abs{\vec{k}} = \sqrt{
- 4M_l^2 + s + t 
} $ (see Eqs. (\ref{k0_tChannel})$-$(\ref{absk_uChannel})). Thus, this calculation is dominated by evaluations of quark-antiquark polarization functions which have a finite absolute value of external 3-momentum $\abs{\vec{k}}$. As discussed in previous sections (and in the Appendix \ref{numericalResultsB0}), the different regularization procedures lead to equal results if $\abs{\vec{k}} = 0$ but differ when considering $\abs{\vec{k}} \geq 0$. Firstly, this is caused because the quark-antiquark polarization function in the 3-momentum sphere intersection regularization has contributions coming from the one fermion, one loop integral, $\mathcal{A}$, regularized differently from the integral $A$ used in the usual 3-momentum regularization. Secondly, as explained before, in the usual 3-momentum regularization the two fermion, one loop integral, $B_0$ is evaluated with a change of variables which breaks certain symmetries of the integral when $\abs{\vec{k}} \geq 0$. In the regularization proposed here, that is not the case, resulting in a different $B_0$ function for finite $\abs{\vec{k}}$. The union between these features explains the different behavior observed in this figure and others that will be analyzed later. Focusing on the results with the 3-momentum sphere intersection regularization (Fig. \ref{sigma_uuuu_udud_Yama}), we can observe a minimum for both processes, right after the reaction threshold energy. After this minimum, the cross sections increase, almost linearly with the center of mass energy, with the cross section for the $u d \to u d $ process being slightly larger, for the same temperature, when compared with the $u u \to u u$ process due to the flavor factors that multiply the different meson contributions, as noted in Ref. \cite{Rehberg:1996vd}. Furthermore, the effect of temperature is to overall decrease the cross sections for a given value of center of mass energy. We note that the minimum center of mass energy for each process (large dots) also decrease with increasing temperature due to the larger restoration of chiral symmetry. The effect of increasing temperature is known to restore chiral symmetry and decrease the quark effective masses towards their bare values and the threshold energies for the reaction to occur, on the center of mass frame, depends only on the outgoing particles masses. Additionally, in the $us\to us$ process considering the simple 3-momentum regularization, Fig. \ref{sigma_usus_ssssKlev}, there is a peak like structure not visible in the case of the new 3-momentum sphere intersection regularization, Fig. \ref{sigma_usus_ssssYama}. In Ref. \cite{Rehberg:1996vd}, this structure was also observed and the authors claimed that its origin is a cutoff artifact. Our calculation seems to more or less agree with such statement: since we do not observe such peak, we can root the origin of such structure to regularization artifacts but not specifically to the presence of a cutoff since the new regularization procedure, proposed in this work, also employs the use of a 3-momentum cutoff.

Next, we turn or attention to Fig. \ref{sigma_udbarudbar_uubaruubar_uubarddbar_uubarssbar} where we present the results for the cross section of quark-antiquark processes with only light quarks in the initial state. This requirement includes the following quark-antiquark processes: $u \bar{d}\to u \bar{d}$, $u \bar{u} \to u \bar{u}$, $u \bar{u} \to d \bar{d}$ and $u \bar{u} \to s \bar{s}$. Again, we show results for $T=215~\mathrm{MeV}$ in the top panels, $T=250~\mathrm{MeV}$ in the bottom panels whilst comparing the results of using both the usual 3-momentum regularization, given in the left panels (Figs. \ref{sigma_udbarudbar_uubaruubar_uubarddbar_uubarssbarT0215Klev} and \ref{sigma_udbarudbar_uubaruubar_uubarddbar_uubarssbarT0250Klev}), with the results of using the regularization proposed in this work, shown in the right panels (Figs. \ref{sigma_udbarudbar_uubaruubar_uubarddbar_uubarssbarT0215Yama} and \ref{sigma_udbarudbar_uubaruubar_uubarddbar_uubarssbarT0250Yama}). The most striking feature of these results is that both regularization give almost exactly the same results, with very tiny differences observed at larger center of mass energies, for both temperatures. This can be traced back to the fact the quark-antiquark processes depend on both the $s$- and $t$-channels. In the case of meson propagators in the $s$-channel, they are always evaluated with zero external 3-momentum, $\abs{ \vec{k} }^{(s)}=0$ and  $ k_0^{(s)} = \sqrt{s} $. Since both regularization yield the same results at zero external 3-momentum, it implies that all contributions to the differential cross section coming from the $s$-channel are identical in both regularization schemes. In the case of meson propagators in the $t$-channel, such is not the case since they are evaluated for $ k_0^{(t)} = 0 $ and 
$\abs{ \vec{k} }^{(t)}=\sqrt{-t}$. However, the integration range over $t$ for processes with $m_1=m_2$ and $m_3=m_4$, are zero exactly at $s=s_\mathrm{min}$ and increase with increasing values of center of mass energy, $\sqrt{s}$. Thus, for small values of $\sqrt{s}$ only small values of external 3-momentum are considered and only in the $t$-channel, with the contributions from the $s$-channel being all for zero external 3-momentum. So, this behavior explains why, for these processes, the cross sections are almost identical in both regularization schemes, becoming slightly different only with increasing center of mass energies. The large peaks observed in this figures is due the fact that these cross sections are being evaluated near the Mott temperatures of the pions and kaons, meaning their behavior will be dominated by the existence of these resonances at certain values of center of mass energy \cite{Rehberg:1996vd}.

\begin{figure*}[ht!]
\begin{subfigure}[b]{0.4\textwidth}
\includegraphics[width=\textwidth]
{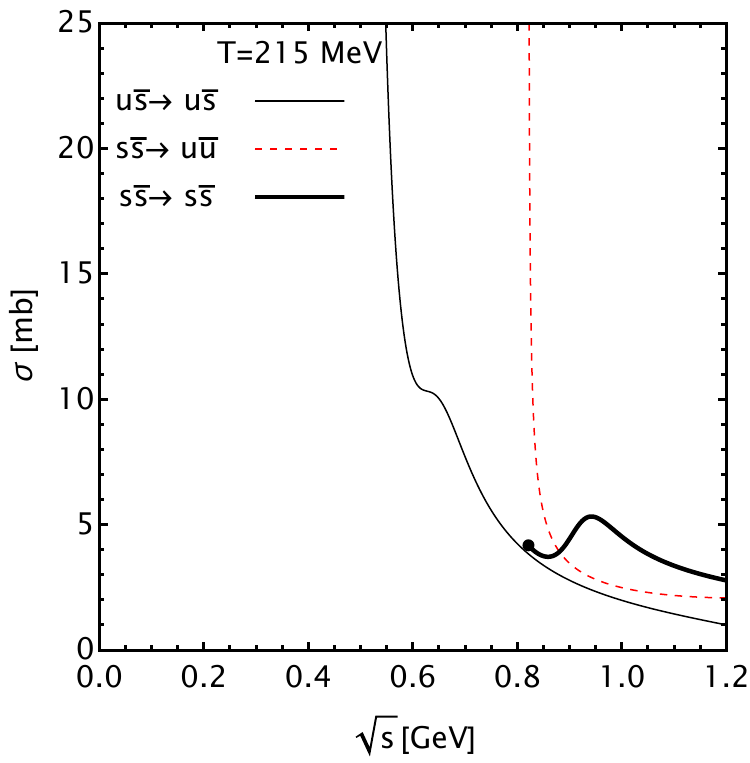}
\caption{}
\label{sigma_usbarusbar_ssbaruubar_ssbarssbarT0215Klev}
\end{subfigure}
\begin{subfigure}[b]{0.4\textwidth}
\includegraphics[width=\textwidth]
{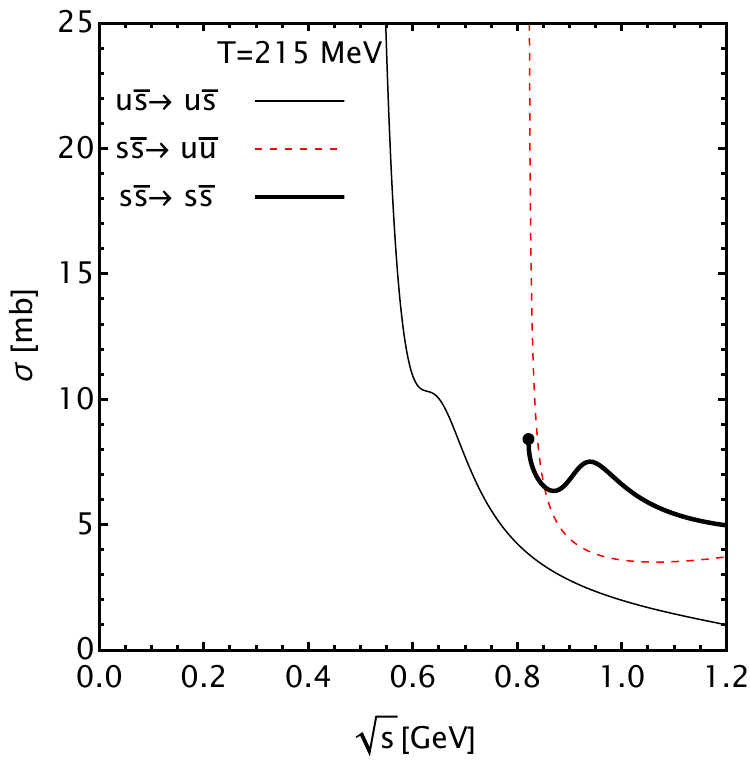}
\caption{}
\label{sigma_usbarusbar_ssbaruubar_ssbarssbarT0215Yama}
\end{subfigure}
\\
\begin{subfigure}[b]{0.4\textwidth}
\includegraphics[width=\textwidth]
{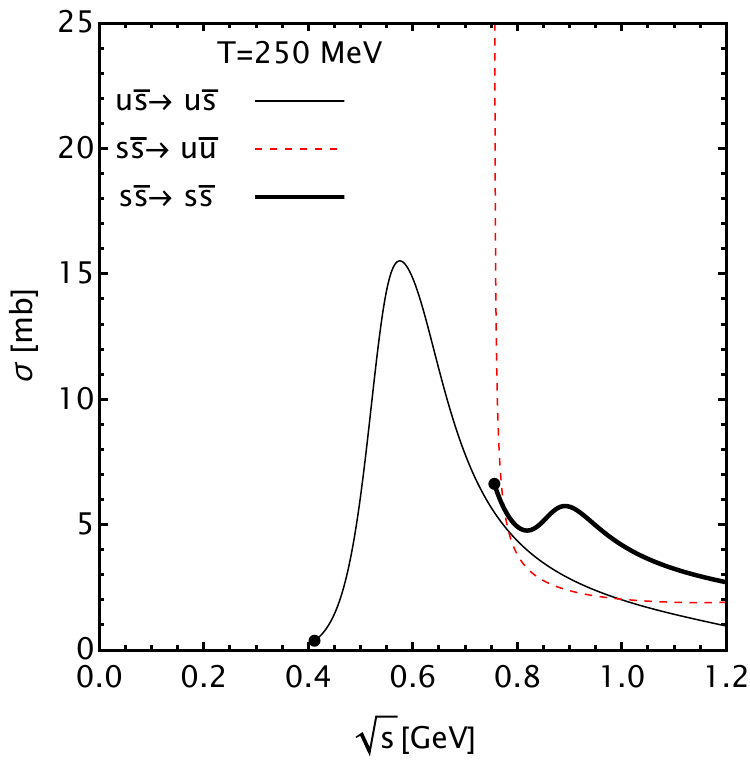}
\caption{}
\label{sigma_usbarusbar_ssbaruubar_ssbarssbarT0250Klev}
\end{subfigure}
\begin{subfigure}[b]{0.4\textwidth}
\includegraphics[width=\textwidth]
{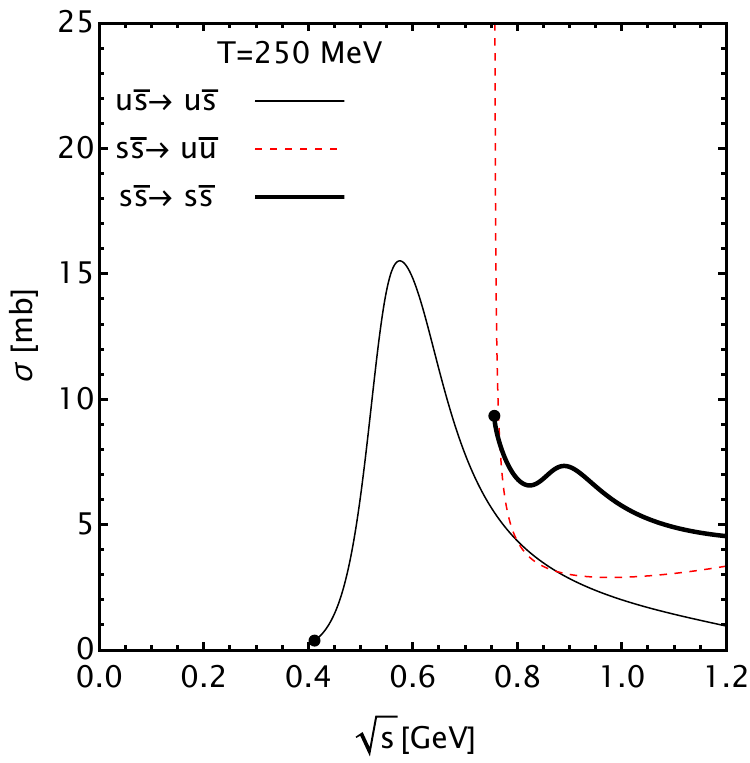}
\caption{}
\label{sigma_usbarusbar_ssbaruubar_ssbarssbarT0250Yama}
\end{subfigure}
\caption{Cross sections ($\left[\sigma\right]=\mathrm{mb}$), as a function of the center of mass energy ($\left[\sqrt{s}\right]=\mathrm{GeV}$), for the $u \bar{s} \to u \bar{s}$, $s \bar{s} \to u \bar{u}$ and $s \bar{s} \to s \bar{s}$ processes, for $T=215~\mathrm{MeV}$ in the top and $T=250~\mathrm{MeV}$ in the bottom panels. On the left-hand side panels, \ref{sigma_usbarusbar_ssbaruubar_ssbarssbarT0215Klev} and \ref{sigma_usbarusbar_ssbaruubar_ssbarssbarT0250Klev}, the usual 3-momentum regularization proposed in Ref. \cite{Rehberg:1995nr} is used whereas, in the right-hand side panel, \ref{sigma_usbarusbar_ssbaruubar_ssbarssbarT0215Yama} and \ref{sigma_usbarusbar_ssbaruubar_ssbarssbarT0250Yama}, the 3-momentum sphere intersection regularization, proposed in this work is used. The large dots mark the minimum center of mass energy requirements.
}
\label{sigma_usbarusbar_ssbaruubar_ssbarssbar}
\end{figure*}

Finally, in Fig. \ref{sigma_usbarusbar_ssbaruubar_ssbarssbar}, we present a comparison of cross sections for quark-antiquark processes involving a strange antiquark in the initial state. The processes shown in this figure are: $u\bar{s} \to u\bar{s}$, $s\bar{s} \to u\bar{u}$ and $s\bar{s} \to s\bar{s}$. The first observation, present in both regularization schemes is the diverging nature of the  $s\bar{s} \to u \bar{u}$ cross section at the energy threshold. As pointed out in Ref. \cite{Rehberg:1996vd}, this is a kinematic singularity caused by the exothermic nature of this particular quark-antiquark process. Here, the cross section for process $s\bar{s} \to s \bar{s}$ is quite different for different regularizations but is extremely similar for the other two processes. Again, the resonant structure observed is due to the proximity to the Mott temperatures of the kaon resonances.

\subsubsection{Results at finite temperature and chemical potential in the 3-momentum sphere intersection regularization scheme}

In this section we show the results of evaluating quark-quark, quark-antiquark and antiquark-antiquark processes at finite (fixed) temperature and (increasing) quark chemical potential, as a function of the center of mass energy, $\sqrt{s}$, allowing the study of the influence of finite density effects on the cross sections. For the temperature we used the previous value of $T=250~\mathrm{MeV}$ while, for the chemical potential, we used 6 different values, $\mu=\qty{0,100,200,300,400,500,600}~\mathrm{MeV}$. Additionally, we consider equal quark chemical potentials, $\mu=\mu_u=\mu_d=\mu_s$. In this case we will have 16 different processes instead of the 11 different processes at $\mu=0$. The extra processes that have to be calculated are: $\bar{u}\,\bar{u} \to \bar{u}\,\bar{u}$, $\bar{u}\,\bar{d} \to \bar{u}\,\bar{d}$, $\bar{u}\,\bar{s} \to \bar{u}\,\bar{s}$, $\bar{s}\,\bar{s} \to \bar{s}\,\bar{s}$ and $s \bar{u} \to s \bar{u}$. These processes are degenerate to their particle analogues at zero chemical potential (when switching quarks to antiquarks and antiquark to quarks) however, when considering finite chemical potential, they will be different.

\begin{figure*}[t!]
\begin{subfigure}[b]{0.4\textwidth}
\includegraphics[width=\textwidth]
{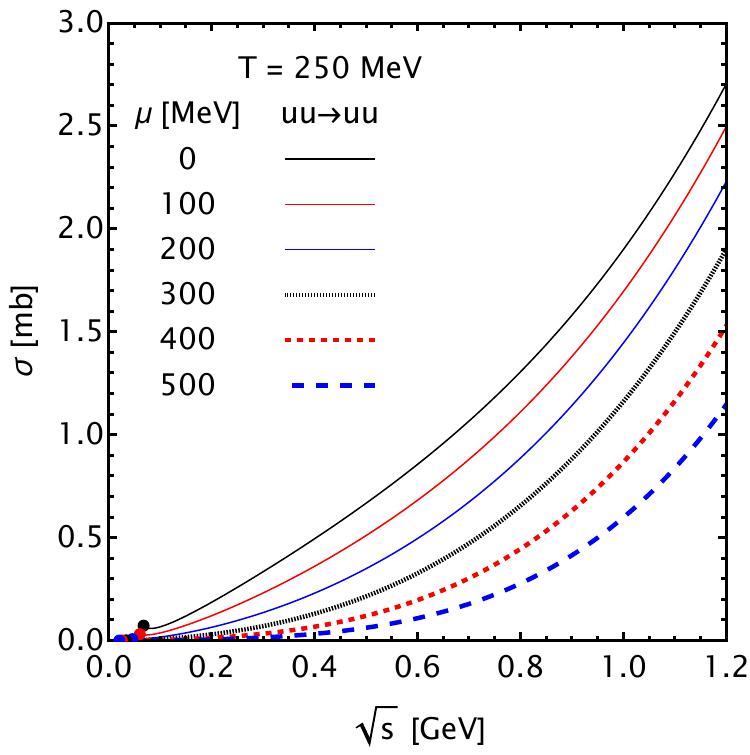}
\caption{}
\label{sigma_uuuu_finit_chem_pot}
\end{subfigure}
\begin{subfigure}[b]{0.4\textwidth}
\includegraphics[width=\textwidth]
{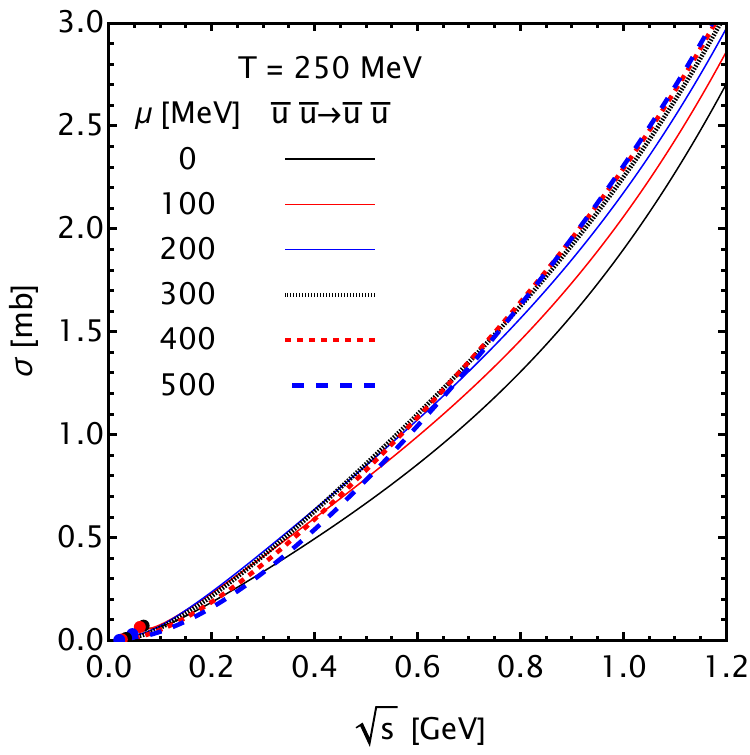}
\caption{}
\label{sigma_ubarubarubarubar_finit_chem_pot}
\end{subfigure}
\\
\begin{subfigure}[b]{0.4\textwidth}
\includegraphics[width=\textwidth]
{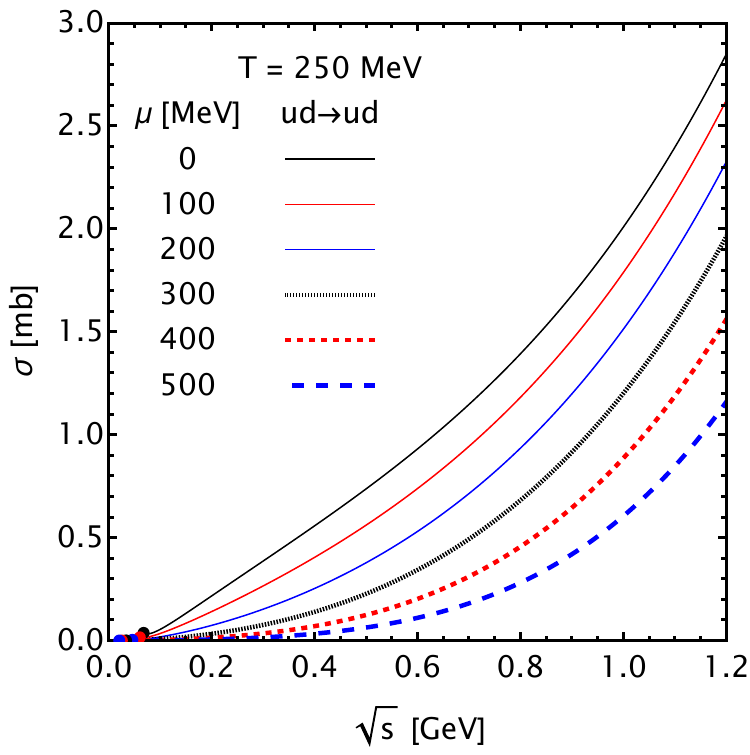}
\caption{}
\label{sigma_udud_finit_chem_pot}
\end{subfigure}
\begin{subfigure}[b]{0.4\textwidth}
\includegraphics[width=\textwidth]
{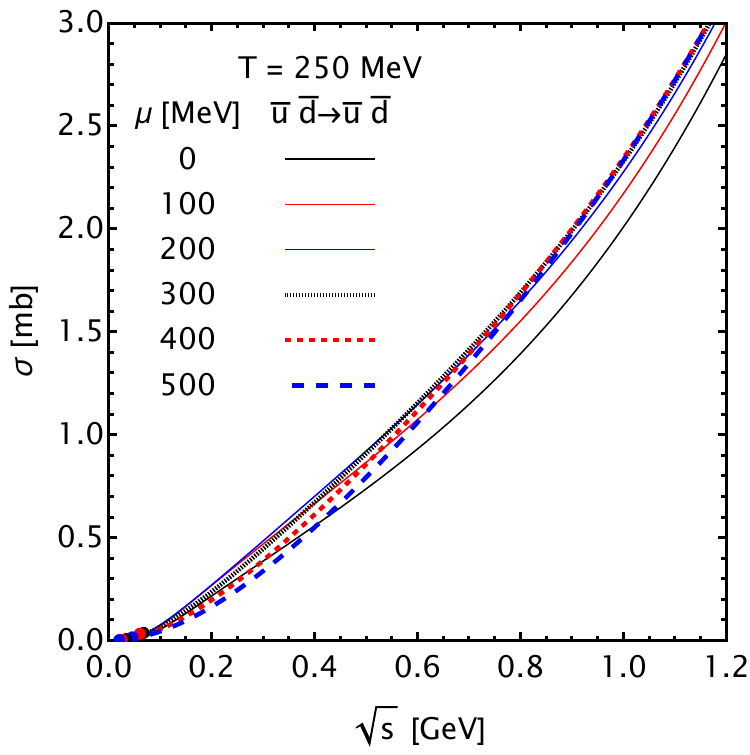}
\caption{}
\label{sigma_ubardbarubardbar_finite_chem_pot}
\end{subfigure}
\caption{
Cross sections ($\left[\sigma\right]=\mathrm{mb}$), as a function of the center of mass energy ($\left[\sqrt{s}\right]=\mathrm{GeV}$), for the $uu \to uu$, $\bar{u}\,\bar{u} \to \bar{u}\,\bar{u}$, $ud \to ud$, $\bar{u}\,\bar{d} \to \bar{u}\,\bar{d}$ processes, for $T=250~\mathrm{MeV}$ and increasing chemical potentials. The large dots mark the minimum center of mass energy requirements for a given processes to occur. Here, the 3-momentum sphere intersection regularization is used.
}
\label{sigma_quarkquark_antiquarkantiquark_finite_chem_pot}
\end{figure*}

\begin{figure*}[t!]
\begin{subfigure}[b]{0.4\textwidth}
\includegraphics[width=\textwidth]
{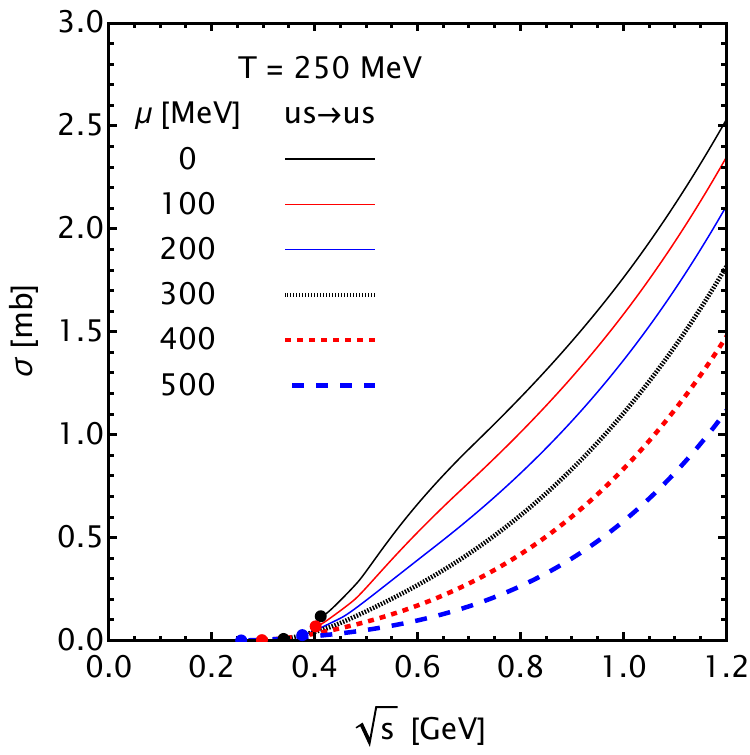}
\caption{}
\label{sigma_usus_finite_chem_pot}
\end{subfigure}
\begin{subfigure}[b]{0.4\textwidth}
\includegraphics[width=\textwidth]
{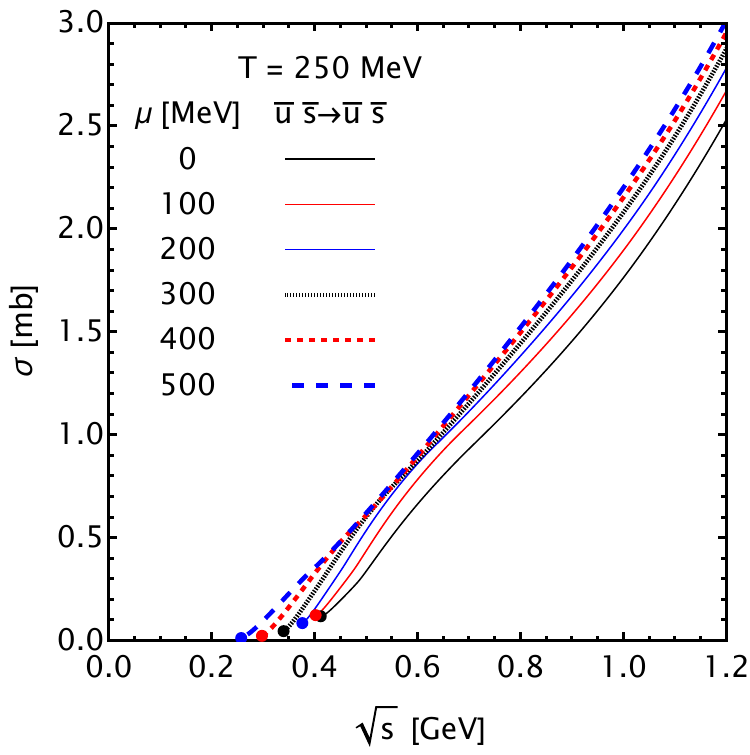}
\caption{}
\label{sigma_ubarsbarubarsbar_finite_chem_pot}
\end{subfigure}
\\
\begin{subfigure}[b]{0.4\textwidth}
\includegraphics[width=\textwidth]
{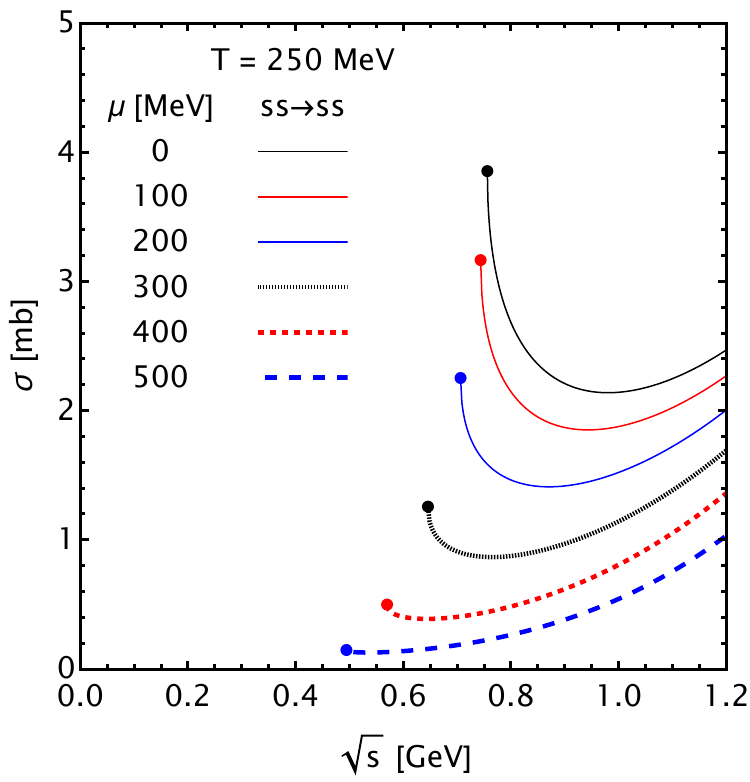}
\caption{}
\label{sigma_ssss_finite_chem_pot}
\end{subfigure}
\begin{subfigure}[b]{0.4\textwidth}
\includegraphics[width=\textwidth]
{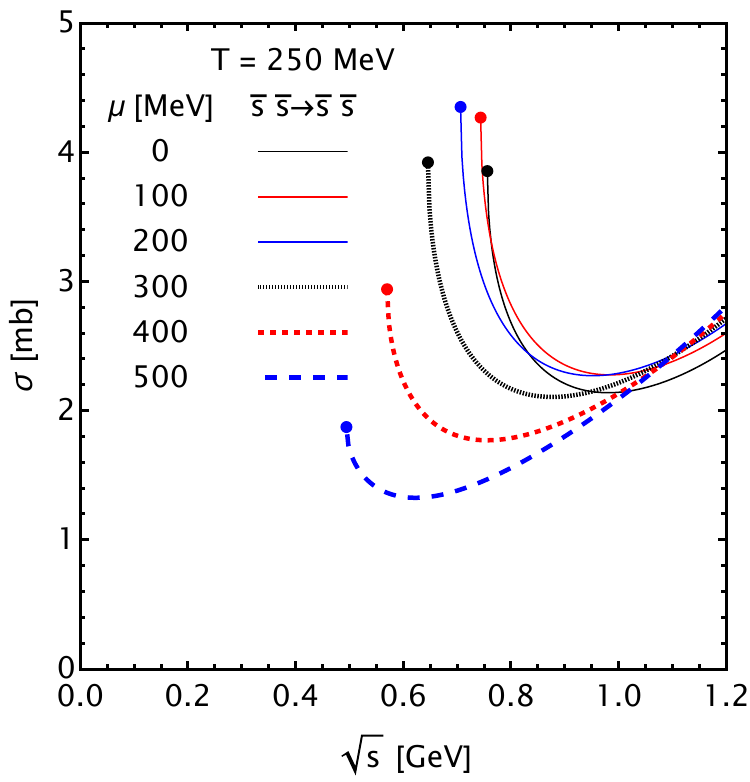}
\caption{}
\label{sigma_sbarsbarsbarsbar_finite_chem_pot}
\end{subfigure}
\caption{
Cross sections ($\left[\sigma\right]=\mathrm{mb}$), as a function of the center of mass energy ($\left[\sqrt{s}\right]=\mathrm{GeV}$), for the $us \to us$, $\bar{u}\,\bar{s} \to \bar{u}\,\bar{s}$, $ss \to ss$, $\bar{s}\,\bar{s} \to \bar{s}\,\bar{s}$ processes, for $T=250~\mathrm{MeV}$ and increasing chemical potentials. The large dots mark the minimum center of mass energy requirements for a given processes to occur. Here, the 3-momentum sphere intersection regularization is used.
}
\label{sigma_quarkquark_antiquarkantiquark_strangeness_finite_chem_pot}
\end{figure*}

We start by presenting all the quark-quark and antiquark-antiquark cross sections in Figs. \ref{sigma_quarkquark_antiquarkantiquark_finite_chem_pot} and \ref{sigma_quarkquark_antiquarkantiquark_strangeness_finite_chem_pot}. When considering finite chemical potential and, consequently, finite baryon density, the medium is characterized by an asymmetry between the number of particles and antiparticles. Such asymmetry is responsible, for instance, for the difference observed in the meson behavior of certain meson states at finite chemical potential and the masses of the charged kaons and their scalar particles are different in the presence of non-zero baryon density \cite{Costa:2019bua}. Since the cross sections are highly dependent on the meson propagators and, taking into account that finite chemical potential already breaks the symmetry between some charged mesons, it is natural to expect it to also break the symmetry between processes which, at zero chemical potential, are identical. Hence, in these figures, one can observe the effect of breaking the symmetry between charged mesons at finite density. Of course, the observed differences between quark-quark and antiquark-antiquark at finite chemical potential is also connected to the Pauli blocking factor considered to evaluate the cross section, see Eq. (\ref{cross_section_definition}).

In Fig. \ref{sigma_uuuu_finit_chem_pot} and \ref{sigma_ubarubarubarubar_finit_chem_pot} we show the results for the processes $uu \to uu$ and $\bar{u}\,\bar{u} \to \bar{u}\,\bar{u}$. As expected, at zero chemical potential, these processes yield the same result however, at finite chemical potential, they have different behaviors. As the chemical potential increases, the cross section for the $uu \to uu$ process consistently decreases across all values of the center of mass energy. Conversely, for the $\bar{u}\,\bar{u} \to \bar{u}\,\bar{u}$ process, the cross section demonstrates an increase with chemical potential for large center of mass energy values. However, for small energies, the behavior of the cross section becomes more complex and does not follow a straightforward trend. The $ud \to ud$ cross section and $\bar{u}\,\bar{d} \to \bar{u}\,\bar{d}$ cross sections, as well as the for the $us \to us$ and $\bar{u}\,\bar{s} \to \bar{u}\,\bar{s}$ processes, shown in Figs. \ref{sigma_udud_finit_chem_pot}, \ref{sigma_ubardbarubardbar_finite_chem_pot}, \ref{sigma_usus_finite_chem_pot} and \ref{sigma_ubarsbarubarsbar_finite_chem_pot} respectively, share the same overall behavior that was previously described about processes involving only the $u$ quark.

The minimum center of mass energies for these processes to occur, represented by the colored dots in Fig. \ref{sigma_quarkquark_antiquarkantiquark_finite_chem_pot}, are independent of the considered process and are equal for a specific chemical potential. This is a consequence of the isospin symmetry considered in this study which yields equal effective quark masses for the up and down quarks. Naturally, processes involving the strange quark have a higher threshold energy due to the higher strange quark effective mass. As expected from the restoration of chiral symmetry in the medium, the threshold energies for all processes gets smaller with increasing chemical potential since the quarks effective masses tend to their respective bare values. This feature is common to all results at finite chemical potential.

We highlight that, in our calculation, the quark-quark process $us \to us$ and the equivalent process $su \to su$ are identical. From the physics point of view this is obvious however, from the practical point of view this is only possible because the 3-momentum sphere intersection regularization does not ruin this feature, as explained before. If one uses the usual 3-momentum regularization, at finite chemical potential, the cross sections for the processes $us \to us$ and $su \to su$ will be different.

\begin{figure*}[t!]
\begin{subfigure}[b]{0.32\textwidth}
\includegraphics[width=\textwidth]
{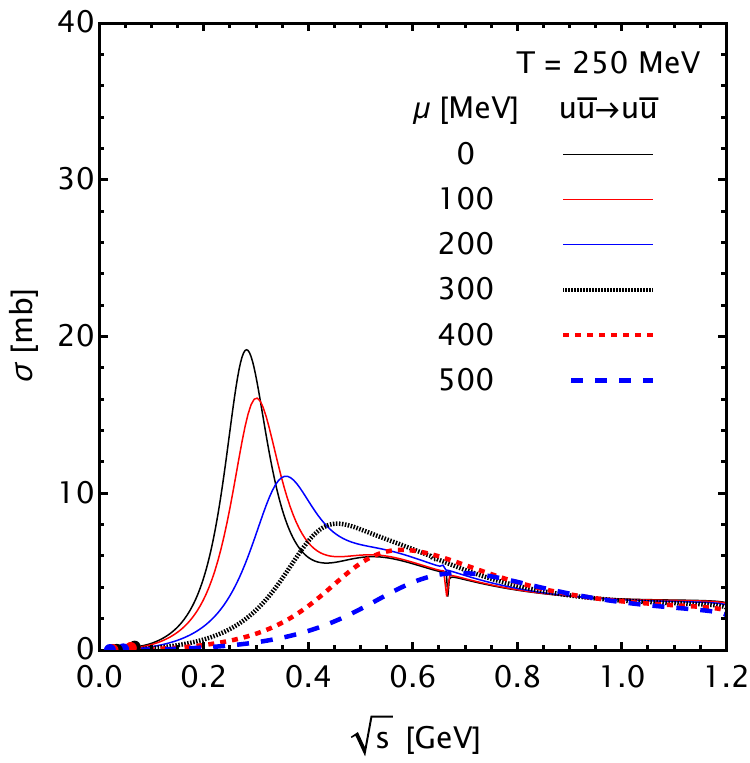}
\caption{}
\label{sigma_uubaruubar_finite_chem_pot}
\end{subfigure}
\begin{subfigure}[b]{0.32\textwidth}
\includegraphics[width=\textwidth]
{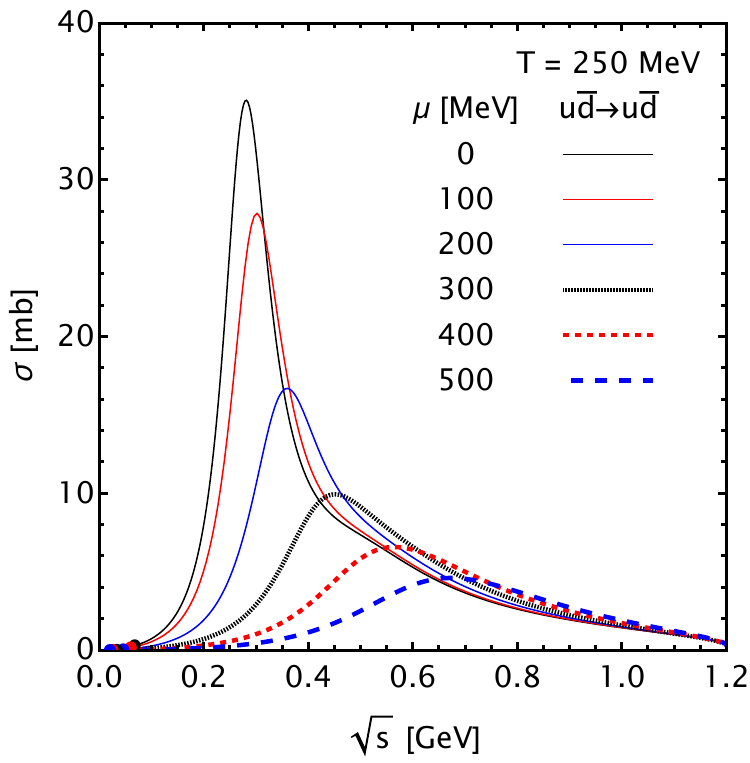}
\caption{}
\label{sigma_udbarudbar_finite_chem_pot}
\end{subfigure}
\begin{subfigure}[b]{0.32\textwidth}
\includegraphics[width=\textwidth]
{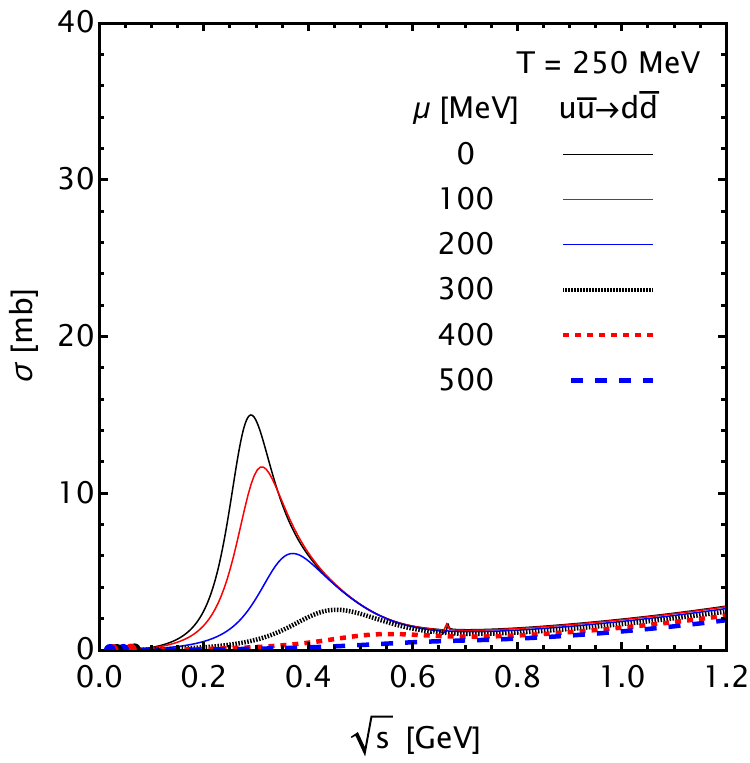}
\caption{}
\label{sigma_uubarddbar_finite_chem_pot}
\end{subfigure}
\\
\begin{subfigure}[b]{0.32\textwidth}
\includegraphics[width=\textwidth]
{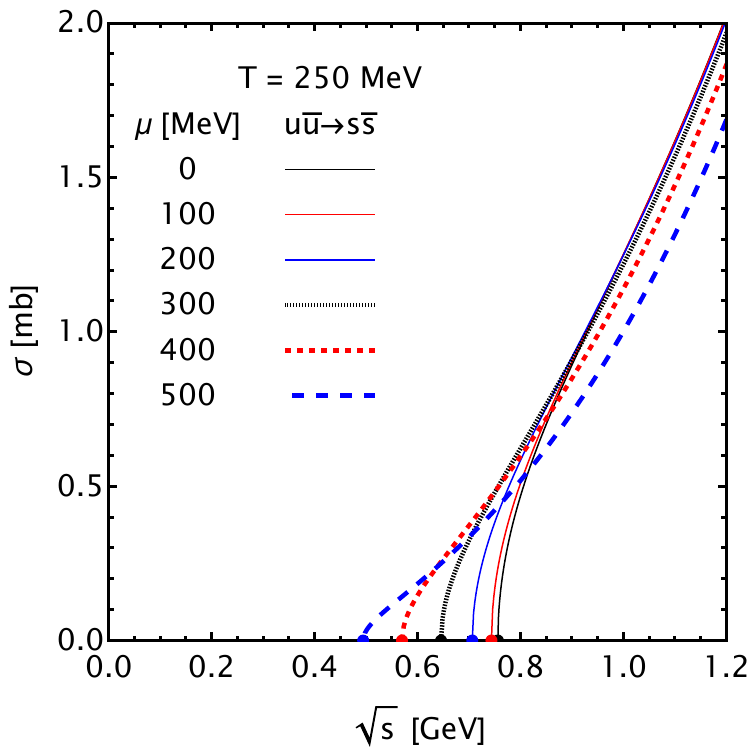}
\caption{}
\label{sigma_uubarssbar_finite_chem_pot}
\end{subfigure}
\begin{subfigure}[b]{0.32\textwidth}
\includegraphics[width=\textwidth]
{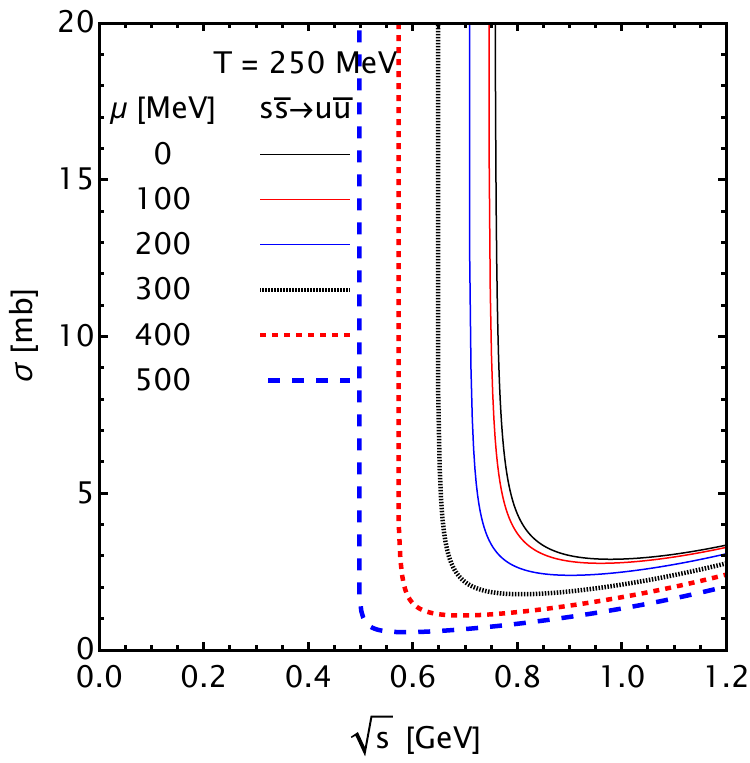}
\caption{}
\label{sigma_ssbaruubar_finite_chem_pot}
\end{subfigure}
\begin{subfigure}[b]{0.32\textwidth}
\includegraphics[width=\textwidth]
{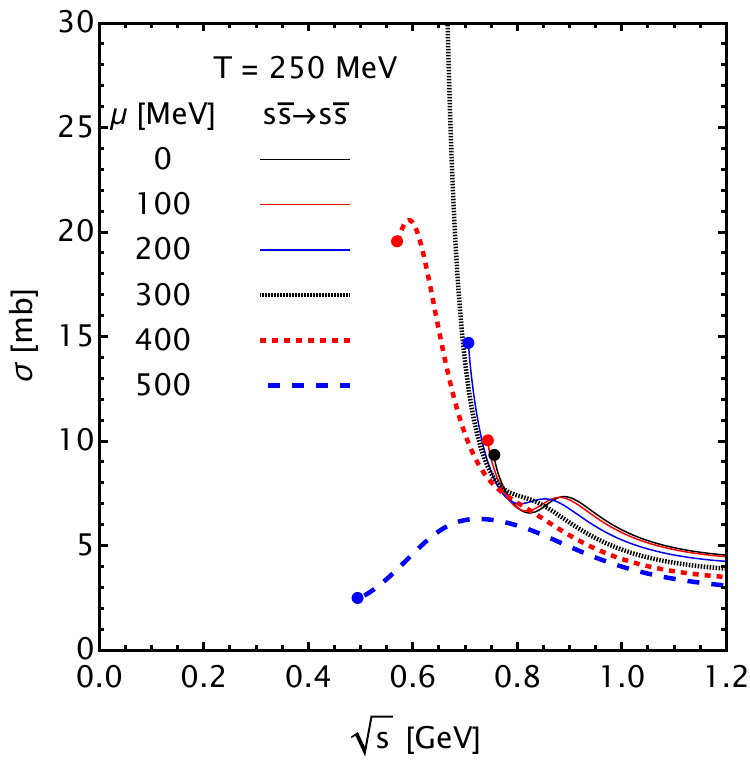}
\caption{}
\label{sigma_ssbarssbar_finite_chem_pot}
\end{subfigure}
\caption{
Cross sections ($\left[\sigma\right]=\mathrm{mb}$), as a function of the center of mass energy ($\left[\sqrt{s}\right]=\mathrm{GeV}$), for the $u\bar{u} \to u\bar{u}$, $u\bar{d} \to u\bar{d}$, $u\bar{u} \to d\bar{d}$, $u\bar{u} \to s\bar{s}$, $s\bar{s} \to u\bar{u}$ and $s\bar{s} \to s\bar{s}$ processes, for $T=250~\mathrm{MeV}$ and increasing chemical potentials. The large dots mark the minimum center of mass energy requirements for a given processes to occur. Here, the 3-momentum sphere intersection regularization is used.
}
\label{sigma_quarkantiquark_zero_strangeness_chem_pot}
\end{figure*}

\begin{figure*}[t!]
\begin{subfigure}[b]{0.4\textwidth}
\includegraphics[width=\textwidth]
{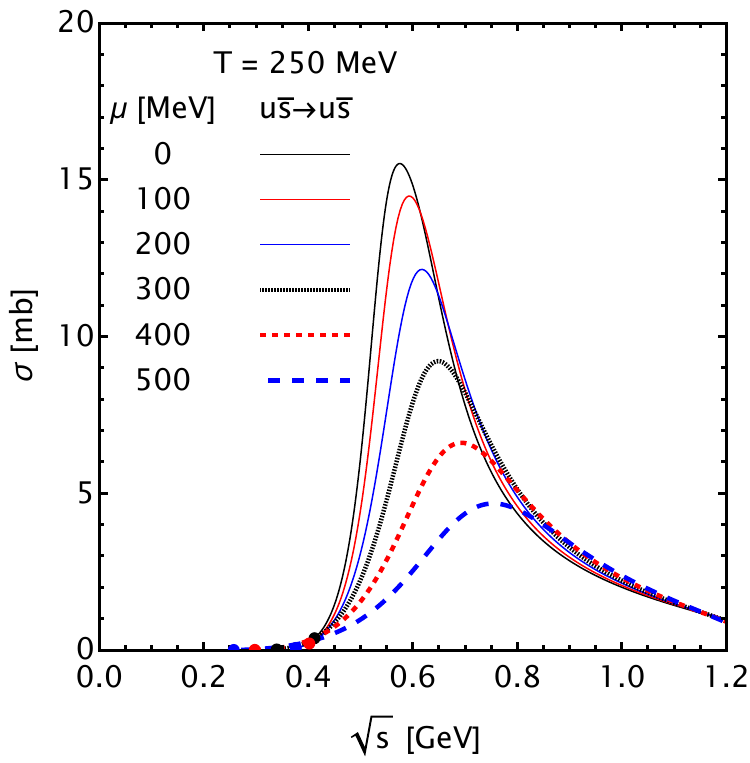}
\caption{}
\label{sigma_usbarusbar_finite_chem_pot}
\end{subfigure}
\begin{subfigure}[b]{0.4\textwidth}
\includegraphics[width=\textwidth]
{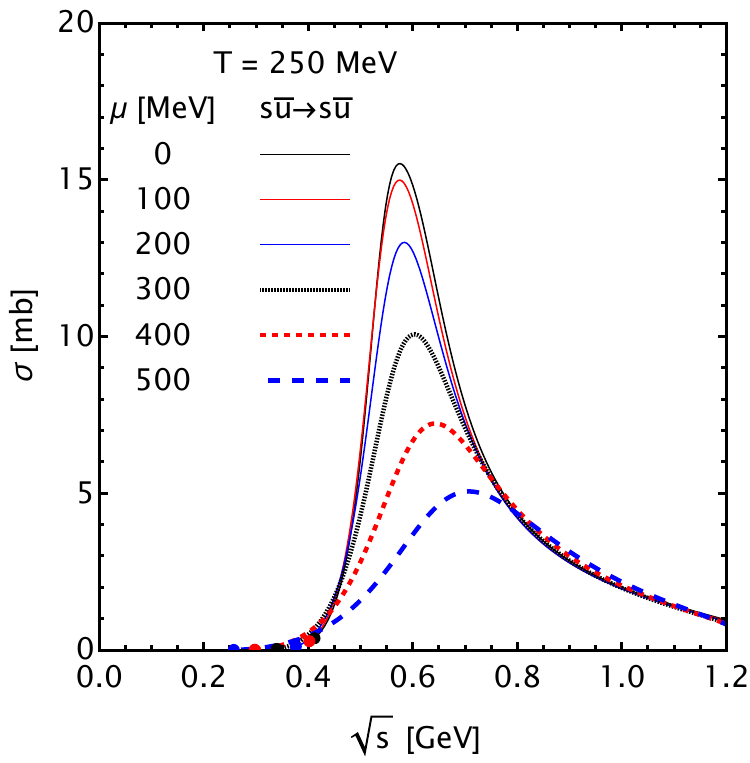}
\caption{}
\label{sigma_subarsubar_finite_chem_pot}
\end{subfigure}
\caption{
Cross sections ($\left[\sigma\right]=\mathrm{mb}$), as a function of the center of mass energy ($\left[\sqrt{s}\right]=\mathrm{GeV}$), for the $u\bar{s} \to u\bar{s}$ and $s\bar{u} \to s\bar{u}$ processes, for $T=250~\mathrm{MeV}$ and increasing chemical potentials. The large dots mark the minimum center of mass energy requirements for a given processes to occur. Here, the 3-momentum sphere intersection regularization is used.
}
\label{sigma_quarkantiquark_strangeness_chem_pot}
\end{figure*}

We now turn our attention to the quark-antiquark cross sections. In Fig. \ref{sigma_quarkantiquark_zero_strangeness_chem_pot} we have the processes with zero net strangeness,which include: $u\bar{u} \to u\bar{u}$, $u\bar{d} \to u\bar{d}$, $u\bar{u} \to d\bar{d}$, $u\bar{u} \to s\bar{s}$, $s\bar{s} \to u\bar{u}$ and $s\bar{s} \to s\bar{s}$. The cross sections of quark-antiquark processes with strangeness are presented in Fig. \ref{sigma_quarkantiquark_strangeness_chem_pot}, and include: $u \bar{s} \to u \bar{s}$ and $s \bar{u} \to s \bar{u}$ processes. As in the case of zero chemical potential, several of these quantities display a resonance structure due to the proximity of the values of temperature and chemical potential to the Mott temperature of the pions and kaons. Of course, when considering finite density, for every chemical potential one can find the respective Mott temperature. Thus, when considering in-medium systems, the melting point of the meson bound states is defined by a point in the space of possible thermodynamic variables, in this case, there will be a Mott temperature and a Mott chemical potential. The study of meson behavior in the medium, within the context of the model used here, is beyond the scope of this work and is left as future work. However, in the range of chemical potentials considered here, the Mott temperature of pions and kaons decreases with increasing temperature. Such effect can be clearly seen in some of the cross sections of Figs. \ref{sigma_quarkantiquark_zero_strangeness_chem_pot} and \ref{sigma_quarkantiquark_strangeness_chem_pot}. The peak caused by the close proximity to the Mott temperatures of the pion and kaons gets smaller with increasing chemical potential. This can be observed in the cross sections of the following processes: $u\bar{u} \to u\bar{u}$, $u\bar{d} \to u\bar{d}$, $u\bar{u} \to d\bar{d}$, $u \bar{s} \to u \bar{s}$ and $s \bar{u} \to s \bar{u}$.

The kinematic divergence present in the exothermic cross section of the process $s\bar{s} \to u\bar{u}$ is present for all considered chemical potentials however, it moves to smaller values of $\sqrt{s}$. The reduction of the strange quark mass, driven by the mechanism of restoration of chiral symmetry, is responsible for this behavior since the threshold energy for this process to occur gets smaller with increasing chemical potential.

The cross section for the $u \bar{s} \to u \bar{s}$ and $s \bar{u} \to s \bar{u}$ are very similar. As mentioned, they also display a peak structure related to the proximity to meson melting points. However, the peaks are not located at the same points. These processes, in the $s$-channel, are mediated by the charged kaons, see Table \ref{s_channel_prop_qqbar}. While $u \bar{s} \to u \bar{s}$ involves the positively charged $K^+$ , the process $s \bar{u} \to s \bar{u}$ exchanges a negatively charged $K^-$. Again, at finite density these meson are not degenerate, displaying different propagators, masses, widths and consequently, different melting points (labeled by a Mott chemical potential and temperature). Thus, in each process, the peak structure is located near its respective charged kaon melting point and their difference in essence, amounts to the charged kaons possessing slightly different melting points as chemical potential increases.

It is expected that the analyzing the effective meson-quark couplings at finite temperature and chemical potential \cite{Costa:2008dp}, could provide additional information regarding the fact that cross sections for quark-quark processes gets smaller with chemical potential while, in general, the antiquark-antiquark cross sections gets larger. However, such study goes beyond the scope of the present work.

\section{Conclusions}
\label{conclusions}

In this work we derived the one and two fermion line integrals from a general NJL model. We considered that the NJL Lagrangian contained the usual Dirac contribution for fermions fields and different scalar-pseudoscalar quark-quark interactions, contained in $\mathcal{L}_{\mathrm{int}}$. The formalism developed here is applicable to any number of scalar-pseudoscalar multi-quark interactions. For instance, one could consider the usual four quark scalar-pseudoscalar interaction, the 't Hooft interaction, 8-quark interactions or explicit chiral symmetry breaking interactions. In order to derive the one and two fermion line integrals, we performed the linear and quadratic expansions of the product between quark bilinear operators in the mean field approximation. From these expansions one can extract thermodynamics quantities and meson propagators. The latter were then used to evaluate quark-quark and quark-antiquark cross sections at finite temperature and chemical potential. Furthermore, the meson propagators obtained in this work are equivalent to the ones used in other works where they are derived from the Bethe-Salpeter equation in the random phase approximation.

In order to perform these tasks, we introduced a new approach to control the diverging nature of the model, the so-called 3-momentum sphere intersection regularization. This new procedure is essential in order to maintain the symmetries of the model. For instance, it ensures that the cross sections for the processes $u s \to u s $ and $s u \to s u $ are the same for all values of temperature and chemical potential, which does not happen in the usual 3-momentum regularization.

As a practical implementation of the developed formalism, and to demonstrate the differences between the 3-momentum sphere intersection regularization and the usual 3-momentum regularization, we considered the standard $SU(3)$ NJL model, with four and six quark interactions \cite{Rehberg:1995kh}, to evaluate quark-quark and quark-antiquark cross sections. We observed quantitative and qualitative differences when comparing quark-quark cross sections in both schemes. For quark-antiquark cross sections however, the results are similar. This is expected since, from the kinematic point of view, the differential cross sections for quark-antiquark processes do not have major dependencies on the absolute value of the external 3-momentum, $\abs{\vec{k}}$, meaning that both regularization schemes will give very similar results. This is a feature of the 3-momentum sphere intersection regularization: it depends on the absolute value of the external 3-momentum, $\abs{\vec{k}}$. Since we are regularizing the integral by evaluating it in the region of integration corresponding to the intersection of two spheres with centers distanced $\abs{\vec{k}}$ and radii $\Lambda$, when $\abs{\vec{k}}=0$, both spheres completely intersect and we end up with the usual 3-momentum regularization.

Thus, not only this new approach to the 3-momentum regularization scheme yields different quantitative and qualitative results for the quark-quark cross sections, it also provides a way to correctly account for the symmetries of the system. These points are very important when using the cross sections evaluated from this model to obtain the quark relaxation time and use it to predict the in-medium transport properties of the quark plasma. The impact of the regularization scheme proposed in this work on the evaluation of transport coefficients like shear/bulk viscosities and thermal/electric conductivities is currently being analyzed and being prepared for future publications.

As future work, we would like to use the Pauli-Villars regularization in order to regularize the model. We are specially interested on studying the differences to the cross sections, between using the 3-momentum sphere intersection regularization, used in this work, and applying the Pauli-Villars regularization. As discussed, it is extremely important to use a regularization procedure which, like the one presented here, is able to preserve the symmetries between meson and antimesons.

Another extension of this work is to include the effect of magnetic fields in the calculation of the one and two fermion line integrals. Such extension can be very important to study systems under the presence of intense magnetic fields such as the case of heavy ion collision experiments or to study the matter inside magnetars. In such case, the Pauli-Villars regularization may be an extremely useful approach in order to study the effects of an external magnetic field. When considering a finite magnetic field, the 3-momentum integration is broken down in to an integration over the momenta aligned with the magnetic field, usually $p_z$, and the two dimensional integration over the transverse momenta, $p_x-p_y$, which is replaced by a sum over discrete values, the Landau levels. If considering a 3-momentum cutoff, the transverse component of the momentum should also be restricted, i.e., cutting off the sum. However, when dealing with an integration from 0 to infinity, as is the case in the Pauli-Villars regularization, this sum can be performed analytically, greatly simplifying the calculations.

We also wish to include in the model a coupling of the quark fields to an external static background of gluons. In such extended model, one is able to also study the deconfinement transition by studying the Polyakov loop order parameter at finite temperature and chemical potential. Such improved models are named Polyakov$-$Nambu$-$Jona-Lasinio (PNJL) models. In this work we used the pion Mott temperature as an indication for the system to be in the regime of deconfined matter. However, using the PNJL model would allow for a better description of the deconfinement transition and a better qualitative and quantitative indication of the presence of deconfined matter. Not only that, the PNJL is known to yield better qualitative results when compared to lattice QCD calculations, yielding a better qualitative prediction of the quark-quark and quark-antiquark cross sections. One can even use the so-called entangled-PNJL model where the coupling constants of the model are also dependent on the Polyakov loop order parameter.

Finally, the formalism developed here can also be extended to allow for the inclusion of pseudoscalar condensation at finite temperature and chemical potential, e.g., pion and kaon condensation \cite{Barducci:2004nc,Xia:2013caa}. In asymmetric matter, such as the one created in heavy-ion collisions experiments or present inside neutron stars, the isospin chemical potential can be sufficiently large for meson condensation to occur \cite{Albrecht:2010eg,Florkowski:1995hv,Khunjua:2019nnv,Andersen:2007qv,Xia:2013caa}.

\section*{Acknowledgments}

This work was supported by a research grant under Project No. PTDC/FIS-NUC/29912/2017, funded by national funds through FCT (Fundação para a Ciência e a Tecnologia, I.P, Portugal) and co-financed by the European Regional Development Fund (ERDF) through the Portuguese Operational Program for Competitiveness and Internationalization, COMPETE 2020, by the project CERN/FIS-PAR/0040/2019 and by national funds from FCT, within the Projects No. UIDB/04564/2020 and No. UIDP/04564/2020.

\appendix

\section{Linear expansion of the product between $N$ commuting operators}
\label{linear_expansion_product_N_operators}

An operator, $\operator_i$, can be written as its own expectation value, plus a small perturbation $\delta \operator_i$ around it \cite{Pereira:2021xxv}:
\begin{align}
\operator_i 
& = 
\expval*{ \operator_i } + 
\qty( \operator_i - \expval*{ \operator_i } ) 
=
\expval*{ \operator_i } + \delta \operator_i .
\label{MF_operator}
\end{align}
The perturbation is defined as:
\begin{align}
\delta \operator_i 
= 
\operator_i - \expval*{ \operator_i } .
\label{MF_operator_pertubation}
\end{align}  
Using these expressions, we can write the product between two commuting operators, $\operator_1$ and $\operator_2$, as:
\begin{align*}
\operator_1 \operator_2 
& = 
\qty( \expval*{ \operator_1 } + \delta \operator_1 ) 
\qty( \expval*{ \operator_2 } + \delta \operator_2 ) 
\\
& = 
\expval*{ \operator_1 } \expval*{ \operator_2 } + 
\expval*{ \operator_1 } \delta \operator_2 + 
\delta \operator_1 \expval*{ \operator_2 } +  
\delta \operator_1 \delta \operator_2 .
\numberthis
\label{product2Operators}
\end{align*}
The so-called linear product between these two operators, $\Linear{ \operator_1 \operator_2  }$, is defined by neglecting the quadratic term $\delta\hat{\mathcal{O}}_1 \delta\hat{\mathcal{O}}_2$ and then, re-writing the perturbation, $\delta \operator_i$, using its definition (see Eq. (\ref{MF_operator_pertubation})). Thus, one can write:
\begin{align}
\Linear{ \operator_1 \operator_2 }
& =  
\operator_1 \expval*{ \operator_2 } + 
\operator_2 \expval*{ \operator_1 }
- \expval*{ \operator_1 } \expval*{ \operator_2 } .
\label{linear_2_operator_product}
\end{align}

The process to define the linear expansion of the product between three commuting operators is exactly the same. Using Eq. (\ref{MF_operator}), we can write the product between three operators as:
\begin{align*}
\operator_1 \operator_2 \operator_3
& = 
\qty( \expval*{ \operator_1 } + \delta \operator_1 ) 
\qty( \expval*{ \operator_2 } + \delta \operator_2 ) 
\qty( \expval*{ \operator_3 } + \delta \operator_3 ) 
\\
& =
\expval*{ \operator_1 } \expval*{ \operator_2 } \expval*{ \operator_3 } + 
\expval*{ \operator_2 } \expval*{ \operator_3 } \delta \operator_1 + \expval*{ \operator_1 } \expval*{ \operator_3 } \delta \operator_2 + \expval*{ \operator_1 } \expval*{ \operator_2 } \delta \operator_3 
\\
&
+ \expval*{ \operator_3 } \delta \operator_1 \delta \operator_2 
+ \expval*{ \operator_2 } \delta \operator_1 \delta \operator_3 
+ \expval*{ \operator_1 } \delta \operator_2 \delta \operator_3 + \delta \operator_1 \delta \operator_2 \delta \operator_3 .
\numberthis
\label{product_3_operators}
\end{align*}
In this case, one neglects the terms proportional to both $(\delta \operator)^2$ and $(\delta \operator)^3$. Then, using Eq. (\ref{MF_operator_pertubation}), one gets:
\begin{align}
\Linear{ \operator_1 \operator_2 \operator_3 }
& =
-2 
\expval*{\operator_1} 
\expval*{\operator_2} 
\expval*{\operator_3} + 
\operator_1 \expval*{\operator_2} \expval*{\operator_3} + \expval*{\operator_1} \operator_2 \expval*{\operator_3} + \expval*{\operator_1} \expval*{\operator_2} \operator_3 .
\label{linear_3_operator_product}
\end{align}


This procedure can be performed to products with any number of operators. Thus, the linear product between $N$ operators can be obtained, by the three following operations: firstly, writing each operator as in Eq. (\ref{MF_operator}), secondly, neglecting terms proportional to $( \delta \operator )^n$ with $n\geq2$, thirdly, writing the perturbations using the definition stated in Eq. (\ref{MF_operator_pertubation}).

Interestingly, one can obtain the linear product between $N$ operators iteratively, by elevating the linear product between two operators, $\Linear{ \operator_1 \operator_2 }$ (explicitly given in Eq. (\ref{linear_2_operator_product})), to a definition. For completeness, one must also consider that the linear expansion of an operator or a constant (products of mean fields) is the operator or constant itself, $\Linear{\operator} = \operator$ and $\Linear{\mathrm{cte}} = \mathrm{cte}$. Using these definitions, one can build the linear expansion of any number of operators by iteratively using the following defining property:
\begin{align}
\Linear{ \operator_1\operator_2 \ldots \operator_N }
=
\Linear{ 
\Linear{\operator_1\operator_2 \ldots \operator_{N-1}}
\operator_N }.
\label{linear_N_operator_product_definition}
\end{align}
Here, the particular set of $N-1$ operators chosen to be inside the nested linear operation does not matter. Recursively, one can continue breaking down the innermost linear product, creating an iterative set of operations until the defining case of two operators is reached: $\Linear{ \operator_1 \operator_2 }$. Explicitly, one can write:
\begin{align}
\Linear{ \operator_1\operator_2 \ldots \operator_N }
=
\Linear{ 
\Linear{
\Linear{ \ldots
\Linear{ \operator_1 \operator_2 }
\operator_3 } \operator_4 \ldots
\operator_{N-1} }
\operator_N } .
\end{align}

For instance, following these definitions, the linear expansion of the product between three commuting operators is given by:
\begin{align*}
\Linear{ \operator_1 \operator_2 \operator_3 }
& =
\Linear{ \Linear{ \operator_1 \operator_2 } \operator_3 }
=
\Linear{ \Linear{ \operator_1 \operator_3 } \operator_2 }
=
\Linear{ \Linear{ \operator_2 \operator_3 } \operator_1 } .
\numberthis
\end{align*}
Here, it is highlighted the fact that the particular combination of two operators chosen in the inner linear operation does not matter. By substituting the definition of $\Linear{ \operator_1 \operator_2 }$ in the above (see Eq. (\ref{linear_2_operator_product})), one recovers the linear product between three operators obtained in Eq. (\ref{linear_3_operator_product}). For instance, for four operators, one can write:
\begin{align*}
\Linear{ \operator_1 \operator_2 \operator_3 \operator_4 }
& = 
\Linear{ \Linear{ \operator_1 \operator_2 \operator_3 } \operator_4 } 
\\
& =
\Linear{ \Linear{ \Linear{ \operator_1 \operator_2 } \operator_3 } \operator_4 } .
\numberthis
\end{align*}
Here, we only wrote one particular choice for the nested linerization operations. There are several other possible permutations among the operators that give the same result.

Due to its recursive nature, applying the definition given in Eq. (\ref{linear_N_operator_product_definition}) to get the linear product between $N$ operators can be quite cumbersome. Fortunately, the linear product between $N=n+1$ operators, with $n \geq 1$, can also be obtained via the following formula:
\begin{align}
\Linear{ \prod_{i=1}^{n+1} \operator_i } & =  
\qty(
\sum_{i=1}^{n+1} \frac{ \operator_i }{ \expval*{ \operator_i } }
- n  
)
\prod_{j=1}^{n+1} \expval*{ \operator_j } .
\label{linear_exp_Noperator}
\end{align}
This formula can be proved by mathematical induction. First, one considers the base case and show that the proposition is valid for $n=1$. Substituting $n=1$, in Eq. (\ref{linear_exp_Noperator}), it yields the linear 2-operator product given in Eq. (\ref{linear_2_operator_product}). This establishes the validity of the formula for the $n=1$ case. The next step is to assume the validity of the formula written in Eq. (\ref{linear_exp_Noperator}) for a given $n=m$, and then show that it also holds for the case $n=m+1$. Thus, one has to show that the following equality holds:
\begin{align}
\Linear{ \prod_{i=1}^{m+2} \operator_i } & =  
\qty(
\sum_{i=1}^{m+2} \frac{ \operator_i }{ \expval*{ \operator_i } }
- \qty(m+1)
)
\prod_{j=1}^{m+2} \expval*{ \operator_j } .
\label{inductive_step}
\end{align}
The right-hand side of this equation can be written as:
\begin{align*}
\RHS
& =
\qty(
\sum_{i=1}^{m+1} 
\frac{ \operator_i }{ \expval*{ \operator_i } } 
+
\frac{ \operator_{m+2}  }{ \expval*{ \operator_{m+2}  } }
- 
\qty(m+1)
)
\prod_{j=1}^{m+2} \expval*{ \operator_j } .
\numberthis
\end{align*}
Now, consider the left-hand side of Eq. (\ref{inductive_step}). Writing the $m+2$ contribution in the product explicitly and using the defining property of the linear product between $N$ operators given in Eq. (\ref{linear_N_operator_product_definition}), we can write the left-hand side as:
\begin{align*}
\LHS 
& = 
\Linear{  \qty(\prod_{i=1}^{m+1} \operator_i ) \operator_{m+2} }
\\
& = 
\Linear{
\Linear{ \prod_{i=1}^{m+1} \operator_i } 
\operator_{m+2}
}
\\
& = 
\qty(
\sum_{i=1}^{m+1} 
\frac{ 1 }{ \expval*{ \operator_i } } 
\Linear{ \operator_i \operator_{m+2} }
- m \operator_{m+2}
)
\prod_{j=1}^{m+1} \expval*{ \operator_j }  .
\numberthis
\end{align*}
Here, we assumed that Eq. (\ref{linear_exp_Noperator}) is valid for some $m$ and used the fact that $\Linear{\operator_{m+2}} = \operator_{m+2}$. The first term can be expanded using the base case, i.e., the result for the linear product of two operators given in Eq. (\ref{linear_2_operator_product}). We write:
\begin{align*}
\LHS 
& = 
\qty(
\sum_{i=1}^{m+1} 
\frac{ 1 }{ \expval*{ \operator_i } } 
\qty(
\frac{ \operator_i }{ \expval*{ \operator_i } } +
\frac{ \operator_{m+2} }{ \expval*{ \operator_{m+2} } } 
- 1  )
\expval*{ \operator_i } \expval*{ \operator_{m+2} } 
- m \operator_{m+2}
)
\prod_{j=1}^{m+1} \expval*{ \operator_j }
\\
& = 
\qty(
\sum_{i=1}^{m+1} 
\frac{ \operator_i }{ \expval*{ \operator_i } } 
+
\frac{ \operator_{m+2} }{ \expval*{ \operator_{m+2} } } 
- 
\qty(m+1) 
)
\prod_{j=1}^{m+2} \expval*{ \operator_j }
\\
& = \RHS .
\numberthis
\end{align*}
Since both sides of Eq. (\ref{inductive_step}) are equal, the formula also holds for $n=m+1$, completing the inductive step. Hence, by mathematical induction, Eq. (\ref{linear_exp_Noperator}) is valid for $n \geq 1$.

\section{Quadratic expansion of the product between $N$ commuting operators}
\label{quadratic_expansion_product_N_operators}

Similarly to the linear product between $N$ commuting operators, the quadratic product between $N$ commuting operators can be be obtained by writing each operator as in Eq. (\ref{MF_operator}), neglecting terms proportional to cubic perturbations or higher, $( \delta \operator )^n$ for $n\geq 3$, and then re-writing the perturbations using Eq. (\ref{MF_operator_pertubation}). 

Consider the product between three operators, $\operator_1$, $\operator_2$ and $\operator_3$. It was explicitly written out in Eq. (\ref{product_3_operators}). Neglecting the term $\delta\hat{\mathcal{O}}_1 \delta\hat{\mathcal{O}}_2 \delta\hat{\mathcal{O}}_3$ and re-writing the perturbations, yields its quadratic expansion:
\begin{align*}
\Quadratic{ \operator_1 \operator_2 \operator_3 }
& =  
\operator_1 \operator_2 \expval*{ \operator_3 } +
\operator_3 \operator_1 \expval*{ \operator_2 } +
\operator_2 \operator_3 \expval*{ \operator_1 } 
\\
& \quad \quad
-\operator_1 \expval*{ \operator_2 } \expval*{ \operator_3 } 
-\operator_3 \expval*{ \operator_1 } \expval*{ \operator_2 } 
-\operator_2 \expval*{ \operator_3 } \expval*{ \operator_1 } 
+
\expval*{ \operator_1 } 
\expval*{ \operator_2 } 
\expval*{ \operator_3 } .
\numberthis
\label{quadratic_3_operator_product}
\end{align*}

Following the same steps, the quadratic product between four operators can be written as:
\begin{align*}
\Quadratic{ 
\operator_1
\operator_2
\operator_3
\operator_4
}
= &
- 2 \expval*{\operator_2} \expval*{\operator_3} \expval*{\operator_4} \operator_1 
- 2 \expval*{\operator_1} \expval*{\operator_3} \expval*{\operator_4} \operator_2 
\\
& 
- 2 \expval*{\operator_1} \expval*{\operator_2} \expval*{\operator_4} \operator_3 
- 2 \expval*{\operator_1} \expval*{\operator_2} \expval*{\operator_3} \operator_4 
\\
& 
+ \expval*{\operator_2} \expval*{\operator_4} \operator_1 \operator_3 
+ \expval*{\operator_1} \expval*{\operator_4} \operator_2 \operator_3
+ \expval*{\operator_2} \expval*{\operator_3} \operator_1 \operator_4 
\\
& 
+ \expval*{\operator_1} \expval*{\operator_3} \operator_2 \operator_4 
+ \expval*{\operator_1} \expval*{\operator_2} \operator_3 \operator_4
+ \expval*{\operator_3} \expval*{\operator_4} \operator_1 \operator_2 
\\
& 
+ 3 \expval*{\operator_1} \expval*{\operator_2} \expval*{\operator_3} \expval*{\operator_4} .
\numberthis
\label{quadratic_4_operator_product}
\end{align*}
In order to write this expression, the terms proportional to both $\delta\hat{\mathcal{O}}_1 \delta\hat{\mathcal{O}}_2 \delta\hat{\mathcal{O}}_3$ and $\delta\hat{\mathcal{O}}_1 \delta\hat{\mathcal{O}}_2 \delta\hat{\mathcal{O}}_3 \delta\hat{\mathcal{O}}_4$ were neglected, after each operator in the product $\operator_1 \operator_2 \operator_3 \operator_4$ was written using Eq. (\ref{MF_operator}).

In a similar way to the linear expansion of the product between $N$ operators described in the previous section, one can obtain the quadratic product between $N$ operators iteratively, by taking the quadratic product between three operators, $\Quadratic{\operator_1 \operator_2 \operator_3}$, as a definition. Additionally, one must consider the following statements: $\Quadratic{\operator_1 \operator_2} = \operator_1 \operator_2$, $\Quadratic{\operator} = \operator$ and $\Quadratic{\mathrm{cte}} = \mathrm{cte}$. Using these constraints, the quadratic expansion of the product of any number of operators can be obtained recursively using the following defining property:
\begin{align}
\Quadratic{ \operator_1\operator_2 \ldots \operator_N }
=
\Quadratic{ 
\Quadratic{\operator_1\operator_2 \ldots \operator_{N-1}}
\operator_N }.
\label{quadratic_N_operator_product_definition}
\end{align}
The particular set of $N-1$ operators, chosen to be inside the nested quadratic operation, does not matter. Recursively, one can continue breaking down the innermost quadratic product, creating an iterative set of operations until the defining case is reached: $\Quadratic{ \operator_1 \operator_2 \operator_3 }$. Explicitly, one can write:
\begin{align}
\Quadratic{ \operator_1\operator_2 \ldots \operator_N }
=
\Quadratic{ 
\Quadratic{
\Quadratic{ \ldots
\Quadratic{ \operator_1 \operator_2 \operator_3 }
\operator_4 } \operator_5 \ldots
\operator_{N-1} }
\operator_N } .
\end{align}
As an example, the quadratic product between four operators, can be obtained by writing:
\begin{align*}
\Quadratic{ \operator_1 \operator_2 \operator_3 \operator_4 }
& = 
\Quadratic{ \Quadratic{ \operator_1 \operator_2 \operator_3 } \operator_4 } .
\numberthis
\end{align*}
Substituting $\Quadratic{ \operator_1 \operator_2 \operator_3 }$ by its definition, given in Eq. (\ref{quadratic_3_operator_product}), one recovers the expression given in Eq. (\ref{quadratic_4_operator_product}).

Instead of using the recursive approach laid out above, the quadratic product between $N=n+2$ operators, with $n \geq 1$, can be conveniently obtained using the following formula:
\begin{align}
\Quadratic{ \prod_{i=1}^{n+2} \operator_i }
& =  
\qty(
\frac{1}{2}
\sum_{i=1}^{n+2}
\sum_{j=1}^{n+2}
\frac{ \operator_i  }{ \expval*{ \operator_i } }
\frac{ \operator_j }{ \expval*{ \operator_j } }
\qty( 1 - \delta_{ij} )
- n  
\sum_{i=1}^{n+2}
\frac{ \operator_i }{ \expval*{ \operator_i } }
+\frac{n}{2} \qty(n+1)
)
\prod_{k=1}^{n+2} \expval*{ \operator_k } .
\label{quadratic_exp_Noperator}
\end{align}
As for the linear product expansion, this formula can be proved by mathematical induction. First we consider the base case and show that the expression is valid for $n=1$, i.e., for the quadratic expansion of the product between three operators. Using this formula one can readily obtain the result presented in Eq. (\ref{quadratic_3_operator_product}), validating the formula for the base case with $n=1$. Next, assuming that the formula holds for $n=m$, one must demonstrate that it is also valid for $n=m+1$. Hence, we have to show that the following relations holds ($n=m+1$ case):
\begin{align}
\Quadratic{ \prod_{i=1}^{m+3} \operator_i }
& =  
\qty(
\frac{1}{2}
\sum_{i=1}^{m+3}
\sum_{j=1}^{m+3}
\frac{ \operator_i  }{ \expval*{ \operator_i } }
\frac{ \operator_j }{ \expval*{ \operator_j } }
\qty( 1 - \delta_{ij} )
- \qty(m+1)  
\sum_{i=1}^{m+3}
\frac{ \operator_i }{ \expval*{ \operator_i } }
+\frac{\qty(m+1)}{2} \qty(m+2)
)
\prod_{k=1}^{m+3} \expval*{ \operator_k } .
\label{inductive_step_quadratic_expansion}
\end{align}
Let us focus on the right-hand side of Eq. (\ref{inductive_step_quadratic_expansion}). Separating the $m+3$ contribution in the sums, we can write:
\begin{align*}
\RHS 
 = &
\bigg(
\frac{1}{2}
\sum_{i=1}^{m+2}
\sum_{j=1}^{m+2}
\frac{ \operator_i  }{ \expval*{ \operator_i } }
\frac{ \operator_j }{ \expval*{ \operator_j } }
\qty( 1 - \delta_{ij} )
+
\frac{ \operator_{m+3} }{ \expval*{ \operator_{m+3} } }
\sum_{i=1}^{m+2}
\frac{ \operator_i  }{ \expval*{ \operator_i } }
\\
& \qquad
- \qty(m+1)  
\sum_{i=1}^{m+2}
\frac{ \operator_i }{ \expval*{ \operator_i } }
- \qty(m+1)  
\frac{ \operator_{m+3} }{ \expval*{ \operator_{m+3} } }
+\frac{\qty(m+1)}{2} \qty(m+2)
\bigg)
\prod_{k=1}^{m+3} \expval*{ \operator_k } .
\numberthis
\label{inductive_step_RHS}
\end{align*}

Consider the left-hand side of Eq. (\ref{inductive_step_quadratic_expansion}). One can separate the $i=m+3$ contribution to the product and, using the defining property of Eq. (\ref{quadratic_N_operator_product_definition}), write:
\begin{align*}
\LHS 
& = 
\Quadratic{ 
\qty(\prod_{i=1}^{m+2} \operator_i )
\operator_{m+3} 
}
\\
& = 
\Quadratic{ 
\Quadratic{\prod_{i=1}^{m+2} \operator_i }
\operator_{m+3} }
\\
& =
\Bigg(
\frac{1}{2}
\sum_{i=1}^{m+2}
\sum_{j=1}^{m+2}
\frac
{ \Quadratic{\operator_i \operator_j \operator_{m+3}} }
{ \expval*{ \operator_i } \expval*{ \operator_j } \expval*{ \operator_{m+3} } }
\qty( 1 - \delta_{ij} )
\\
& \qquad \qquad \qquad 
- m  
\sum_{i=1}^{m+2}
\frac{ \operator_i }{ \expval*{ \operator_i } }
\frac{ \operator_{m+3}  }{ \expval*{ \operator_{m+3} } }
+
\frac{m}{2} \qty(m+1)
\frac{ \operator_{m+3}  }{ \expval*{ \operator_{m+3} } }
\Bigg)
\prod_{k=1}^{m+3} \expval*{ \operator_k } .
\numberthis
\end{align*}
Here, we applied the base case (see Eq.(\ref{quadratic_exp_Noperator})). Defining the first term inside the brackets by $t$, we can write it as:
\begin{align}
t
& =
\frac{1}{2}
\sum_{i=1}^{m+2}
\sum_{j=1}^{m+2}
\frac
{ \Quadratic{\operator_i \operator_j \operator_{m+3}} }
{ \expval*{ \operator_i } \expval*{ \operator_j } \expval*{ \operator_{m+3} } }
\qty( 1 - \delta_{ij} ).
\end{align}
Using Eq. (\ref{quadratic_3_operator_product}), we can write $t$ as:
\begin{align*}
t
=
\frac{1}{2}
\sum_{i=1}^{m+2}
\sum_{j=1}^{m+2}
\frac{ 1  }{ \expval*{ \operator_i } }
\frac{ 1 }{ \expval*{ \operator_j } }
\frac{ 1  }{ \expval*{ \operator_{m+3} } }
\qty( 1 - \delta_{ij} )
\Bigg(
&
\operator_i \operator_j \expval*{ \operator_{m+3} } +
\operator_{m+3} \operator_i \expval*{ \operator_j } +
\operator_j \operator_{m+3} \expval*{ \operator_i } 
\\
&  \quad 
-\operator_i \expval*{ \operator_j } \expval*{ \operator_{m+3} } 
-\operator_{m+3} \expval*{ \operator_i } \expval*{ \operator_j } 
\\
& \qquad \quad
-\operator_j \expval*{ \operator_{m+3} } \expval*{ \operator_i } 
+
\expval*{ \operator_i } 
\expval*{ \operator_j } 
\expval*{ \operator_{m+3} } 
\Bigg) .
\numberthis
\end{align*}
After some straightforward algebra, we can write $t$ as:
\begin{align*}
t
 =
\frac{1}{2}
\sum_{i=1}^{m+2}
\sum_{j=1}^{m+2}
\frac{ \operator_i  }{ \expval*{ \operator_i } }
\frac{ \operator_j }{ \expval*{ \operator_j } }
\qty( 1 - \delta_{ij} )
& +
\frac{ \operator_{m+3}  }{ \expval*{ \operator_{m+3} } }
\qty(m+1) \sum_{i=1}^{m+2} \frac{ \operator_i  }{ \expval*{ \operator_i } } 
\\
-
\qty(m+1) \sum_{i=1}^{m+2} \frac{ \operator_i  }{ \expval*{ \operator_i } } 
& -
\frac{1}{2}
\qty(m+2)\qty(m+1)
\frac{ \operator_{m+3}  }{ \expval*{ \operator_{m+3} } }
+
\frac{1}{2}
\qty(m+2)\qty(m+1) .
\numberthis
\end{align*}
Where we have used, $
\sum_{i,j=1}^{m+2}
\qty( 1 - \delta_{ij} )
=
\qty(m+2)\qty(m+1)$ and $
\sum_{i,j=1}^{m+2}
\frac{ \operator_i  }{ \expval*{ \operator_i } }
\qty( 1 - \delta_{ij} )
=
\qty(m+1) \sum_{i=1}^{m+2} \frac{ \operator_i  }{ \expval*{ \operator_i } }
$. Finally, the left-hand side of Eq. (\ref{inductive_step_quadratic_expansion}) can be written as:
\begin{align*}
\LHS 
 =
& \bigg(
\frac{1}{2}
\sum_{i=1}^{m+2}
\sum_{j=1}^{m+2}
\frac{ \operator_i  }{ \expval*{ \operator_i } }
\frac{ \operator_j }{ \expval*{ \operator_j } }
\qty( 1 - \delta_{ij} )
+
\frac{ \operator_{m+3}  }{ \expval*{ \operator_{m+3} } }
\sum_{i=1}^{m+2} \frac{ \operator_i  }{ \expval*{ \operator_i } } 
\\
& \qquad
-
\qty(m+1) \sum_{i=1}^{m+2} \frac{ \operator_i  }{ \expval*{ \operator_i } } 
-
\qty(m+1)
\frac{ \operator_{m+3}  }{ \expval*{ \operator_{m+3} } }
+
\frac{\qty(m+1)}{2}
\qty(m+2)
\bigg)
\prod_{k=1}^{m+3} \expval*{ \operator_k } 
\\
= &
\RHS .
\numberthis
\end{align*}
Hence, the left-hand side is equal to the right-hand side given in Eq. (\ref{inductive_step_RHS}), the formula also holds for $n=m+1$, completing the inductive step. Hence, by mathematical induction, Eq. (\ref{quadratic_exp_Noperator}) is valid for $n \geq 1$.

\section{Defining the region of integration for the $B_0$ and $\mathcal{A}$ functions after changing variables}
\label{integration_region_B0COV}

In order to complete the change of variables performed in the two fermion, one loop integral (in the $\abs{\vec{k}}>0$ case) defined in Eqs. (\ref{E_B0_COV})$-$(\ref{B0scattGeneralDef}), one must find the limits of integration. The case of equal masses was performed in Ref. \cite{Yamazaki:2012ux}. By construction, the following equality holds, $E[M_j,\vec{p}-\vec{k}]^2-E\qty[M_i,\vec{p}]^2 = \vec{k}^2  - 2 \abs{\vec{p}} \abs{\vec{k}} \cos \qty[\theta] - M_i^2 + M_j^2= 2 E \varepsilon$, with $E \qty[ M, \vec{p}]$ defined in Eq. (\ref{energyEquation}) while, $E$ and $\varepsilon$, are the new integration variables, defined in Eqs. (\ref{E_B0_COV}) and (\ref{epsilon_B0_COV}), respectively. Solving this equation with respect to  $\cos \qty[\theta] $ and squaring both sides, yields:
\begin{align}
\cos \qty[\theta]^2
=
\frac
{
\qty( \vec{k}^2 - M_i^2 + M_j^2 - 2 E \varepsilon )^2
}{
4 \vec{p}^2 \vec{k}^2 
} .
\end{align}
In order to have a right-hand side which depends only on the new integration variables and the external 3-momentum, $\abs{\vec{k}}$, we can solve $E \qty[ M_i, \vec{p}]= \sqrt{ \vec{p}^2 + M_i^2 } = \qty( E - \frac{\varepsilon}{2} )$ for $\vec{p}^2$ (see Eqs. (\ref{E_B0_COV}) and (\ref{epsilon_B0_COV})) and substitute in the above equation. 
Since, in the original integration bounds, the angle $\theta \in \qty[0,\pi]$, it implies that $\cos^2[\theta] \leq 1 $. This allows us to define the following inequality:
\begin{align}
P \qty[ \varepsilon, E, M_i, M_j, \abs{\vec{k}} ]
=
\qty( \vec{k}^2 - M_i^2 + M_j^2 - 2 E \varepsilon )^2
-
4 \vec{k}^2 \qty( \qty( E - \frac{\varepsilon}{2} )^2 - M_i^2 )
\leq 0 .
\label{integrationRegionPolynomial}
\end{align}
In the above, for simplicity, we have defined the function $P \qty[ \varepsilon, E, M_i, M_j, \abs{\vec{k}} ]$. It is easier to obtain the integration region if we write the function $P \qty[ \varepsilon, E, M_i, M_j, \abs{\vec{k}} ]$ as a polynomial in $\varepsilon$, with coefficients which depend on the remaining variables, $E$, $M_i$, $M_j$ and $\vec{k}^2$. The resulting inequality is quadratic in $\varepsilon$:
\begin{align}
P \qty[ \varepsilon, E, M_i, M_j, \abs{\vec{k}} ]
=
C_0 \qty[ E, M_i, M_j, \abs{\vec{k}} ]
+
C_1 \qty[ E, M_i, M_j, \abs{\vec{k}} ] \varepsilon
+
C_2 \qty[ E, M_i, M_j, \abs{\vec{k}} ] \varepsilon^2
\leq 0 ,
\end{align}
where,
\begin{align}
C_0 \qty[ E, M_i, M_j, \abs{\vec{k}} ]
& = 
\vec{k}^4 + 2 \vec{k}^2 \qty( M_i^2 + M_j^2 -2 E^2  ) +\qty( M_i^2 - M_j^2 )^2 ,
\\
C_1 \qty[ E, M_i, M_j, \abs{\vec{k}} ] 
& = 
4 E \qty( M_i^2 - M_j^2 ) ,
\\
C_2 \qty[ E, M_i, M_j, \abs{\vec{k}} ]
& =
4 E^2 - \vec{k}^2 .
\end{align}
The concavity of the inequality can be found by checking the sign of the coefficient in front of the term proportional to $\varepsilon^2$, $C_2 \qty[ E, M_i, M_j, \abs{\vec{k}} ]$. One can show that this coefficient is always bigger than zero due to the so-called triangle inequality: in a normed vector space $V$, one can write the triangle inequality: $ \abs{ \vec{x} + \vec{y} } \leq \abs{\vec{x}} + \abs{\vec{y}}$, for all $\vec{x}$, $\vec{y}$ that belongs to $V$. Trivially, we can write $\abs{\vec{k}}=\abs{ \qty(\vec{k} - \vec{p}) + \vec{p} }$ and, making use of the triangle inequality, it follows that:
\begin{align*}
\abs{\vec{k}} 
& \leq
\abs{ \vec{k} - \vec{p} } + \abs{\vec{p} } 
\\
\Rightarrow
\vec{k}^2
& \leq
\qty( \abs{ \vec{k} - \vec{p} } )^2 + 
\qty( \abs{\vec{p} }  )^2 +
2 \abs{ \vec{k} - \vec{p} } \abs{ \vec{p} } .
\numberthis
\end{align*}
One can always write $\abs{ \vec{p} } \leq \sqrt{ \vec{p}^2 + M_i^2 }=E\qty[M_i,\vec{p}]$ and $\abs{ \vec{k} - \vec{p} } \leq \sqrt{ \qty(\vec{k-p})^2 + M_j^2 }=E[M_j,\vec{p}-\vec{k}]$, since $M_i$ and $M_j$ are real and $M_i,M_j > 0$. The above inequality then becomes:
\begin{align*}
\vec{k}^2
& <
\qty( 
E[M_j,\vec{p}-\vec{k}] 
+ 
E\qty[M_i,\vec{p}]
)^2
=
4 E^2
\\
\Rightarrow
0
& <
4 E^2 - \vec{k}^2.
\numberthis
\end{align*}
Thus, the coefficient of the quadratic term, $C_2 \qty[ E, M_i, M_j, \abs{\vec{k}} ]
 =
4 E^2 - \vec{k}^2$, is always positive, meaning that it corresponds to a parabola with the concavity turned upwards. Hence, for every fixed value of $E$, the region of integration is an upwards parabola. Since we are interested in the region delimited by the values of $\varepsilon$ and $E$ for which  $P \qty[ \varepsilon, E, M_i, M_j, \abs{\vec{k}} ] \leq 0$, this implies that the region of integration corresponds to the interior of the parabola (the region between the two zeros).

To finish determining the integration region, we need to apply the restriction coming from the 3-momentum cutoff in the 3-momentum sphere intersection regularization (see Section \ref{regularization}). These restrictions can readily be found by demanding that, $\abs{\vec{p}} \leq \Lambda$ and $\abs{\vec{p}-\vec{k}} \leq \Lambda$. For the first constraint, we get:
\begin{align*}
\abs{\vec{p}} \leq \Lambda
\Rightarrow &
\qty( E - \frac{\varepsilon}{2} )^2 \leq 
\Lambda^2 + M_i^2
\Rightarrow 
-\sqrt{\Lambda^2 + M_i^2 } \leq E - \frac{\varepsilon }{2}\leq \sqrt{ \Lambda^2 + M_i^2}.
\numberthis
\label{cutoffConstraint1}
\end{align*}
For the second constraint, one gets:
\begin{align*}
\abs{\vec{p}-\vec{k}} & \leq \Lambda
\Rightarrow
\qty( E + \frac{\varepsilon}{2} )^2 \leq \Lambda^2 + M_j^2
\Rightarrow
-\sqrt{\Lambda ^2 + M_j^2 } \leq E + \frac{\varepsilon }{2}\leq \sqrt{ \Lambda^2 + M_j^2 } .
\numberthis
\label{cutoffConstraint2}
\end{align*}
At this point, one specific choice of integration order must be chosen in order to evaluate the integral. As previously discussed it is quite useful to utilize both integration orders since, in certain terms, one can analytically perform the innermost integration when using the order $\int \dd{ \varepsilon } \int \dd{ E }$ while, for other terms, this analytical integration can only be performed for the other integration order, $\int \dd{ E } \int \dd{ \varepsilon }$.

It will also be quite useful to evaluate the integral $R_{ij}$, defined by:
\begin{align}
R_{ij} \qty[M_i, M_j, \abs{\vec{k}} ]
=
\frac{ \abs{\vec{k}} }{ 2\pi }
\int_{ \mathbb{I}_{\abs{\vec{k}}}^\Lambda }
\frac{ \dd[3]{\vec{p}}  }{ E\qty[M_i,\vec{p}] E[M_j,\vec{p}-\vec{k}] } .
\label{Rij_integral_definition}
\end{align}
Where the integration is to be made considering the 3-momentum sphere intersection regularization. Changing to the $E$ and $\varepsilon$ variables (see Eqs. (\ref{E_B0_COV}) and (\ref{epsilon_B0_COV}) and (\ref{jacobianCOV_knot0})), one obtains:
\begin{align}
R_{ij} \qty[\Lambda,M_i, M_j, \abs{\vec{k}} ]
= 
\int_{ \mathbb{I}_{\abs{\vec{k}}}^\Lambda }
\frac{\dd{\phi}}{2\pi}
\dd{ E } \dd{ \varepsilon } .
\label{Rij_integral_Eepsilon_variables}
\end{align}
Hence, one immediately finds that $2\pi R_{ij}$ corresponds to the volume of the integration region delimited by the 3-momentum sphere intersection regularization in the new variables $E$ and $\varepsilon$. Evaluating this quantity will be quite useful in order to: $(1)$ check that the change of variables is valid; $(2)$ verify that both integration orders, $\int \dd{ \varepsilon } \int \dd{ E }$   and $\int \dd{ E } \int \dd{ \varepsilon }$, yield the same results. In a later section we will also show how to evaluate this integral ($R_{ij}$), without making the change of variables, allowing for a multiple numerical cross check for the validity of the proposed integration procedure.

\subsection{ Integration order $\dd{ \varepsilon } \dd{ E }$ }

In this section we will consider the $\dd{ \varepsilon } \dd{ E }$ integration order, hence we will integrate first over the $E$ variable. In other words, we write the integration of a function $f(E,\varepsilon)$, as:
\begin{align}
\int_{\mathbb{I}_{\abs{\vec{k}}}^\Lambda}
\frac{\dd{\phi}}{2\pi}
\dd{ \varepsilon } \dd{ E }
f(E,\varepsilon)
=
\int_{ \depsilondEepsilonMin }^{ \depsilondEepsilonMax }
\dd{\varepsilon}
\int_{ \depsilondEEMin }^{ \depsilondEEMax } 
\dd{E} 
f(E,\varepsilon)
\end{align}
and define the different functions, $\depsilondEEMin$, $\depsilondEEMax$, $\depsilondEepsilonMin$ and $\depsilondEepsilonMax$, which encompass the integration region and are functions of the masses ($M_i$ and $M_j$), the absolute value of the external 3-momentum ($\abs{\vec{k}}$), and the 3-momentum cutoff, $\Lambda$. In the special case in which $f(E,\varepsilon)=1$, one obviously gets the integral $R_{ij}$, see Eq. (\ref{Rij_integral_Eepsilon_variables}).

To achieve this, we start by solving Eq. (\ref{integrationRegionPolynomial}) with respect to $E$. This yields two solutions ($E_\mathrm{min}^\pm$):
\begin{equation*}
E_\mathrm{min}^\pm
\qty[M_i, M_j, \abs{\vec{k}}, \varepsilon]
=
\begin{cases}
-
\frac
{ ( M_i^2 - M_j^2 )}
{ 4 \varepsilon }
-
\frac
{ \varepsilon  \qty( M_i^2 + M_j^2 )}
{ 2 ( M_i^2 - M_j^2 ) } ,
&\text{if} \;
\varepsilon = \pm \abs{\vec{k}} ,
\\
\frac{
\varepsilon  ( M_i^2 - M_j^2 )
\pm
\sqrt{
\vec{k}^2 
\qty( \vec{k}^2 + ( M_i - M_j )^2 -\varepsilon^2) 
\qty( \vec{k}^2 + ( M_i + M_j )^2 -\varepsilon^2) 
} 
}
{
2 
\qty( \abs{\vec{k}} - \varepsilon )
\qty( \abs{\vec{k}} + \varepsilon )
}  ,
&\text{otherwise}.
\end{cases}
\numberthis
\end{equation*}
In the special case in which $M_i=M_j=M$ one gets:
\begin{align}
E_\mathrm{min}^\pm
\qty[M, M, \abs{\vec{k}}, \varepsilon]
=
\pm
\frac{
\sqrt{
\vec{k}^2 
\qty( \vec{k}^2 -\varepsilon^2) 
\qty( \vec{k}^2 + 4 M^2 -\varepsilon^2) 
} 
}
{
2 
\qty( \abs{\vec{k}} - \varepsilon )
\qty( \abs{\vec{k}} + \varepsilon )
} .
\end{align}

From the restrictions imposed by the cutoff in Eqs. (\ref{cutoffConstraint1}) and (\ref{cutoffConstraint2}), one can further constrain the integration region. One can define the following function from solving theses equation with  respect to $E$:
\begin{align}
E_{ \Lambda_i }^\pm
[\Lambda , M_i, \varepsilon] 
& = 
+\frac{\varepsilon }{2}
\pm
\sqrt{ \Lambda^2 + M_i^2 } ,
\\
E_{ \Lambda_j }^\pm
[\Lambda ,  M_j, \varepsilon]
& = 
-\frac{\varepsilon }{2}
\pm
\sqrt{ \Lambda^2 + M_j^2 } .
\end{align}
By definition, the variable $E$ is positive (see Eqs. (\ref{E1_def}), (\ref{E2_def}) and (\ref{E_B0_COV})). Hence, we are interested only on the functions, $E_{ \Lambda_i }^+$ and $E_{ \Lambda_j }^+$. Furthermore, these functions intersect at a certain point, $\qty(\varepsilon_{\Lambda_s},E_{\Lambda_s})$. Such point is the solution to the equation, $E_{ \Lambda_i }^+ = E_{ \Lambda_j }^+$ and it is given by:
\begin{align}
\varepsilon_{\Lambda_s} \qty[\Lambda, M_i, M_j]
& =
\sqrt{ \Lambda^2 + M_j^2 }
-
\sqrt{ \Lambda^2 + M_i^2 } ,
\label{epsilonLambdaSwitch}
\\
E_{ \Lambda_s } \qty[\Lambda, M_i, M_j]
& = 
\frac{1}{2} 
\qty( \sqrt{ \Lambda^2 + M_i^2 } + \sqrt{ \Lambda^2 + M_j^2} ) .
\label{ELambdaSwitch}
\end{align}
Also, at this point, the restriction coming from the cutoff demands one to switch from one branch, to another. This can be more easily understood after a graphical analysis. Thus, we can unify the constraint coming from the 3-momentum cutoff in the following equation:
\begin{equation*}
E_{ \Lambda }
[\Lambda , M_i, M_j, \varepsilon] 
=
\begin{cases}
E_{ \Lambda_j }^+
[\Lambda , M_j, \varepsilon] ,
&\text{if} \;
\varepsilon > \varepsilon_{\Lambda_s} \qty[\Lambda, M_i, M_j] ,
\\
E_{ \Lambda_i }^+
[\Lambda , M_i, \varepsilon] ,
&\text{otherwise}.
\end{cases}
\numberthis
\end{equation*}

At this point we can define the lower bound of the $\dd{E}$ integration, $\depsilondEEMin$. It is simply given by:
\begin{align}
\depsilondEEMin
\qty[\Lambda, M_i, M_j, \abs{\vec{k}}, \varepsilon]
=
E_\mathrm{min}^+
\qty[M_i, M_j, \abs{\vec{k}}, \varepsilon] .
\label{depsilondE_EMin}
\end{align}
The upper bound of the $\dd{E}$ integration, $\depsilondEEMax$, however, is not so simple. When dealing with equal fermion masses ($M_i=M_j=M$), the constraint coming from the cutoff defined previously, completely fixes the upper bound. However, when dealing with different masses, this is not true and a switch from $E_{ \Lambda }$ to the function $E_\mathrm{min}^-$, can occur. This is also better understood after a visual analysis of the interplay between the functions. In order to define $\depsilondEEMax$ we will need to find other quantities of interest where some of the previously mentioned quantities intersect. Firstly, consider a possible intersection between the function $E_\mathrm{min}^-$ and $E_{ \Lambda_i }^+$, i.e., solutions of the equation, $E_\mathrm{min}^-\qty[M_i, M_j, \abs{\vec{k}}, \varepsilon_{\gamma_1}]=E_{ \Lambda_i }^+
[\Lambda , M_i, \varepsilon_{\gamma_1}]$. This results in four different intersection points, with only one being meaningful to the integration region (the other ones are outside the region). Its $\qty(\varepsilon_{\gamma_1},E_{\gamma_1})$ values are given by:
\begin{align}
\varepsilon_{\gamma_1}
\qty[ \Lambda, M_i, M_j, \abs{\vec{k}}]
& =
\sqrt{ \qty( \abs{\vec{k}} - \Lambda )^2 + M_j^2 }
-
\sqrt{ \Lambda^2 + M_i^2 } ,
\label{epsilonGamma1}
\\
E_{\gamma_1}
\qty[ \Lambda, M_i, M_j, \abs{\vec{k}}]
& =
\frac{1}{2}
\qty(
\sqrt{ \qty( \abs{\vec{k}} - \Lambda )^2 + M_j^2 }
+
\sqrt{ \Lambda^2 + M_i^2 }
) .
\label{EGamma1}
\end{align}
Likewise, the solution of interest to the equation $E_\mathrm{min}^-\qty[M_i, M_j, \abs{\vec{k}}, \varepsilon_{\gamma_2}]=E_{ \Lambda_j }^+
[\Lambda , M_j, \varepsilon_{\gamma_2}]$ is:
\begin{align}
\varepsilon_{\gamma_2}
\qty[ \Lambda, M_i, M_j, \abs{\vec{k}}]
& =
\sqrt{ \Lambda^2 + M_j^2 } 
-
\sqrt{ \qty( \abs{\vec{k}} - \Lambda )^2 + M_i^2 } ,
\label{epsilonGamma2}
\\
E_{\gamma_2}
\qty[ \Lambda, M_i, M_j, \abs{\vec{k}}]
& =
\frac{1}{2}
\qty(
\sqrt{ \Lambda^2 + M_j^2 }
+
\sqrt{ \qty( \abs{\vec{k}} - \Lambda )^2 + M_i^2 }
) .
\label{EGamma2}
\end{align}
Secondly, the equations $E_\mathrm{min}^+=E_\mathrm{min}^-$ are equal in four different points, with two of them being useful for our purposes, $\varepsilon_{\alpha_1}$ and $\varepsilon_{\alpha_2}$. They can be calculated to yield:
\begin{align}
\varepsilon_{\alpha_1}
\qty[ M_i, M_j, \abs{\vec{k}} ]
& =
-\sqrt{ \vec{k}^2 +( M_i - M_j )^2 }
=
- \varepsilon_{\alpha_2}
\qty[ M_i, M_j, \abs{\vec{k}} ] .
\label{EMinMinusEqualEMinPlusIntersectionEpsilon}
\end{align}
To each of these $\varepsilon$ points there are corresponding $E$ values, given by:
\begin{align}
E_{\alpha_1}
\qty[ M_i, M_j, \abs{\vec{k}} ]
& =
\frac{1}{2}
\frac{
\qty( M_i + M_j )
}
{
\qty( M_i - M_j )
}
\sqrt{ \vec{k}^2 + ( M_i - M_j )^2 }
=
- E_{\alpha_2}
\qty[ M_i, M_j, \abs{\vec{k}} ] .
\end{align}
From the above, one can observe that, $E_{\alpha_{1}},E_{\alpha_{2}} \to \infty$ for the degenerate mass case, meaning that the two equations, $E_\mathrm{min}^+$ and $E_\mathrm{min}^-$, are never equal. However, for different masses, there will exist two switching points, symmetric with respect to $\varepsilon$ (see Eq. (\ref{EMinMinusEqualEMinPlusIntersectionEpsilon})). Thus, as previously aluded to, increasing the difference between the masses, causes $E_{\alpha_1}$ and $E_{\alpha_2}$ to decrease and the upper bound, $\depsilondEEMax$, ceases to be defined by $E_{ \Lambda }$ switching to $E_\mathrm{min}^-$. In the case where $M_i>M_j$, one can write:
\begin{equation*}
E_\mathrm{max}^{(M_i>M_j)}
\qty[\Lambda, M_i, M_j, \abs{\vec{k}}, \varepsilon]
=
\begin{cases}
E_\mathrm{min}^-
\qty[M_i, M_j, \abs{\vec{k}}, \varepsilon] ,
& 
\begin{array}{l}
\text{if} \;
\varepsilon < \varepsilon_{\gamma_1}
\qty[ \Lambda, M_i, M_j, \abs{\vec{k}}] \; \text{and}
\\
\phantom{\text{if} \;}E_{\alpha_1}
\qty[ M_i, M_j, \abs{\vec{k}} ] < E_{\gamma_1}
\qty[ \Lambda, M_i, M_j, \abs{\vec{k}}]  ,
\end{array}
\\
E_{ \Lambda }
[\Lambda , M_i, M_j, \varepsilon]   ,
&\;\text{otherwise}.
\end{cases}
\numberthis
\end{equation*}
In the opposite mass case, $M_j>M_i$, the upper bound is:
\begin{equation*}
E_\mathrm{max}^{(M_j>M_i)}
\qty[\Lambda, M_i, M_j, \abs{\vec{k}}, \varepsilon]
=
\begin{cases}
E_\mathrm{min}^-
\qty[M_i, M_j, \abs{\vec{k}}, \varepsilon] ,
& 
\begin{array}{l}
\text{if} \;
\varepsilon > \varepsilon_{\gamma_2}
\qty[ \Lambda, M_i, M_j, \abs{\vec{k}}] \; \text{and}
\\
\phantom{\text{if} \;}E_{\alpha_2}
\qty[ M_i, M_j, \abs{\vec{k}} ] < E_{\gamma_2}
\qty[ \Lambda, M_i, M_j, \abs{\vec{k}}]  ,
\end{array}
\\
E_{ \Lambda }
[\Lambda , M_i, M_j, \varepsilon]   ,
&\;\text{otherwise}.
\end{cases}
\numberthis
\end{equation*}
As explained before, for the degenerate mass case, the upper bound is trivially given by the cutoff restriction only, i.e., $E_{ \Lambda }$. Gathering all this information, one can write the upper bound of the $\dd{E}$ integration, $\depsilondEEMax$, as:
\begin{equation*}
\depsilondEEMax
\qty[\Lambda, M_i, M_j, \abs{\vec{k}}, \varepsilon]
=
\begin{cases}
E_\mathrm{max}^{(M_i>M_j)}
\qty[\Lambda, M_i, M_j, \abs{\vec{k}}, \varepsilon] ,
&\text{if} \;
M_i > M_j ,
\\
E_\mathrm{max}^{(M_j>M_i)}
\qty[\Lambda, M_i, M_j, \abs{\vec{k}}, \varepsilon] ,
&\text{if} \;
M_i < M_j ,
\\
E_{ \Lambda }
[\Lambda , M_i, M_j, \varepsilon] ,
&\text{otherwise}.
\end{cases}
\numberthis
\label{depsilondE_EMax}
\end{equation*}

Finally, taking into account all the intersections of interest, one can write the lower bound of the $\dd{\varepsilon}$ integration, $\depsilondEepsilonMin$, as:
\begin{equation*}
\depsilondEepsilonMin
\qty[\Lambda, M_i, M_j, \abs{\vec{k}}]
=
\begin{cases}
\varepsilon_{\alpha_1}
\qty[ M_i, M_j, \abs{\vec{k}} ] ,
& 
\begin{array}{l}
\text{if} \;
M_i > M_j \; \text{and}
\\
\phantom{\text{if} \;}
E_{ \Lambda }
[\Lambda , M_i, M_j, 
\varepsilon_{\alpha_1}
\qty[ M_i, M_j, \abs{\vec{k}} ]
] 
\geq 
E_{\alpha_1}
\qty[ M_i, M_j, \abs{\vec{k}} ] ,
\end{array}
\\
\varepsilon_{\gamma_1}
\qty[ \Lambda, M_i, M_j, \abs{\vec{k}}] ,
&\;\text{otherwise} .
\end{cases}
\numberthis
\label{depsilondE_epsilonMin}
\end{equation*}
The upper bound, on the other hand, is given by:
\begin{equation*}
\depsilondEepsilonMax
\qty[\Lambda, M_i, M_j, \abs{\vec{k}}]
=
\begin{cases}
\varepsilon_{\alpha_2}
\qty[ M_i, M_j, \abs{\vec{k}} ] ,
& 
\begin{array}{l}
\text{if} \;
M_i < M_j \; \text{and}
\\
\phantom{\text{if} \;}E_{ \Lambda }
[\Lambda , M_i, M_j, 
\varepsilon_{\alpha_2}
\qty[ M_i, M_j, \abs{\vec{k}} ]
] 
\geq 
E_{\alpha_2}
\qty[ M_i, M_j, \abs{\vec{k}} ] ,
\end{array}
\\
\varepsilon_{\gamma_2}
\qty[ \Lambda, M_i, M_j, \abs{\vec{k}}] ,
&\;\text{otherwise} .
\end{cases}
\numberthis
\label{depsilondE_epsilonMax}
\end{equation*}
One can check that the size of the integration region gets smaller with increasing $\abs{\vec{k}}$. This is a consequence of the regularization used in this work. As explained earlier, increasing the value of $\abs{\vec{k}}$ is equivalent to increasing the distance between the center of two 3-momentum spheres of radii $\Lambda$. For $\abs{\vec{k}}=2\Lambda$ the spheres no longer intersect and the integration region should be zero. Hence, to improve the numerical implementation of the calculation, one can additionally include a condition to only calculate the integral if $0<\abs{\vec{k}}<2\Lambda$.

\begin{figure*}[ht!]
\begin{subfigure}[b]{0.32\textwidth}
\includegraphics[width=\textwidth]
{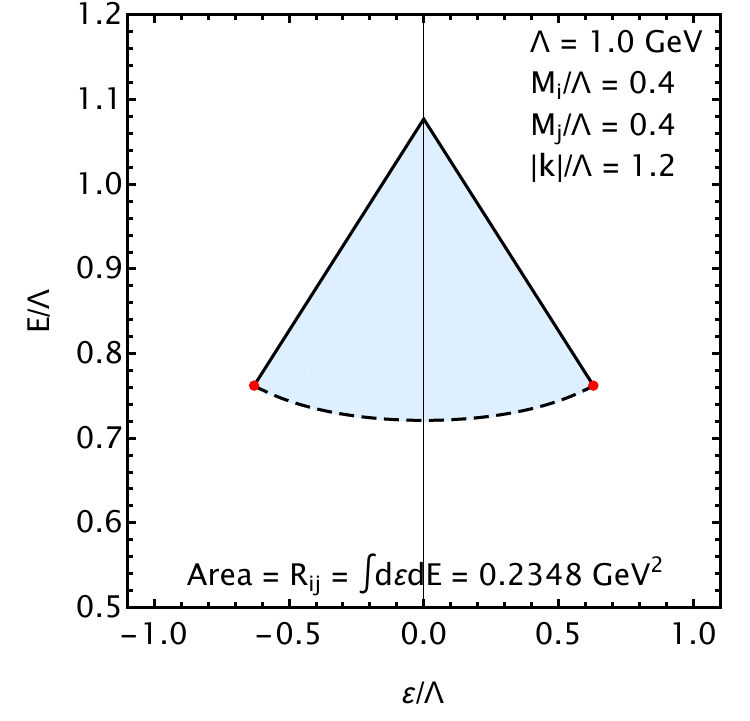}
\caption{$M_i=M_j$}
\label{depsilondΕ_IntegrationRegionM1equalM2}
\end{subfigure}
\begin{subfigure}[b]{0.32\textwidth}
\includegraphics[width=\textwidth]
{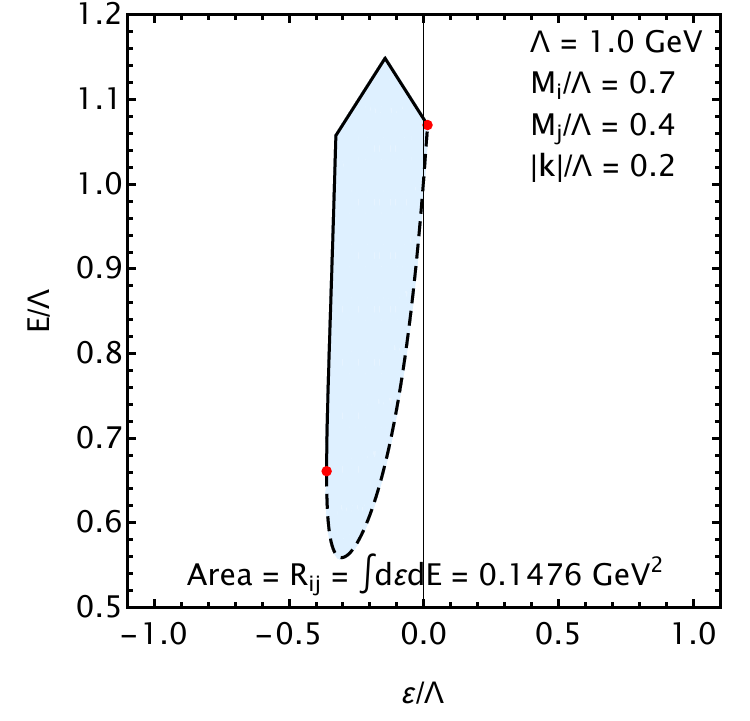}
\caption{$M_i>M_j$}
\label{depsilondΕ_IntegrationRegionM1largerM2}
\end{subfigure}
\begin{subfigure}[b]{0.32\textwidth}
\includegraphics[width=\textwidth]
{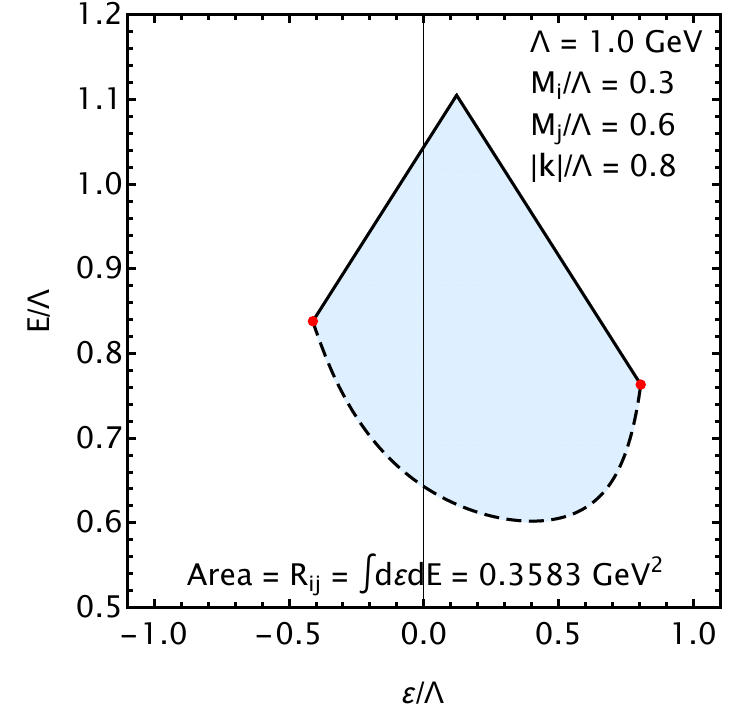}
\caption{$M_i<M_j$}
\label{depsilondΕ_IntegrationRegionM2largerM1}
\end{subfigure}
\caption{
Region of the integration $\int_{ \depsilondEepsilonMin }^{  \depsilondEepsilonMax }
\dd{ \varepsilon } \int_{ \depsilondEEMin }^{ \depsilondEEMax }  \dd{E} $, for fixed values of $\Lambda$, $\abs{\vec{k}}$ and different values of $M_i$, $M_j$, with $M_i=M_j$ in the left panel, $M_i>M_j$ in the middle and $M_j>M_i$ in the right panel. The full black line is the upper limit of the integration over the variable $E$, $\depsilondEEMax
\qty[\Lambda, M_i, M_j, \abs{\vec{k}}, \varepsilon]$ (Eq. (\ref{depsilondE_EMax})) while, the dashed black line, is the inferior bound, $\depsilondEEMin
\qty[\Lambda, M_i, M_j, \abs{\vec{k}},\varepsilon]$ (Eq. (\ref{depsilondE_EMin})). The red dots represent the minimum and maximum values of the integration over $\varepsilon$, $\depsilondEepsilonMin
\qty[\Lambda, M_i, M_j, \abs{\vec{k}}]$ and $\depsilondEepsilonMax
\qty[\Lambda, M_i, M_j, \abs{\vec{k}}]$, respectively (Eqs. (\ref{depsilondE_epsilonMin}) and (\ref{depsilondE_epsilonMax})). In each plot we also give the values for the area of each integration region, which corresponds to evaluating the $R_{ij}$ integral, defined in Eq. (\ref{Rij_integral_definition}).
}
\label{depsilondΕ_integrationRegions}
\end{figure*}

In Fig. \ref{depsilondΕ_integrationRegions} we show the region of integration $\int_{ \depsilondEepsilonMin }^{  \depsilondEepsilonMax }
\dd{ \varepsilon } \int_{ \depsilondEEMin }^{ \depsilondEEMax }  \dd{E} $, for fixed values of $\Lambda$, $\abs{\vec{k}}$, $M_i$ and $M_j$. We consider the three possible scenarios of $M_i=M_j$, $M_i>M_j$ and $M_i<M_j$, in the left, middle and right panels, respectively. For simplicity, we fix the cutoff to $\Lambda=1~\mathrm{GeV}$ and consider different values of $\abs{\vec{k}}$ and masses. The goal is to provide different scenarios for easier reproduction of these results. We also show the values for the different areas resulting from integrating the region (see Eq. (\ref{Rij_integral_Eepsilon_variables})). As discussed before, calculating these areas corresponds to evaluating the $R_{ij}$ integral first defined in Eq (\ref{Rij_integral_definition}).

\subsection{ Integration order $\dd{ E } \dd{ \varepsilon }$ }

In this section we will consider the $\dd{ E } \dd{ \varepsilon }$ integration order, meaning that one first integrates over the $\varepsilon$ variable. We write:
\begin{align}
\int_{\mathbb{I}_{\abs{\vec{k}}}^\Lambda}
\frac{\dd{\phi}}{2\pi}
\dd{ E } \dd{ \varepsilon }
f(E,\varepsilon)
=
\int_{ \dEdepsilonEMin }^{ \dEdepsilonEMax } 
\dd{E} 
\int_{ \dEdepsilonepsilonMin }^{ \dEdepsilonepsilonMin } 
\dd{\varepsilon}
f(E,\varepsilon) .
\end{align}
Again, if $f(E,\varepsilon)=1$, one recovers the integral $R_{ij}$, see Eq. (\ref{Rij_integral_Eepsilon_variables}).

For this case, the starting point is finding the solutions to Eq. (\ref{integrationRegionPolynomial}) with respect to $\varepsilon$. This yields two solutions ($\varepsilon_\mathrm{min}^\pm$):
\begin{equation*}
\varepsilon_\mathrm{min}^\pm
\qty[M_i, M_j, \abs{\vec{k}}, E]
=
\begin{cases}
-
\frac
{ ( M_i^2 - M_j^2 )}
{ 4 E }
-
\frac
{ 2 E \qty( M_i^2 + M_j^2 )}
{ ( M_i^2 - M_j^2 ) } ,
&\text{if} \;
E = \pm \nicefrac{ \abs{\vec{k}} }{ 2 } ,
\\
\frac
{
-2 E \qty( M_i^2 - M_j^2 )
\pm \sqrt{ 
\vec{k}^2 
\qty( 4 E^2 - \vec{k}^2 - ( M_i - M_j )^2 ) 
\qty( 4 E^2 - \vec{k}^2 - ( M_i + M_j )^2 )}
}
{
4 E^2 - \vec{k}^2
} ,
&\text{otherwise}.
\end{cases}
\numberthis
\end{equation*}
For equal masses, $M_i=M_j=M$, one gets:
\begin{align}
\varepsilon_\mathrm{min}^\pm
\qty[M, M	, \abs{\vec{k}}, E]
=
\pm
\frac
{
\sqrt{ 
\vec{k}^2 
\qty( 4 E^2 - \vec{k}^2 ) 
\qty( 4 E^2 - \vec{k}^2 - 4M^2 )}
}
{
4 E^2 - \vec{k}^2
}  .
\end{align}

As before, the cutoff restrictions to the integration region can be obtained by solving Eqs. (\ref{cutoffConstraint1}) and (\ref{cutoffConstraint2}) with  respect to $\varepsilon$. It yields:
\begin{align}
\varepsilon_{\Lambda_i }^\pm
[\Lambda , M_i, E] 
& = 
2 E \pm 2 \sqrt{ \Lambda^2 + M_i^2 } ,
\\
\varepsilon_{ \Lambda_j }^\pm
[\Lambda ,  M_j, E]
& = 
-2 E \pm 2 \sqrt{ \Lambda^2 + M_j^2 } .
\end{align}

The upper and lower bounds of the $\dd{\varepsilon}$ integration, $\dEdepsilonepsilonMax$ and $\dEdepsilonepsilonMin$ can be defined from these equations. Starting with the upper bound, it is defined by the functions $\varepsilon_\mathrm{min}^+$ and $\varepsilon_{\Lambda_j }^+$ which intersect at the point $\qty(E_{\gamma_2},\varepsilon_{\gamma_2})$, previously defined in Eqs. (\ref{epsilonGamma2}) and (\ref{EGamma2}). Hence, the upper bound of the $\dd{\varepsilon}$ integration, $\dEdepsilonepsilonMax$, can be defined as:
\begin{equation*}
\dEdepsilonepsilonMax
\qty[\Lambda, M_i, M_j, \abs{\vec{k}},E]
=
\begin{cases}
\varepsilon_{\Lambda_j }^+
[\Lambda , M_j, E]  ,
&\text{if} \;
E \geq E_{\gamma_2}
\qty[ \Lambda, M_i, M_j, \abs{\vec{k}}]  ,
\\
\varepsilon_\mathrm{min}^+
\qty[M_i, M_j, \abs{\vec{k}}, E] ,
&\text{otherwise}.
\end{cases}
\numberthis
\label{dEdepsilon_epsilonMax}
\end{equation*}
Similarly, the lower bound is defined by the functions $\varepsilon_\mathrm{min}^-$ and $\varepsilon_{\Lambda_i }^-$. These functions intersect at the point $\qty(\varepsilon_{\gamma_1},E_{\gamma_1})$, defined in Eqs. (\ref{epsilonGamma1}) and (\ref{EGamma1}). Thus, we can write:
\begin{equation*}
\dEdepsilonepsilonMin
\qty[\Lambda, M_i, M_j, \abs{\vec{k}},E]
=
\begin{cases}
\varepsilon_{\Lambda_i }^-
[\Lambda , M_i, E]  ,
&\text{if} \;
E \geq E_{\gamma_1}
\qty[ \Lambda, M_i, M_j, \abs{\vec{k}}] ,
\\
\varepsilon_\mathrm{min}^-
\qty[M_i, M_j, \abs{\vec{k}}, E] ,
&\text{otherwise}.
\end{cases}
\numberthis
\label{dEdepsilon_epsilonMin}
\end{equation*}

As with the integration order considered in the previous section, $\dd{ \varepsilon } \dd{ E }$, depending if one is considering equal or different fermion masses, the lower bound of the $\dd{E}$ integration, $\dEdepsilonepsilonMin$ may be defined by the intersection of the functions $\varepsilon_\mathrm{min}^-$ with $\varepsilon_\mathrm{min}^+$, by the intersection of $\varepsilon_\mathrm{min}^+$ with $\varepsilon_{\Lambda_i }^-$ or by the intersection of $\varepsilon_\mathrm{min}^-$ with $\varepsilon_{\Lambda_j }^+$. When the fermion masses are equal, the lower bound is given by the first intersection mentioned, i.e., when $\varepsilon_\mathrm{min}^-=\varepsilon_\mathrm{min}^+$. This occurs at the point $\qty(E_\alpha,\varepsilon_\alpha)$ defined by:
\begin{align}
E_\alpha \qty[ M_i, M_j, \abs{\vec{k}} ]
& = 
\frac{1}{2} \sqrt{ \vec{k}^2 + \qty( M_i + M_j )^2 } ,
\\
\varepsilon_\alpha  \qty[ M_i, M_j, \abs{\vec{k}} ]
& =
-
\frac
{ \qty( M_i^2 - M_j^2 ) \sqrt{ \vec{k}^2 + ( M_i + M_j)^2 } }
{ \qty( M_i + M_j )^2 } .
\end{align}
If $M_i>M_j$, the lower bound of the $\dd{E}$ integration can also be located at the above intersection point, $E_\alpha$. However, for certain values of the parameters $\Lambda$ $M_i$, $M_j$ and $\abs{\vec{k}}$, the lower bound of the $\dd{E}$ integration can occur at the intersection between $\varepsilon_\mathrm{min}^+$ and $\varepsilon_{\Lambda_i }^-$. This intersection also occurs at the point $\qty(\varepsilon_{\gamma_1},E_{\gamma_1})$, defined in Eqs. (\ref{epsilonGamma1}) and (\ref{EGamma1}). Hence, we can write:
\begin{equation*}
E_\mathrm{min}^{(M_i>M_j)}
\qty[\Lambda, M_i, M_j, \abs{\vec{k}}]
=
\begin{cases}
E_{\gamma_1}
\qty[ \Lambda, M_i, M_j, \abs{\vec{k}}] ,
& 
\begin{array}{l}
\text{if} \;
\varepsilon_{\Lambda_i }^-
[\Lambda , M_i, \varepsilon_\alpha  \qty[ M_i, M_j, \abs{\vec{k}} ]] > \varepsilon_\alpha  \qty[ M_i, M_j, \abs{\vec{k}} ]  ,
\end{array}
\\
E_\alpha \qty[ M_i, M_j, \abs{\vec{k}} ]   ,
&\;\text{otherwise}.
\end{cases}
\numberthis
\end{equation*}
Likewise, for the opposite mass case, in which $M_j>M_i$, one can realize that the lower bound of the integration can occur at the point $E_\alpha$ or, it may occur at the intersection between $\varepsilon_\mathrm{min}^-$ and $\varepsilon_{\Lambda_j }^+$, defined by the point $\qty(E_{\gamma_2},\varepsilon_{\gamma_2})$ given in Eqs. (\ref{epsilonGamma2}) and (\ref{EGamma2}). Thus, we write:
\begin{equation*}
E_\mathrm{min}^{(M_j>M_i)}
\qty[\Lambda, M_i, M_j, \abs{\vec{k}}]
=
\begin{cases}
E_{\gamma_2}
\qty[ \Lambda, M_i, M_j, \abs{\vec{k}}] ,
& 
\begin{array}{l}
\text{if} \;
\varepsilon_{\Lambda_j }^+
[\Lambda , M_i, \varepsilon_\alpha  \qty[ M_i, M_j, \abs{\vec{k}} ]] < \varepsilon_\alpha  \qty[ M_i, M_j, \abs{\vec{k}} ]  ,
\end{array}
\\
E_\alpha \qty[ M_i, M_j, \abs{\vec{k}} ]   ,
&\;\text{otherwise}.
\end{cases}
\numberthis
\end{equation*}
Collecting these conditions, we can finally define:
\begin{equation*}
\dEdepsilonEMin
\qty[\Lambda, M_i, M_j, \abs{\vec{k}}]
=
\begin{cases}
E_\mathrm{min}^{(M_i>M_j)}
\qty[\Lambda, M_i, M_j, \abs{\vec{k}}] ,
&\text{if} \;
M_i > M_j ,
\\
E_\mathrm{min}^{(M_j>M_i)}
\qty[\Lambda, M_i, M_j, \abs{\vec{k}} ] ,
&\text{if} \;
M_i < M_j ,
\\
E_\alpha \qty[ M_i, M_j, \abs{\vec{k}} ]  ,
&\text{otherwise}.
\end{cases}
\numberthis
\label{dEdepsilon_EMin}
\end{equation*}
The upper bound of the $\dd{E}$ integration is much simpler to define. It is defined by the intersection between the cutoff restrictions, which takes place when $\varepsilon_{\Lambda_i }^-=\varepsilon_{\Lambda_j }^+$, in the formerly defined point $\qty(E_{ \Lambda_s }, \varepsilon_{\Lambda_s } )$ given in Eqs. (\ref{epsilonLambdaSwitch}) and (\ref{ELambdaSwitch}). In such a way, the upper limit is then simply given by:
\begin{align}
\dEdepsilonEMax
\qty[\Lambda, M_i, M_j, \abs{\vec{k}}]
=
E_{ \Lambda_s } \qty[\Lambda, M_i, M_j]  .
\label{dEdepsilon_EMax}
\end{align}

\begin{figure*}[ht!]
\begin{subfigure}[b]{0.32\textwidth}
\includegraphics[width=\textwidth]
{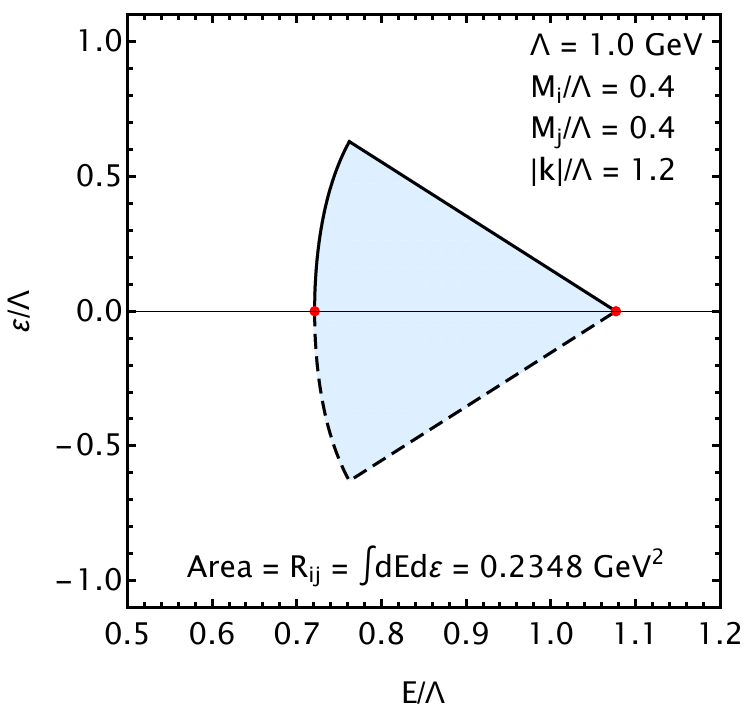}
\caption{$M_i=M_j$}
\label{dΕdepsilon_IntegrationRegionM1equalM2}
\end{subfigure}
\begin{subfigure}[b]{0.32\textwidth}
\includegraphics[width=\textwidth]
{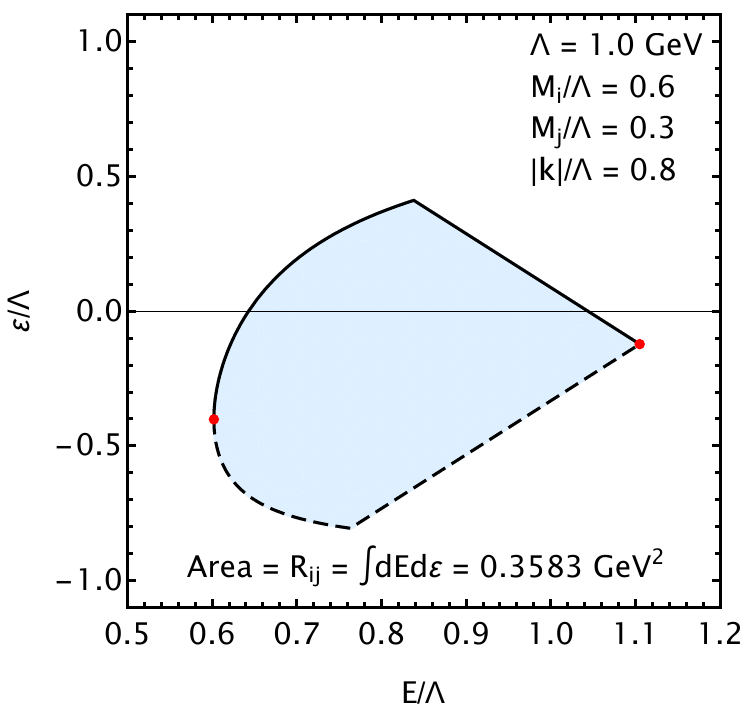}
\caption{$M_i>M_j$}
\label{dΕdepsilon_IntegrationRegionM1largerM2}
\end{subfigure}
\begin{subfigure}[b]{0.32\textwidth}
\includegraphics[width=\textwidth]
{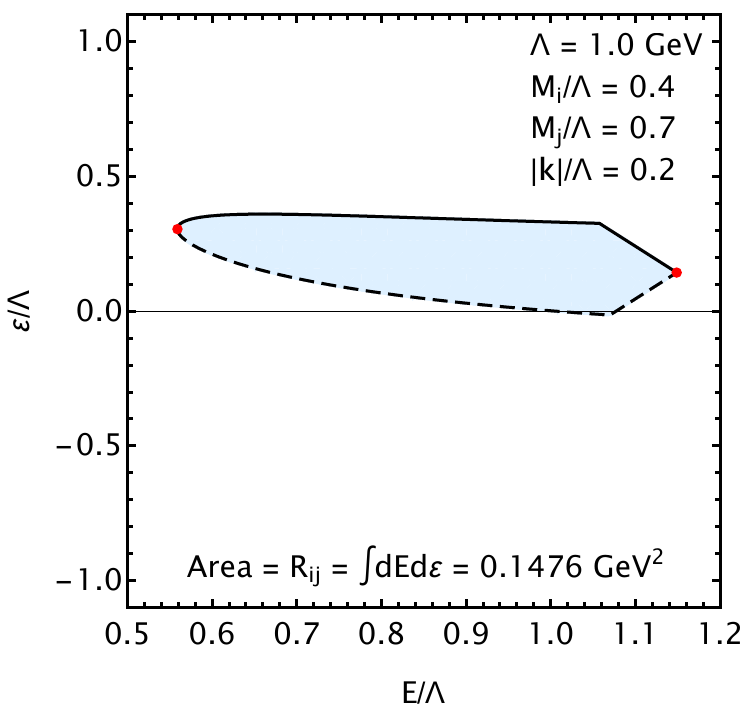}
\caption{$M_i<M_j$}
\label{dΕdepsilon_IntegrationRegionM2largerM1}
\end{subfigure}
\caption{
Region of the integration $\int_{ \dEdepsilonEMin }^{ \dEdepsilonEMax }  \dd{E} \int_{ \dEdepsilonepsilonMin }^{  \dEdepsilonepsilonMax }
\dd{ \varepsilon }$, for fixed values of $\Lambda$, $\abs{\vec{k}}$ and different values of $M_i$, $M_j$, with $M_i=M_j$ in the left panel, $M_i>M_j$ in the middle and $M_j>M_i$ in the right panel. The full black line is the upper limit of the integration over the variable $\varepsilon$, $\dEdepsilonepsilonMax
\qty[\Lambda, M_i, M_j, \abs{\vec{k}},E]$ (Eq. (\ref{dEdepsilon_epsilonMax})) while, the dashed black line, is the inferior bound, $\dEdepsilonepsilonMin
\qty[\Lambda, M_i, M_j, \abs{\vec{k}},E]$ (Eq. (\ref{dEdepsilon_epsilonMin})). The red dots represent the minimum and maximum values of the integration over $E$, $\dEdepsilonEMin
\qty[\Lambda, M_i, M_j, \abs{\vec{k}}]$ and $\dEdepsilonEMax
\qty[\Lambda, M_i, M_j, \abs{\vec{k}}]$, respectively ((Eq. (\ref{dEdepsilon_EMin})) and (Eq. (\ref{dEdepsilon_EMax}))). In each plot we also give the values for the area of each integration region, which corresponds to evaluating the $R_{ij}$ integral, defined in Eq. (\ref{Rij_integral_definition}).
}
\label{dΕdepsilon_integrationRegions}
\end{figure*}

In Fig. \ref{dΕdepsilon_integrationRegions} we show the region of integration $\int_{ \dEdepsilonEMin }^{ \dEdepsilonEMax }  \dd{E} \int_{ \dEdepsilonepsilonMin }^{  \dEdepsilonepsilonMax }
\dd{ \varepsilon }$, for fixed values of $\Lambda$, $\abs{\vec{k}}$, $M_i$ and $M_j$. Again, we consider the three possible scenarios of $M_i=M_j$, $M_i>M_j$ and $M_i<M_j$, in the left, middle and right panels, respectively. It is now easy to check that one gets exactly the same areas (or, identically, values for the $R_{ij}$ integral) using this integration order, for the same set of parameters, when comparing with the other possible integration order whose areas are shown in Fig. \ref{depsilondΕ_integrationRegions}.

\subsection{Calculating the integral $R_{ij}$ without changing variables}
\label{integration_region_Rij}

In this section we show how to evaluate the integral $R_{ij}$, defined in Eq. (\ref{Rij_integral_definition}), without changing to the $E$ and $\varepsilon$ variables. This is useful for numerical cross checks, as well as evaluating the vacuum limit of the one fermion line integral in the 3-momentum sphere intersection regularization, $\mathcal{A}$.

As discussed in Section \ref{regularization}, using the 3-momentum sphere intersection regularization amounts to performing an integral over the region defined by the intersection between two spheres, $\mathbb{I}_{\abs{\vec{k}}}^\Lambda$. One of the spheres is centered at the origin while the other is distanced $\abs{\vec{k}}$ from the first, in the $p_z$ axis, $\mathbb{I}_{\abs{\vec{k}}}^\Lambda =\mathbb{S}_0^\Lambda \cap \mathbb{S}_{\abs{\vec{k}}}^\Lambda$. Such region can be observed in Fig. \ref{integration_region}. The major difficulty of evaluating the integral $R_{ij}$ over this 3-dimensional region lays in the fact that it is not mirror-symmetric in the $p_z$ axis. In order to make the region symmetrical we are allowed to consider, in Eq. (\ref{Rij_integral_definition}), the following change of variables, $\vec{p} \to \vec{p} + \nicefrac{\vec{k}}{2}$. With this transformation, we are allowed to write the $R_{ij}$ integral as:
\begin{align}
R_{ij} \qty[\Lambda,M_i, M_j, \abs{\vec{k}} ]
=
\frac{ \abs{\vec{k}} }{ 2\pi }
\int_{ {\mathbb{I}^\p}_{\abs{\vec{k}}}^\Lambda }
\frac{ \dd[3]{\vec{p}}  }
{ 
E\qty[M_i, \vec{p} + \frac{\vec{k}}{2}] 
E[M_j,\vec{p} - \frac{\vec{k}}{2} ] 
} .
\end{align}
Where, the new region of integration, is defined by the intersection of the two spheres distanced $\abs{\vec{k}}$ from each other, with one centered at $\qty(p_x,p_y,p_z)=\qty(0,0,\nicefrac{\abs{\vec{k}}}{2})$ and the other centered at $\qty(p_x,p_y,p_z)=\qty(0,0,-\nicefrac{\abs{\vec{k}}}{2})$ i.e., ${\mathbb{I}^\p}_{\abs{\vec{k}}}^\Lambda = \mathbb{S}_{\nicefrac{\abs{\vec{k}}}{2}}^\Lambda \cap \mathbb{S}_{-\nicefrac{\abs{\vec{k}}}{2}}^\Lambda $. This new integration region is mirror-symmetric with respect to all axis, allowing for a simpler parametrization of the integration bounds.

\begin{figure*}[ht!]
\begin{subfigure}[b]{0.4\textwidth}
\includegraphics[width=\textwidth]
{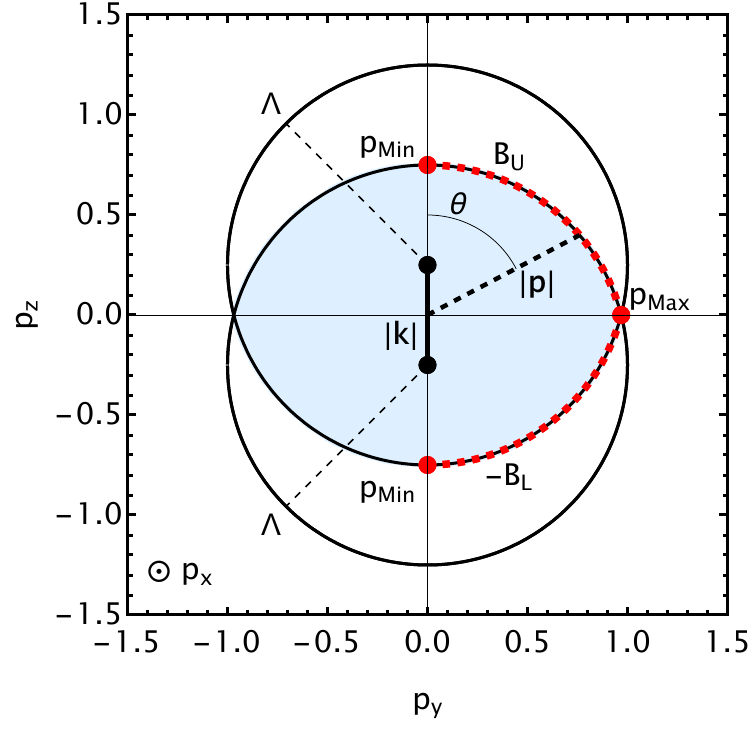}
\end{subfigure}
\caption{
Region of integration in the 3-momentum sphere intersection regularization scheme used to regularize the $\mathcal{A}$ integral.
}
\label{AIntegralIntegrationRegion}
\end{figure*}

In Fig. \ref{AIntegralIntegrationRegion} we show the 2-dimensional cross section of the intersection ${\mathbb{I}^\p}_{\abs{\vec{k}}}^\Lambda$, in the $p_y-p_z$ plane. Consider the more general case of an integral, $I$, with an integrand, $F$, which is independent of the azimuthal angle, $\phi$:
\begin{align}
I = 
\int_{ {\mathbb{I}^\p}_{\abs{\vec{k}}}^\Lambda }
\dd[3]{\vec{p}}
F [\abs{\vec{p}},\theta] .
\label{general_integrand_p_theta_3MSIR}
\end{align}
This integral can be separated into two parts: one corresponding to the upper hemisphere and the other to the lower hemisphere of the intersection. Consider also the functions $B_U$ and $B_L$ which parametrize the boundary of the upper and lower hemispheres, respectively (see Fig. \ref{AIntegralIntegrationRegion}). Hence, we can write, $I = I_U + I_L$, where:
\begin{align}
\frac{I_U}{2\pi} & = 
\int_0^{ p_{\mathrm{Max}} } \dd{\abs{\vec{p}}}
\vec{p}^2
\int_0^{ \nicefrac{\pi}{2} } \dd{\theta}
\sin [\theta]
F [\abs{\vec{p}},\theta] 
-
\int_{ p_{\mathrm{Min}} }^{ p_{\mathrm{Max}} } \dd{\abs{\vec{p}}}
\vec{p}^2
\int_0^{ B_U } \dd{\theta}
\sin [\theta]
F [\abs{\vec{p}},\theta] 
,
\\
\frac{I_L}{2\pi} & = 
\int_0^{ p_{\mathrm{Max}} } \dd{\abs{\vec{p}}}
\vec{p}^2
\int_0^{ \nicefrac{\pi}{2} } \dd{\theta}
\sin [\theta]
F [\abs{\vec{p}},\theta] 
-
\int_{ p_{\mathrm{Min}} }^{ p_{\mathrm{Max}} } \dd{\abs{\vec{p}}}
\vec{p}^2
\int_{ B_L }^{\pi} \dd{\theta}
\sin [\theta]
F [\abs{\vec{p}},\theta] 
.
\end{align}
In the above, we first consider the integral of a sphere of radius $p_{\mathrm{Max}}$ (the distance from the origin  to the point where both boundaries intersect, $B_U=B_L$, which occurs when $\theta=\nicefrac{\pi}{2}$) and than subtract the extra volume coming from the boundaries, $B_U$ and $B_L$ , up to this maximum sphere. The quantity $p_{\mathrm{Min}}$ is the distance from the origin to the boundaries when $\theta=0$. The quantities $B_U$, $B_L$, $p_{\mathrm{Min}}$ and $p_{\mathrm{Max}}$, will be defined later. For simplicity, we apply the change of variables, $\alpha=\cos[\theta]$ and change the integration bounds accordingly to get:
\begin{align*}
-\frac{I}{2\pi} 
& = 
\int_0^{ p_{\mathrm{Max}} } \dd{\abs{\vec{p}}}
\vec{p}^2
\int_{ 1 }^{ -1 } \dd{\alpha}
F [\abs{\vec{p}},\alpha] 
\\
& -
\int_{ p_{\mathrm{Min}} }^{ p_{\mathrm{Max}} } \dd{\abs{\vec{p}}}
\vec{p}^2
\int_{ 1 }^{ B_U } \dd{\alpha}
F [\abs{\vec{p}},\alpha] 
\\
& -
\int_{ p_{\mathrm{Min}} }^{ p_{\mathrm{Max}} } \dd{\abs{\vec{p}}}
\vec{p}^2
\int_{ B_L }^{ -1 } \dd{\alpha}
F [\abs{\vec{p}},\alpha]  .
\numberthis
\end{align*}
This expression can be further simplified by breaking the $\dd{\abs{\vec{p}}}$ integral in the first term at the point $p_{\mathrm{Min}}$ and by rearranging the integral. Finally, we can write the integral as:
\begin{align}
I
& = 
-2 \pi
\qty(
\int_0^{ p_{\mathrm{Min}} } \dd{\abs{\vec{p}}}
\vec{p}^2
\int_{ 1 }^{ -1 } \dd{\alpha}
F [\abs{\vec{p}},\alpha] 
+
\int_{ p_{\mathrm{Min}} }^{ p_{\mathrm{Max}} } \dd{\abs{\vec{p}}}
\vec{p}^2
\int_{ B_U }^{ B_L } \dd{\alpha}
F [\abs{\vec{p}},\alpha] 
) .
\end{align}
At this point we define the functions that define the integration bounds. For the different hemispheres, the equations which define the boundaries are given by:
\begin{align}
\vec{p}^2 \sin[\theta]^2 \cos[\phi]^2
+
\vec{p}^2 \sin[\theta]^2 \sin[\phi]^2
+
\qty( \abs{\vec{p}} \cos[\theta] \pm \frac{\abs{\vec{k}}}{2} )^2
=
\Lambda^2 .
\end{align}
Where the plus sign is for the upper hemisphere and the minus sign for the lower hemisphere. Solving these equations with respect to $\cos[\theta]$, yields:
\begin{align}
\cos[\theta]
=
\pm
\frac
{\Lambda ^2 - \nicefrac{\vec{k}^2}{4} - \vec{p}^2}
{\abs{\vec{k}} \abs{\vec{p}}} .
\label{boundary_condition}
\end{align}
Again, the plus sign is for the upper region while the minus sign is for the lower hemisphere. Thus, in the $\alpha=\cos[\theta]$ variable we can find write $B_U$ and $B_L$ as:
\begin{align}
B_U \qty[\Lambda, \abs{\vec{k}}, \abs{\vec{p}}] 
= 
\frac
{\Lambda ^2 - \nicefrac{\vec{k}^2}{4} - \vec{p}^2}
{\abs{\vec{k}} \abs{\vec{p}}}
= 
- B_L \qty[\Lambda, \abs{\vec{k}}, \abs{\vec{p}}] .
\end{align}
The quantities $p_{\mathrm{Min}}$ and $p_{\mathrm{Max}}$ can be derived by plugging $\theta=0$ and $\theta=\nicefrac{\pi}{2}$ in Eq. (\ref{boundary_condition}), to yield:
\begin{align}
p_{\mathrm{Min}} \qty[\Lambda, \abs{\vec{k}}]
& = 
\Lambda -\frac{ \abs{\vec{k}} }{2} , 
\label{pMin_general_integrand_p_theta_3MSIR}
\\
p_{\mathrm{Max}} \qty[\Lambda, \abs{\vec{k}}]
& = 
\sqrt{ \Lambda ^2 - \frac{ \vec{k}^2 }{4} } .
\label{pMax_general_integrand_p_theta_3MSIR}
\end{align}

We are finally ready to evaluate the integral $R_{ij}$. Applying the tools developed above, we can write $F [\abs{\vec{p}},\theta] = 1/\qty( 
E\qty[M_i, \vec{p} + \frac{\vec{k}}{2}] 
E[M_j,\vec{p} - \frac{\vec{k}}{2} ] 
) $. Hence, we can explicitly write:
\begin{align*}
R_{ij} \qty[\Lambda,M_i, M_j, \abs{\vec{k}} ]
=
& - \abs{\vec{k}}
\int_0^{ p_{\mathrm{Min}} } \dd{\abs{\vec{p}}}
\vec{p}^2
\int_{ 1 }^{ -1 } 
\frac{ \dd{\alpha} }
{ 
\sqrt{ \vec{p}^2 + \frac{\vec{k}^2}{4} + \vec{p} \abs{\vec{k}} \cos[\theta] + M_i^2 }
\sqrt{ \vec{p}^2 + \frac{\vec{k}^2}{4} - \vec{p} \abs{\vec{k}} \cos[\theta] + M_j^2 }
}
\\
& - \abs{\vec{k}}
\int_{ p_{\mathrm{Min}} }^{ p_{\mathrm{Max}} } \dd{\abs{\vec{p}}}
\vec{p}^2
\int_{ B_U }^{ B_L } 
\frac{ \dd{\alpha} }
{ 
\sqrt{ \vec{p}^2 + \frac{\vec{k}^2}{4} + \vec{p} \abs{\vec{k}} \cos[\theta] + M_i^2 }
\sqrt{ \vec{p}^2 + \frac{\vec{k}^2}{4} - \vec{p} \abs{\vec{k}} \cos[\theta] + M_j^2 }
}
.
\numberthis
\end{align*}
We can calculate the innermost integrals analytically:
\begin{align*}
\int 
\dd{\alpha}
&
\frac{ \abs{\vec{k}} \vec{p}^2 }
{ 
\sqrt{ \vec{p}^2 + \frac{\vec{k}^2}{4} + \abs{\vec{p}} \abs{\vec{k}} \cos[\theta] + M_i^2 }
\sqrt{ \vec{p}^2 + \frac{\vec{k}^2}{4} - \abs{\vec{p}} \abs{\vec{k}} \cos[\theta] + M_j^2 }
}
\\
& \qquad \qquad \qquad \qquad \qquad \qquad \qquad =
-2 \abs{\vec{p}}
\arcsin
\qty[
\frac{
\sqrt{
\frac{\vec{k}^2}{4} + M_j^2 + \vec{p}^2 - \alpha \abs{\vec{k}} \abs{\vec{p}} }
}{
\sqrt{2} 
\sqrt{
\frac{\vec{k}^2}{4} + 
\frac{ M_i^2 + M_j^2 }{2}  + \vec{p}^2   }
}
] 
+ \mathrm{cte} .
\numberthis
\end{align*}
Using this primitive, we are finally able to write $R_{ij}$ as:
\begin{align*}
R_{ij} \qty[\Lambda,M_i, M_j, \abs{\vec{k}} ]
& = 
2
\sumeta
\eta
\int_0^{ \Lambda -\frac{ \abs{\vec{k}} }{2} } \dd{\abs{\vec{p}}}
\abs{\vec{p}}
\arcsin
\qty[
\frac{
\sqrt{
\frac{\vec{k}^2}{4} + M_j^2 + \vec{p}^2 + \eta \abs{\vec{k}} \abs{\vec{p}} }
}{
\sqrt{2} 
\sqrt{
\frac{\vec{k}^2}{4} + 
\frac{ M_i^2 + M_j^2 }{2}  + \vec{p}^2  }
}
]
\\
&
+ 2
\sumeta
\eta
\int_{ \Lambda -\frac{ \abs{\vec{k}} }{2} }^{ \sqrt{ \Lambda ^2 - \frac{ \vec{k}^2 }{4} } } 
\dd{\abs{\vec{p}}}
\abs{\vec{p}}
\arcsin
\qty[
\frac{
\sqrt{
M_j^2 + \eta \Lambda^2 + (1-\eta)(\vec{p}^2 + \frac{\vec{k}^2}{4})  }
}{
\sqrt{2} 
\sqrt{
\frac{\vec{k}^2}{4} + 
\frac{ M_i^2 + M_j^2 }{2}  + \vec{p}^2   }
}
] .
\numberthis
\end{align*}
In Table \ref{Rij_values} we give some results from solving the integral which can then be directly compared to the results obtained when using the variables $E$ and $\varepsilon$, given in Figs. \ref{depsilondΕ_integrationRegions} and \ref{dΕdepsilon_integrationRegions}.

\begin{table*}[ht!]
\begin{center}
\begin{tabular*}{0.52\textwidth}{@{\extracolsep{\fill}}ccccc@{}}
\toprule
\multicolumn{1}{c}{$\Lambda~\qty[\text{GeV}]$ } & 
\multicolumn{1}{c}{$M_i~\qty[\text{GeV}]$ } & 
\multicolumn{1}{c}{$M_j~\qty[\text{GeV}]$ }&
\multicolumn{1}{c}{$\abs{\vec{k}}~\qty[\text{GeV}]$ } &
\multicolumn{1}{c}{$R_{ij}~\qty[\text{GeV}^2]$} 
\\
\midrule
$1.0$ & $0.4$ & $0.4$ & $1.2$ & $0.2348$ 
\\
$1.0$ & $0.3$ & $0.6$ & $0.8$ & $0.3583$ 
\\
$1.0$ & $0.6$ & $0.3$ & $0.8$ & $0.3583$ 
\\
$1.0$ & $0.4$ & $0.7$ & $0.2$ & $0.1476$ 
\\
$1.0$ & $0.7$ & $0.4$ & $0.2$ & $0.1476$ 
\\
$1.0$ & $1.2$ & $0.2$ & $0.5$ & $0.2334$ 
\\
$1.0$ & $0.2$ & $1.2$ & $0.5$ & $0.2334$ 
\\
\bottomrule
\end{tabular*}
\end{center}
\caption{
Values for the $R_{ij}$ integral, defined in Eq. (\ref{Rij_integral_definition}), for different values of the cutoff, $\Lambda$, effective masses, $M_i$ and $M_j$ and absolute value of the 3-momentum, $\abs{\vec{k}}$. The $R_{ij}$ integral is identical to the area defined by the integral $\int_{ \depsilondEepsilonMin }^{  \depsilondEepsilonMax }
\dd{ \varepsilon } \int_{ \depsilondEEMin }^{ \depsilondEEMax }  \dd{E} $ and $\int_{ \dEdepsilonEMin }^{ \dEdepsilonEMax }  \dd{E} \int_{ \dEdepsilonepsilonMin }^{  \dEdepsilonepsilonMax }
\dd{ \varepsilon }$
} (see Eq. (\ref{Rij_integral_Eepsilon_variables}) and Figs. \ref{depsilondΕ_integrationRegions} and \ref{dΕdepsilon_integrationRegions} for the evaluation of the $R_{ij}$ using the $E$ and $\varepsilon$ variables).
\label{Rij_values}
\end{table*}

\newpage
\section{Numerical results for the two fermion line integral, $B_0$}
\label{numericalResultsB0}

In this section we show numerical results for the two fermion line integral, $B_0$, by choosing a particular value for the cutoff, $\Lambda$, and constant values for the fermion masses, $M_i$ and $M_j$. This means we are not supposing any model for the underlying fermions dynamics, meaning that the fermion masses are not temperature and/or chemical potential dependent. We chose to perform such study in order to isolate the temperature and chemical potential dependence of the $B_0$ function. We show several results for this function, in a diverse set of scenarios, in order to simplify the task of reproducibility. As already stated, the calculation of this integral is extremely important to studies involving mesons masses, the phase diagram of strongly interacting matter and the transport properties of the quark-gluon plasma in the relaxation time approximation, in the context of the NJL model and similar models.

\begin{figure*}[ht!]
\begin{subfigure}[b]{0.4\textwidth}
\includegraphics[width=\textwidth]
{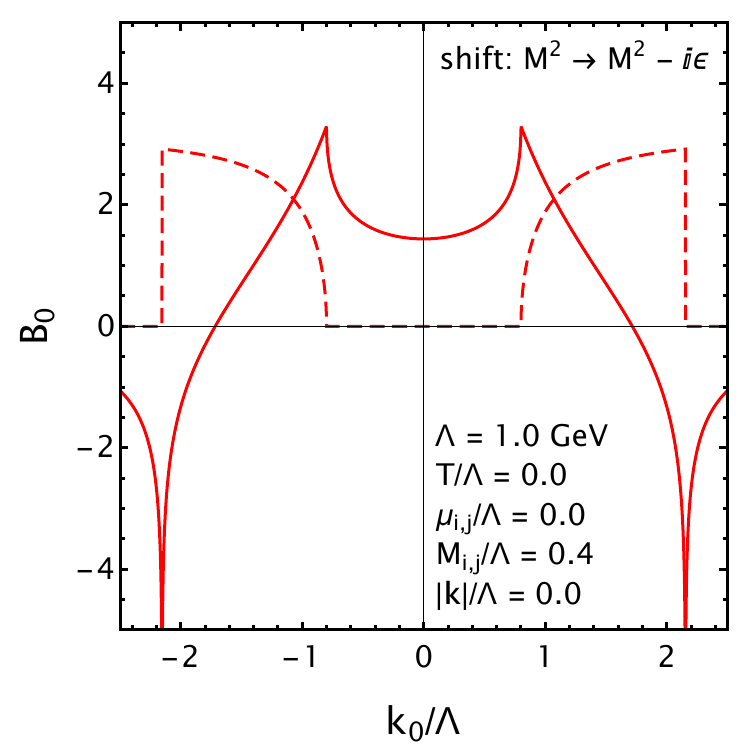}
\label{plotB0vsk0T0Cpi0Cpj0L1Mi04Mj04k0_M_shift}
\end{subfigure}
\begin{subfigure}[b]{0.4\textwidth}
\includegraphics[width=\textwidth]
{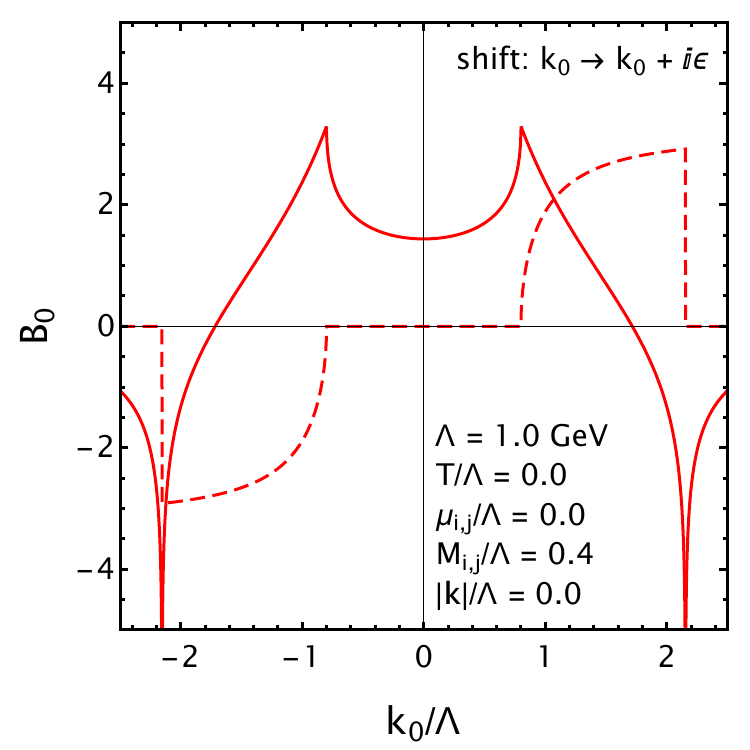}
\label{plotB0vsk0T0Cpi0Cpj0L1Mi04Mj04k0_k0_shift}
\end{subfigure}
\caption{$B_0$ integral, in the vacuum, as a function of the zero component of the external momentum, normalized by the cutoff, $k_0/\Lambda$. The cutoff is fixed to $\Lambda
=1~\mathrm{GeV}$, the fermion masses are equal and fixed to $M_i=M_j=0.4 \Lambda$ and the magnitude of the external 3-momentum, $\abs{\vec{k}}=0.0$. The full and dashed lines correspond to the real and imaginary parts part of the $B_0$ integral, respectively. In the left panel it was considered the mass shift, $M^2 \to M^2-i\epsilon$ while, in the right panel, the external momentum shift was considered, $k_0 \to k_0 + i\epsilon$.  }
\label{plotB0vsk0T0Cpi0Cpj0L1Mi04Mj04k0_different_shifts}
\end{figure*}

In Fig. \ref{plotB0vsk0T0Cpi0Cpj0L1Mi04Mj04k0_different_shifts}, we show the result for the real and imaginary parts of the $B_0$ function, in the vacuum, as a function of the zero component of the external momentum, normalized by the cutoff, $k_0/\Lambda$. We fixed the 3-momentum cutoff, $\Lambda=1~\mathrm{GeV}$, the fermion masses are equal and fixed to $M_i=M_j=0.4 \Lambda$ and the magnitude of the external 3-momentum is zero, $\abs{\vec{k}}=0.0$.

In Ref. \cite{Rehberg:1995nr}, other values for the above parameters are chosen, yielding a slightly different result. However, when using the same parameter set, we were able to obtain exactly the same results presented in Ref. \cite{Rehberg:1995nr}. As already discussed, our approach only differs from the one proposed in that work if one is considering the case with $\abs{\vec{k}} > 0$, being exactly the same if $\abs{\vec{k}}=0$.

As discussed previously, one must consider a pole shift in order to apply the Sokhotski–Plemelj theorem. The way in which this shift is done, only changes the behavior of the imaginary parts of the $B_0$ integral, as shown in Eqs. (\ref{symm_symmetric}) and (\ref{symm_antisymmetric}). In each panel of this figure, we show the result for a specific shift: on the left the mass shift is used, $M^2 \to M^2 - i \epsilon$ while, on the right, the external momentum shift is used, $k_0 \to k_0 + i\epsilon$. As predicted, using the mass shift yields a $\Im[B_0]$ function that is symmetric with respect to $k_0$ while, using the external momentum shift, $\Im[B_0]$ is antisymmetric. By considering the opposite sign in the shifts, $M^2 \to M^2 + i \epsilon$ and $k_0 \to k_0 - i\epsilon$, gives an overall minus factor in the imaginary parts, yielding a total of four possible different  scenarios for the imaginary part of $B_0$. In the following numerical results, we always use the mass shift, $M^2 \to M^2 - i \epsilon$.

Although the specific choice for the pole shift changes the properties of the imaginary part, it does not change the results for the cross sections in the NJL model, for example. Indeed, in order to study cross sections in the NJL model, the meson propagators are calculated which are dependent on the quark-antiquark polarization function, as discussed in earlier sections of this work. These quantities are calculated using the $B_0$ function, and display the same symmetry properties as the $B_0$ integral. However, in the differential cross section calculation, one only needs products of meson propagators with the complex conjugate of other meson propagators, rendering the calculation invariant under a specific overall sign of the imaginary part. For more details see Section \ref{cross_sections}.

\begin{figure*}[t!]
\begin{subfigure}[b]{0.4\textwidth}
\includegraphics[width=\textwidth]
{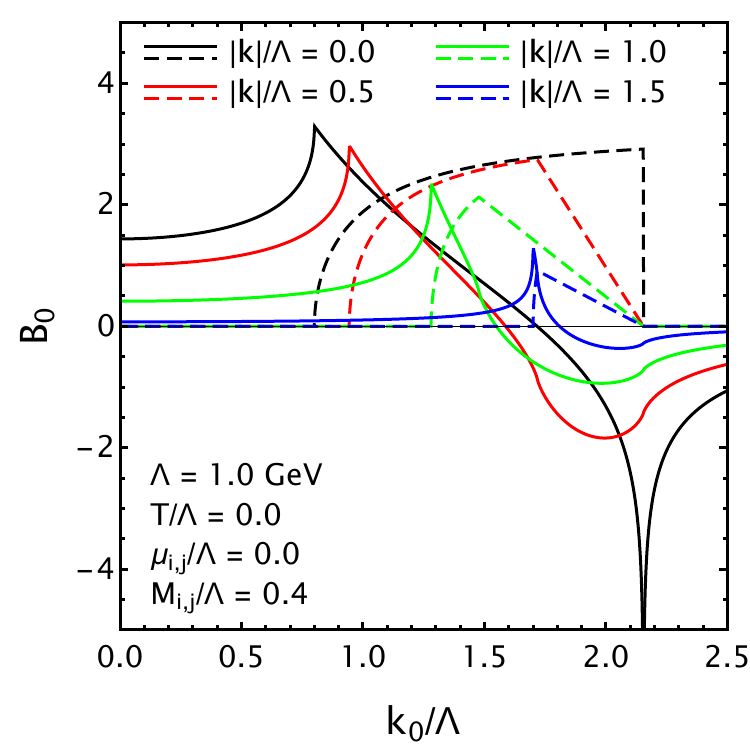}
\label{plotB0vsk0T0Cpi0Cpj0L1Mi04Mj04differentk}
\end{subfigure}
\begin{subfigure}[b]{0.4\textwidth}
\includegraphics[width=\textwidth]
{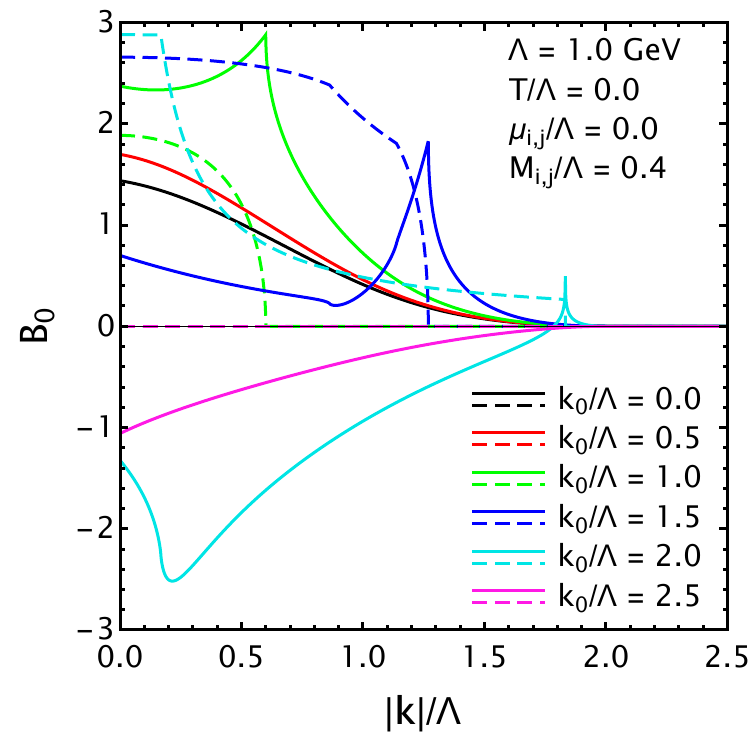}
\label{plotB0vskT0Cpi0Cpj0L1Mi04Mj04differentw}
\end{subfigure}
\caption{$B_0$ integral, in the vacuum, as a function of the zero component of the external momentum, normalized by the cutoff, $k_0/\Lambda$ (left panel) and as a function of the magnitude of the external 3-momentum, normalized by the cutoff, $\abs{\vec{k}}/\Lambda$ (right panel). The full and dashed lines correspond to the real and imaginary parts part of the $B_0$ integral, respectively. The cutoff is fixed to $\Lambda
=1~\mathrm{GeV}$, the fermion masses are equal and fixed to $M_i=M_j=0.4 \Lambda$.  }
\label{plotB0vsk0T0Cpi0Cpj0L1Mi04Mj04differentwk}
\end{figure*}

In Fig. \ref{plotB0vsk0T0Cpi0Cpj0L1Mi04Mj04differentwk}, the $B_0$ function is shown as a function of the external momentum, in the vacuum.
On the left panel, it is shown as a function of the zero component of the external momentum, $k_0/\Lambda$, while, on the right panel, it is shown as a function of the magnitude of the external 3-momentum, $\abs{\vec{k}}/\Lambda$. In the left panel, four different values for $\abs{\vec{k}}$ are considered and analogously, in the right panel, six values $k_0$ are fixed, making the information in both plots complement each other. In the calculation of $B_0$ as a function of $k_0$, we only show results for positive $k_0$ since the function is symmetric with respect to $k_0$.

From the results of the $B_0$ as a function of $\abs{\vec{k}}$ (right panel) one can see the effect of the 3-momentum sphere intersection regularization. The absolute value of the external 3-momentum, $\abs{\vec{k}}$, is the distance between two spheres, each with radii, $\Lambda$. At, $\abs{\vec{k}}=2\Lambda$ the spheres no longer intersect and the $B_0$ function should be zero. Indeed that is the observed behavior in the right panel of Fig. \ref{plotB0vsk0T0Cpi0Cpj0L1Mi04Mj04differentwk}. In fact this is observed in all the calculations shown in this work, even at finite temperature and finite chemical potential, as we will see later. Moreover, we can observe that the presence of an imaginary part is not constant, certain scenarios do not have a finite imaginary part.

In the results concerning the dependence on $k_0$ (see left panel), we have fixed $\abs{\vec{k}}$ to four specific values: $\abs{\vec{k}}/\Lambda=\qty{0.0,0.5,1.0,1.5}$. We do not consider larger values because, for $\abs{\vec{k}}\geq 2 \Lambda$, within our approach, the $B_0$ function is automatically zero, due to the regularization proposed in this work. Interestingly, we can also observe from this plot that the starting $k_0$ point for a finite imaginary part, $\Im[B_0]$, is $\abs{\vec{k}}$ dependent while, the ending point is not. Also, differently from the previous case, at $k_0=2\Lambda$ the function remains finite. This is not surprising since the regularization is made only over the loop's 3-momentum while, the integration over the zero component of the momentum inside the loop is made using the Matusbara formalism, with the sum calculated from negative to positive infinity (see Eq. (\ref{B0def})).

\begin{figure*}[t!]
\begin{subfigure}[b]{0.4\textwidth}
\includegraphics[width=\textwidth]
{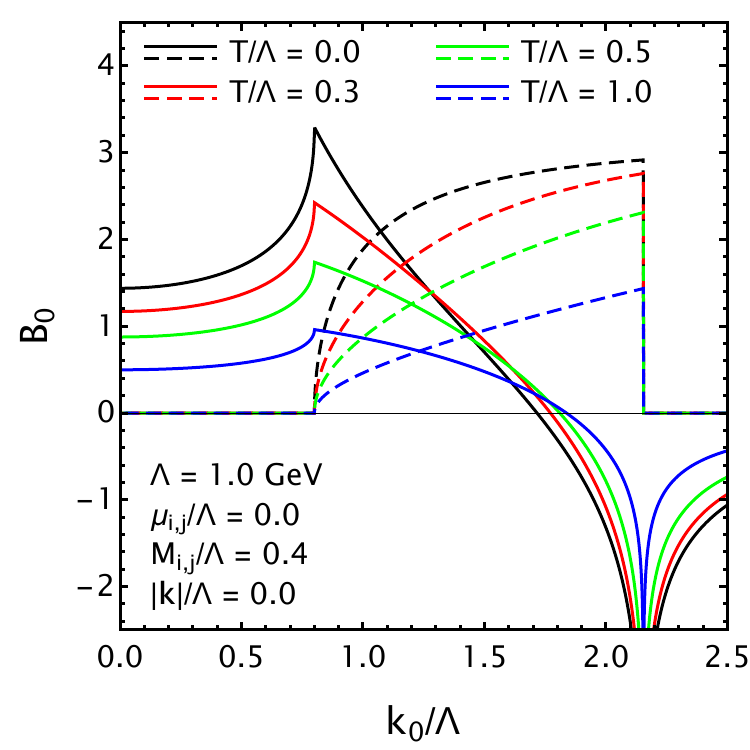}
\label{plotB0vsk0Cpi0Cpj0L1Mi04Mj04k00differentT}
\end{subfigure}
\begin{subfigure}[b]{0.4\textwidth}
\includegraphics[width=\textwidth]
{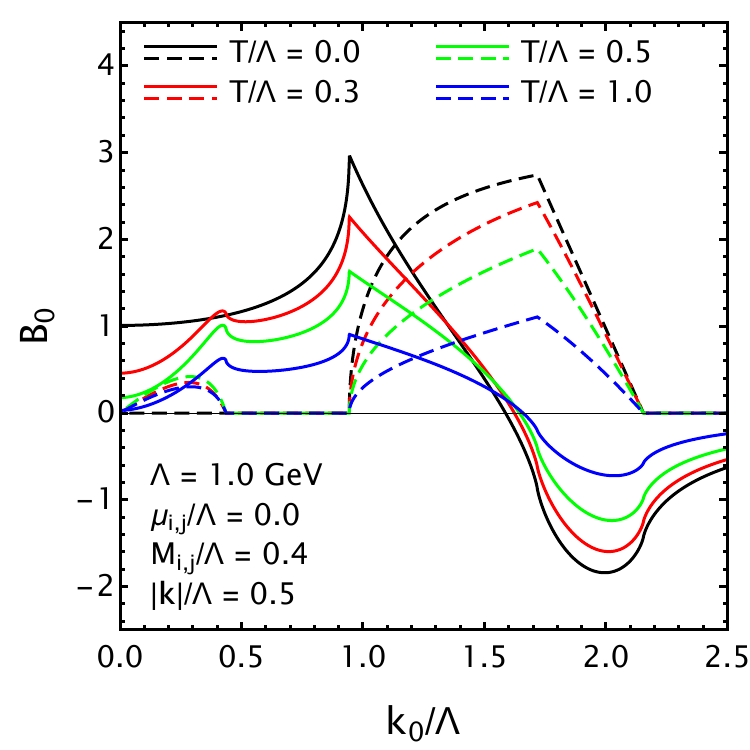}
\label{plotB0vsk0Cpi0Cpj0L1Mi04Mj04k05differentT}
\end{subfigure}
\\
\begin{subfigure}[b]{0.4\textwidth}
\includegraphics[width=\textwidth]
{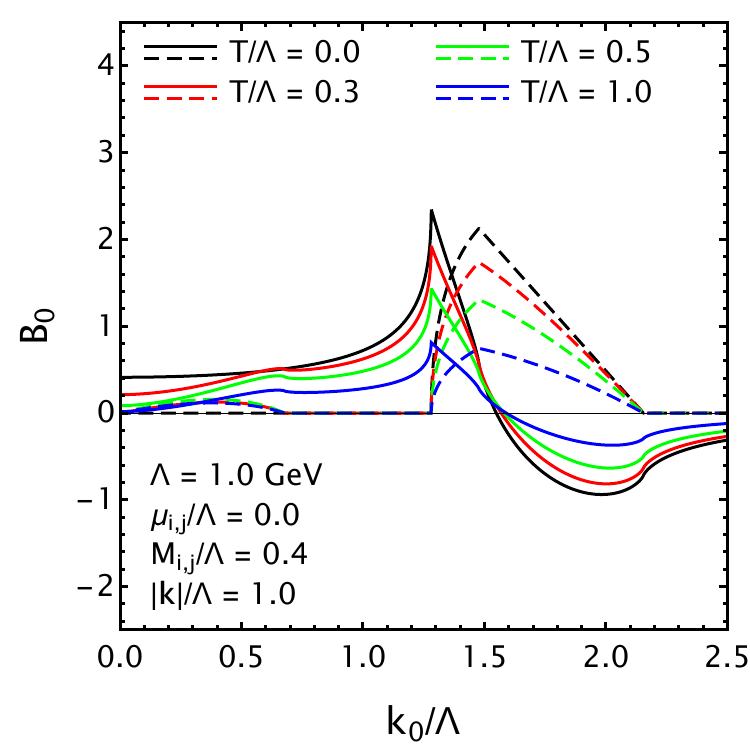}
\label{plotB0vsk0Cpi0Cpj0L1Mi04Mj04k10differentT}
\end{subfigure}
\begin{subfigure}[b]{0.4\textwidth}
\includegraphics[width=\textwidth]
{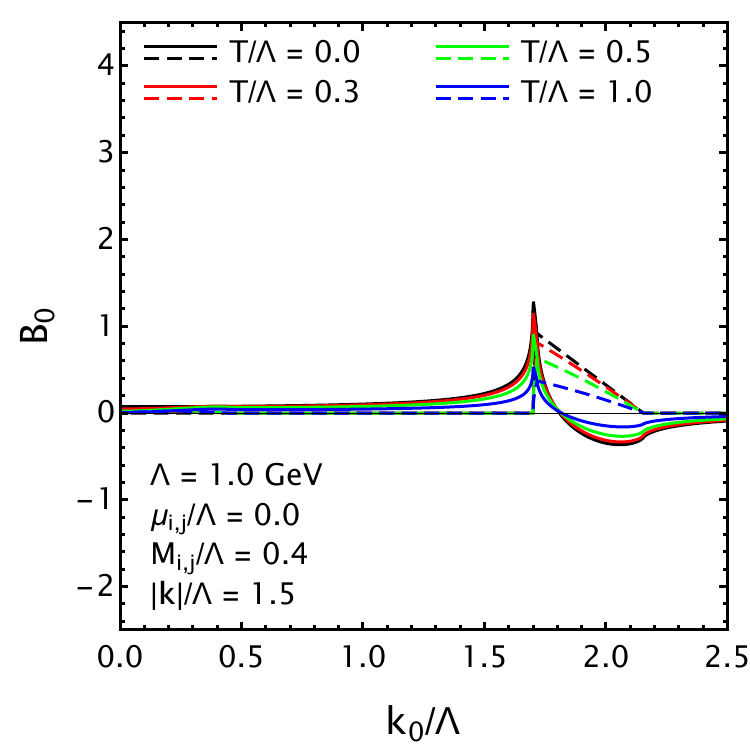}
\label{plotB0vsk0Cpi0Cpj0L1Mi04Mj04k15differentT}
\end{subfigure}
\caption{$B_0$ integral, for zero chemical potential, as a function of the zero component of the external momentum, normalized by the cutoff, $k_0/\Lambda$. The full and dashed lines correspond to the real and imaginary parts part of the $B_0$ integral, respectively. The cutoff is fixed to $\Lambda
=1~\mathrm{GeV}$, the fermion masses are equal and fixed to $M_i=M_j=0.4 \Lambda$. In each panel, we fix the magnitude of the external 3-momentum, $\abs{\vec{k}}/\Lambda=\qty{0.0,0.5,1.0,1.5}$ and the temperature was fixed to different values, corresponding to lines with different colors.}
\label{plotB0vsk0Cpi0Cpj0L1Mi04Mj04k1differentkT}
\end{figure*}

\begin{figure*}[t!]
\begin{subfigure}[b]{0.4\textwidth}
\includegraphics[width=\textwidth]
{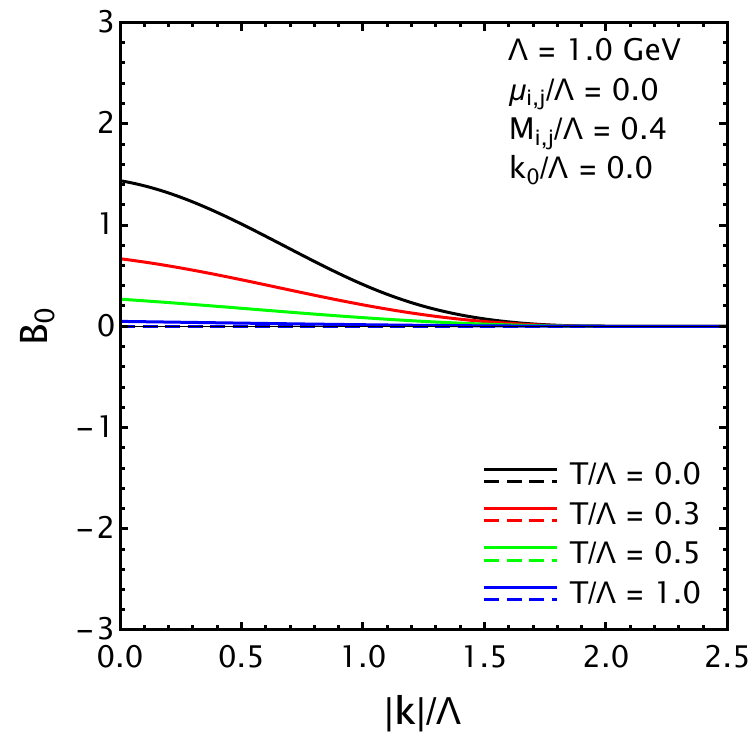}
\label{plotB0vskCpi0Cpj0L1Mi04Mj04w00differentT}
\end{subfigure}
\begin{subfigure}[b]{0.4\textwidth}
\includegraphics[width=\textwidth]
{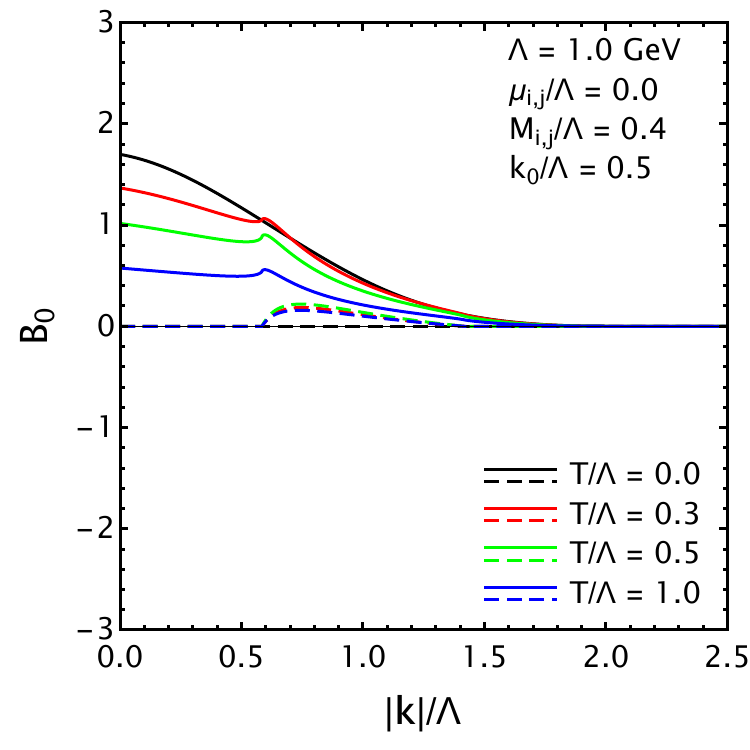}
\label{plotB0vskCpi0Cpj0L1Mi04Mj04w05differentT}
\end{subfigure}
\\
\begin{subfigure}[b]{0.4\textwidth}
\includegraphics[width=\textwidth]
{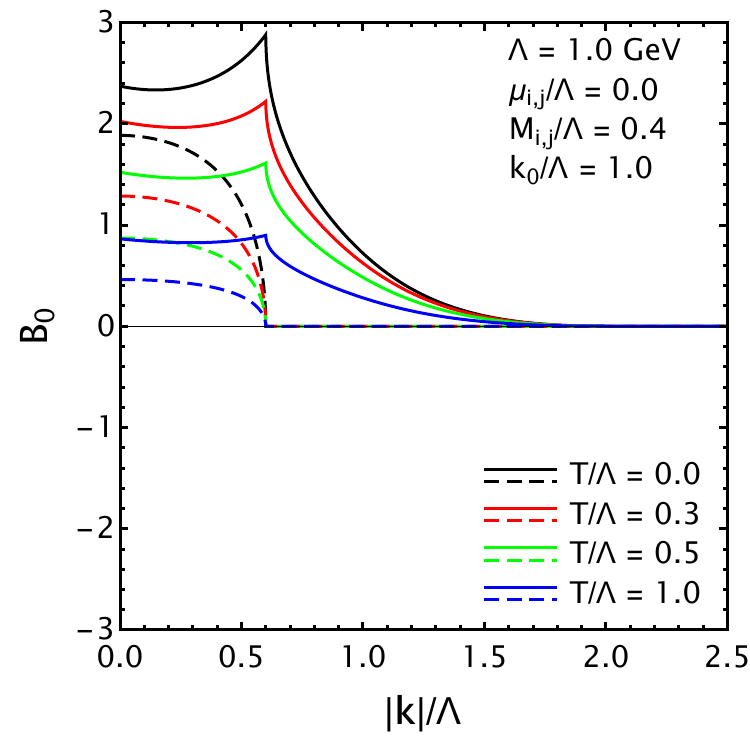}
\label{plotB0vskCpi0Cpj0L1Mi04Mj04w10differentT}
\end{subfigure}
\begin{subfigure}[b]{0.4\textwidth}
\includegraphics[width=\textwidth]
{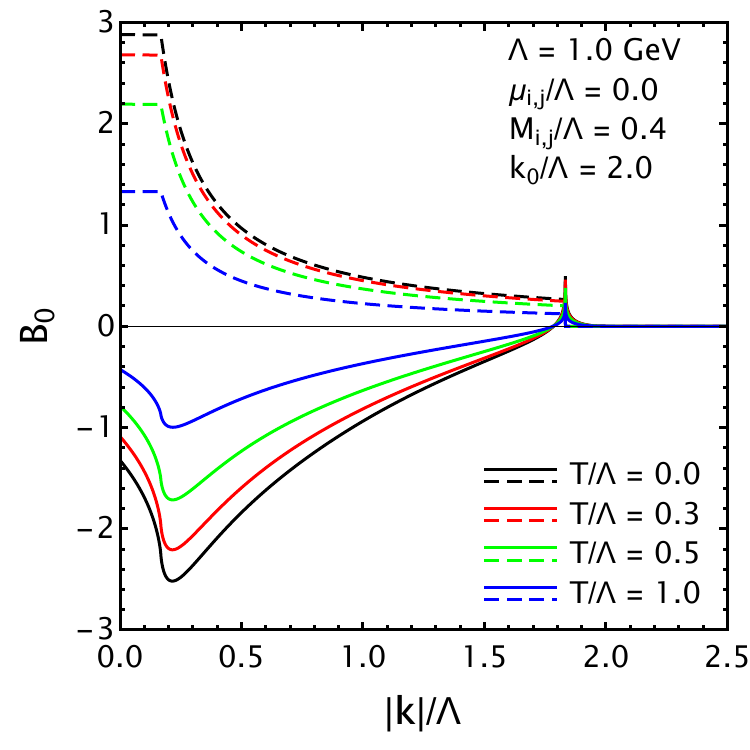}
\label{plotB0vskCpi0Cpj0L1Mi04Mj04w20differentT}
\end{subfigure}
\caption{$B_0$ integral, for zero chemical potential, as a function of the magnitude of the external 3-momentum, normalized by the cutoff, $\abs{\vec{k}}/\Lambda$. The full and dashed lines correspond to the real and imaginary parts part of the $B_0$ integral, respectively. The cutoff is fixed to $\Lambda
=1~\mathrm{GeV}$, the fermion masses are equal and fixed to $M_i=M_j=0.4 \Lambda$. In each panel, we fix the magnitude of the zero component of the external momentum, $k_0/\Lambda=\qty{0.0,0.5,1.0,2.0}$ and the temperature was fixed to different values, corresponding to lines with different colors.}
\label{plotB0vskCpi0Cpj0L1Mi04Mj04differentwT}
\end{figure*}

In Figs. \ref{plotB0vsk0Cpi0Cpj0L1Mi04Mj04k1differentkT} and \ref{plotB0vskCpi0Cpj0L1Mi04Mj04differentwT}, we show several results at finite temperature and zero chemical potential ($\mu_i=\mu_j=0$). In the first, $B_0$  is displayed as a function of $k_0$ while, in the latter, as a function of $\abs{\vec{k}}$. Following the previous pattern, the cutoff and masses are still fixed to $\Lambda
=1~\mathrm{GeV}$ and $M_i=M_j=0.4 \Lambda$, respectively.
In Fig. \ref{plotB0vsk0Cpi0Cpj0L1Mi04Mj04k1differentkT}, specific values for the ratio $\abs{\vec{k}}/ \Lambda$ are fixed while, in Fig. \ref{plotB0vskCpi0Cpj0L1Mi04Mj04differentwT}, some values of $k_0$ are fixed instead.

For finite $\abs{\vec{k}}$ (observe Fig. \ref{plotB0vsk0Cpi0Cpj0L1Mi04Mj04k1differentkT}), the starting and ending points for a finite imaginary part, are not temperature dependent. Indeed, in each panel of this figure, the imaginary part stars and ends at the same values of $k_0$. This can be understood from the equations which define the imaginary parts: they can be non-zero exclusively inside precisely defined kinematic regions, which are not temperature or chemical potential dependent, see Eqs. (\ref{imagPartPair}) and (\ref{imagPartScat}).

When comparing the zero with the finite temperature cases, we can observe that the there is an extra imaginary contribution, which does not exist at zero temperature, present at low values of $k_0$, for finite values of $\abs{\vec{k}}$. This extra contribution to the imaginary part, not present in the vacuum, is caused by the presence of a finite contribution coming from the scattering part of the function, $\Im[B_{0,\scat}]$. In the vacuum only the $\Im[B_{0,\pair}]$ is non-zero, since only pair creation and pair annihilation processes are possible. However, at finite temperature, the scattering channel opens. In these figures we can also observe the effect of the regularization, independent of the temperature, at $\abs{\vec{k}}=2\Lambda$, the $B_0$ function is zero.

\begin{figure*}[ht!]
\begin{subfigure}[b]{0.4\textwidth}
\includegraphics[width=\textwidth]
{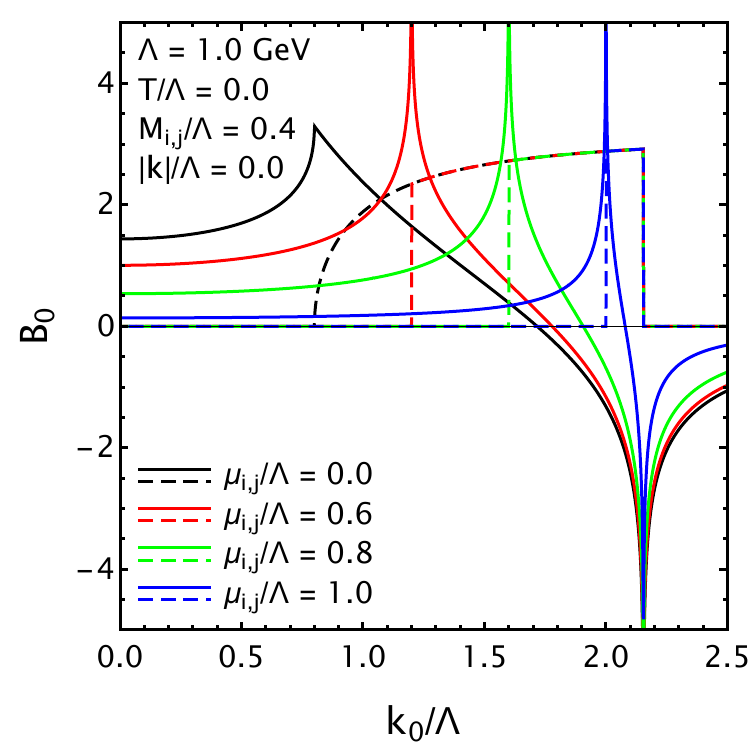}
\label{plotB0vsk0T0L1Mi04Mj04k00differentCp}
\end{subfigure}
\begin{subfigure}[b]{0.4\textwidth}
\includegraphics[width=\textwidth]
{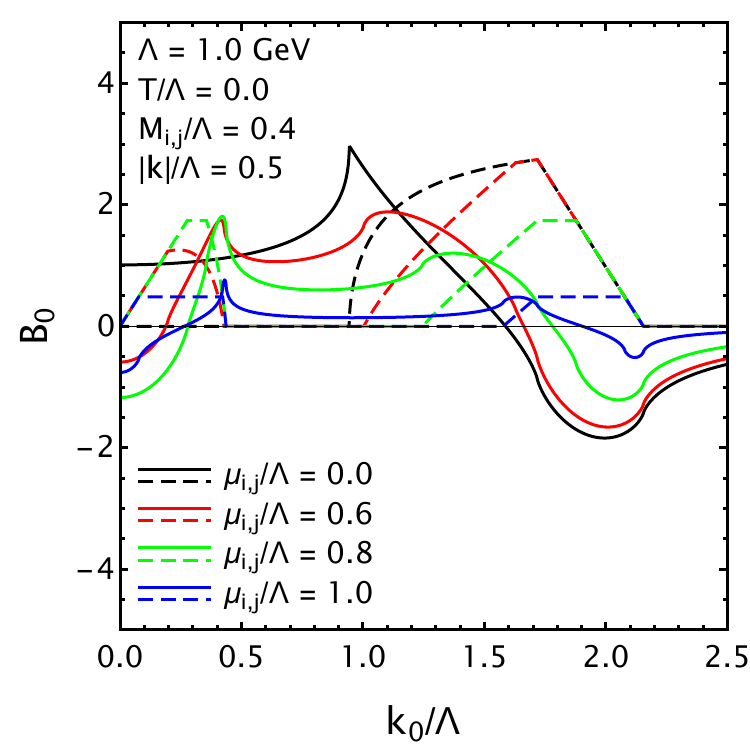}
\label{plotB0vsk0T0L1Mi04Mj04k05differentCp}
\end{subfigure}
\\
\begin{subfigure}[b]{0.4\textwidth}
\includegraphics[width=\textwidth]
{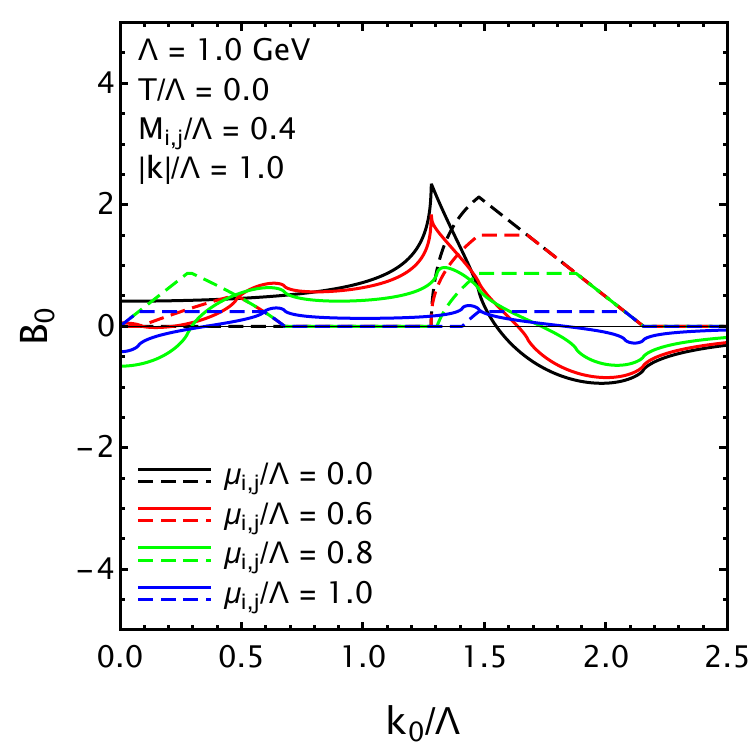}
\label{plotB0vsk0T0L1Mi04Mj04k10differentCp}
\end{subfigure}
\begin{subfigure}[b]{0.4\textwidth}
\includegraphics[width=\textwidth]
{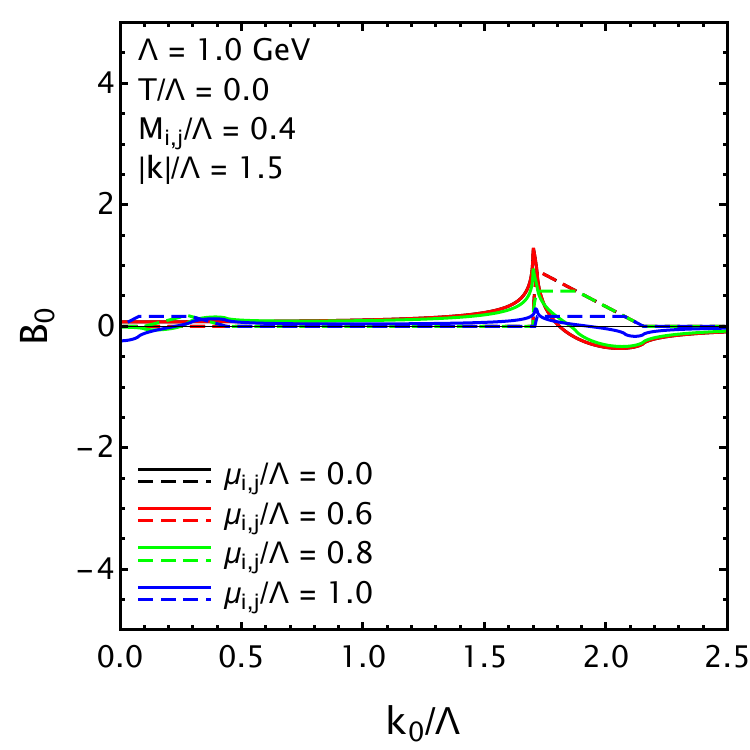}
\label{plotB0vsk0T0L1Mi04Mj04k15differentCp}
\end{subfigure}
\caption{$B_0$ integral, for zero temperature, as a function of the zero component of the external momentum, normalized by the cutoff, $k_0/\Lambda$. The full and dashed lines correspond to the real and imaginary parts part of the $B_0$ integral, respectively. The cutoff is fixed to $\Lambda
=1~\mathrm{GeV}$, the fermion masses are equal and fixed to $M_i=M_j=0.4 \Lambda$. In each panel, we fix the magnitude of the external 3-momentum, $\abs{\vec{k}}/\Lambda=\qty{0.0,0.5,1.0,1.5}$ and the chemical potentials are equal and fixed to different values, corresponding to lines with different colors.}
\label{plotB0vsk0T0L1Mi04Mj04k00differentkCp}
\end{figure*}

\begin{figure*}[ht!]
\begin{subfigure}[b]{0.4\textwidth}
\includegraphics[width=\textwidth]
{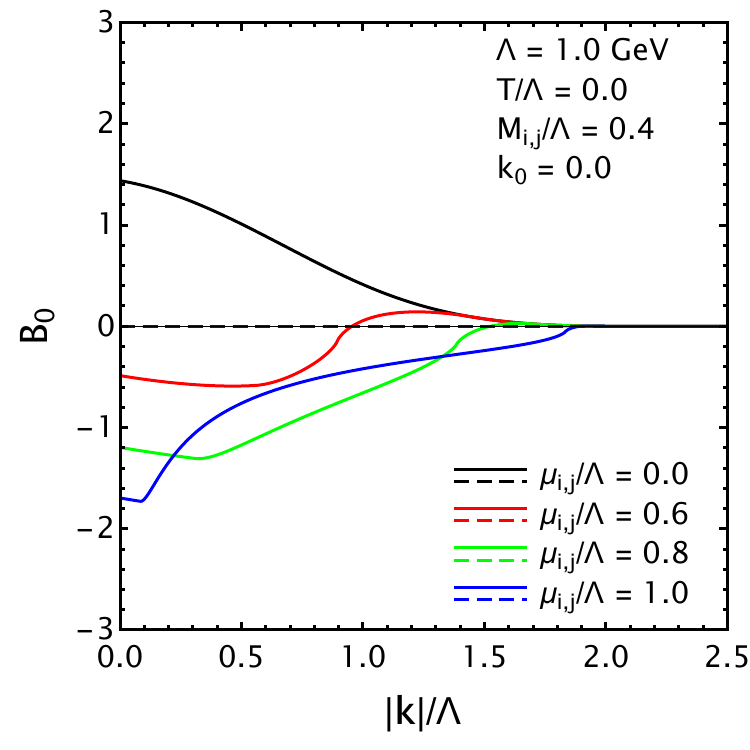}
\label{plotB0vskT0L1Mi04Mj04w00differentCp}
\end{subfigure}
\begin{subfigure}[b]{0.4\textwidth}
\includegraphics[width=\textwidth]
{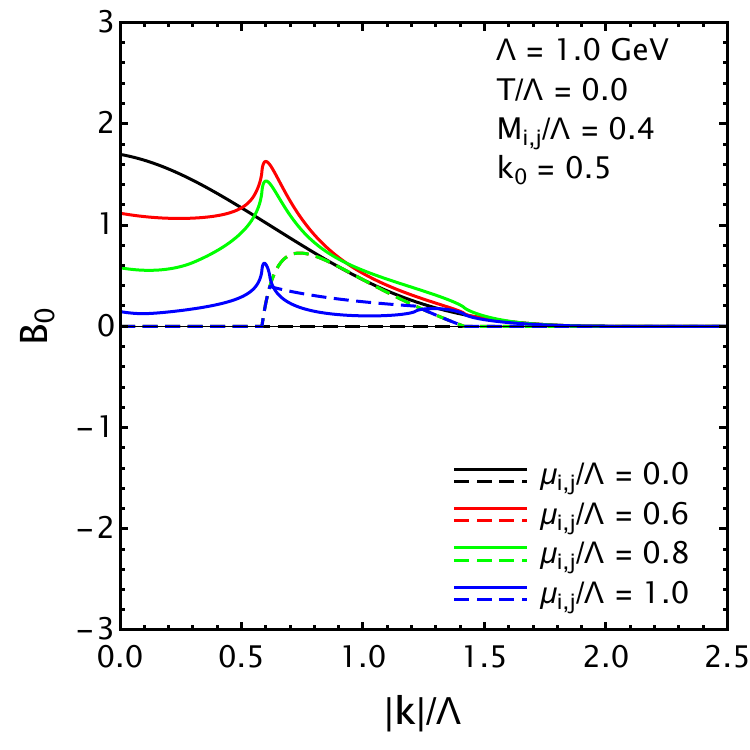}
\label{plotB0vskT0L1Mi04Mj04w05differentCp}
\end{subfigure}
\\
\begin{subfigure}[b]{0.4\textwidth}
\includegraphics[width=\textwidth]
{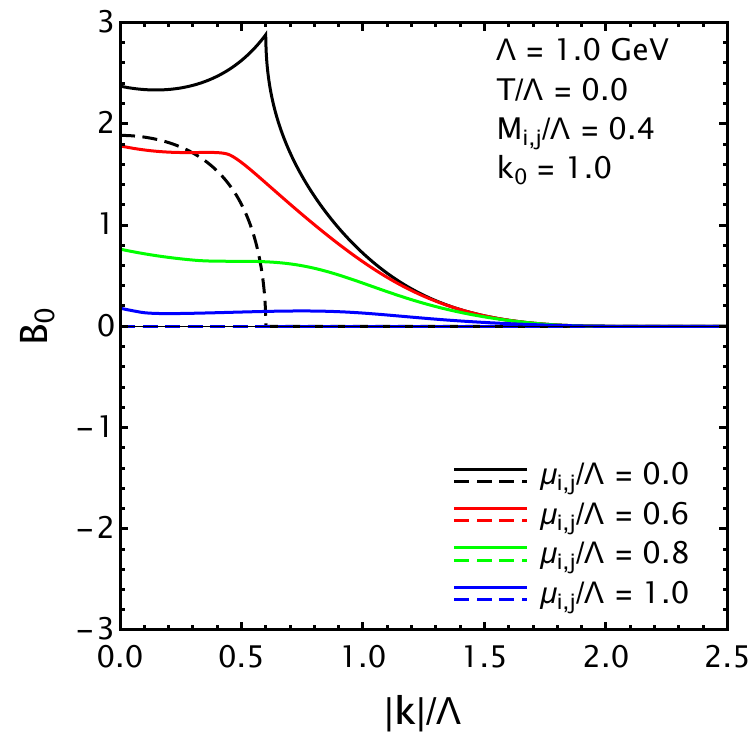}
\label{plotB0vskT0L1Mi04Mj04w10differentCp}
\end{subfigure}
\begin{subfigure}[b]{0.4\textwidth}
\includegraphics[width=\textwidth]
{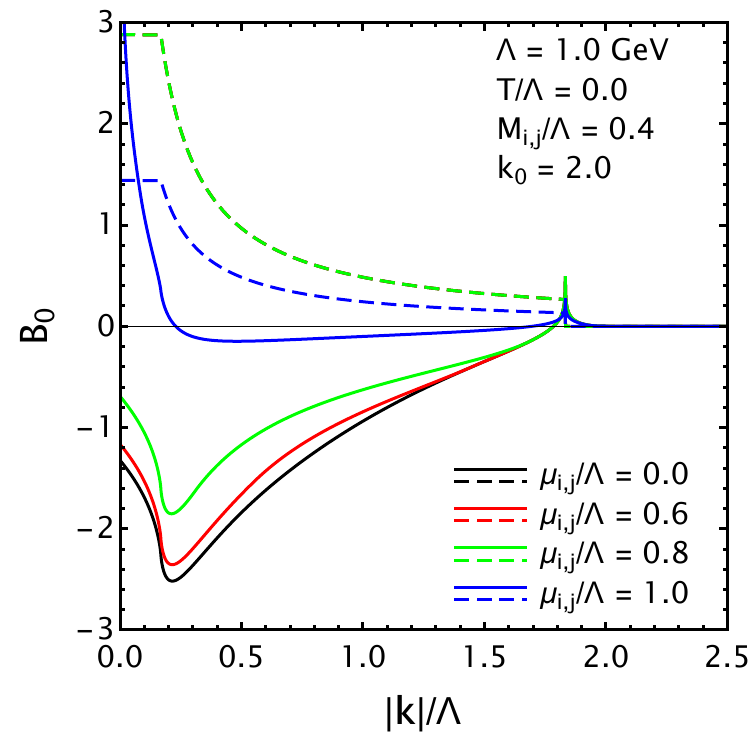}
\label{plotB0vskT0L1Mi04Mj04w20differentCp}
\end{subfigure}
\caption{$B_0$ integral, for zero temperature, as a function of the magnitude of the external 3-momentum, normalized by the cutoff, $\abs{\vec{k}}/\Lambda$. The full and dashed lines correspond to the real and imaginary parts part of the $B_0$ integral, respectively. The cutoff is fixed to $\Lambda
=1~\mathrm{GeV}$, the fermion masses are equal and fixed to $M_i=M_j=0.4 \Lambda$. In each panel, we fix the magnitude of the zero component of the external momentum, $k_0/\Lambda=\qty{0.0,0.5,1.0,2.0}$ and the chemical potentials are equal and fixed to different values, corresponding to lines with different colors.}
\label{plotB0vskT0L1Mi04Mj04w20differentwCp}
\end{figure*}

Next, we analyse the complementary physical scenario from the one discussed above, considering zero temperature and finite chemical potential. For simplicity, we considered degenerate chemical potentials, i.e., $\mu_i=\mu_j$. However, the formalism presented is perfectly suited to study cases with different chemical potentials, $\mu_i \neq \mu_j$. Such scenario arise in the study of neutron star matter and Heavy-Ion Collision experiments where asymmetric nuclear/quark matter, is present. Following the previous analysis, we show the $B_0$ integral as a function of $k_0/ \Lambda$ and $\abs{\vec{k} }/ \Lambda$, in Figs. \ref{plotB0vsk0T0L1Mi04Mj04k00differentkCp} and Fig. \ref{plotB0vskT0L1Mi04Mj04w20differentwCp}, respectively.

At zero temperature, only for values of chemical potential higher than the fermions masses  one gets results that differ from the vacuum result. This characteristic is to be expected since at zero temperature, $T=0$, the Fermi-Dirac distributions present in the $B_0$ integral act as step-functions. Hence, below a certain energy regime, only the vacuum contributes: only when the momenta is higher the Fermi's momenta, for a fixed chemical potential, does this momentum contributes to the momenta integration. This feature is sometimes termed by Silver Blaze property \cite{Cohen:2003kd,Otto:2022jzl}.

Previously, when discussing Fig. \ref{plotB0vsk0Cpi0Cpj0L1Mi04Mj04k1differentkT}, we argued that the values of $k_0$, which define the intervals in which the imaginary part is finite, are neither temperature or chemical potential dependent. However, Fig. \ref{plotB0vsk0T0L1Mi04Mj04k00differentkCp} seems to contradict this statement, since increasing the chemical potential drives the onset of the imaginary part to larger values of $k_0$. However, this is a feature of the Eqs. (\ref{GPair}) and (\ref{GScat}) themselves and not caused by the restriction imposed by the endpoints of the integration. In these figures, as expected, we continue to observe the general effect of the regularization, with $B_0=0$ for $\abs{\vec{k}} \geq 2 \Lambda$.

\begin{figure*}[t!]
\begin{subfigure}[b]{0.4\textwidth}
\includegraphics[width=\textwidth]
{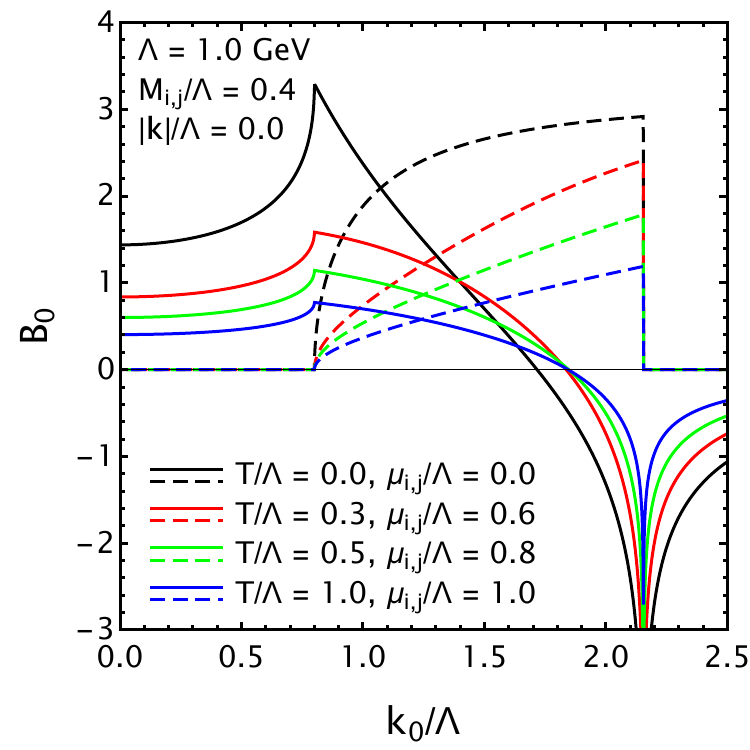}
\label{plotB0vsk0L1Mi04Mj04k00differentTCp}
\end{subfigure}
\begin{subfigure}[b]{0.4\textwidth}
\includegraphics[width=\textwidth]
{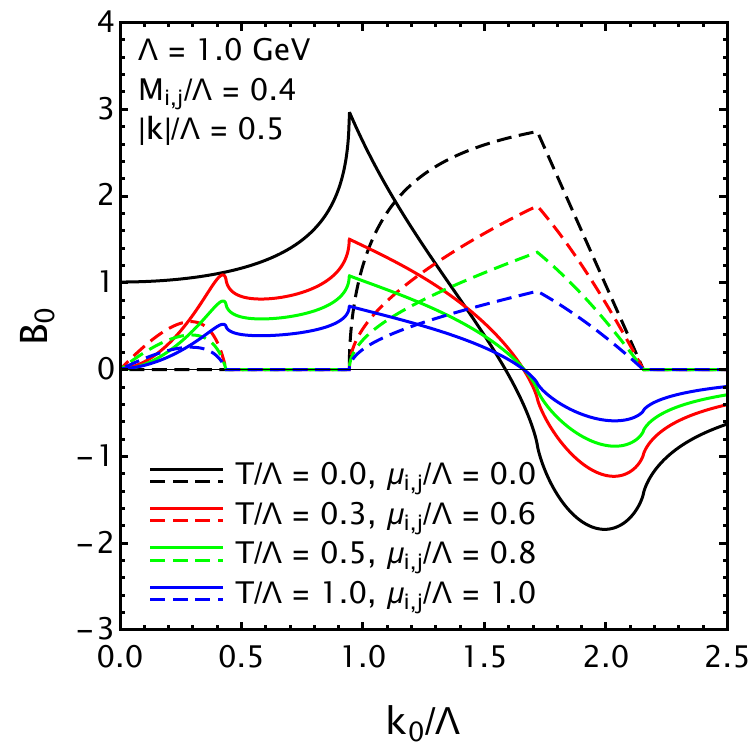}
\label{plotB0vsk0L1Mi04Mj04k05differentTCp}
\end{subfigure}
\\
\begin{subfigure}[b]{0.4\textwidth}
\includegraphics[width=\textwidth]
{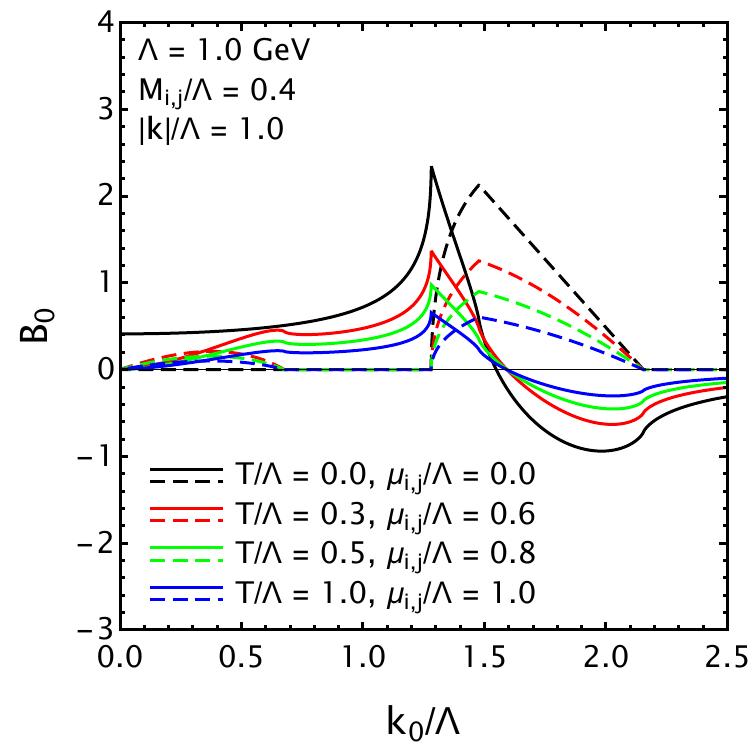}
\label{plotB0vsk0L1Mi04Mj04k10differentTCp}
\end{subfigure}
\begin{subfigure}[b]{0.4\textwidth}
\includegraphics[width=\textwidth]
{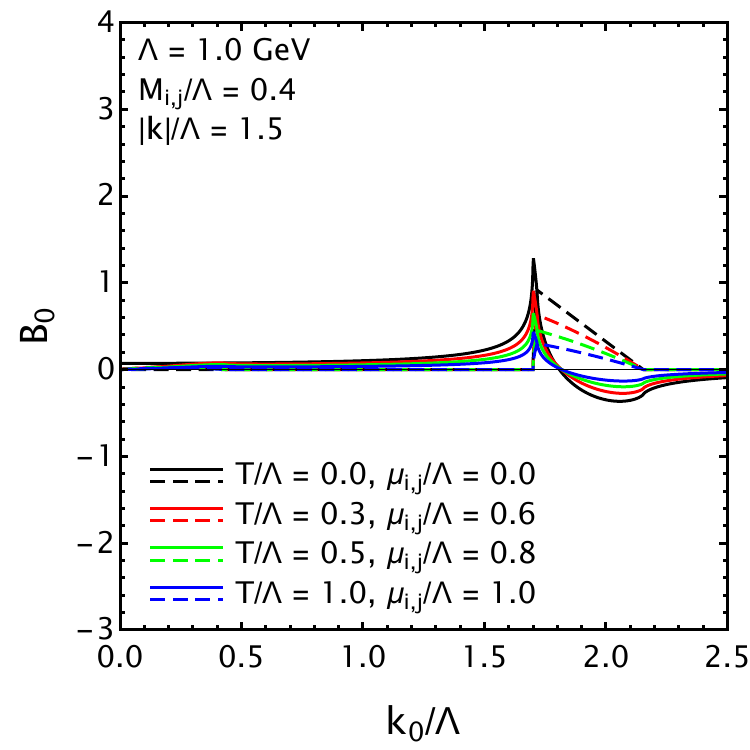}
\label{plotB0vsk0L1Mi04Mj04k15differentTCp}
\end{subfigure}
\caption{$B_0$ integral, as a function of the zero component of the external momentum, normalized by the cutoff, $k_0/\Lambda$. The full and dashed lines correspond to the real and imaginary parts part of the $B_0$ integral, respectively. The cutoff is fixed to $\Lambda
=1~\mathrm{GeV}$, the fermion masses are equal and fixed to $M_i=M_j=0.4 \Lambda$. In each panel, we fix the magnitude of the external 3-momentum, $\abs{\vec{k}}/\Lambda=\qty{0.0,0.5,1.0,1.5}$ and the temperature and chemical potentials are fixed to different values, corresponding to lines with different colors.}
\label{plotB0vsk0L1Mi04Mj04k15differentkTCp}
\end{figure*}

\begin{figure*}[t!]
\begin{subfigure}[b]{0.4\textwidth}
\includegraphics[width=\textwidth]
{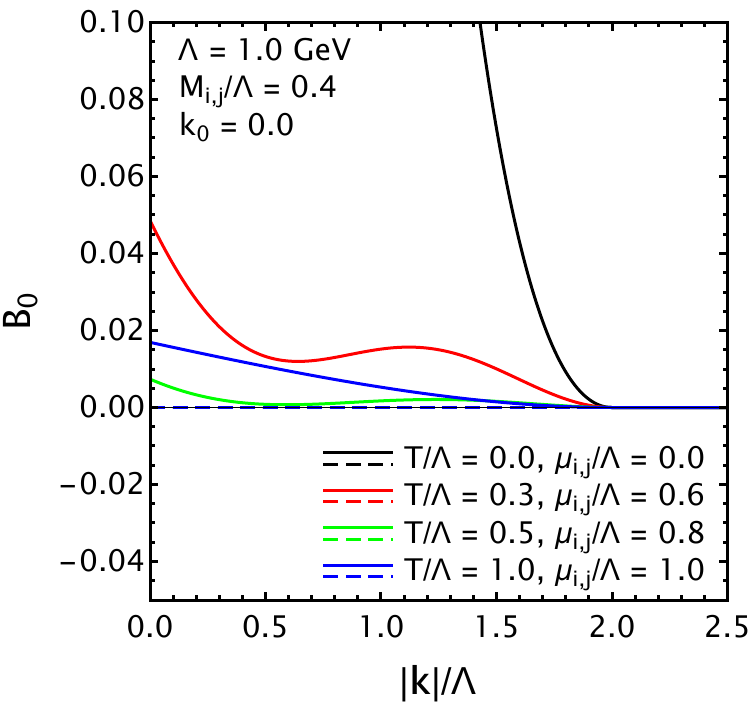}
\label{plotB0vskL1Mi04Mj04w00differentTCp}
\end{subfigure}
\begin{subfigure}[b]{0.4\textwidth}
\includegraphics[width=\textwidth]
{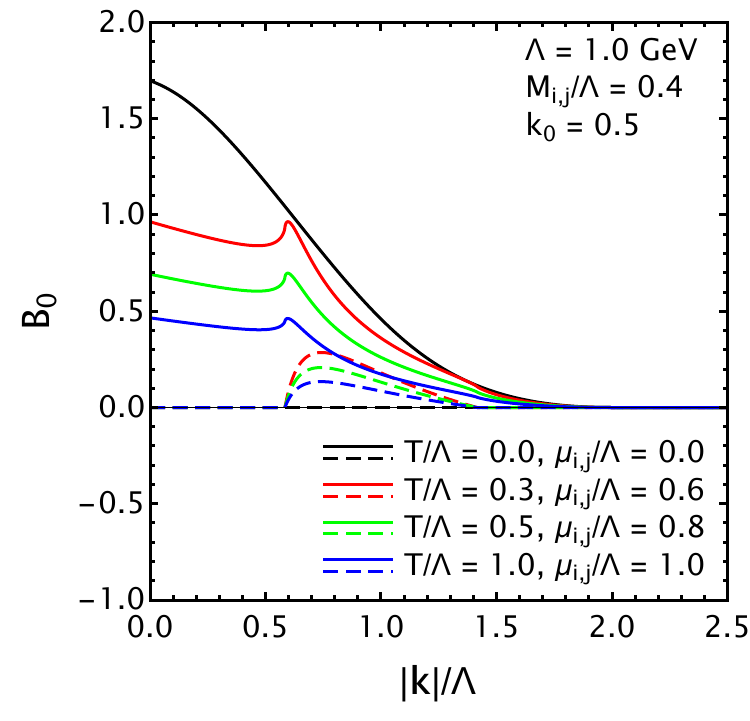}
\label{plotB0vskL1Mi04Mj04w05differentTCp}
\end{subfigure}
\\
\begin{subfigure}[b]{0.4\textwidth}
\includegraphics[width=\textwidth]
{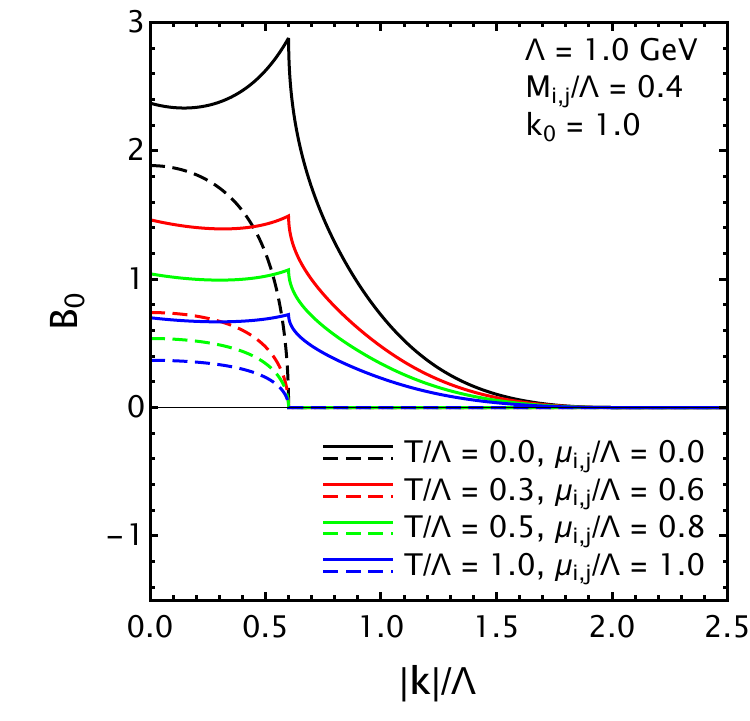}
\label{plotB0vskL1Mi04Mj04w10differentTCp}
\end{subfigure}
\begin{subfigure}[b]{0.4\textwidth}
\includegraphics[width=\textwidth]
	{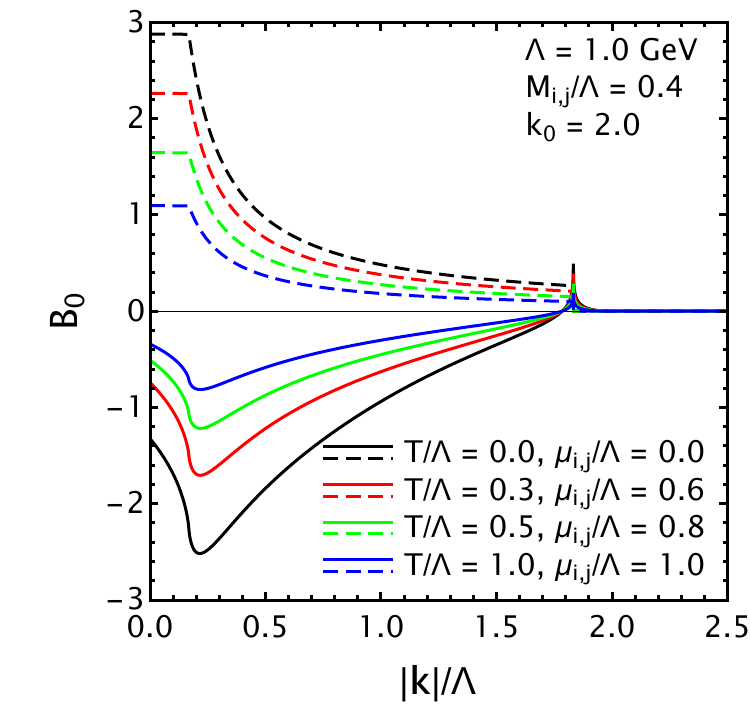}
\label{plotB0vskL1Mi04Mj04w15differentTCp}
\end{subfigure}
\caption{$B_0$ integral, for zero temperature, as a function of the magnitude of the external 3-momentum, normalized by the cutoff, $\abs{\vec{k}}/\Lambda$. The full and dashed lines correspond to the real and imaginary parts part of the $B_0$ integral, respectively. The cutoff is fixed to $\Lambda
=1~\mathrm{GeV}$, the fermion masses are equal and fixed to $M_i=M_j=0.4 \Lambda$. In each panel, we fix the magnitude of the zero component of the external momentum, $k_0/\Lambda=\qty{0.0,0.5,1.0,2.0}$ and the temperature and chemical potentials are fixed to different values, corresponding to lines with different colors.}
\label{plotB0vskT0L1Mi04Mj04differentwCp}
\end{figure*}

Lastly, in Figs. \ref{plotB0vsk0L1Mi04Mj04k15differentkTCp} and \ref{plotB0vskT0L1Mi04Mj04differentwCp}, we show the completely in-medium scenario, where finite temperature and finite degenerate chemical potential are considered. The general behavior is very similar to the ones found in the previously discussed scenarios, presenting only quantitative differences.

In all scenarios considered in this section, a general pattern can be observed in the real part of the integral, $\Re[B_0]$: it changes behavior when there is a switch on/off of the imaginary part, be it as a function of $k_0$ or as a function of $\abs{\vec{k}}$. This can be understood taking into account the nature of the integration for a specific set of values of $\{k_0,\abs{\vec{k}}\}$. For the imaginary contribution to be non-zero, there must exist a singularity inside the integration bounds. Whenever one is dealing with a pair of values for $\{k_0,\abs{\vec{k}}\}$, which lays outside the integration bounds, there should not be any singularity within the integration. When the opposite happens, there is a singular point inside the integration region and the now improper integral is assigned a value by the application of the Cauchy-Principal value.

In the case where there is a model for the fermion masses, which encodes the temperature and chemical potential dependence of the fermion masses, the interpretation of the results gets more complicated. For instance, in the case of the NJL model, one foresees that as temperature increases, at zero chemical potential, the quark effective mass decreases. A similar pattern is expect at finite chemical potential. In such cases, the behavior of the $B_0$ function as a function of temperature and chemical potential is different and outside the scope of this section. Here our aim was to simply analyze the thermodynamic structure of the two fermion line integral $B_0$ by itself.

\bibliography{bib}

\end{document}